\documentclass[12pt]{article}

\usepackage[a4paper,text={16.8cm,22.4cm}]{geometry}
\usepackage{amsmath,amsfonts,slashed,amssymb,tikz,bm,psfrag,graphicx,color,dsfont}
\usepackage{multicol}
\usepackage{float}

\RequirePackage[sort&compress,square,comma,numbers]{natbib}
\allowdisplaybreaks
\begin{document}

\begin{titlepage}

\begin{flushright}
\normalsize
July 25, 2019
\end{flushright}

\vspace{0.1cm}
\begin{center}
\Large\bf
Precision calculations of $B \to V$ form factors in QCD
\end{center}

\vspace{0.5cm}
\begin{center}
{\bf Jing Gao$^{a,b}$, Cai-Dian L\"{u}$^{a,b}$, Yue-Long Shen$^{c}$,
Yu-Ming Wang$^{d}$,  Yan-Bing Wei$^{d}$  } \\
\vspace{0.7cm}
{\sl   ${}^a$ \, Institute of High Energy Physics, CAS, P.O. Box 918(4) Beijing 100049,  China \\
 ${}^b$ \,  School of Physics, University of Chinese Academy of Sciences, Beijing 100049,  China \\
 ${}^c$ \, College of Information Science and Engineering,
Ocean University of China, Songling Road 238, Qingdao, 266100 Shandong, P.R. China \\
${}^d$ School of Physics, Nankai University, 300071 Tianjin, China \,
}
\end{center}

\vspace{0.2cm}
\begin{abstract}

Applying the vacuum-to-$B$-meson correlation functions
with an interpolating current for the light vector meson we construct the light-cone sum rules (LCSR)
for the  ``effective" form factors $\xi_{\parallel}(n \cdot p)$, $\xi_{\perp}(n \cdot p)$,
$\Xi_{\parallel}(\tau, n \cdot p)$ and $\Xi_{\perp}(\tau, n \cdot p)$,
defined by the corresponding hadronic  matrix elements in soft-collinear effective theory (SCET),
entering  the leading-power factorization formulae for QCD form factors
responsible for  $B \to V \ell \bar \nu_{\ell}$ and $B \to V \ell \bar \ell$ decays
at large hadronic recoil at next-to-leading-order in QCD.
The evanescent-operator  approach for the perturbative matching of the
effective operators from $\rm {SCET_{I}} \to {HQET}$ is employed in the determination of the hard-collinear functions
entering the SCET factorization formulae for the vacuum-to-$B$-meson correlation functions.
The light-quark mass effect for the local SCET form factors $\xi_{\parallel}(n \cdot p)$ and $\xi_{\perp}(n \cdot p)$
is also computed from the LCSR method with the $B$-meson light-cone distribution amplitude
$\phi_B^{+}(\omega, \mu)$  at ${\cal O}(\alpha_s)$.
Furthermore, the subleading power corrections to $B \to V$ form factors from the higher-twist
$B$-meson light-cone distribution amplitudes are also computed with the same method  at tree level up to the twist-six accuracy.
Employing the two different models for the $B$-meson light-cone distribution amplitudes consistent with QCD equations of motion,
we observe  that  the higher-twist corrections to $B \to V$ form factors are dominated by the two-particle
twist-five distribution amplitude $g_B^{-}(\omega, \mu)$, in analogy to the previous observation for $B \to P$ form factors.
Having at our disposal the LCSR predictions for $B \to V$ form factors, we further perform new determinations of the CKM
matrix element $|V_{ub}|$ from the semileptonic $B \to \rho \, \ell \, \bar \nu_{\ell}$ and
$B \to \omega \, \ell \, \bar  \nu_{\ell}$  decays,
and predict the normalized differential branching fractions and
the $q^2$-binned $K^{\ast}$ longitudinal polarization fractions of the exclusive rare
$B \to K^{\ast} \, \nu_{\ell} \, \bar \nu_{\ell}$ decays.

\end{abstract}

\vfil

\end{titlepage}

\section{Introduction}

Precision calculations of $B \to V$ form factors are indispensable
for the determinations of  the CKM matrix elements from the semileptonic  $B \to V \, \ell  \, \bar \nu_{\ell}$ decays
and the radiative penguin $B \to V \gamma$ decays and for the theory descriptions of the
electroweak penguin $B \to V \ell \bar \ell$ decays as well as the hadronic two-body $B$-meson  decays
in QCD. In the low hadronic recoil region, the unquenched lattice QCD calculations of $B \to K^{\ast}$
form factors have been performed \cite{Horgan:2013hoa,Horgan:2015vla}
(see references therein for discussions on the earlier calculations with the quenched approximation)
by employing the MILC Collaboration gauge-field ensembles with an improved staggered quark action \cite{Bazavov:2009bb}.
In the large hadronic recoil region, distinct analytical QCD methods have been developed for the systematic
calculations of the heavy-to-light $B$-meson decay form factors with the aid of the heavy quark expansion.

QCD factorization formulae for heavy-to-light form factors at large recoil were originally proposed
in \cite{Beneke:2000wa} at leading power in $\Lambda/m_{Q}$, where both the soft contribution satisfying
the large-recoil symmetry relations and the hard spectator scattering effect violating the symmetry relations
were shown to appear simultaneously in contrast to the hard-collinear factorization for the pion-photon form factor.
With the advent of soft-collinear effective theory (SCET) factorization properties of heavy-to-light form factors
can be addressed by integrating out the hard and hard-collinear fluctuations of the QCD matrix elements one after the other.
Implementing the first-step matching procedure for the QCD current $\bar \psi \, \Gamma_i \, Q$ will give rise to the so-called
${\rm A0}$-type and ${\rm B1}$-type ${\rm SCET}_{\rm I}$  operators \cite{Bauer:2002aj,Beneke:2002ph,Beneke:2003pa}, both of which can contribute to heavy-to-light form factors at leading power  in $\Lambda/m_{Q}$. Performing the perturbative matching of the effective currents
from ${\rm SCET}_{\rm I} \to {\rm SCET}_{\rm II}$  indicates that the soft-collinear factorization for
the ${\rm A0}$-type matrix elements cannot be achieved, due to the emergence of end-point divergences appearing in the
convolution integrals of the hard-collinear functions and the light-cone distribution amplitudes.
By contrast, the non-local form factors defined by the ${\rm B1}$-type ${\rm SCET}_{\rm I}$ operators can be further expressed
as the convolution of the jet functions and the hadronic distribution amplitudes \cite{Beneke:2003pa}.
It is then evident that the theory predictions of  heavy-to-light form factors from the SCET factorization formulae
cannot be made without the knowledge of the standard light-cone distribution amplitudes and the matrix elements
of the ${\rm A0}$-type ${\rm SCET}_{\rm I}$ operators.

Applying the dispersion relations and perturbative QCD factorization theorems for the vacuum-to-vector-meson
correlation functions, the QCD light-cone sum rules (LCSR) for  $B \to V$ form factors can be readily constructed
\cite{Ball:1997rj,Ball:1998kk,Ball:2004rg,Straub:2015ica} with the parton-hadron duality ansatz
and the narrow-width approximation for the vector mesons (see \cite{Meissner:2013hya,Hambrock:2015aor,Cheng:2017sfk} for further discussions
and \cite{Cheng:2018ouz} for the sum-rule construction with the helicity form-factor scheme).
Alternatively, the QCD LCSR  for heavy-to-light $B$-meson decay form factors can be
derived from the vacuum-to-$B$-meson correlation functions,
following the analogous strategies, at leading order (LO) \cite{Khodjamirian:2005ea,Khodjamirian:2006st,Khodjamirian:2010vf,Khodjamirian:2012rm}
and next-to-leading order (NLO) \cite{Wang:2015vgv,Wang:2017jow,Lu:2018cfc} in the strong coupling $\alpha_s$,
where the factorization formulae for the correlation functions under discussion were established with the
diagrammatic approach and the strategy of regions \cite{Beneke:1997zp,Smirnov:2002pj}.
Constructing the LCSR for the ${\rm SCET}_{\rm I}$ matrix elements entering the QCD factorization formulae of
heavy-to-light $B$-meson decay form factors has been achieved in \cite{DeFazio:2005dx,DeFazio:2007hw}
employing the vacuum-to-$B$-meson correlation functions.
Compared to the QCD factorization approach,  the LCSR calculations of heavy-to-light $B$-meson form factors
depend on the duality assumption of either the light-meson channel or the $B$-meson channel.

Yet another factorization approach to compute heavy-to-light $B$-meson form factors at large recoil
has been developed to regularize the rapidity divergences of the ${\rm A0}$-type SCET matrix elements
by the intrinsic transverse momenta of the soft and collinear partons involved
in the hard scattering processes  \cite{Keum:2000wi,Lu:2000em}.
Perturbative QCD corrections to the short-distance matching coefficient functions entering the
transverse-momentum-dependent (TMD) factorization formulae of several hard exclusive processes
of phenomenological interest \cite{Nandi:2007qx,Li:2010nn,Li:2012nk} have been accomplished at leading-twist accuracy.
The  Sudakov and threshold resummation of enhanced logarithms entering the  TMD wavefunctions have been
performed for the $B$-meson \cite{Li:2012md} and for the pion \cite{Li:2013xna}
with the Collins-Soper-Sterman (CSS) formalism \cite{Collins:1981uk,Collins:1984kg,Collins:2011zzd}.
In addition, constructing the factorization-compatible definitions of the TMD wavefunctions
free of the rapidity pinch singularities has been discussed in \cite{Li:2014xda,Wang:2015qqr},
where the non-dipolar off-light-cone  Wilson lines were introduced in the unsubtracted TMD pion wavefunctions to reduce
the soft  subtraction functions.
However, a definite power counting scheme for all the momentum modes involved in the exclusive $B$-meson decays
still needs to be constructed for the TMD factorization approach to clarify the conceptual differences between
the perturbative QCD (PQCD) framework \cite{Keum:2000wi,Lu:2000em}
and the QCD factorization approach \cite{Beneke:2000wa,Beneke:2003pa}
and to develop the TMD factorization for hard exclusive processes into a systematic theoretical framework.

Applying the SCET factorization for the QCD $B \to V$ form factors at large  recoil,
we aim at computing the hadronic matrix elements of both the ${\rm A0}$-type and ${\rm B}$-type ${\rm SCET}_{\rm I}$  operators
by constructing the corresponding sum rules  from the vacuum-to-$B$-meson correlation functions,
in analogy to the prescriptions developed in \cite{DeFazio:2005dx,DeFazio:2007hw}.
The major new  improvements of the present paper can be summarized as follows.

\begin{itemize}

\item{We establish the factorization formulae of  the vacuum-to-$B$-meson correlation functions
defined with an interpolating current for the vector meson and an effective weak current in ${\rm SCET}_{\rm I}$
at one loop using the evanescent approach \cite{Dugan:1990df,Herrlich:1994kh}, instead of substituting the light-cone projector of the
$B$-meson for evaluating the corresponding ${\rm SCET}_{\rm I}$ diagrams prior to performing the loop-momentum integration.
In addition, QCD resummation of enhanced logarithms of $m_b/\Lambda$ entering the ${\rm A0}$-type and ${\rm B}$-type
hard matching coefficients for the weak current $\bar \psi \, \Gamma_i \, Q$ is accomplished
at next-to-leading-logarithmic (NLL) and leading-logarithmic  (LL) accuracy
with the standard renormalization-group (RG) formalism \cite{Hill:2004if,Beneke:2005gs}.}

\item{Applying the SCET representations of the vacuum-to-$B$-meson correlation functions
in the presence of the subleading-power Lagrangian ${\cal L}_{\xi m}^{(1)}$,
we construct the sum rules for the light-quark mass contributions to the ${\rm A0}$-type  SCET form factors
at tree level with the power counting scheme $m \sim \Lambda$.
We demonstrate explicitly that the flavour SU(3)-symmetry breaking effects for the longitudinal
vector meson form factors are not suppressed by powers of $\Lambda/m_b$ and evidently preserve the
large recoil symmetry relations for the soft contributions to the semileptonic $B \to V$ form factors. }

\item{We compute the subleading power corrections to $B \to V$ form factors from the higher-twist
$B$-meson light-cone distribution amplitudes (LCDA) with the aid of the LCSR technique  up to the twist-six accuracy.
In particular, we employ a complete parametrization of the light-ray matrix element
$\langle 0 | \bar q_{\alpha}(z_1 \, \bar n) \, g_s \, G_{\mu \nu} (z_2 \, \bar n) \, h_{v \, \beta}(0 ) |\bar B_v  \rangle$
in heavy-quark effective theory (HQET) defining eight independent invariant functions presented in \cite{Braun:2017liq}
instead of four  three-particle distribution amplitudes proposed in \cite{Kawamura:2001jm} in the light-cone limit.
In an attempt to understand the systematic uncertainty of the LCSR predictions for  $B \to V$ form factors, we apply
two distinct models for the two-particle and three-particle $B$-meson LCDA, satisfying the classical QCD equations of motion
and the corresponding asymptotic behaviours at small quark and gluon momenta determined by the conformal spins of the soft fields,
as constructed in \cite{Braun:2017liq,Lu:2018cfc}.  }

\item{For the sake of understanding the long-standing discrepancy of the form-factor ratio
${\cal R}_1 = \left [ {(m_B +m_V) /  m_B} \right ] \,  T_1(q^2) / V(q^2)$
predicted by the QCD sum rule technique with the vector-meson LCDA \cite{Ball:2004rg}
and by the QCD factorization approach \cite{Beneke:2000wa,Beneke:2005gs,Beneke:2008ei},
we carry out a detailed comparison of the various terms contributing to the factorization formula of
the  ratio ${\cal R}_1$ with their counterparts in the framework of the LCSR with the $B$-meson distribution amplitudes
and identify the dominating QCD mechanisms responsible for the above-mentioned  discrepancy.}

\end{itemize}

The outline of this paper is as follows. We will review the ${\rm SCET (hc, c, s)}$ representations
of  $B \to V$ form factors at large hadronic recoil from integrating out the  hard-scale fluctuations,
which express the seven QCD form factors  in terms of the four ${\rm SCET}_{\rm I}$ matrix elements,
$\xi_{a}(n \cdot p)$ and  $\Xi_{a}(\tau, n \cdot p)$ (with $a = \parallel, \perp$),
as well as the perturbatively calculable short-distance coefficients at leading power in $\Lambda/m_b$
in section \ref{section: SCET-I factorization}.
Constructing the SCET sum rules for these ``effective" form factors with the $B$-meson distribution amplitudes
at leading twist accuracy will be presented in section \ref{section: SCET sum rules at leading twist}
with a detailed demonstration of the factorization formulae for the vacuum-to-$B$-meson correlation functions
at NLO in $\alpha_s$, where we pay particular attention to the infrared subtractions for deriving the master formulae
of the hard-collinear matching functions in the presence of the evanescent operators.
We proceed to compute the subleading power corrections
to $B \to V$ form factors from both the two-particle and three-particle higher-twist $B$-meson LCDA
with the LCSR  method at tree level up to the twist-six accuracy in section \ref{section: SCET rum rules at higher twist},
where the operator identities between the two-body and three-body light-ray HQET operators
at classical level are employed to reduce the resulting sum rules.
Phenomenological aspects of the newly derived LCSR for $B \to V$ form factors will be explored in
section \ref{section: numerical analysis}, including the numerical impacts of the subleading-power corrections
in the heavy-quark expansion, an exploratory comparison of our predictions
of the form-factor ratios with the SCET factorization calculations,
the extrapolations of our results toward large momentum transfer with the $z$-series parametrization,
the exclusive determinations of the CKM matrix elements $|V_{ub}|$ from the partial branching fractions
of $B \to \rho \, \ell \, \bar \nu_{\ell}$ and $B \to \omega \, \ell \, \bar \nu_{\ell}$,
and the  $q^2$-binned distributions of the branching fractions
as well as  the $K^{\ast}$ longitudinal polarized fractions of the
flavour-changing-neutral-current (FCNC) induced $B \to K^{\ast} \, \nu_{\ell} \, \bar \nu_{\ell}$ decays.
A summary of our main observations and concluding remarks on the future development will be displayed in
section \ref{section: summary}.
We further collect the explicit expressions for the ${\rm A0}$-type and ${\rm B}$-type hard functions
from matching the QCD weak current  $\bar \psi \, \Gamma_i \, Q$ onto ${\rm SCET}_{\rm I}$
at NLO and LO in $\alpha_s$, respectively,  in Appendix \ref{appendix: hard functions}.
Two phenomenological models for the two-particle and three-particle $B$-meson distribution amplitudes,
up to the twist-six accuracy, employed in the numerical computations of the semileptonic $B \to V$ form factors
are collected in Appendix \ref{appendix: B-meson DAs}.

\section{QCD factorization for $B \to V$ form factors}
\label{section: SCET-I factorization}

The purpose of this section is, following closely \cite{Hill:2004if,Becher:2004kk,Beneke:2005gs},
to summarize the soft-collinear factorization formulae for $B \to V$ form factors
at large hadronic recoil  in ${\rm SCET}_{\rm I}$ by
integrating out the strong interaction dynamics at the hard scale $m_b$,
for the sake of establishing the  theoretical framework  for computing the
SCET matrix elements from the LCSR method.
The resulting ${\rm SCET (hc, c, s)}$ representations for the QCD
heavy-to-light form factors are given by \cite{Beneke:2003pa,Lange:2003pk,Bauer:2002aj}
\begin{eqnarray}
F^{B \to V}_{i}(n \cdot  p) = C_i^{\rm (A0)}(n \cdot  p) \, \xi_a(n \cdot p)
+ \int d \tau \, C_i^{\rm (B1)}(\tau, n \cdot p) \, \Xi_a(\tau, n \cdot p) \,, \,\,\,
(a \, = \, \parallel,  \,\,  \perp)\,,
\label{SCET-I factorization formula for B to V FFs}
\end{eqnarray}
where the seven $B \to V$ form factors are expressed in terms of the four ``effective"
form factors in ${\rm SCET_{I}}$ at leading power in the heavy quark expansion.
The hard matching coefficients for both the ${\rm A0}$-type and ${\rm B1}$-type SCET currents have been
computed at one-loop accuracy \cite{Hill:2004if,Bauer:2000yr,Beneke:2004rc}.
As we aim at computing the semileptonic $B \to V$ form factors with the aid of the
factorization formula (\ref{SCET-I factorization formula for B to V FFs}) at ${\cal O}(\alpha_s)$,
we will need the perturbative matching functions $ C_i^{\rm (A0)}(n \cdot  p)$ at
NLO in QCD and the ${\rm B1}$-type hard functions $C_i^{\rm (B1)}(\tau, n \cdot p)$ at tree level
as displayed in Appendix \ref{appendix: hard functions}.
The ``effective" form factors $\xi_{a}(n \cdot p)$ and $\Xi_{a}(\tau, n \cdot p)$
are defined by the hadronic matrix elements of
the corresponding ${\rm SCET_{I}}$ operators \cite{Beneke:2005gs}
\begin{eqnarray}
&& \langle V(p, \epsilon^{\ast}) | \left (\bar \xi \, W_c \right) \,
\gamma_5 \, h_v \,   | \bar B_v \rangle  =
- n \cdot p \, (\epsilon^{\ast} \cdot v) \, \xi_{\parallel}(n \cdot p) \,, \nonumber \\
&& \langle V(p, \epsilon^{\ast}) | \left (\bar \xi \, W_c \right) \,
\gamma_5 \, \gamma_{\mu \perp} \, h_v \,   | \bar B_v \rangle  =
- n \cdot p \, (\epsilon^{\ast}_{\mu} - \epsilon^{\ast} \cdot v \, \bar n_{\mu}) \,
\xi_{\perp}(n \cdot p) \,, \nonumber \\
&& \langle V(p, \epsilon^{\ast}) | \left (\bar \xi \, W_c \right) \,
\gamma_5 \, \left (W_c^{\dagger} \,\, i \, \not \! \! D_{c \perp} \, W_c \right)(r n) \, h_v \,   | \bar B_v \rangle
=  - n \cdot p  \, m_b \, \epsilon^{\ast} \cdot v \,
\int_0^1 \, d \tau \, e^{i \, \tau \, n \cdot p \, r} \, \Xi_{\parallel}(\tau, \, n \cdot p)  \,, \nonumber \\
&& \langle V(p, \epsilon^{\ast}) | \left (\bar \xi \, W_c \right) \,
\gamma_5 \, \gamma_{\mu \perp} \,
\left (W_c^{\dagger} \,\, i \, \not \! \! D_{c \perp} \, W_c \right)(r n) \, h_v \,   | \bar B_v \rangle  \nonumber \\
&& =  - n \cdot p  \, m_b \,  (\epsilon^{\ast}_{\mu} - \epsilon^{\ast} \cdot v \, \bar n_{\mu}) \,
\int_0^1 \, d \tau \, e^{i \, \tau \, n \cdot p \, r} \, \Xi_{\perp}(\tau, \, n \cdot p)  \,,
\label{Definition: SCET-I form factors}
\end{eqnarray}
where  the light-cone Wilson line is introduced to restore the collinear gauge invariance \cite{Beneke:2003pa,Beneke:2002ni}
\begin{eqnarray}
W_c(x) =  {\rm P} \, {\rm exp}  \,
\left [ i \, g_s \, \int_{-\infty}^{0} \, d s \, n \cdot A_c(x + s \, n) \right ] \,.
\label{collinear Wilson line}
\end{eqnarray}

The QCD matrix elements of the heavy-to-light currents $\bar \psi \, \Gamma_i \, Q$
are parameterized by  the semileptonic  $B \to V$ form factors in the standard way
\cite{Beneke:2000wa} \footnote{Notice that there is an obvious misprint in the definition
of the tensor form factor $T_1(q^2)$ presented in \cite{Beneke:2005gs}.}
\begin{eqnarray}
&& c_V \,  \langle V(p, \epsilon^{\ast}) | \bar q \, \gamma_{\mu} \, b | \bar B (p+q) \rangle
= - {2 \, i \,   V(q^2) \over m_B + m_V} \,
\epsilon_{\mu \nu \rho \sigma} \, \epsilon^{\ast \, \nu}\,
p^{\rho} \, q^{\sigma} \,, \nonumber \\
&& c_V \, \langle V(p, \epsilon^{\ast}) | \bar q \, \gamma_{\mu} \, \gamma_5 \, b | \bar B (p+q) \rangle
=  {2 \, m_V \, \epsilon^{\ast} \cdot q  \over q^2} \, q_{\mu} \, A_0(q^2) \nonumber \\
&& \hspace{5.5 cm}  + (m_B +m_V)\, \left[\epsilon^{\ast}_{\mu} - {\epsilon^{\ast} \cdot q \over q^2}  \, q_{\mu} \right ]
 \, A_1(q^2)    \nonumber  \\
&&  \hspace{5.5 cm}  - {\epsilon^{\ast} \cdot q  \over m_B +m_V} \, \left [ (2 \, p + q)_{\mu}
- {m_B^2 - m_V^2 \over q^2}  \, q_{\mu}\right ] \, A_2(q^2) \,, \nonumber \\
&& c_V \, \langle V(p, \epsilon^{\ast}) | \bar q \, i \, \sigma_{\mu \nu} \, q^{\nu}\, b | \bar B (p+q) \rangle
=  2 \, i \, T_1(q^2) \, \epsilon_{\mu \nu \rho \sigma} \, \epsilon^{\ast \, \nu}\,
p^{\rho} \, q^{\sigma} \,, \nonumber \\
&& c_V \, \langle V(p, \epsilon^{\ast}) | \bar q \, i \, \sigma_{\mu \nu} \, \gamma_5 \, q^{\nu}\, b | \bar B (p+q) \rangle
= T_2(q^2) \, \left [ (m_B^2 -m_V^2) \, \epsilon^{\ast}_{\mu}
-  (\epsilon^{\ast} \cdot q)  \, (2 \, p + q)_{\mu} \right ]  \nonumber  \\
&& \hspace{6.5 cm}  + \,  T_3(q^2) \,  (\epsilon^{\ast} \cdot q)\,
\left [q_{\mu} - {q^2  \over m_B^2 -m_V^2} \,  (2 \, p + q)_{\mu} \right ],
\end{eqnarray}
with the convention $\epsilon_{0123}=-1$. We have introduced the factor $c_V$
to account for the flavour structure of vector mesons
with $c_V = \pm \sqrt{2}$ for the $\rho^0$ and $\omega$
(obviously, $\sqrt{2}$ for $b \to u$ transition and $-\sqrt{2}$ for the $b \to d$ transition)
and $c_V = 1$ otherwise.
At maximal hadronic recoil $q^2=0$ there exist  two  relations for
the above-mentioned $B \to V$ form factors in QCD
\begin{eqnarray}
{m_B + m_V \over 2 \, m_V} \, A_1(0) - {m_B - m_V \over 2 \, m_V} \, A_2(0) = A_0(0) \,,
\qquad T_1(0) = T_2(0) \,,
\end{eqnarray}
which are free of both radiative and power corrections.
Employing the SCET representation of the QCD heavy-to-light current
(see \cite{Beneke:2005gs} for the explicit expressions of the A-type and B-type SCET currents)
\begin{eqnarray}
(\bar \psi \, \Gamma_i \, Q)(0) &=& \int d {\hat s} \, \sum_{j} \, \tilde{C}_{i j}^{(\rm A0)}(\hat s) \,
O_{j}^{(\rm A0)}(s; 0) + \int d  {\hat s} \, \sum_{j} \, \tilde{C}_{i j \mu}^{(\rm A1)}(\hat s) \,
O_{j}^{(\rm A1) \mu}(s; 0) \nonumber \\
&& +  \int d  {\hat s_1} \, \int d  {\hat s_2} \, \sum_{j} \, \tilde{C}_{i j \mu}^{(\rm B1)}(\hat s_1, \hat s_2) \,
O_{j}^{(\rm B1) \mu}(s_1, s_2; 0) + ...    \,,
\end{eqnarray}
and performing the $\hat{s}$ and $\hat{s}_1$ integrations for the resulting SCET matrix elements
\begin{eqnarray}
&& \langle V(p, \epsilon^{\ast}) | \int d {\hat s} \,  \tilde{C}_{i j}^{(\rm A0)}(\hat s) \,
O_{j}^{(\rm A0)}(s; 0)  | \bar B_v \rangle \nonumber \\
&& =  C_{i j}^{(\rm A0)} \left ({n \cdot p \over m_b}, \mu \right ) \,
\langle V(p, \epsilon^{\ast}) | (\bar \xi \, W_c)(0) \, \Gamma_j^{\prime} \, h_v(0)| \bar B_v \rangle \,, \nonumber \\
&& \langle V(p, \epsilon^{\ast}) | \int d {\hat s_1} \, d {\hat s_2} \,  \tilde{C}_{i j}^{(\rm B1)}(\hat s_1, \hat s_2) \,
O_{j}^{(\rm B1)}(s_1, s_2; 0)  | \bar B_v \rangle \nonumber \\
&&  =  {n \cdot p   \over m_b}  \, \int d \tau \, C_{i j \mu}^{(\rm B1)} \left ({n \cdot p \, \bar \tau \over m_b},
 {n \cdot p \, \tau \over m_b},\mu \right )  \,
\, \int {d r \over 2 \, \pi} \, e^{-i \, \tau \, n \cdot p \, r}  \, \nonumber \\
&& \hspace{0.5 cm} \langle V(p, \epsilon^{\ast}) | (\bar \xi \, W_c)(0) \,
\left (W_c^{\dagger} \,\, i \,  D_{c \perp}^{\mu} \, W_c \right)(r n)
\, \Gamma_j^{\prime} \, h_v(0)| \bar B_v \rangle \,,
\end{eqnarray}
we can readily derive the ${\rm SCET_{I}}$ factorization formulae for the QCD form factors
at leading power in the heavy quark expansion
\begin{eqnarray}
&& {m_B \over m_B + m_V} \, V(n \cdot p)  =  C_V^{(\rm A0)} \, \left ({n \cdot p \over m_b}, \mu \right ) \,
\xi_{\perp}(n \cdot p)  \nonumber \\
&& \hspace{4 cm} + \int_0^1  d \tau \, C_V^{(\rm B1)} \,
\left ({n \cdot p \, \bar \tau \over m_b},
 {n \cdot p \, \tau \over m_b},\mu \right )  \, \Xi_{\perp}(\tau, n \cdot p) \,,  \nonumber \\
&& {2 \, m_V \over n \cdot p} \, A_0(n \cdot p)  =  C_{f_0}^{(\rm A0)} \, \left ({n \cdot p \over m_b}, \mu \right ) \,
\xi_{\parallel}(n \cdot p)   + \int_0^1  d \tau \, C_{f_0}^{(\rm B1)} \,
\left ({n \cdot p \, \bar \tau \over m_b},
 {n \cdot p \, \tau \over m_b},\mu \right )  \, \Xi_{\parallel}(\tau, n \cdot p) \,,  \nonumber  \\
&& {m_B + m_V \over n \cdot p} \, A_1(n \cdot p)  =  C_V^{(\rm A0)} \, \left ({n \cdot p \over m_b}, \mu \right ) \,
\xi_{\perp}(n \cdot p) \nonumber \\
&& \hspace{4 cm} + \int_0^1  d \tau \, C_V^{(\rm B1)} \,
\left ({n \cdot p \, \bar \tau \over m_b},
 {n \cdot p \, \tau \over m_b},\mu \right )  \, \Xi_{\perp}(\tau, n \cdot p) \,,  \nonumber  \\
&& {m_B + m_V \over n \cdot p} \, A_1(n \cdot p)
- {m_B - m_V \over m_B} \, A_2(n \cdot p) \nonumber \\
&& =  C_{f_+}^{(\rm A0)} \, \left ({n \cdot p \over m_b}, \mu \right ) \,
\xi_{\parallel}(n \cdot p)   + \int_0^1  d \tau \, C_{f_+}^{(\rm B1)} \,
\left ({n \cdot p \, \bar \tau \over m_b},
 {n \cdot p \, \tau \over m_b},\mu \right )  \, \Xi_{\parallel}(\tau, n \cdot p) \,,  \nonumber  \\
&& T_1(n \cdot p)  =  C_{T_1}^{(\rm A0)} \, \left ({n \cdot p \over m_b}, \mu \right ) \,
\xi_{\perp}(n \cdot p)  + \int_0^1  d \tau \, C_{T_1}^{(\rm B1)} \,
\left ({n \cdot p \, \bar \tau \over m_b},
 {n \cdot p \, \tau \over m_b},\mu \right )  \, \Xi_{\perp}(\tau, n \cdot p) \,,  \nonumber \\
&& {m_B \over n \cdot p} \,T_2(n \cdot p)  =  C_{T_1}^{(\rm A0)} \, \left ({n \cdot p \over m_b}, \mu \right ) \,
\xi_{\perp}(n \cdot p)  + \int_0^1  d \tau \, C_{T_1}^{(\rm B1)} \,
\left ({n \cdot p \, \bar \tau \over m_b},
 {n \cdot p \, \tau \over m_b},\mu \right )  \, \Xi_{\perp}(\tau, n \cdot p) \,,  \nonumber \\
&& {m_B \over n \cdot p} \,T_2(n \cdot p) - T_3(n \cdot p) \nonumber \\
&& =  C_{f_T}^{(\rm A0)} \, \left ({n \cdot p \over m_b}, \mu \right ) \,
\xi_{\parallel}(n \cdot p)  + \int_0^1  d \tau \, C_{f_T}^{(\rm B1)} \,
\left ({n \cdot p \, \bar \tau \over m_b},
 {n \cdot p \, \tau \over m_b},\mu \right )  \, \Xi_{\parallel}(\tau, n \cdot p) \,.
\label{SCET-I factorization formulae}
\end{eqnarray}
The coefficient functions $C_{i j}^{(\rm A0)}$ and $C_{i j \mu}^{(\rm B1)}$ are obtained from the
Fourier transformations of the position-space coefficient functions
$\tilde{C}_{i j}^{(\rm A0)}$ and $\tilde{C}_{i j \mu}^{(\rm B1)}$ \cite{Beneke:2003pa}.
It is evident that only five independent combinations of ${\rm A0}$- and ${\rm B1}$-type  SCET operators
appear in the factorization formulae for the seven different $B \to V$ form factors,
implying the two additional relations \cite{Beneke:2000wa,Burdman:2000ku}
\begin{eqnarray}
{m_B \over m_B + m_V} \, V(n \cdot p)  = {m_B + m_V \over n \cdot p} \, A_1(n \cdot p) \,,
\qquad  T_1(n \cdot p) =  {m_B \over n \cdot p} \,T_2(n \cdot p) \,,
\label{exact form factor relations at LP}
\end{eqnarray}
which are fulfilled to all orders in perturbative expansion at leading power in $\Lambda/m_b$.

\section{The $B$-meson  LCSR for the SCET $B \to V$ form factors}
\label{section: SCET sum rules at leading twist}

In this section we turn to construct the SCET sum rules for the ``effective" form factors
$\xi_{a}(n \cdot p)$ and  $\Xi_{a}(\tau, n \cdot p)$ (with $a = \parallel, \perp$)
entering the factorization formulae (\ref{SCET-I factorization formulae}) for the QCD $B \to V$ form factors
at one-loop accuracy. To this end, we will first demonstrate the soft-collinear factorization theorems for the
corresponding vacuum-to-$B$-meson correlation functions with an interpolating current for the
collinear vector meson at leading power in the heavy-quark expansion.
We further place particular attention to the treatment of evanescent operators in dimensional regularization
for the determination of the perturbative matching coefficients from  $\rm {SCET_{I}} \to {HQET}$.
The summation of parametrically large logarithms  $\ln ({m_b / \Lambda})$ appearing in the hard functions
in front of both A0- and B-type  $\rm {SCET_{I}}$ operators is achieved at
NLL  and LL accuracy, respectively,
by employing the RG formalism in momentum space.

\subsection{The $B$-meson  LCSR for $\xi_{\parallel}(n \cdot p)$}

Following the standard strategy we start with the construction of the vacuum-to-$B$-meson
correlation function
\begin{eqnarray}
\Pi_{\nu, \|}(p, q) = \int d^4 x \, e^{i p \cdot x} \, \langle 0 |
{\rm T}  \left \{j_{\nu}(x),  \,\, \left (\bar \xi \, W_c \right)(0) \,
\gamma_5 \, h_v(0) \,    \right \}   | \bar B_v \rangle \,,
\label{correlation function for xi-L}
\end{eqnarray}
where the local QCD current $j_{\nu}$ interpolates current for the longitudinal polarization state
of the collinear vector meson
\begin{eqnarray}
j_{\nu}(x) = \bar q^{\prime}(x) \, \gamma_{\nu} \, q(x) \,.
\end{eqnarray}
The SCET representation of the QCD interpolating current can be obtained following the prescriptions
described in \cite{Beneke:2002ph}
\begin{eqnarray}
j_{\nu} = j_{\xi \xi, \nu}^{(0)} + j_{\xi \xi, \perp \, \nu}^{(1)} \,
+ j_{\xi q_s, \parallel \, \nu}^{(2)}  + j_{\xi q_s, \perp \, \nu}^{(2)} + ... \,,
\end{eqnarray}
where the explicit expressions of the effective currents are given by
\begin{eqnarray}
j_{\xi \xi, \nu}^{(0)} &=& \bar \xi \, {\not \! n \over 2} \, \xi \,\, \bar n_{\nu} \,,  \nonumber \\
j_{\xi \xi, \perp \,  \nu}^{(1)}  &=&   \bar \xi \, \gamma_{\nu \perp} \,
{1 \over i \, n \cdot D_c} \, i \not \! \! D_{c \perp}\,  {\not \! n \over 2} \, \xi
+ \bar \xi \, i \not \! \! D_{c \perp}\,
{1 \over i \, n \cdot D_c} \, \gamma_{\nu \perp} \,  {\not \! n \over 2} \, \xi   \,,  \nonumber \\
j_{\xi q_s, \parallel \, \nu}^{(2)} &=& \left ( \bar \xi \, W_c \,  {\not \! n \over 2} \, Y_s^{\dagger} \, q_s
+ \bar q_s \, Y_s \,  {\not \! n \over 2} \,  W_c^{\dagger} \, \xi \right ) \,\, \bar n_{\nu} \,, \nonumber \\
j_{\xi q_s, \perp \, \nu}^{(2)} &=&  \bar \xi \, W_c \,   \gamma_{\perp \nu}  \, Y_s^{\dagger}\, q_s
+ \bar q_s   \,   Y_s \,  \gamma_{\perp \nu} \, W_c^{\dagger} \, \xi \,.
\end{eqnarray}
To maintain the collinear and soft gauge invariance both the collinear Wilson line
defined in (\ref{collinear Wilson line}) and the following light-like Wilson line
\begin{eqnarray}
Y_s(x) &=&  {\rm P \,\, exp} \left [i \, g_s \, \int_{-\infty}^{0} \, ds \,  \bar n \cdot A_s(x + s \bar n) \right ] \,,
\end{eqnarray}
is introduced for the SCET currents in a general gauge.
It is then straightforward to identify the leading-power contribution to the correlation function (\ref{correlation function for xi-L})
\begin{eqnarray}
&& \Pi_{\nu, \|}(p, q) \nonumber \\
&& = \int d^4 x \, e^{i p \cdot x} \,
 \langle 0 | {\rm T}  \left \{ j_{\xi q_s, \parallel \, \nu}^{(2)}(x),  \,\, \left (\bar \xi \, W_c \right)(0) \,
\gamma_5 \, h_v(0) \,    \right \}   | \bar B_v \rangle \nonumber \\
&& + \int d^4 x \, e^{i p \cdot x} \, \int d^4 y \,
 \langle 0 | {\rm T}  \left \{ j_{\xi \xi, \nu}^{(0)}(x),  \,\, i \, {\cal L}_{\xi q_s}^{(2)}(y),  \,\,
\left (\bar \xi \, W_c \right)(0) \, \gamma_5 \, h_v(0) \,    \right \}   | \bar B_v \rangle  \nonumber \\
&& + \int d^4 x \, e^{i p \cdot x} \, \int d^4 y \, \int d^4 z \,
 \langle 0 | {\rm T}  \left \{ j_{\xi \xi, \nu}^{(0)}(x),  \,\, i \, {\cal L}_{\xi q_s}^{(1)}(y),  \,\,
i \, {\cal L}_{\xi m}^{(1)}(z),  \,\,
\left (\bar \xi \, W_c \right)(0) \, \gamma_5 \, h_v(0) \,    \right \}   | \bar B_v \rangle  \nonumber \\
&& \equiv \Pi_{\nu, \|}^{A}(p, q)  + \Pi_{\nu, \|}^{B}(p, q) + \Pi_{\nu, \|}^{C}(p, q) \,,
\label{def£ºxi-L-A-B-C}
\end{eqnarray}
where the third term $\Pi_{\nu, \|}^{C}$ takes into account the light-quark mass effect.
The multipole expanded SCET Lagrangian up to the ${\cal O}(\lambda^2)$ accuracy \cite{Beneke:2002ni}
have been derived with the position-space formalism \cite{Beneke:2002ph}
\begin{eqnarray}
{\cal L}_{\xi}^{(0)} &=& \bar \xi \, \left ( i \, \bar n \cdot D
+ i \not \! \! D_{\perp c} \,\, {1 \over i \, n \cdot D_c} \, i \not \! \! D_{\perp c} \right ) \,
{\not \! n \over 2 }  \,\, \xi  \,,  \nonumber \\
{\cal L}_{\xi m}^{(1)} &=& m \,\, \bar \xi \, \left [ i \not \! \! D_{\perp c},  \, {1 \over i \, n \cdot D_c}   \right ] \,
{\not \! n \over 2 }  \,\, \xi  \,, \nonumber \\
{\cal L}_{\xi m}^{(2)} &=& -m^2 \,\, \bar \xi \,  {1 \over i \, n \cdot D_c}  \,
{\not \! n \over 2 }  \,\, \xi  \nonumber   \,, \\
{\cal L}_{\xi q_s}^{(1)} &=& \bar q_s \,  W_c^{\dag} \,\,  i \not \! \! D_{\perp c}\,\, \xi
- \bar \xi \,\,   i \not \! \! \overleftarrow{D}_{\perp c} \,\,  W_c \, q_s,   \nonumber   \\
{\cal L}_{\xi q_s}^{(2)} &=& \bar q_s \,  W_c^{\dag} \,\,   \left ( i \, \bar n \cdot D
+ i \not \! \! D_{\perp c} \,\, {1 \over i \, n \cdot D_c} \, i \not \! \! D_{\perp c} \right ) \,\,
{\not \! n \over 2 }  \,\, \xi   \nonumber \\
&& -  \, \bar \xi \, {\not \! n \over 2 } \,\,  \left ( i \, \bar n \cdot \overleftarrow{D}
+ \, i \not \! \! \overleftarrow{D}_{\perp c} \,\, {1 \over i \, n \cdot \overleftarrow{D}_c} \,
i \not \! \! \overleftarrow{D}_{\perp c} \right ) \,\,
W_c \,\, q_s   \nonumber \\
&& + \,  \bar q_s  \, \overleftarrow{D}_s^{\mu} \, x_{\perp \mu} \,  W_c^{\dag} \,\,  i \not \! \! D_{\perp c}\,\, \xi
- \bar \xi  \,  i \not \! \! \! \overleftarrow{ D}_{\perp c} \,\,  W_c \, x_{\perp \mu}  \, D_s^{\mu} \,   q_s \,.
\end{eqnarray}
Our major objective is then to perform the perturbative matching of the ${\rm SCET}_{\rm I}$ correlation functions
$\Pi_{\nu, \|}^{i}$ as defined in (\ref{def£ºxi-L-A-B-C}) onto ${\rm SCET}_{\rm II}$
\begin{eqnarray}
\Pi_{\nu, \|}^{i}(p, q) = {\tilde{f}_B(\mu) \, m_B \over 2}\,
\sum_{m = \pm } \, \int_0^{+\infty} \, d \omega \, J_{\|, m}^{i}
\left ({\mu^2 \over n \cdot p \, \omega}, {\omega \over \bar n \cdot p} \right )  \,
\phi_B^{m}(\omega, \mu) \,\, \bar n_{\nu},  \,\, \, (i=A, B, C)
\label{SCET factorization formula L}
\end{eqnarray}
at one-loop accuracy, by integrating out the hard-collinear fluctuations from the scale $\sqrt{m_b \, \Lambda}$.

\subsubsection{SCET factorization for $\Pi_{\nu, \|}^{A}(p, q)$}

The hard-collinear functions $J_{\|, m}^{A} $ entering the SCET factorization formula (\ref{SCET factorization formula L})
can be determined by investigating the partonic matrix element
\begin{eqnarray}
F_{\|}^{A} = \int d^4 x \, e^{i p \cdot x} \,
\left \langle 0 \left | {\rm T}  \left \{ \bar q_s(x) \, Y_s \,  {\not \! n \over 2} \,  W_c^{\dagger} \, \xi(x),
\,\, \left (\bar \xi \, W_c \right)(0) \, \gamma_5 \, h_v(0) \,    \right \}   \right | \bar q_s(k) \, h_v \right \rangle \,.
\label{correlator: F-L-A}
\end{eqnarray}
Evaluating the tree-level contribution to the SCET amplitude $F_{\|}^{A}$ leads to
\begin{eqnarray}
F_{\|, \, \rm LO}^{A} = - {i \over \bar n \cdot p - \omega + i0} \,\,
\bar q_s(k) \, {\not \! n \over 2} \, \gamma_5 \,  h_v
=  - {i \over \bar n \cdot p - \omega^{\prime} + i0} \ast
\langle O_{\|, \,-} (\omega, \omega^{\prime})  \rangle^{(0)}  \,,
\end{eqnarray}
where we have introduced the convention $\omega= \bar n \cdot k$  and the asterisk
indicates the convolution integration over the variable $\omega^{\prime}$.
The light-cone  matrix element $\langle O_{\|, \, -} (\omega, \omega^{\prime}) \rangle$
is defined as
\begin{eqnarray}
\langle O_{\|, \, -} (\omega, \omega^{\prime})  \rangle
= \langle 0 |O_{\|, \, -} (\omega^{\prime}) | \bar q_s(k) \, h_v  \rangle
=\bar q_s(k) \, {\not \! n \over 2} \, \gamma_5 \,  h_v  \,\,
\delta(\omega - \omega^{\prime}) + {\cal O}(\alpha_s),
\end{eqnarray}
where the HQET operator $O_{\|, \, -} (\omega^{\prime})$  in the  momentum space reads
\begin{eqnarray}
O_{\|, \, -} (\omega^{\prime}) = {1 \over 2 \, \pi} \, \int d t \, e^{i \, t \, \omega^{\prime}} \,
\left ( \bar q_s Y_s \right )(t \, \bar n) \, \,  {\not \! n \over 2} \, \gamma_5 \,\,
\left (Y_s^{\dag} h_v \right ) (0)\,.
\end{eqnarray}
The light-ray effective operator $ O_{\|, \, + }(\omega^{\prime})$ can be defined in
an analogous way
\begin{eqnarray}
O_{\|, \, +} (\omega^{\prime}) = {1 \over 2 \, \pi} \, \int d t \, e^{i \, t \, \omega^{\prime}} \,
\left ( \bar q_s Y_s \right )(t \, \bar n) \, \,  {\not \! \bar  n \over 2} \, \gamma_5 \,\,
\left (Y_s^{\dag} h_v \right ) (0)\,.
\end{eqnarray}
Implementing the perturbative matching relation for the matrix element $F_{\|}^{A}$
\begin{eqnarray}
F_{\|}^{A}  = (- i) \, \sum_{m=\pm} \, J_{\|, \, m}^{A}
\left ({\mu^2 \over n \cdot p \, \omega^{\prime}}, {\omega^{\prime} \over \bar n \cdot p} \right )  \,
\ast \langle O_{\|, \,m} \, (\omega, \omega^{\prime})  \rangle  \,,
\label{matching relation for F-L-A}
\end{eqnarray}
we can readily derive the tree-level short-distance functions
\begin{eqnarray}
J_{\|, \, -}^{A, \, (0)} = {1 \over \bar n \cdot p - \omega^{\prime} + i0}\,,
\qquad J_{\|, \, +}^{A, \, (0)} = 0 \,.
\end{eqnarray}
Employing the definition of the two-particle $B$-meson LCDA in the coordinate space \cite{Grozin:1996pq,Beneke:2000wa}
\begin{eqnarray}
&& \langle 0 | \left ( \bar q_s Y_s \right )_{\beta}(t \, \bar n) \, \, \,
\left ( Y_s^{\dag} h_v\right )_{\alpha}(0) | \bar B_v \rangle \nonumber \\
&& = - {i \, \tilde{f}_B(\mu) \, m_B \over 4} \,
\left \{ {1 + \not \! v \over 2} \,  \left [ 2 \, \tilde{\phi}_B^{+}(t, \mu)
+ \left (\tilde{\phi}_B^{-}(t, \mu) - \tilde{\phi}_B^{+}(t, \mu) \right ) \, \not \! n  \, \right ]
\, \gamma_5 \right \}_{\alpha \beta }  \,,
\label{Definition of B-meson LCDA}
\end{eqnarray}
the resulting SCET factorization formula for the correlation function $\Pi_{\nu, \|}^{A}$  is
\begin{eqnarray}
\Pi_{\nu, \|}^{A}(p, q) = {\tilde{f}_B(\mu) \, m_B \over 2}\,
\, \int_0^{+\infty} \, d \omega \, J_{\|, -}^{A, (0)}
\left ({\mu^2 \over n \cdot p \, \omega}, {\omega \over \bar n \cdot p} \right )  \,
\phi_B^{-}(\omega, \mu) \,\, \bar n_{\nu} + {\cal O}(\alpha_s) \,.
\end{eqnarray}

\begin{figure}
\begin{center}
\includegraphics[width=1.0 \columnwidth]{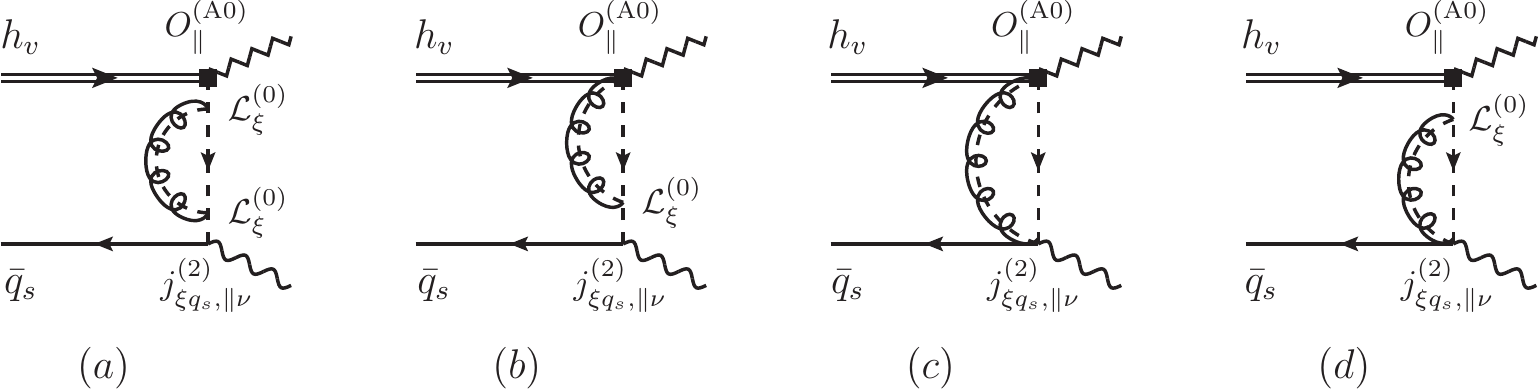}
\vspace*{0.1cm}
\caption{Diagrammatical representation of the vacuum-to-$B$-meson correlation function
$\Pi_{\nu, \|}^{A}(p, q)$  defined with  the ${\rm A0}$-type SCET operator
$O_{\|}^{\rm {(A0)}}= \left (\bar \xi \, W_c \right) \, \gamma_5 \, h_v$
and the power suppressed interpolating current $j_{\xi q_s, \parallel \, \nu}^{(2)}$ at  one loop.}
\label{fig: correlator for F-L-A}
\end{center}
\end{figure}

We proceed to determine the NLO contribution to the jet function $ J_{\|, \pm}^{A}$ by
expanding the matching relation (\ref{matching relation for F-L-A}) up to the ${\cal O}(\alpha_s)$ accuracy.
To this end, we will need to evaluate the one-loop ${\rm SCET}_{\rm I}$ diagrams
presented in figure \ref{fig: correlator for F-L-A} with the
subleading power SCET Feynman rules collected in \cite{Beneke:2018rbh}.
The self-energy correction to the hard-collinear quark propagator displayed
in  the diagram (a) of figure \ref{fig: correlator for F-L-A}
can be readily written as \cite{Wang:2015ndk}
\begin{eqnarray}
F_{\|, \, {\rm NLO}}^{A, (a)} = -{\alpha_s \, C_F \over 4 \, \pi} \,
\left [ {1 \over \epsilon}  + \ln {\mu^2 \over n \cdot p \, (\omega- \bar n \cdot p) - i 0} + 1 \right ]  \,
F_{\|, \, {\rm LO}}^{A} \,.
\end{eqnarray}
Obviously, the NLO correction from the hard-collinear Wilson lines
presented in the diagram (c) of figure \ref{fig: correlator for F-L-A}
yields vanishing contribution due to $n^2=0$.
One can further verify that the hard-collinear corrections displayed in
the diagrams (b) and (d) of figure \ref{fig: correlator for F-L-A}
give rise to the identical  results
\begin{eqnarray}
F_{\|, \, {\rm NLO}}^{A, (b)} &=& F_{\|, \, {\rm NLO}}^{A, (d)}
= - {2 \, g_s^2 \, C_F \over \bar n \cdot p - \omega} \,\,
\bar q_s(k) \, {\not \! n \over 2} \, \gamma_5 \,  h_v  \, \nonumber \\
&& \times \, \int {d^D l \over  (2 \pi)^D} \,
{n \cdot (p+l) \over [n \cdot (p+l) \, \bar n \cdot (p-k+l) + l_{\perp}^2 + i 0]
[ n \cdot l + i 0] [l^2 + i 0] }  \,,
\end{eqnarray}
which can be evaluated straightforwardly with dimensional regularization scheme
\begin{eqnarray}
F_{\|, \, {\rm NLO}}^{A, (b)} &=& F_{\|, \, {\rm NLO}}^{A, (d)}
= {\alpha_s \, C_F \over 2 \, \pi} \,\,
\bigg  \{ {1 \over \epsilon^2}   + {1 \over \epsilon} \,
\left [  \ln {\mu^2 \over n \cdot p \, (\omega- \bar n \cdot p)} + 1 \right ]
+ {1 \over 2} \, \ln^2 {\mu^2 \over n \cdot p \, (\omega- \bar n \cdot p)} \nonumber \\
&& +  \, \ln {\mu^2 \over n \cdot p \, (\omega- \bar n \cdot p)}
- {\pi^2 \over 12} + 2  \bigg \}   \,\, F_{\|, \, {\rm LO}}^{A}  \,.
\end{eqnarray}
Adding up different pieces together and applying the matching condition (\ref{matching relation for F-L-A})
leads to the hard-collinear functions at the one-loop accuracy
\begin{eqnarray}
J_{\|, \, -}^{A, \, (1)} &=&  J_{\|, \, -}^{A, \, (0)} \,
\bigg \{ 1 +  {\alpha_s \, C_F \over 4 \, \pi}  \,
\bigg  [ {4 \over \epsilon^2}   + {1 \over \epsilon} \,
\left ( 4 \,  \ln {\mu^2 \over n \cdot p \, (\omega- \bar n \cdot p) } + 3  \right )
+ 2 \, \ln^2 {\mu^2 \over n \cdot p \, (\omega- \bar n \cdot p)} \nonumber \\
&& + \, 3 \, \ln {\mu^2 \over n \cdot p \, (\omega- \bar n \cdot p)}
- {\pi^2 \over 3} + 7  \bigg ]  \bigg \} \,, \nonumber \\
 J_{\|, \, +}^{A, \, (1)} &=& 0 \,,
\label{jet functions: L-A}
\end{eqnarray}
which are in precise agreement with the results presented in \cite{DeFazio:2007hw}.

\subsubsection{SCET factorization for $\Pi_{\nu, \|}^{B}(p, q)$}

\begin{figure}
\begin{center}
\includegraphics[width=0.8 \columnwidth]{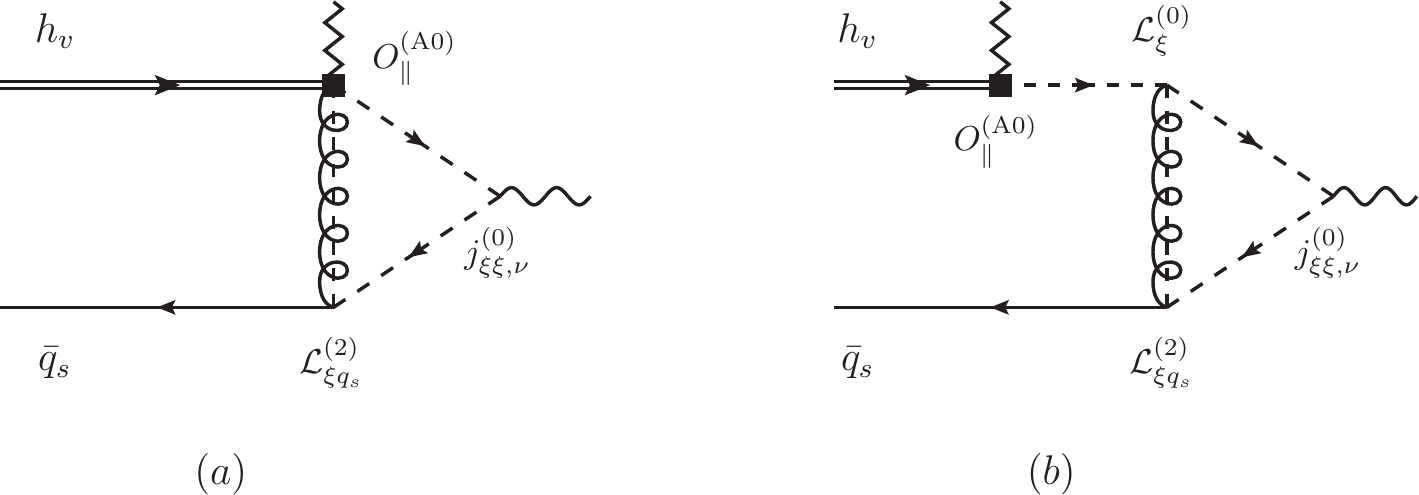}
\vspace*{0.1cm}
\caption{Diagrammatical representation of the vacuum-to-$B$-meson correlation function
$\Pi_{\nu, \|}^{B}(p, q)$  defined with  the ${\rm A0}$-type SCET operator
$O_{\|}^{\rm {(A0)}}= \left (\bar \xi \, W_c \right) \, \gamma_5 \, h_v$,
the leading power interpolating current $j_{\xi \xi, \nu}^{(0)}$
and the subleading power SCET Lagrangian ${\cal L}_{\xi q_s}^{(2)}$.}
\label{fig: correlator for F-L-B}
\end{center}
\end{figure}

Along the same vein, the jet function $J_{\|, \, \pm}^{B}$ can be determined by
performing the perturbative factorization  for  the partonic matrix element
\begin{eqnarray}
F_{\|}^{B} = \int d^4 x \, e^{i p \cdot x} \, \int d^4 y \,
 \left \langle 0 \left | {\rm T}  \left \{ \bar \xi(x) \, {\not \! n \over 2} \, \xi(x),
  \,\, i \, {\cal L}_{\xi q_s}^{(2)}(y),  \,\,
\left (\bar \xi \, W_c \right)(0) \, \gamma_5 \, h_v(0) \,    \right \}   \right | \bar q_s(k) \, h_v  \right \rangle,
\hspace{0.5 cm}
\end{eqnarray}
taking advantage of the matching relation in analogy to (\ref{matching relation for F-L-A})
\begin{eqnarray}
F_{\|}^{B}  = (- i) \, \sum_{m=\pm} \, J_{\|, \, m}^{B}
\left ({\mu^2 \over n \cdot p \, \omega^{\prime}}, {\omega^{\prime} \over \bar n \cdot p} \right )  \,
\ast \langle O_{\|, \,m} \, (\omega, \omega^{\prime})  \rangle  \,.
\label{matching relation for F-L-B}
\end{eqnarray}
Evaluating the diagram (a) in figure \ref{fig: correlator for F-L-B} with the SCET Feynman rules leads to
\begin{eqnarray}
F_{\|, \rm {LO}}^{B, (a)} &=& -2  \, g_s^2 \, C_F \, \bar q_s(k) \, {\not \! n \over 2} \, \gamma_5 \,  h_v \nonumber \\
&& \times \,  \int {d^D l \over (2 \pi)^D} \,
{n \cdot (p-l) \over [n \cdot l \, \bar n \cdot (l+k) + l_{\perp}^2 + i 0]
[ n \cdot (p-l) \, \bar n \cdot (p-l-k) + l_{\perp}^2  + i 0] [l^2 + i 0] } \, \nonumber \\
&=& {\alpha_s \, C_F \over 2 \, \pi} \, {\bar n \cdot p - \omega \over \omega} \,
\left  [ {1 \over \epsilon} + \ln {\mu^2 \over n \cdot p \, (\omega- \bar n \cdot p)}
+ {1 \over 2} \, \left ( { \bar n \cdot p - \omega \over \bar n \cdot p} \right )  + 1 \right ] \nonumber \\
&&   \times \, \ln \left ( { \bar n \cdot p - \omega \over \bar n \cdot p} \right )  \,
 F_{\|, \, {\rm LO}}^{A}.
 \label{FL-B-a: result}
\end{eqnarray}
We can proceed to write down the QCD amplitude for the diagram (b) in figure \ref{fig: correlator for F-L-B}
\begin{eqnarray}
F_{\|, \rm {LO}}^{B, (b)} &=&  g_s^2 \, C_F \, \int {d^D l  \over (2 \pi)^D} \, \int {d^D L  \over (2 \pi)^D}
\frac{n \cdot L \,\, n \cdot (p+L) \,\, n \cdot (p+L+l)}
{[L^2 + i 0] [(p+L)^2 + i 0] [(p+L+l)^2 + i 0] [l^2+ i 0]}  \nonumber \\
&&  \bar q_s(k) \, \bigg \{ (2 \, \pi)^4 \, \delta^4(l+L+k)  \,
\left [\bar n^{\alpha} + \gamma_{\perp}^{\alpha} \, {\not \! L_{\perp} \over n \cdot L}
+ {n^{\alpha}  \over n \cdot l} \, {L^2 \over n \cdot L}  \right ]  \, {\not \! n \over 2}  \nonumber \\
&& + \, k_{\perp \, \beta} \, \partial_{\perp}^{\beta} \,
\left [  (2 \, \pi)^4 \, \delta^4(l+L+k) \right ] \,
\left [ \gamma_{\perp}^{\alpha} - { \not \! L_{\perp} \over n \cdot L} \, n^{\alpha} \right ] \,  \bigg \}  \nonumber \\
&& \left [ \bar n_{\alpha} +  {\gamma_{\alpha \, \perp} \, (\not \! L_{\perp} + \not l_{\perp}) \over n \cdot (p+L+l)}
+ {\not \! L_{\perp}  \,  \gamma_{\alpha \perp} \over n \cdot (p+L)}
- {L_{\perp}^2 +  \not \! L_{\perp}  \, \not  l_{\perp}  \over n \cdot (p+L) \, n \cdot (p+L+l)} \, n_{\alpha}  \right ]
\, {\not \!  \bar n \over 2} \, \gamma_5 \, h_v \,, \hspace{0.5 cm}
\end{eqnarray}
which can be further evaluated with two distinct approaches by identifying the transverse derivative
$\partial_{\perp}^{\beta}$ acting on the Dirac $\delta$-function as $\partial / \partial k_{\perp \beta}$ and
$\partial / \partial L_{\perp \beta}$, respectively.
We have verified explicitly that both the two calculational methods lead to the identical results
\begin{eqnarray}
F_{\|, \rm {LO}}^{B, (b)} &=&  {\alpha_s \, C_F \over 4 \, \pi} \,
\bigg \{ - {2 \over \epsilon^2 } + {1 \over \epsilon} \,
\left [ - 2 \, \ln \left ( {\mu^2 \over n \cdot p \, (\omega - \bar n \cdot p)} \right )
+ \, 2 \,\, {\ln (1 + \eta) \over \eta} - 3 \right ] \nonumber \\
&& - \ln^2 \left ( {\mu^2 \over n \cdot p \, (\omega - \bar n \cdot p)} \right )
+ \ln \left ( {\mu^2 \over n \cdot p \, (\omega - \bar n \cdot p)} \right )  \,
\left [  2 \,\, {\ln (1 + \eta) \over \eta} - 3 \right ] \nonumber \\
&&  +  \, {1 \over \eta} \, \ln^2(1+\eta)  + \left ({4 \over \eta} + 1 \right ) \, \ln (1+\eta)
+ {\pi^2 \over 6} - 8 \bigg \} \,\, F_{\|, \, {\rm LO}}^{A}  \,,
\label{result of F-L-B}
\end{eqnarray}
where we have defined $\eta= - \omega / \bar n \cdot p$.
Applying the matching condition for the matrix element $F_{\|}^{B}$
presented in (\ref{matching relation for F-L-B}) we can readily derive the hard-collinear
function $J_{\|, \, \pm}^{B}$ at tree level
\begin{eqnarray}
J_{\|, \, -}^{B} &=&  {\alpha_s \, C_F \over 4 \, \pi}  \, J_{\|, \, -}^{A, \, (0)} \,
\bigg \{   - {2 \over \epsilon^2}   + {1 \over \epsilon} \,
\left [ - 2 \,  \ln \left ( {\mu^2 \over n \cdot p \, (\omega- \bar n \cdot p) } \right )  - 2 \, \ln (1+ \eta) - 3  \right ]
\nonumber \\
&& - \,  \ln^2 \left ( {\mu^2 \over n \cdot p \, (\omega- \bar n \cdot p) } \right )
+  \ln \left ( {\mu^2 \over n \cdot p \, (\omega- \bar n \cdot p) } \right ) \,
\left [ - 2 \, \ln (1+ \eta) - 3 \right ] - \ln^2 (1+\eta) \nonumber \\
&& + \,  \left ({2 \over \eta} -1 \right ) \, \ln (1+ \eta)
+ {\pi^2 \over 6} - 8  \bigg \} \,, \nonumber \\
 J_{\|, \, +}^{B} &=& 0 \,,
\label{jet functions: L-B}
\end{eqnarray}
which are again in complete agreement with the previous calculations displayed in \cite{DeFazio:2007hw}.

\subsubsection{SCET factorization for $\Pi_{\nu, \|}^{C}(p, q)$}

\begin{figure}
\begin{center}
\includegraphics[width=0.4 \columnwidth]{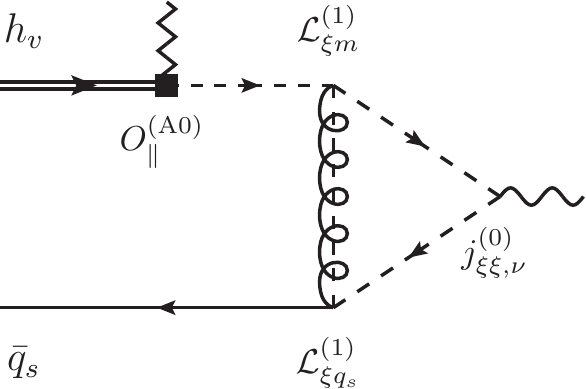}
\vspace*{0.1cm}
\caption{Diagrammatical representation of the vacuum-to-$B$-meson correlation function
$\Pi_{\nu, \|}^{B}(p, q)$  defined with  the ${\rm A0}$-type SCET operator
$O_{\|}^{\rm {(A0)}}= \left (\bar \xi \, W_c \right) \, \gamma_5 \, h_v$,
the leading power interpolating current $j_{\xi \xi, \nu}^{(0)}$
and the subleading power SCET Lagrangians ${\cal L}_{\xi q_s}^{(1)}$
and ${\cal L}_{\xi m}^{(1)}$.}
\label{fig: correlator for F-L-C}
\end{center}
\end{figure}

We are now in a position to compute the jet function $J_{\|, \, \pm}^{C}$
entering the SCET factorization formula (\ref{SCET factorization formula L})
by inspecting the QCD matrix element
\begin{eqnarray}
F_{\|}^{C} &=& \int d^4 x \, e^{i p \cdot x} \, \int d^4 y \, \int d^4 z \,  \nonumber  \\
&&  \left  \langle 0 \left | {\rm T}  \left \{\bar \xi(x) \, {\not \! n \over 2} \, \xi(x),
\,\, i \, {\cal L}_{\xi q_s}^{(1)}(y),  \,\,
i \, {\cal L}_{\xi m}^{(1)}(z),  \,\,
\left (\bar \xi \, W_c \right)(0) \, \gamma_5 \, h_v(0) \,    \right \}
\right | \bar q_s(k) \, h_v \right \rangle  \,,
\end{eqnarray}
which can be further matched onto the light-ray HQET operators defining the $B$-meson LCDA
\begin{eqnarray}
F_{\|}^{C}  = (- i) \, \sum_{m=\pm} \, J_{\|, \, m}^{C}
\left ({\mu^2 \over n \cdot p \, \omega^{\prime}}, {\omega^{\prime} \over \bar n \cdot p} \right )  \,
\ast \langle O_{\|, \,m} \, (\omega, \omega^{\prime})  \rangle  \,.
\label{matching relation for F-L-C}
\end{eqnarray}
The LO amplitude of $F_{\|}^{C}$ can be obtained by computing the diagram in figure \ref{fig: correlator for F-L-C}
\begin{eqnarray}
F_{\|, \rm{LO}}^{C} &=& - {m \over n \cdot p}  \, {g_s^2 \, C_F \over \bar n \cdot p - \omega} \,\,
\bar q_s(k) \, {\not \! \bar n \over 2} \, \gamma_5 \,  h_v \,  \nonumber \\
&& \times \,  \int {d^D l \over (2 \pi)^D} \,
{(D-2) \,  (n \cdot l)^2 \over [n \cdot l \, \bar n \cdot (l+k) + l_{\perp}^2 + i 0]
[ n \cdot (p-l) \, \bar n \cdot (p-l-k) + l_{\perp}^2  + i 0] [l^2 + i 0] } \, \nonumber \\
&=&  {m \over \omega}  \, {\alpha_s \, C_F \over 4 \, \pi} \, {i \over \bar n \cdot p - \omega + i 0} \,
\ln \left ( {\bar n \cdot p - \omega \over \bar n \cdot p} \right ) \,
\bar q_s(k) \, {\not \! \bar n \over 2} \, \gamma_5 \,  h_v,
\label{result of F-L-C}
\end{eqnarray}
where $m$ indicates the mass of the collinear quark produced from the weak decay of the heavy quark.
It is evident that the light-quark mass effect defined by $F_{\|}^{C}$ is not suppressed by any powers of $\Lambda/m_b$
in the heavy quark expansion compared with the SCET matrix elements $F_{\|}^{A}$  and $F_{\|}^{B}$ .
The resulting jet functions at tree level $J_{\|, \, \pm}^{C}$ are given by
\begin{eqnarray}
J_{\|, \, +}^{C} &=&  -  {m \over \omega}  \,
{1 \over \bar n \cdot p - \omega + i0} \,  {\alpha_s \, C_F \over 4 \, \pi}  \,
\ln \left ( {\bar n \cdot p - \omega \over \bar n \cdot p} \right ) \,, \nonumber \\
J_{\|, \, -}^{C} &=& 0 \,,
 \label{jet functions: L-C}
\end{eqnarray}
which are consistent with the results derived from the diagrammatic factorization approach
for the corresponding vacuum-to-$B$-meson correlation function in QCD \cite{Lu:2018cfc}.

Plugging the obtained jet functions (\ref{jet functions: L-A}),
(\ref{jet functions: L-B}) and (\ref{jet functions: L-C}) into the factorization formula
(\ref{SCET factorization formula L}) and employing the decomposition of  $\Pi_{\nu, \|}$
defined in (\ref{def£ºxi-L-A-B-C}) yields
\begin{eqnarray}
\Pi_{\nu, \|}(p, q) &=& {\tilde{f}_B(\mu) \, m_B \over 2}\,
\, \int_0^{+\infty} \, {d \omega \over \bar n \cdot p - \omega + i0} \,
\bigg \{  \left [ 1 + {\alpha_s \, C_F \over 4 \, \pi}  \, \hat{J}_{\|,-}^{(\rm A0)}
\left ({\mu^2 \over n \cdot p \, \omega}, {\omega \over \bar n \cdot p} \right ) \right ] \,
\phi_B^{-}(\omega, \mu) \,\,  \nonumber \\
&&  + \left [ \, {\alpha_s \, C_F \over 4 \, \pi}  \, \hat{J}_{\|,+}^{(m)}
\left ({\mu^2 \over n \cdot p \, \omega}, {\omega \over \bar n \cdot p} \right ) \right ] \,
\phi_B^{+}(\omega, \mu)  \bigg \} \,\,  \bar n_{\nu} \,,
\label{Final SCET factorization formula for Pi-L}
\end{eqnarray}
where the normalized one-loop jet functions $\hat{J}_{\|,-}^{(\rm A0)}$ and $\hat{J}_{\|,+}^{(m)}$ read
\begin{eqnarray}
\hat{J}_{\|,-}^{(\rm A0)} &=&  \ln^2 \left ( {\mu^2 \over n \cdot p \, (\omega- \bar n \cdot p) } \right )
- 2 \, \ln \left ( {\mu^2 \over n \cdot p \, (\omega- \bar n \cdot p) } \right ) \, \ln(1+\eta)
-\ln^2 (1+\eta) \nonumber \\
&& + \left ({2 \over \eta} -1 \right ) \,\ln(1+\eta)  - {\pi^2 \over 6}  - 1 \,, \nonumber \\
\hat{J}_{\|,+}^{(m)} &=&  - {m \over \omega}  \, \ln \left ( {\bar n \cdot p - \omega \over \bar n \cdot p} \right )   \,.
\end{eqnarray}
To facilitate the construction of the SCET sum rules for the effective form factor $\xi_{\|}(n \cdot p)$,
we need to work out the  dispersion representation of the factorization formula (\ref{Final SCET factorization formula for Pi-L})
by computing the spectral functions of the various convolution  integrations over the variable $\omega$
\begin{eqnarray}
\Pi_{\nu, \|}(p, q) &=& - {\tilde{f}_B(\mu) \, m_B \over 2}\,
\, \int_0^{+\infty} \, {d \omega^{\prime} \over \omega^{\prime}  - \bar n \cdot p - i0} \,
\left [ \phi_{B, \rm{eff}}^{-}(\omega^{\prime}, \mu)
+ \phi_{B, m}^{+}(\omega^{\prime}, \mu) \right ]  \,  \bar n_{\nu}  \,,
\label{dispersipon form for Pi-L}
\end{eqnarray}
where the effective $B$-meson ``distribution amplitudes" are introduced to describe both the
hard-collinear and soft fluctuations  \cite{Wang:2015vgv,Lu:2018cfc}
\begin{eqnarray}
\phi_{B, \rm{eff}}^{-}(\omega^{\prime}, \mu) &=& \phi_{B}^{-}(\omega^{\prime}, \mu)
+ {\alpha_s \, C_F \over 4 \, \pi} \,
\bigg \{ \int_0^{\omega^{\prime}}  d \omega \,
\left [ {2 \over \omega-\omega^{\prime}} \,
\left ( \ln {\mu^2 \over n \cdot p \, \omega^{\prime}}
- 2 \, \ln {\omega^{\prime} - \omega \over \omega^{\prime}} \right )  \right ]_{\oplus} \,
\phi_{B}^{-}(\omega, \mu) \nonumber \\
&& - \int_{\omega^{\prime}}^{\infty} \, d \omega \,
\left [  \ln^2 {\mu^2 \over n \cdot p \, \omega^{\prime}}
- \left (2 \, \ln {\mu^2 \over n \cdot p \, \omega^{\prime}}  + 3 \right ) \,
\ln {\omega-\omega^{\prime} \over \omega^{\prime}}
+ 2 \, \ln {\omega \over \omega^{\prime}}  + {\pi^2 \over 6} - 1 \right ] \, \nonumber  \\
&& \times \,  {d \phi_{B}^{-}(\omega, \mu) \over d\omega} \bigg \}   \,, \nonumber \\
\phi_{B, m}^{+}(\omega^{\prime}, \mu) &=& {\alpha_s \, C_F \over 4 \, \pi} \,
m \, \int_{\omega^{\prime}}^{\infty}  d \omega  \, \ln{\omega - \omega^{\prime} \over \omega^{\prime}} \,\,
 {d \over d\omega} \, {\phi_{B}^{+}(\omega, \mu)  \over \omega} \,.
 \label{spectral function for F-L-A0}
\end{eqnarray}
The plus function appearing in (\ref{spectral function for F-L-A0}) is defined in the standard way \cite{Lange:2003pk}
\begin{eqnarray}
\int_0^{\omega^{\prime}} \,\,  d \omega \, [f(\omega,\omega^{\prime} )]_{\oplus} \,\, g( \omega)
= \int_0^{\omega^{\prime}} \,\,  d \omega \, f(\omega,\omega^{\prime} ) \,\,
[ g( \omega) - g(\omega^{\prime})] \,.
\end{eqnarray}

Matching the spectral representation of the factorization formula (\ref{dispersipon form for Pi-L})
for the vacuum-to-$B$-meson correlation function $\Pi_{\nu, \|}$
with the corresponding hadronic dispersion relation
\begin{eqnarray}
&& \Pi_{\nu, \|}(p, q) \nonumber \\
&& =  \left [- {f_{V, \|}  \, m_V \over m_V^2/n \cdot p -\bar n \cdot p - i 0}  \,
\left ( {n \cdot p \over 2 \, m_V} \right )^2 \, \xi_{\|}(n \cdot p)
+ \int_{\omega_s}^{+\infty} \, {d \omega^{\prime} \over \omega^{\prime}  - \bar n \cdot p - i0} \,\,
\rho^{h}_{\|}(\omega^{\prime}, n \cdot p) \right ] \, \bar n_{\nu} \,, \hspace{0.8 cm}
\end{eqnarray}
we can readily derive the NLO sum rules for the SCET form factor  $\xi_{\|}(n \cdot p)$
\begin{eqnarray}
&& \xi_{\|, \, \rm {NLO}}(n \cdot p) \nonumber  \\
&& = 2 \, {\tilde{f}_B(\mu)  \over f_{V, \|} } \,
{ m_B \, m_V \over (n \cdot p)^2} \, \int_{0}^{\omega_s} \, d \omega^{\prime} \,
{\rm exp} \left [ - {n \cdot p \, \omega^{\prime} - m_V^2 \over n \cdot p \, \omega_M} \right ] ¡¢£¬
\left [ \phi_{B, \rm{eff}}^{-}(\omega^{\prime}, \mu) +  \phi_{B, m}^{+}(\omega^{\prime}, \mu)\right ] \,.
\hspace{0.5 cm}
\end{eqnarray}
The scale-independent longitudinal decay constant of the vector meson is defined as follows
\begin{eqnarray}
c_V \, \langle V(p, \epsilon^{\ast}) | j_{\nu} | 0 \rangle = - i \, f_{V, \| } \,\,  m_V \,\,  \epsilon^{\ast}_{\nu}(p) \,.
\end{eqnarray}
The HQET $B$-meson decay constant $\tilde{f}_B(\mu)$ will be expressed in terms of  the QCD decay constant $f_B$
at one loop \cite{Beneke:2011nf}
\begin{eqnarray}
\tilde{f}_B(\mu) =  \left [ 1- {\alpha_s \, C_F  \over 4 \, \pi} \,
\left (3 \ln {\mu \over m_b} + 2 \right ) \right ]^{-1} \, f_B =  K^{-1}(m_b, \mu) \, f_B \,.
\label{HQET relation for fB}
\end{eqnarray}
Taking the factorization scale $\mu$ as a hard-collinear scale $\mu_{hc} \sim \sqrt{\Lambda \, m_b}$
will introduce the enhanced logarithms of $m_b/\mu$ in the perturbative matching coefficient $K(m_b, \mu)$,
whose resummation  at the NLL accuracy  can be achieved with the standard RG approach.
Solving the two-loop RG evolution equation for the HQET decay constant $\tilde{f}_B(\mu)$
\cite{Ji:1991pr,Broadhurst:1991fz} leads to
\begin{eqnarray}
\tilde{f}_B(\mu) = U_2(\mu_{h2}, \mu) \,  \tilde{f}_B(\mu_{h2})   \,,
\end{eqnarray}
where the explicit expression of the evolution function $U_2(\mu_{h2}, \mu)$ has been
derived in \cite{Beneke:2011nf,Wang:2016qii}.
Since the soft scale $\mu_0$ entering the initial condition of the $B$-meson distribution amplitudes
$\phi_B^{\pm}(\omega, \mu_0)$ is numerically comparable to the hard-collinear scale
$\mu_{hc} \simeq 1.5 \, {\rm GeV}$, we not perform the NLL resummation of
the parametrically large logarithms of $\mu / \mu_0$ by applying the Lange-Neubert evolution equation
at two loops \cite{Braun:2019wyx}.
It is then straightforward to write down the resummation improved SCET sum rules for $\xi_{\|}(n \cdot p)$
\begin{eqnarray}
\xi_{\|, \, \rm {NLL}}(n \cdot p) &=& 2 \, {U_2(\mu_{h2}, \mu) \, \tilde{f}_B(\mu_{h2})  \over f_{V, \|} } \,
{ m_B \, m_V \over (n \cdot p)^2} \, \nonumber \\
&&  \times \, \int_{0}^{\omega_s} \, d \omega^{\prime} \,
{\rm exp} \left [ - {n \cdot p \, \omega^{\prime} - m_V^2 \over n \cdot p \, \omega_M} \right ] \,
\left [ \phi_{B, \rm{eff}}^{-}(\omega^{\prime}, \mu) +  \phi_{B, m}^{+}(\omega^{\prime}, \mu)\right ] \,.
\label{SCET sum rules for xi-L}
\end{eqnarray}

\subsection{The $B$-meson  LCSR for $\Xi_{\parallel}(\tau, n \cdot p)$}
\label{section: SCET factorization for Xi-L}

\begin{figure}
\begin{center}
\includegraphics[width=0.4 \columnwidth]{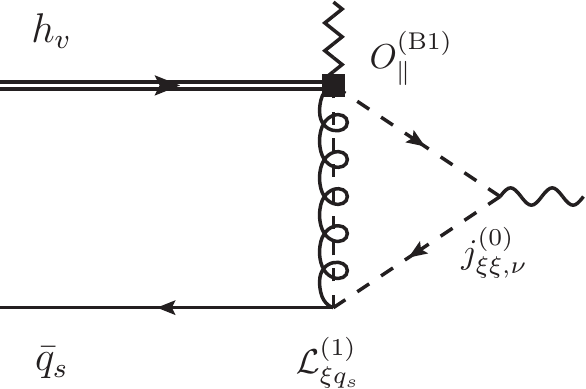}
\vspace*{0.1cm}
\caption{Diagrammatical representation of the vacuum-to-$B$-meson correlation function
$\widetilde{\Pi}_{\nu, \|}(p, q, \tau)$  defined with  the ${\rm B1}$-type SCET operator
$O_{\|}^{\rm{(B1)}} = (\bar \xi \, W_c)(0) \,
\gamma_5 \,\, (W_c^{\dagger} \,\, i  {\not  \! \!  D_{c, \,\perp}} \,\, W_c)(r \, n) \,\, h_v(0)$,
the leading power interpolating current $j_{\xi \xi, \nu}^{(0)}$ for the longitudinal polarized vector meson
and the subleading power SCET Lagrangian ${\cal L}_{\xi q_s}^{(1)}$.}
\label{fig: correlator for F-tilde-L}
\end{center}
\end{figure}

We aim at constructing the SCET  sum rules for the $\rm {B1}$-type  non-local form factor
$\Xi_{\parallel}(\tau, n \cdot p)$ from the following correlation function
\begin{eqnarray}
\widetilde{\Pi}_{\nu, \|}(p, q, \tau) &=& { n \cdot p \over 2\, \pi} \, \int d^4 x \, e^{i p \cdot x} \,
 \int d r \, e^{-i n \cdot p \, \tau \, r}  \, \nonumber \\
&&  \langle 0 | {\rm T}  \left \{j_{\nu}(x),  \,\, \left (\bar \xi \, W_c \right)(0) \,
\gamma_5 \,\, (W_c^{\dagger} \,\, i  \not \! \!  D_{c, \,\perp} \, W_c)(r \, n) \,\, h_v(0) \,    \right \}   | \bar B_v \rangle \,,
\label{correlation function for Xi-L}
\end{eqnarray}
the leading power contribution of which can be readily identified as
\begin{eqnarray}
\widetilde{\Pi}_{\nu, \|}(p, q, \tau)
&=& { n \cdot p \over 2\, \pi} \,  \int d^4 x  \, e^{i p \cdot x} \, \int d^4 y \,
\int d r \, e^{-i n \cdot p \, \tau \, r}  \, \nonumber \\
&& \langle 0 | {\rm T}  \left \{j_{\xi \xi, \nu}^{(0)}(x), \,\, i \, {\cal L}_{\xi q_s}^{(1)}(y), \,\, \left (\bar \xi \, W_c \right)(0) \,
\gamma_5 \,\, (W_c^{\dagger} \,\, i  \not \!  D_{c, \,\perp} \, W_c)(r \, n) \,\, h_v(0) \,    \right \}   | \bar B_v \rangle \,.
\hspace{0.5 cm}
\end{eqnarray}
The soft-collinear factorization formula for the vacuum-to-$B$-meson correlation function $\widetilde{\Pi}_{\nu, \|}$
can be obtained by integrating out the hard-collinear dynamics
\begin{eqnarray}
\widetilde{\Pi}_{\nu, \|}(p, q, \tau) = {\tilde{f}_B(\mu) \, m_B \over 2}\,
\sum_{m = \pm } \, \int_0^{+\infty} \, d \omega \, \tilde{J}_{\|, m}
\left ({\mu^2 \over n \cdot p \, \omega}, {\omega \over \bar n \cdot p}, \tau \right )  \,
\phi_B^{m}(\omega, \mu) \,\, \bar n_{\nu}.  \,\, \,
\label{factorization formula for Pi-tilde-L}
\end{eqnarray}
The short-distance matching coefficient functions $\tilde{J}_{\|, m}$ can be determined by
investigating the SCET matrix element
\begin{eqnarray}
\tilde{F}_{\|}(p, q, \tau)
&=& { n \cdot p \over 2\, \pi} \,  \int d^4 x  \, e^{i p \cdot x} \, \int d^4 y \,
\int d r \, e^{-i n \cdot p \, \tau \, r}  \, \bigg \langle 0 \bigg | {\rm T}  \bigg \{\bar \xi(x) \, {\not \! n \over 2} \, \xi(x), \,\,
i \, {\cal L}_{\xi q_s}^{(1)}(y), \,\, \nonumber \\
&&  \left (\bar \xi \, W_c \right)(0) \,
\gamma_5 \,\, (W_c^{\dagger} \,\, i  \not \! \! D_{c \,\perp} \, W_c)(r \, n) \,\, h_v(0) \,    \bigg \}
\bigg | \bar q_s(k) \,  h_v \bigg \rangle \,.
\end{eqnarray}
Evaluating the tree-level diagram displayed in figure \ref{fig: correlator for F-tilde-L}
with the SCET Feynman rules yields
\begin{eqnarray}
\tilde{F}_{\|}(p, q, \tau) &=& g_s^2 \, C_F \, \, \bar q_s(k) \, {\not \! \bar n \over 2} \, \gamma_5 \,  h_v \nonumber \\
&&  \hspace{-0.8 cm} \int{d^D l \over (2 \pi)^D } \,
{(D-2) \,  n \cdot l \, n \cdot (p-l) \, \delta(\tau - n \cdot l / n \cdot p)
\over [n \cdot l \, \bar n \cdot (l+k) + l_{\perp}^2 + i 0]
[ n \cdot (p-l) \, \bar n \cdot (p-l-k) + l_{\perp}^2  + i 0] [l^2 + i 0] } \,,
\hspace{0.8 cm}
\end{eqnarray}
which can be further computed with the contour integration method
\begin{eqnarray}
\tilde{F}_{\|}(p, q, \tau) &=&  \left ( - i \right ) \, {\alpha_s \, C_F \over 2 \, \pi} \, {n \cdot p \over \omega} \,
\ln (1+\eta) \, \left [ (1-\tau) \, \theta(\tau) \, \theta(1-\tau) \right ] \,\,
 \, \bar q_s(k) \, {\not \! \bar n \over 2} \, \gamma_5 \,  h_v \,.
\end{eqnarray}
Applying the matching relation for the SCET matrix element $\tilde{F}_{\|}(p, q, \tau)$
\begin{eqnarray}
\tilde{F}_{\|}(p, q, \tau) = (- i) \, \sum_{m=\pm} \, \tilde{J}_{\|, \, m}
\left ({\mu^2 \over n \cdot p \, \omega^{\prime}}, {\omega^{\prime} \over \bar n \cdot p}, \tau \right )  \,
\ast \langle O_{\|, \,m} \, (\omega, \omega^{\prime})  \rangle \,,
\end{eqnarray}
we obtain the jet functions  entering the factorization formula (\ref{factorization formula for Pi-tilde-L}) at tree level
\begin{eqnarray}
\tilde{J}_{\|, \, +} = {\alpha_s \, C_F \over 2\, \pi} \, {n \cdot p \over \omega} \,
\ln (1+\eta) \, \left [ (1-\tau) \, \theta(\tau) \, \theta(1-\tau) \right ] \,,
\qquad \tilde{J}_{\|, \, -}=0.
\end{eqnarray}

Matching the spectral representation of the SCET factorization formula
(\ref{factorization formula for Pi-tilde-L}) for the correlation function
$\widetilde{\Pi}_{\nu, \|}$
\begin{eqnarray}
\widetilde{\Pi}_{\nu, \|}(p, q, \tau) &=& {\alpha_s \, C_F \over 4 \, \pi} \,
\tilde{f}_B(\mu) \, m_B \, \left [ (1-\tau) \, \theta(\tau) \, \theta(1-\tau) \right ] \, \nonumber \\
&& \times \, \int_{0}^{\infty} {d \omega^{\prime} \over \omega^{\prime}  - \bar n \cdot p - i 0} \,
\left [ \int_{\omega^{\prime}}^{\infty}  d \omega \, {n \cdot p \over \omega} \, \phi_B^{+}(\omega, \mu) \right ] \,
\bar n_{\nu} \,
\end{eqnarray}
with the corresponding hadronic representation
\begin{eqnarray}
\widetilde{\Pi}_{\nu, \|}(p, q, \tau) &=& \bigg [- {f_{V, \|} \,\, m_V \over m_V^2/n \cdot p - \bar n \cdot p - i 0} \,
\left ({n \cdot p \over 2 \, m_V} \right )^2 \, m_b \,\, \Xi_{\|}(\tau, n \cdot p) \nonumber \\
&& + \int_{\omega_s}^{+\infty} \, {d \omega^{\prime} \over \omega^{\prime}  - \bar n \cdot p - i0} \,\,
\tilde{\rho}^{h}_{\|}(\omega^{\prime}, n \cdot p, \tau) \bigg ] \, \bar n_{\nu} \,,
\end{eqnarray}
and employing the parton-hadron duality approximation with the aid of the Borel transformation,
we can derive the  SCET  sum rules for the non-local form factor $\Xi_{\|}(\tau, n \cdot p)$
\begin{eqnarray}
\Xi_{\|}(\tau, n \cdot p) &=& - {\alpha_s \, C_F \over \pi} \,
{U_2(\mu_{h2}, \mu) \, \, \tilde{f}_B(\mu_{h2})  \over f_{V,\|}} \,
{ m_B \, m_V  \over \, n \cdot p \, m_b} \,
\, \left [ (1-\tau) \, \theta(\tau) \, \theta(1-\tau) \right ] \,
\nonumber \\
&& \times \, \int_{0}^{\omega_s} \, d \omega^{\prime} \,
{\rm exp} \left [ - {n \cdot p \, \omega^{\prime} - m_V^2 \over n \cdot p \, \omega_M} \right ]
\int_{\omega^{\prime}}^{\infty}  \, d \omega \, { \phi_B^{+}(\omega, \mu) \over \omega} \,
+ {\cal O}(\alpha_s^2) \,,
\label{SCET sum rules for Xi-L}
\end{eqnarray}
where the summation of the enhanced logarithms of $m_b/\mu$ entering the hard matching coefficient
$K^{-1}(m_b, \mu)$ has been included  at the LL accuracy.

It will be interesting to compare the obtained sum rules for $\Xi_{\|}(\tau, n \cdot p)$
presented in (\ref{SCET sum rules for Xi-L}) with
the direct QCD calculation in the SCET framework.
Integrating the hard-collinear fluctuations,
the ${\rm B}$-type ${\rm SCET}_{\rm I}$ form factor $\Xi_{\|}(\tau, n \cdot p)$ can be further factorized as the
convolution of the short-distance coefficient function and distributions amplitudes of the $B$-meson
and the vector meson in ${\rm SCET}_{\rm II}$
\begin{eqnarray}
\Xi_{\|}^{\rm {SCET}}(\tau, n \cdot p) &=& {2 \, m_V \over n \cdot p} \, {m_B \over 4 \, m_b} \,
\left [ U_2(\mu_{h2}, \mu) \, \, \tilde{f}_B(\mu_{h2}) \right ]  \, f_{V,\|} \,  \nonumber \\
&& \times \, \int_0^{\infty} \, d \omega \, \int_0^1 d v \,
\phi_B^{+}(\omega, \mu) \, \phi_{V, \|}(v, \mu) \,
J_{V, \|}^{\rm {SCET}}(\tau, v, \omega) \,,
\label{SCET factorization of Xi-L}
\end{eqnarray}
where the jet function $J_{V, \|}^{\rm {SCET}}$ has been determined at NLO in $\alpha_s$  \cite{Beneke:2005gs}
by implementing the ultraviolet renormalization and the infrared subtraction with the
dimensional regularization scheme and the evanescent operator approach.
Substituting the tree-level expression of $J_{V, \|}^{\rm {SCET}}$ into the SCET factorization
formula (\ref{SCET factorization of Xi-L}) and employing the asymptotic form of the vector-meson
distribution amplitude $\phi_{V, \|}(v, \mu) = 6 \, v \, (1-v)$ leads to
\begin{eqnarray}
\Xi_{\|, \, \rm {LO}}^{\rm {SCET}}(\tau, n \cdot p) &=& - {3 \, g_s^2 \, C_F \over N_c} \,
{m_B \, m_V \over (n \cdot p)^2 \, m_b} \, \left [ U_2(\mu_{h2}, \mu) \, \, \tilde{f}_B(\mu_{h2}) \right ]
\, f_{V,\|} \,  \, \left [ (1-\tau) \, \theta(\tau) \, \theta(1-\tau) \right ] \, \nonumber \\
&&  \times \, \int_{0}^{\infty}  \, {d \omega \over \omega} \, \phi_B^{+}(\omega, \mu)  \,.
\label{SCET factorization of Xi-L at LL}
\end{eqnarray}
Based upon the power counting scheme for the sum rule parameters
\begin{eqnarray}
\omega_s \sim \omega_M \sim {\cal O} \left ( {\Lambda^2 / m_b} \right )  \,,
\end{eqnarray}
the resummation improved SCET sum rules (\ref{SCET sum rules for Xi-L}) for $\Xi_{\|}(\tau, n \cdot p)$
can be further reduced to
\begin{eqnarray}
\Xi_{\|}(\tau, n \cdot p) &=& - {\alpha_s \, C_F \over \pi} \,
{U_2(\mu_{h2}, \mu) \, \, \tilde{f}_B(\mu_{h2})  \over f_{V,\|}} \,
{ m_B \, m_V  \over \, n \cdot p \, m_b} \,
\, \left [ (1-\tau) \, \theta(\tau) \, \theta(1-\tau) \right ] \,
\nonumber \\
&& \times \,  \int_{0}^{\infty}  \, {d \omega \over \omega} \, \phi_B^{+}(\omega, \mu)  \,\,
\left \{  \omega_M \, \left ( 1 - e^{-\omega_s/\omega_M}  \right )  \,
{\rm exp} \left [  { m_V^2 \over n \cdot p \, \omega_M} \right ]  \right \} \,,
\end{eqnarray}
which can be readily demonstrated to be identical to the tree-level SCET factorization formula
(\ref{SCET factorization of Xi-L at LL}) by employing the QCD sum rules for the vector-meson
decay constant at leading power approximation \cite{Reinders:1984sr}
\begin{eqnarray}
f_{V,\|}^2= {N_c \over 12 \, \pi^2} \, n \cdot p\, \omega_M \,
\left \{  \left ( 1 - e^{-\omega_s/\omega_M}  \right )  \,
{\rm exp} \left [  { m_V^2 \over n \cdot p \, \omega_M} \right ]  \right \}
+ {\cal O}(\alpha_s) \,.
\end{eqnarray}
However, we mention in passing that the advantage of the SCET factorization over the LCSR approach
presented here lies in the fact that it is free of the systematic uncertainty induced by the parton-hadron duality
approximation of the light-meson channel and the perturbative correction to
 the hard-collinear function $\Xi_{\|}(\tau, n \cdot p)$
at ${\cal O}(\alpha_s^{\ell})$ can be determined by computing the ${\rm SCET}_{\rm I}$
Feynman diagrams at $(\ell-1)$ loops instead of evaluating the effective diagrams
for the vacuum-to-$B$-meson correlation function $\widetilde{\Pi}_{\nu, \|}(p, q, \tau)$
at $\ell$ loops.

\subsection{The $B$-meson  LCSR for $\xi_{\perp}(n \cdot p)$}

We proceed to construct the SCET sum rules for the $\rm {A0}$-type form factor  $\xi_{\perp}(n \cdot p)$
by investigating the following vacuum-to-$B$-meson correlation function
\begin{eqnarray}
\Pi_{\mu \nu \rho, \perp}(p, q) = \int d^4 x \, e^{i p \cdot x} \, \langle 0 |
{\rm T}  \left \{j_{\nu \rho}(x),  \,\, \left (\bar \xi \, W_c \right)(0) \,
\gamma_5 \, \gamma_{\mu\, \perp}\, h_v(0) \,    \right \}   | \bar B_v \rangle \,,
\label{correlation function for xi-T}
\end{eqnarray}
where the interpolating current for the transversely polarized  vector meson is given by
\begin{eqnarray}
j_{\nu \rho}(x) = \bar q^{\prime}(x) \, \gamma_{\nu} \, \gamma_{\rho\, \perp} \, q(x)\,.
\end{eqnarray}
Matching the QCD interpolating current $j_{\nu \rho}$ onto ${\rm SCET}_{\rm I}$ yields
\begin{eqnarray}
j_{\nu \rho}(x) = j_{\xi \xi, \, \nu \rho}^{(0)}(x) + j_{\xi \xi, \, \nu \rho}^{(1)}(x)
+ j_{\xi q_s, \, \nu \rho}^{(2)}(x) + ...\,,
\label{SCET representation of the transverse vector current}
\end{eqnarray}
where the resulting effective currents of our interest can be written as
\begin{eqnarray}
j_{\xi \xi, \, \nu \rho}^{(0)} = \bar \xi \,\, {\not \! n \over 2} \,
\gamma_{\rho\, \perp}  \, \xi \,\, \bar n_{\nu} \,,\qquad
j_{\xi q_s, \, \nu \rho}^{(2)} =  \left ( \bar \xi \,\, W_c \,\,  {\not \! n \over 2} \, \gamma_{\rho\, \perp} \, Y_s^{\dag} \,\, q_s
+ \bar q_s \,\, Y_s \,\, {\not \! n \over 2} \, \gamma_{\rho\, \perp} \,\, W_c^{\dag} \,\, \xi \right )  \,\, \bar n_{\nu}  \,.
\end{eqnarray}
The SCET representation of the correlation function (\ref{correlation function for xi-T})
at leading power in the heavy quark expansion can be readily derived as follows
\begin{eqnarray}
\Pi_{\mu \nu \rho, \perp}(p, q) &=& \int d^4 x \, e^{i p \cdot x} \, \left \langle 0 \left |
{\rm T}  \left \{j_{\xi q_s, \, \nu \rho}^{(2)}(x),  \,\, \left (\bar \xi \, W_c \right)(0) \,
\gamma_5 \, \gamma_{\mu\, \perp}\, h_v(0) \,    \right \}   \right | \bar B_v \right \rangle  \nonumber \\
&& + \, \int d^4 x \, e^{i p \cdot x} \, \int d^4 y \,\,  \left \langle 0 \left |
{\rm T}  \left \{j_{\xi \xi, \, \nu \rho}^{(0)}(x),  \,\,i \, {\cal L}_{\xi q_s}^{(2)}(y), \,\,
\left (\bar \xi \, W_c \right)(0) \, \gamma_5 \, \gamma_{\mu\, \perp}\, h_v(0) \,    \right \}  \right | \bar B_v \right \rangle  \,
 \nonumber \\
&& + \, \int d^4 x \, e^{i p \cdot x} \, \int d^4 y \,\, \int d^4 z  \nonumber \\
&& \,\, \hspace{0.4 cm} \left \langle 0 \left | {\rm T}  \left \{j_{\xi \xi, \, \nu \rho}^{(0)}(x),  \,\,i \, {\cal L}_{\xi q_s}^{(1)}(y), \,\,
i \, {\cal L}_{\xi m}^{(1)}(z), \,\,
\left (\bar \xi \, W_c \right)(0) \, \gamma_5 \, \gamma_{\mu\, \perp}\, h_v(0) \,    \right \}  \right | \bar B_v \right \rangle  \,
\nonumber \\
&& \equiv  \Pi_{\mu \nu \rho, \perp}^{A}(p, q) + \Pi_{\mu \nu \rho, \perp}^{B}(p, q)
+ \Pi_{\mu \nu \rho, \perp}^{C}(p, q)  \,.
\label{correlation function for xi-T in SCET}
\end{eqnarray}
Integrating out the hard-collinear dynamics of the vacuum-to-$B$-meson correlation functions
$\Pi_{\mu \nu \rho, \perp}^{i}$  defined in (\ref{correlation function for xi-T in SCET})
 gives rise to the SCET/HQET factorization formulae
\begin{eqnarray}
&& \Pi_{\mu \nu \rho, \perp}^{i}(p, q) \nonumber \\
&& = {\tilde{f}_B(\mu) \, m_B \over 2}\,
\sum_{m = \pm } \, \int_0^{+\infty} \, d \omega \, J_{\perp, m}^{i}
\left ({\mu^2 \over n \cdot p \, \omega}, {\omega \over \bar n \cdot p} \right )  \,
\phi_B^{m}(\omega, \mu) \,\, g_{\mu \rho \, \perp} \,\, \bar n_{\nu},  \,\, \, (i=A, B, C)
\label{SCET factorization formula T}
\end{eqnarray}
where the jet functions $J_{\perp, m}^{i}$ will be determined at ${\cal O}(\alpha_s)$
with the naive dimensional regularization (NDR) scheme for $\gamma_5$, which
anti-commutes with all of the $\gamma$-matrices.

\subsubsection{SCET factorization for $\Pi_{\mu \nu \rho, \perp}^{A}(p, q)$}

Following the strategy presented in section \ref{section: SCET factorization for Xi-L},
the short-distance function $J_{\perp, m}^{A}$ can be obtained by implementing
the perturbative matching $\rm {SCET_{I} \to HQET}$ for the matrix element
\begin{eqnarray}
&& F_{\mu \rho, \perp}^{A}(p, q) \nonumber \\
&& =
\int d^4 x \, e^{i p \cdot x} \, \left \langle 0 \left |
{\rm T}  \left \{\bar q_s(x) \, {\not \! n \over 2} \, \gamma_{\rho\, \perp}  \, \xi(x),
\,\, \left (\bar \xi \, W_c \right)(0) \,
\gamma_5 \, \gamma_{\mu\, \perp}\, h_v(0) \,    \right \}
\right | \bar q_s(k) \, h_v \right \rangle.
\end{eqnarray}
It is straightforward to write down the LO contribution to this  SCET matrix element
\begin{eqnarray}
F_{\mu \rho, \perp, \rm{LO}}^{A}(p, q) &=& - {i \over \bar n \cdot p -\omega + i 0} \,\,
\bar q_s(k) \, \gamma_{\rho \perp} \, \gamma_{\mu \perp} \, {\not \! n \over 2} \, \gamma_5 \, h_v \nonumber \\
&=& - {i \over \bar n \cdot p -\omega^{\prime} + i 0} \ast \langle O_{\mu \rho, \perp}^{(n)}(\omega, \omega^{\prime}) \rangle^{(0)} \,,
\end{eqnarray}
where the HQET matrix element $\langle O_{\mu \rho, \perp}^{(n)}(\omega, \omega^{\prime}) \rangle$
is defined in the standard way
\begin{eqnarray}
\langle O_{\mu \rho, \perp}^{(n)}(\omega, \omega^{\prime}) \rangle
= \langle 0 |  O_{\mu \rho, \perp}^{(n)}(\omega^{\prime}) |  \bar q_s(k) \, h_v \rangle
= \bar q_s(k) \, \gamma_{\rho \perp} \, \gamma_{\mu \perp} \, {\not \! n \over 2} \, \gamma_5 \, h_v \,\,
\delta(\omega - \omega^{\prime}) + {\cal O}(\alpha_s) \,,
\hspace{0.5 cm}
\end{eqnarray}
with the light-ray effective operator in the momentum space
\begin{eqnarray}
 O_{\mu \rho, \perp}^{(n)}(\omega^{\prime})  = {1 \over 2 \, \pi} \, \int d t \, e^{i \, t \, \omega^{\prime}} \,
\left ( \bar q_s Y_s \right )(t \, \bar n) \, \,  \gamma_{\rho \perp} \, \gamma_{\mu \perp} \,
 {\not \! n \over 2} \, \gamma_5  \,\, \left ( Y_s^{\dag} h_v \right ) (0)\,.
\end{eqnarray}
To facilitate the infrared subtraction for the renormalized  matrix element $F_{\mu \rho, \perp}^{A}(p, q)$
beyond the tree level, it will be convenient to introduce the  HQET operator basis
\begin{eqnarray}
 O_{\mu \rho, \perp}^{(n, 1)}(\omega^{\prime})  &=& {1 \over 2 \, \pi} \, \int d t \, e^{i \, t \, \omega^{\prime}} \,
\left ( \bar q_s Y_s \right )(t \, \bar n) \, \,  \left [  \, g_{\rho \mu  \perp} \,
 {\not \! n \over 2}  \,\, \gamma_5 \, \right ] \,\, \left ( Y_s^{\dag} h_v \right )(0) \,, \nonumber \\
 O_{\mu \rho, \perp}^{(n, 2)}(\omega^{\prime})  &=& {1 \over 2 \, \pi} \, \int d t \, e^{i \, t \, \omega^{\prime}} \,
\left ( \bar q_s Y_s \right )(t \, \bar n) \, \, \left [ \,  i \, \epsilon_{\rho \mu \perp} \,
 {\not \! n \over 2}  \, \right ] \,\,  \left ( Y_s^{\dag} h_v \right )(0) \,, \nonumber \\
 O_{\mu \rho, \perp}^{(n, 3)}(\omega^{\prime})  &=& {1 \over 2 \, \pi} \, \int d t \, e^{i \, t \, \omega^{\prime}} \,
\left ( \bar q_s Y_s \right )(t \, \bar n) \, \, \left [ {\not \! n \over 2}  \,\,
\left ( \, { [\gamma_{\rho \perp}, \gamma_{\mu \perp}] \over 2 } \, \gamma_ 5
-  i \, \epsilon_{\rho \mu \perp} \,  \right ) \, \right ] \,\, \left ( Y_s^{\dag} h_v \right )(0) \,,
\end{eqnarray}
where we have employed the following conventions for brevity
\begin{eqnarray}
\epsilon_{\rho \mu \perp}  = {1 \over 2} \, \epsilon_{\rho \mu \alpha \beta} \, n^{\alpha} \, \bar n^{\beta} \,.
\end{eqnarray}
It is apparent that the evanescent operator $O_{\mu \rho, \perp}^{(n, 3)}(\omega^{\prime})$ vanishes in the four dimensional
space-time, however, it may generate the nonvanishing contribution to the perturbative matching coefficient by
mixing into the physical operator under the QCD radiative correction.
Expressing the HQET operator $O_{\mu \rho, \perp}^{(n)}(\omega^{\prime})$ in the given basis
\begin{eqnarray}
O_{\mu \rho, \perp}^{(n)}(\omega^{\prime}) = O_{\mu \rho, \perp}^{(n, 1)}(\omega^{\prime})
+ O_{\mu \rho, \perp}^{(n, 2)}(\omega^{\prime})
+ O_{\mu \rho, \perp}^{(n, 3)}(\omega^{\prime})  \,,
\end{eqnarray}
and employing the matching relation for the SCET matrix element $F_{\mu \rho, \perp}^{A}$
\begin{eqnarray}
F_{\mu \rho, \perp}^{A}(p, q) = (-i) \, \sum_{k=1,2,3}  \, \sum_{m=n, \, \bar n} \,
{\cal J}_{\perp, m}^{A, k} \left ({\mu^2 \over n \cdot p \, \omega^{\prime}}, {\omega^{\prime} \over \bar n \cdot p} \right )  \ast
\langle  O_{\mu \rho, \perp}^{(m, k)}(\omega, \omega^{\prime}) \rangle \,,
\label{matching relation for J-T-A}
\end{eqnarray}
we can readily derive the tree-level jet functions
\begin{eqnarray}
{\cal J}_{\perp, n}^{A, 1, \, (0)} = {\cal J}_{\perp, n}^{A, 2, \, (0)}
={\cal J}_{\perp, n}^{A, 3, \, (0)} = {1 \over \bar n \cdot p - \omega^{\prime} + i 0} \,, \qquad
{\cal J}_{\perp, \bar n}^{A, 1, \, (0)} = {\cal J}_{\perp, \bar n}^{A, 2, \, (0)}
= {\cal J}_{\perp, \bar n}^{A, 3, \, (0)} = 0 \,. \hspace{0.5 cm}
\end{eqnarray}
Taking advantage of the $B$-meson distribution amplitudes defined in (\ref{Definition of B-meson LCDA}),
the factorization formula for the correlation function $\Pi_{\mu \nu \rho, \perp}^{A}(p, q)$ at LO in $\alpha_s$
can be written as
\begin{eqnarray}
\Pi_{\mu \nu \rho, \perp}^{A}(p, q) = {\tilde{f}_B(\mu) \, m_B \over 2}\,
\int_0^{+\infty} \, d \omega \, {\cal J}_{\perp, n}^{A, 1, \, (0)}
\left ({\mu^2 \over n \cdot p \, \omega}, {\omega \over \bar n \cdot p} \right )  \,
\phi_B^{-}(\omega, \mu) \,\, g_{\mu \rho \, \perp} \,\, \bar n_{\nu} +{\cal O}(\alpha_s),
\end{eqnarray}
implying that the jet function $J_{\perp,m}^{A}$ in the SCET factorization formula
(\ref{SCET factorization formula T}) can be constructed
\begin{eqnarray}
J_{\perp,-}^{A, (0)}= {\cal J}_{\perp, n}^{A, 1, (0)}\,,  \qquad
J_{\perp,+}^{A, (0)}= {\cal J}_{\perp, \bar n}^{A, 1, (0)} = 0 \,.
\end{eqnarray}

The NLO contribution to the SCET matrix element $F_{\mu \rho, \perp}^{A}(p, q)$
can be deduced by evaluating the four Feynman diagrams in analogy to those
presented in figure \ref{fig: correlator for F-L-A} with the proper replacement
of the Dirac structures for the effective weak current and the interpolating current
of the vector meson. It is evident from the Wilson-line Feynman rules that the resulting amplitudes of
the one-loop diagrams for $F_{\mu \rho, \perp}^{A}(p, q)$ and $F_{\|}^{A}(p, q)$ are insensitive to the Dirac structures of
the SCET operators defining these two correlation functions.
It is then straightforward to write down the SCET amplitude of $F_{\mu \rho, \perp}^{A}(p, q)$
from (\ref{jet functions: L-A}) at the one-loop accuracy
\begin{eqnarray}
F_{\mu \rho, \perp, \rm{NLO}}^{A}(p, q) &=&
{\alpha_s \, C_F \over 4 \, \pi}  \,
\bigg  [ {4 \over \epsilon^2}   + {1 \over \epsilon} \,
\left ( 4 \,  \ln {\mu^2 \over n \cdot p \, (\omega- \bar n \cdot p) } + 3  \right )
+ 2 \, \ln^2 {\mu^2 \over n \cdot p \, (\omega- \bar n \cdot p)} \nonumber \\
&& + \, 3 \, \ln {\mu^2 \over n \cdot p \, (\omega- \bar n \cdot p)}
- {\pi^2 \over 3} + 7  \bigg ]   \,\, F_{\mu \rho, \perp, \rm{LO}}^{A}(p, q) \, \nonumber \\
&\equiv& (-i) \, \sum_{k=1,2,3}  \,
T_{\perp, n}^{A, k, (1)} \left ({\mu^2 \over n \cdot p \, \omega^{\prime}}, {\omega^{\prime} \over \bar n \cdot p} \right )  \ast
\langle  O_{\mu \rho, \perp}^{(n, k)}(\omega, \omega^{\prime}) \rangle^{(0)}   \,.
\end{eqnarray}
The master formula for the one-loop jet function ${\cal J}_{\perp, n}^{A, 1, \, (1)}$ can be derived by
expanding the matching relation (\ref{matching relation for J-T-A}) at ${\cal O}(\alpha_s)$
\begin{eqnarray}
&& \sum_{k=1,2,3}  \, T_{\perp, n}^{A, k, (1)}
\left ({\mu^2 \over n \cdot p \, \omega^{\prime}}, {\omega^{\prime} \over \bar n \cdot p} \right )  \ast
\langle  O_{\mu \rho, \perp}^{(n, k)}(\omega, \omega^{\prime}) \rangle^{(0)} \nonumber  \\
&&  = \sum_{k=1,2,3}  \,
\bigg [ {\cal J}_{\perp, n}^{A, k, (1)} \left ({\mu^2 \over n \cdot p \, \omega^{\prime}}, {\omega^{\prime} \over \bar n \cdot p} \right )  \ast \langle  O_{\mu \rho, \perp}^{(n, k)}(\omega, \omega^{\prime}) \rangle^{(0)} \nonumber \\
&& \hspace{1.8 cm} + \,  {\cal J}_{\perp, n}^{A, k, (0)}
\left ({\mu^2 \over n \cdot p \, \omega^{\prime}}, {\omega^{\prime} \over \bar n \cdot p} \right )
\ast \langle  O_{\mu \rho, \perp}^{(n, k)}(\omega, \omega^{\prime}) \rangle^{(1)} \bigg ]  \,.
\end{eqnarray}
The ultraviolet renormalized one-loop matrix elements
$\langle  O_{\mu \rho, \perp}^{(n, k)}(\omega, \omega^{\prime}) \rangle^{(1)}$
can be written as
\begin{eqnarray}
\langle  O_{\mu \rho, \perp}^{(n, k)}(\omega, \omega^{\prime}) \rangle^{(1)}
= \sum_j \, \left [M_{k j}^{(1) R} + Z_{k j}^{(1)} \right ] \,
 \langle  O_{\mu \rho, \perp}^{(n, j)}(\omega, \omega^{\prime}) \rangle^{(0)} \,,
 \label{renormalization of HQET operator}
\end{eqnarray}
where the superscript $R$ indicates the infrared regularization scheme implemented in
the computation of the bare matrix elements $M_{k j}^{(1)}$.
Employing the dimensional regularization scheme for both the ultraviolet and infrared divergences,
the bare matrix elements vanish due to the scaleless  one-loop HQET diagrams.
The one-loop jet function ${\cal J}_{\perp, n}^{A, 1, (1)}$ can then be readily determined
by comparing the coefficient of $ \langle  O_{\mu \rho, \perp}^{(n, 1)}(\omega, \omega^{\prime}) \rangle^{(0)}$
\begin{eqnarray}
{\cal J}_{\perp, n}^{A, 1, (1)} =  T_{\perp, n}^{A, 1, (1)} -
\sum_{k=1,2,3}  \, Z_{k 1}^{(1)} \ast {\cal J}_{\perp, n}^{A, k, (0)}  \,.
\label{master formula for J-T-A: V1}
\end{eqnarray}
The infrared subtraction term $ Z_{1 1}^{(1)} \ast {\cal J}_{\perp, n}^{A, 1, (0)}$ removes the soft divergences
of the one-loop amplitude $T_{\perp, n}^{A, 1, (1)}$ such that the resulting short-distance function
${\cal J}_{\perp, n}^{A, 1, (1)}$  is finite as it must be.
Implementing the ultraviolet renormalization for the HQET operator
$O_{\mu \rho, \perp, \rm {bare}}^{(n,2)}(\omega^{\prime})$ yields
\begin{eqnarray}
O_{\mu \rho, \perp}^{(n, 2)}(\omega) =
Z_{2 2}(\omega, \omega^{\prime}) \, \ast \,O_{\mu \rho, \perp, \rm {bare}}^{(n, 2)}(\omega^{\prime}) \,,
\end{eqnarray}
which indicates that the  renormalization constants  $Z_{2 1}^{(1)}$ and $Z_{2 3}^{(1)}$ vanish.
The renormalization constants for the evanescent operator will be determined by applying the
prescription that the infrared finite matrix element
$\langle  O_{\mu \rho, \perp}^{(n, 3)}(\omega, \omega^{\prime}) \rangle$ vanishes with the ultraviolet
divergences treated in dimensional regularization and with the infrared singularities parameterized by the
regulator other than the dimensions of space-time \cite{Dugan:1990df,Herrlich:1994kh}.
Taking advantage of the relation (\ref{renormalization of HQET operator})
and the preceding renormalization scheme  for the evanescent operator we obtain
\begin{eqnarray}
Z_{3 1}^{(1)} = - M_{31}^{(1) \, \rm{off}}  \,.
\label{renormalization of evanescent operator}
\end{eqnarray}
Plugging (\ref{renormalization of evanescent operator}) into (\ref{master formula for J-T-A: V1})
with the vanishing renormalization constant $Z_{2 1}^{(1)}$ gives rise to
\begin{eqnarray}
{\cal J}_{\perp, n}^{A, 1, (1)} =  T_{\perp, n}^{A, 1, (1)} -
Z_{1 1}^{(1)} \, \ast {\cal J}_{\perp, n}^{A, 1, (0)}
+ M_{31}^{(1) \, \rm{off}} \,  \ast {\cal J}_{\perp, n}^{A, 3, (0)} \,.
\label{master formula for J-T-A: V2}
\end{eqnarray}

\begin{figure}
\begin{center}
\includegraphics[width=0.8 \columnwidth]{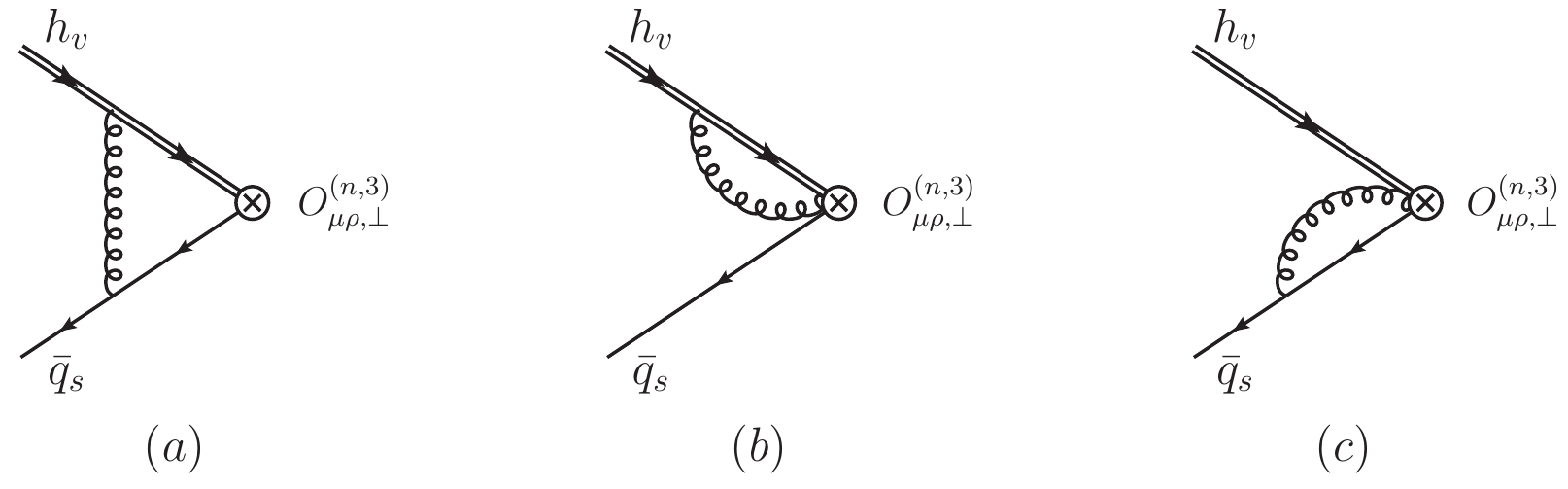}
\vspace*{0.1cm}
\caption{The one-loop HQET diagrams for the renormalization mixing of the
evanescent operator $O_{\mu \rho, \perp}^{(n, 3)}(\omega^{\prime})$
into the physical operator $O_{\mu \rho, \perp}^{(n, 1)}(\omega^{\prime})$.}
\label{fig: evanescent operator at one loop}
\end{center}
\end{figure}

We proceed to compute the one-loop HQET matrix element of the evanescent operator
$O_{\mu \rho, \perp}^{(n, 3)}(\omega^{\prime})$ for the determination of
the bare amplitude $M_{31}^{(1) \, \rm{off}}$.
It is evident that only the diagram (a) in figure \ref{fig: evanescent operator at one loop} can potentially
generate the renormalization mixing of the evanescent
operator $O_{\mu \rho, \perp}^{(n, 3)}(\omega^{\prime})$
into the physical operator $O_{\mu \rho, \perp}^{(n, 1)}(\omega^{\prime})$.
The corresponding one-loop amplitude can be readily derived with the HQET Feynman rules
\begin{eqnarray}
\langle  O_{\mu \rho, \perp}^{(n, 3)}(\omega, \omega^{\prime}) \rangle^{(1, a)}
&=& i \, g_s^2 \, C_F \,\int {d^D l \over (2 \, \pi)^D} \,
{1 \over [(l+k)^2+ i 0] [- v \cdot l + i 0] [l^2+ i 0] } \,\,
\delta(\omega^{\prime}- \omega - \bar n \cdot l) \nonumber \\
&& \bar q_s(k) \, \not \! v  \,\, (\not  l + \not \! k)  \,
\left [ {\not \! n \over 2}  \,\,
\left ( \, { [\gamma_{\rho \perp}, \gamma_{\mu \perp}] \over 2 } \, \gamma_ 5
-  i \, \epsilon_{\rho \mu \perp} \,  \right ) \, \right ]  \,\, h_v\,.
\end{eqnarray}
Performing the loop-momentum integration and employing the classical equation of
motion for the light quark, we observe that the soft gluon correction will not resolve the
dynamical structure of the light-ray HQET operator $O_{\mu \rho, \perp}^{(n, 3)}(\omega^{\prime})$.
Consequently, we obtain
\begin{eqnarray}
M_{31}^{(1) \, \rm{off}} = 0 \,.
\label{result of M31-off}
\end{eqnarray}
Substituting  (\ref{result of M31-off}) into (\ref{master formula for J-T-A: V2})
yields the final result for the one-loop matching coefficient
\begin{eqnarray}
{\cal J}_{\perp, n}^{A, 1, (1)} =  T_{\perp, n}^{A, 1, (1)} -
Z_{1 1}^{(1)} \, \ast {\cal J}_{\perp, n}^{A, 1, (0)} \,,
\label{master formula for J-T-A: V3}
\end{eqnarray}
from which we can further write down its explicit expression
\begin{eqnarray}
J_{\perp,-}^{A, (1)}&=&{\cal J}_{\perp, n}^{A, 1, (1)} =  {\alpha_s \, C_F \over 4 \, \pi}  \,
\bigg  [ 2 \, \ln^2 {\mu^2 \over n \cdot p \, (\omega- \bar n \cdot p)}
 + \, 3 \, \ln {\mu^2 \over n \cdot p \, (\omega- \bar n \cdot p)}
- {\pi^2 \over 3} + 7  \bigg ] \,\, {\cal J}_{\perp, n}^{A, 1, (0)} \,, \hspace{0.5 cm} \nonumber \\
J_{\perp,+}^{A, (1)}&=&{\cal J}_{\perp, \bar n}^{A, 1, (1)} = 0 \,.
\label{result of F-T-A}
\end{eqnarray}
It needs to point out that the absence of the HQET operator mixing between
$O_{\mu \rho, \perp}^{(n, 3)}(\omega^{\prime})$ and  $O_{\mu \rho, \perp}^{(n, 1)}(\omega^{\prime})$
under renormalization arises from the heavy quark spin symmetry, in contrast to the counterpart collinear
operator mixing pattern under the radiative correction \cite{Wang:2017ijn}.

\subsubsection{SCET factorization for $\Pi_{\mu \nu \rho, \perp}^{B}(p, q)$}

\begin{figure}
\begin{center}
\includegraphics[width=0.8 \columnwidth]{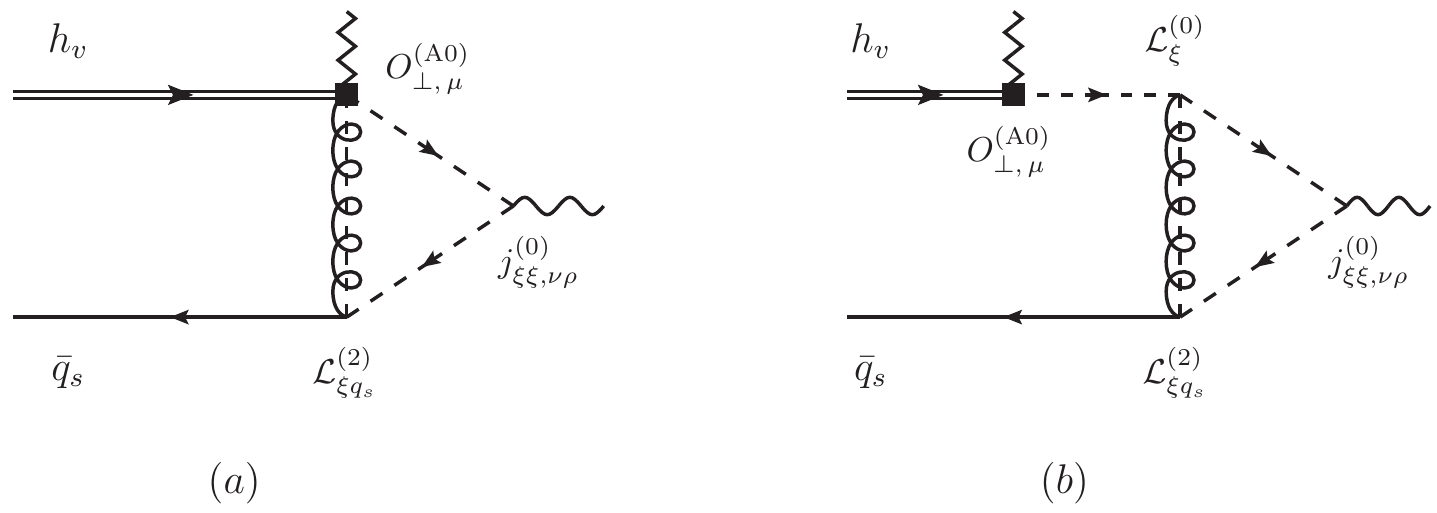}
\vspace*{0.1cm}
\caption{Diagrammatical representation of the vacuum-to-$B$-meson correlation function
$\Pi_{\mu \nu \rho, \perp}^{B}(p, q)$  defined with  the ${\rm A0}$-type SCET operator
$O_{\perp, \, \mu}^{\rm {(A0)}}= \left (\bar \xi \, W_c \right) \,
\gamma_5 \, \gamma_{\mu\, \perp}\, h_v$,
the leading power interpolating current $j_{\xi \xi, \nu \rho}^{(0)}$
and the subleading power SCET Lagrangian ${\cal L}_{\xi q_s}^{(2)}$.}
\label{fig: correlator for F-T-B}
\end{center}
\end{figure}

For the sake of determining the jet functions $J_{\perp, \pm}^{B}$ entering the SCET factorization formula
(\ref{SCET factorization formula T}), we consider the following partonic matrix element
\begin{eqnarray}
F_{\mu \rho, \perp}^{B}(p, q) &=&  \int d^4 x \, e^{i p \cdot x} \, \int d^4 y \,\, \nonumber \\
&& \hspace{-1 cm}  \left \langle 0 \left |
{\rm T}  \left \{\bar \xi(x) \,\, {\not \! n \over 2} \,
\gamma_{\rho\, \perp}  \, \xi(x),  \,\,i \, {\cal L}_{\xi q_s}^{(2)}(y), \,\,
\left (\bar \xi \, W_c \right)(0) \, \gamma_5 \, \gamma_{\mu\, \perp}\, h_v(0) \,    \right \}
\right | \bar q_s(k) \, h_v \right \rangle.
\end{eqnarray}
Applying the SCET Feynman rules we can observe that the contribution from the diagram (a)
displayed in figure \ref{fig: correlator for F-T-B} can be read from the result of the corresponding matrix element
(\ref{FL-B-a: result}) for the effective currents related to the longitudinally polarized vector meson
\begin{eqnarray}
F_{\mu \rho, \perp}^{B, (a)}(p, q) &=& {\alpha_s \, C_F \over 2 \, \pi} \, {\bar n \cdot p - \omega \over \omega} \,
\left  [ {1 \over \epsilon} + \ln {\mu^2 \over n \cdot p \, (\omega- \bar n \cdot p)}
+ {1 \over 2} \,  \left ( { \bar n \cdot p - \omega \over \bar n \cdot p} \right )  + 1 \right ] \nonumber \\
&&   \times \, \ln \left ( { \bar n \cdot p - \omega \over \bar n \cdot p} \right )  \,
F_{\mu \rho, \perp, \rm{LO}}^{A}(p, q).
\end{eqnarray}
Evaluating the $\rm {SCET_I}$ diagram (b) in figure \ref{fig: correlator for F-T-B} leads to
\begin{eqnarray}
F_{\mu \rho, \perp}^{B, (b)}(p, q) &=&  g_s^2 \, C_F \, \int {d^D l  \over (2 \pi)^D} \, \int {d^D L  \over (2 \pi)^D}
\frac{n \cdot L \,\, n \cdot (p+L) \,\, n \cdot (p+L+l)}
{[L^2 + i 0] [(p+L)^2 + i 0] [(p+L+l)^2 + i 0] [l^2+ i 0]}  \nonumber \\
&&  \bar q_s(k) \, \bigg \{ (2 \, \pi)^4 \, \delta^4(l+L+k)  \,
\left [\bar n^{\alpha} + \gamma_{\perp}^{\alpha} \, {\not \! L_{\perp} \over n \cdot L}
+ {n^{\alpha}  \over n \cdot l} \, {L^2 \over n \cdot L}  \right ]  \, {\not \! n \over 2}  \nonumber \\
&& + \, k_{\perp \, \beta} \, \partial_{\perp}^{\beta} \,
\left [  (2 \, \pi)^4 \, \delta^4(l+L+k) \right ] \,
\left [ \gamma_{\perp}^{\alpha} - { \not \! L_{\perp} \over n \cdot L} \, n^{\alpha} \right ] \,  \bigg \}  \nonumber \\
&& \gamma_{\rho \perp} \,
\left [ \bar n_{\alpha} +  {\gamma_{\alpha \, \perp} \, (\not \! L_{\perp} + \not l_{\perp}) \over n \cdot (p+L+l)}
+ {\not \! L_{\perp}  \,  \gamma_{\alpha \perp} \over n \cdot (p+L)}
- {L_{\perp}^2 +  \not \! L_{\perp}  \, \not  l_{\perp}  \over n \cdot (p+L) \, n \cdot (p+L+l)} \, n_{\alpha}  \right ] \nonumber \\
&& \,\, \times \, {\not \!  \bar n \over 2} \, \gamma_5 \, \gamma_{\mu \perp} \,  h_v \,, \hspace{0.5 cm}
\end{eqnarray}
whose result cannot be extracted directly from the counterpart longitudinal matrix element
displayed in (\ref{result of F-L-B}).
Implementing the Dirac algebra reduction in the $D$-dimensional space-time and performing
the loop momentum integration, we obtain
\begin{eqnarray}
F_{\mu \rho, \perp}^{B, (b)}(p, q) &=& {\alpha_s \, C_F \over 4 \, \pi} \,  \,
\bigg \{ - {2 \over \epsilon^2 } + {1 \over \epsilon} \,
\left [ - 2 \, \ln \left ( {\mu^2 \over n \cdot p \, (\omega - \bar n \cdot p)} \right )
+ \, 2 \,\, {\ln (1 + \eta) \over \eta} - 4 \right ] \nonumber \\
&& - \ln^2 \left ( {\mu^2 \over n \cdot p \, (\omega - \bar n \cdot p)} \right )
+ \ln \left ( {\mu^2 \over n \cdot p \, (\omega - \bar n \cdot p)} \right )  \,
\left [  2 \,\, {\ln (1 + \eta) \over \eta} - 4 \right ] \nonumber \\
&&  +  \, {1 \over \eta} \, \ln^2(1+\eta)  + {4 \over \eta}  \, \ln (1+\eta)
+ {\pi^2 \over 6} - 8 \bigg \}  \,\,
F_{\mu \rho, \perp, \rm{LO}}^{A}(p, q).
\end{eqnarray}
Adding up the contributions from these two diagrams  and performing the infrared subtraction
with the evanescent operator approach, the  jet functions  $J_{\perp, \pm}^{B}$ can be  determined as follows
\begin{eqnarray}
J_{\perp, -}^{B} &=&  {\alpha_s \, C_F \over 4 \, \pi}  \,
\bigg  [ - \ln^2 \left ( {\mu^2 \over n \cdot p \, (\omega - \bar n \cdot p)} \right )
+ \ln \left ( {\mu^2 \over n \cdot p \, (\omega - \bar n \cdot p)} \right )  \,
\left [  - 2 \,\, \ln (1 + \eta) - 4 \right ] \nonumber \\
&&  - \ln^2(1+\eta)  + \left ( {2 \over \eta} - 2 \right )  \, \ln (1+\eta)
+ {\pi^2 \over 6} - 8  \bigg ] \,\, {\cal J}_{\perp, n}^{A, 1, (0)}  \,, \nonumber \\
J_{\perp, +}^{B} &=&  0 \,,
\label{result of F-T-B}
\end{eqnarray}
which are consistent with the previous results obtained in \cite{DeFazio:2007hw},
employing the momentum-space projector for the $B$-meson LCDA in the $D$ dimensional
space-time.

\subsubsection{SCET factorization for $\Pi_{\mu \nu \rho, \perp}^{C}(p, q)$}

We proceed to determine the light-quark-mass induced jet functions $J_{\perp, \pm}^{C}$
appearing in the SCET factorization formula (\ref{SCET factorization formula T}) by
investigating the partonic matrix element
\begin{eqnarray}
&& F_{\mu \rho, \perp}^{C}(p, q) =  \int d^4 x \, e^{i p \cdot x} \, \int d^4 y \,\, \int d^4 z  \,\, \nonumber \\
&& \hspace{0.5 cm}
\left \langle 0 \left | {\rm T}  \left \{\bar \xi(x) \,\, {\not \! n \over 2} \,
\gamma_{\rho\, \perp}  \, \xi(x),  \,\,i \, {\cal L}_{\xi q_s}^{(1)}(y), \,\,
i \, {\cal L}_{\xi m}^{(1)}(z), \,\,
\left (\bar \xi \, W_c \right)(0) \, \gamma_5 \, \gamma_{\mu\, \perp}\, h_v(0) \,    \right \}
\right |  \bar q_s(k) \, h_v \right \rangle. \hspace{1.0 cm}
\end{eqnarray}
Computing the tree-level contribution to $F_{\mu \rho, \perp}^{C}$ with the SCET Feynman rules leads to
\begin{eqnarray}
F_{\mu \rho, \perp}^{C}(p, q) &=&   {m \over n \cdot p}  \, {g_s^2 \, C_F \over \bar n \cdot p - \omega} \,\,
\bar q_s(k) \, \gamma_{\rho \perp} \, \gamma_{\mu \perp}\, {\not \! \bar n \over 2} \, \gamma_5 \,  h_v \,
\,  \int {d^D l \over (2 \pi)^D} \, \nonumber \\
&&
{(D-4) \,  (n \cdot l)^2 \over [n \cdot l \, \bar n \cdot (l+k) + l_{\perp}^2 + i 0]
[ n \cdot (p-l) \, \bar n \cdot (p-l-k) + l_{\perp}^2  + i 0] [l^2 + i 0] } \,,
\hspace{0.5 cm}
\end{eqnarray}
which vanishes in the four dimensional space-time in contrast to the result of
the counterpart longitudinal matrix element as displayed in (\ref{result of F-L-C}).
The corresponding jet functions $J_{\perp, \pm}^{C}$ can therefore be determined as
\begin{eqnarray}
J_{\perp, +}^{C} = J_{\perp, -}^{C} = {\cal O}(\alpha_s^2).
\label{result of F-T-C}
\end{eqnarray}
An interesting consequence from such observation is that the light-quark-mass corrections
to the  radiative leptonic decay amplitudes of $B \to \gamma \ell \nu_{\ell}$  and $B_s \to \gamma \ell \bar \ell$
will not give rise to the leading-power contribution  in the heavy quark expansion,
at least, at ${\cal O}(\alpha_s)$.

Substituting the derived jet functions (\ref{result of F-T-A}), (\ref{result of F-T-B})
and (\ref{result of F-T-C}) into the SCET factorization formula (\ref{SCET factorization formula T}),
we can readily write down
\begin{eqnarray}
\Pi_{\mu \nu \rho, \perp}(p, q) &=& {\tilde{f}_B(\mu) \, m_B \over 2}\,
\int_0^{+\infty} \, {d \omega \over \bar n \cdot p - \omega + i 0} \,\,
\left [ 1 + {\alpha_s \, C_F \over 4 \, \pi} \,   \hat{J}_{\perp, -}^{(\rm{A0})}
\left ({\mu^2 \over n \cdot p \, \omega}, {\omega \over \bar n \cdot p} \right )   \right ] \,
\phi_B^{-}(\omega, \mu) \, \nonumber \\
&&   \, g_{\mu \rho \, \perp} \,\, \bar n_{\nu} \, + {\cal O}(\alpha_s^2) \,,
\label{result of F-T in SCET}
\end{eqnarray}
where the renormalized one-loop jet function $\hat{J}_{\perp, -}^{(\rm{A0})}$ reads
\begin{eqnarray}
\hat{J}_{\perp, -}^{(\rm{A0})} &=&  \ln^2 \left ( {\mu^2 \over n \cdot p \, (\omega- \bar n \cdot p) } \right )
- 2 \, \ln \left ( {\mu^2 \over n \cdot p \, (\omega- \bar n \cdot p) } \right ) \, \ln(1+\eta)
- \ln \left ( {\mu^2 \over n \cdot p \, (\omega- \bar n \cdot p) } \right )
\nonumber \\
&& -\ln^2 (1+\eta) + \left ({2 \over \eta} - 2 \right ) \,\ln(1+\eta)  - {\pi^2 \over 6}  - 1 \,.
\end{eqnarray}
The spectral representation of the factorization formula (\ref{result of F-T in SCET}) can be further derived
with the explicit expressions of the various dispersion integrals displayed in \cite{Wang:2015vgv}
\begin{eqnarray}
\Pi_{\mu \nu \rho, \perp}(p, q) = - {\tilde{f}_B(\mu) \, m_B \over 2}\,
\int_0^{+\infty} \, {d \omega^{\prime} \over \omega^{\prime} - \bar n \cdot p - i 0} \,\,
\tilde{\phi}_{B, \rm{eff}}^{-}(\omega^{\prime}, \mu) \,
 \, g_{\mu \rho \, \perp} \,\, \bar n_{\nu} \, + {\cal O}(\alpha_s^2) \,,
\label{spectral representation for F-T-A0}
\end{eqnarray}
where we have introduced the ``effective" $B$-meson distribution amplitude
$\tilde{\phi}_{B, \rm{eff}}^{-}$ for brevity
\begin{eqnarray}
&& \tilde{\phi}_{B, \rm{eff}}^{-}(\omega^{\prime}, \mu) \nonumber \\
&& = \phi_{B}^{-}(\omega^{\prime}, \mu)
+ {\alpha_s \, C_F \over 4 \, \pi} \,
\bigg \{ \int_0^{\omega^{\prime}}  d \omega \,
\left [ {2 \over \omega-\omega^{\prime}} \,
\left ( \ln {\mu^2 \over n \cdot p \, \omega^{\prime}}
- 2 \, \ln {\omega^{\prime} - \omega \over \omega^{\prime}} -{1 \over 2} \right )  \right ]_{\oplus} \,
\phi_{B}^{-}(\omega, \mu) \nonumber \\
&& \hspace{0.5 cm} - \int_{\omega^{\prime}}^{\infty} \, d \omega \,
\bigg [  \ln^2 {\mu^2 \over n \cdot p \, \omega^{\prime}}
- \ln {\mu^2 \over n \cdot p \, \omega^{\prime}}
- \left (2 \, \ln {\mu^2 \over n \cdot p \, \omega^{\prime}}  + 4 \right ) \,
\ln {\omega-\omega^{\prime} \over \omega^{\prime}}
+ 2 \, \ln {\omega \over \omega^{\prime}} \nonumber  \\
&& \hspace{2.5 cm} + \, {\pi^2 \over 6} - 1 \bigg ] \,\,
 {d \phi_{B}^{-}(\omega, \mu) \over d\omega} \bigg \} \,.
 \label{spectral function for F-T-A0}
\end{eqnarray}

Matching the SCET factorization formula (\ref{spectral representation for F-T-A0})
with the hadronic dispersion relation for the vacuum-to-$B$-meson correlation function $\Pi_{\mu \nu \rho, \perp}$
\begin{eqnarray}
&& \Pi_{\mu \nu \rho, \perp}(p, q) \nonumber \\
&& =  \left [- {f_{V, \perp}(\nu)  \, n \cdot p \over m_V^2/n \cdot p -\bar n \cdot p - i 0}  \,
{\xi_{\perp}(n \cdot p) \over 2} \,
+ \int_{\omega_s}^{+\infty} \, {d \omega^{\prime} \over \omega^{\prime}  - \bar n \cdot p - i0} \,\,
\rho^{h}_{\perp}(\omega^{\prime}, n \cdot p) \right ]  \, g_{\mu \rho \, \perp} \,\, \bar n_{\nu} \,, \hspace{0.8 cm}
\end{eqnarray}
and implementing the NLL resummation for the enhanced logarithms of $\mu/\mu_0$,
we obtain the following SCET sum rules for the form factor $\xi_{\perp}(n \cdot p)$ at ${\cal O}(\alpha_s)$
\begin{eqnarray}
\xi_{\perp}(n \cdot p)=  {U_2(\mu_{h2}, \mu) \, \tilde{f}_B(\mu_{h2})  \over f_{V, \perp}(\nu) } \,
{ m_B \over n \cdot p} \,  \int_{0}^{\omega_s} \, d \omega^{\prime} \,
{\rm exp} \left [ - {n \cdot p \, \omega^{\prime} - m_V^2 \over n \cdot p \, \omega_M} \right ] \,\,
 \tilde{\phi}_{B, \rm{eff}}^{-}(\omega^{\prime}, \mu)  \,.
\label{LCSR for xi-T}
\end{eqnarray}
The renormalization-scale dependent transverse decay constant of the vector meson $f_{V, \perp}$ is defined as follows
\begin{eqnarray}
c_V \,\, \langle V(p, \epsilon^{\ast})| j_{\nu \rho} |  0 \rangle
= - \, i \, f_{V, \perp}(\nu) \,\, p_{\nu} \,\, \epsilon^{\ast}_{\rho \perp}(p) \,,
\end{eqnarray}
where the corresponding RG evolution equation can be written as
\begin{eqnarray}
{d  \over d \ln \nu} \, f_{V, \perp}(\nu) = \gamma_{T}(\alpha_s) \,\, f_{V, \perp}(\nu) \,, \qquad
\gamma_{T}(\alpha_s)= \sum_{n=0}^{\infty} \, \left ( {\alpha_s \over 4\, \pi} \right )^{n+1} \, \gamma_{T, n} \,,
\label{RGE for fVT}
\end{eqnarray}
with the first two expansion coefficients \cite{Bell:2010mg}
\begin{eqnarray}
\gamma_{T, 0}= - 2 \, C_F \,, \qquad
\gamma_{T, 1}=  C_F \, \left [  19 \, C_F - {257 \over 9} \, C_A
+ {52 \over 9} \, \left (n_l+1 \right )\, T_F \right ]\,.
\end{eqnarray}
It is evident from the definitions of the $\rm{A0}$-type SCET form factors
(\ref{Definition: SCET-I form factors}) that $\xi_{\perp}(n \cdot p)$ is independent
of the QCD renormalization scale $\nu$ of the transverse decay constant $f_{V, \perp}(\nu)$.
In order to demonstrate such argument from the obtained $B$-meson LCSR displayed in (\ref{LCSR for xi-T}),
we need to distinguish the renormalization scale $\nu$ for the tensor interpolating current of the vector
meson $j_{\nu \rho}$ from the factorization scale $\mu$ that is related to the RG evolution of the light-cone
HQET operators. To this end, we decompose the one-loop jet function $\hat{J}_{\perp, -}^{(\rm{A0})}$
entering the factorization formula (\ref{result of F-T in SCET}) into the following two pieces
\begin{eqnarray}
\hat{J}_{\perp, -}^{(\rm{A0})} \left ({\mu^2 \over n \cdot p \, \omega}, {\omega \over \bar n \cdot p}, \nu \right )
=\hat{J}_{\perp, -}^{(\rm{A0})} \left ({\mu^2 \over n \cdot p \, \omega}, {\omega \over \bar n \cdot p} \right )
+ \delta \hat{J}_{\perp, -}^{(\rm{A0})}  \left ({\mu^2 \over n \cdot p \, \omega}, {\omega \over \bar n \cdot p}, \nu \right ) \,,
\end{eqnarray}
where the QCD scale $\nu$ dependence of the second term on the right-hand side is determined by
the RG evolution equation for the tensor current
\begin{eqnarray}
{d  \over d \ln \nu} \, \delta \hat{J}_{\perp, -}^{(\rm{A0})}
\left ({\mu^2 \over n \cdot p \, \omega}, {\omega \over \bar n \cdot p}, \nu \right )
= \gamma_{T}(\alpha_s) \,\, \delta \hat{J}_{\perp, -}^{(\rm{A0})}
\left ({\mu^2 \over n \cdot p \, \omega}, {\omega \over \bar n \cdot p}, \nu \right )   \,.
\label{RGE for delta-J}
\end{eqnarray}
Employing the consistency condition of the newly defined  function for $\nu=\mu$
\begin{eqnarray}
\delta \hat{J}_{\perp, -}^{(\rm{A0})}
\left ({\mu^2 \over n \cdot p \, \omega}, {\omega \over \bar n \cdot p}, \nu=\mu \right )
=0 \,,
\end{eqnarray}
it is straightforward to write down the solution to (\ref{RGE for delta-J})
\begin{eqnarray}
\delta \hat{J}_{\perp, -}^{(\rm{A0})}
\left ({\mu^2 \over n \cdot p \, \omega}, {\omega \over \bar n \cdot p}, \nu \right )
= {\alpha_s(\mu)\over 4\, \pi} \, \gamma_{T, 0}\, \ln \left ({\nu \over \mu}\right )
+ {\cal O}(\alpha_s^2) \,.
\end{eqnarray}
The resulting hard-collinear function $\hat{J}_{\perp, -}^{(\rm{A0})}$ is therefore given by
\begin{eqnarray}
&& \hat{J}_{\perp, -}^{(\rm{A0})} \left ({\mu^2 \over n \cdot p \, \omega}, {\omega \over \bar n \cdot p}, \nu \right ) \nonumber \\
&& = \ln^2 \left ( {\mu^2 \over n \cdot p \, (\omega- \bar n \cdot p) } \right )
- 2 \, \ln \left ( {\mu^2 \over n \cdot p \, (\omega- \bar n \cdot p) } \right ) \, \ln(1+\eta) \nonumber \\
&& \hspace{0.5 cm} - \ln \left ( {\nu^2 \over n \cdot p \, (\omega- \bar n \cdot p) } \right )
-\ln^2 (1+\eta) + \left ({2 \over \eta} - 2 \right ) \,\ln(1+\eta)  - {\pi^2 \over 6}  - 1 \,.
\end{eqnarray}
As a consequence,  the ``effective" distribution amplitude
entering the SCET sum rules for $\xi_{\perp}(n \cdot p)$ with the two scales $\mu$ and $\nu$
distinct from each other is given by
\begin{eqnarray}
&& \tilde{\phi}_{B, \rm{eff}}^{-}(\omega^{\prime}, \mu, \nu) \nonumber \\
&& = \phi_{B}^{-}(\omega^{\prime}, \mu)
+ {\alpha_s \, C_F \over 4 \, \pi} \,
\bigg \{ \int_0^{\omega^{\prime}}  d \omega \,
\left [ {2 \over \omega-\omega^{\prime}} \,
\left ( \ln {\mu^2 \over n \cdot p \, \omega^{\prime}}
- 2 \, \ln {\omega^{\prime} - \omega \over \omega^{\prime}} -{1 \over 2} \right )  \right ]_{\oplus} \,
\phi_{B}^{-}(\omega, \mu) \nonumber \\
&& \hspace{0.5 cm} - \int_{\omega^{\prime}}^{\infty} \, d \omega \,
\bigg [  \ln^2 {\mu^2 \over n \cdot p \, \omega^{\prime}}
- \ln {\nu^2 \over n \cdot p \, \omega^{\prime}}
- \left (2 \, \ln {\mu^2 \over n \cdot p \, \omega^{\prime}}  + 4 \right ) \,
\ln {\omega-\omega^{\prime} \over \omega^{\prime}}
+ 2 \, \ln {\omega \over \omega^{\prime}} \nonumber  \\
&& \hspace{2.5 cm} + \, {\pi^2 \over 6} - 1 \bigg ] \,\,
 {d \phi_{B}^{-}(\omega, \mu) \over d\omega} \bigg \} \,.
 \label{spectral function for F-T-A0 with different mu and nu}
\end{eqnarray}

\subsection{The $B$-meson  LCSR for $\Xi_{\perp}(\tau, n \cdot p)$}

Now we turn to construct the sum rules for the $\rm{B}$-type SCET form factor $\Xi_{\perp}(\tau, n \cdot p)$
with the following vacuum-to-$B$-meson correlation function
\begin{eqnarray}
\widetilde{\Pi}_{\mu \nu \rho, \perp}(p, q, \tau) &=& { n \cdot p \over 2\, \pi} \, \int d^4 x \, e^{i p \cdot x} \,
 \int d r \, e^{-i n \cdot p \, \tau \, r}  \, \nonumber \\
&&  \langle 0 | {\rm T}  \left \{j_{\nu \rho}(x),  \,\, \left (\bar \xi \, W_c \right)(0) \,
\gamma_5 \,\, \gamma_{\mu \perp} \,\,
(W_c^{\dagger} \,\, i  \not \! \!  D_{c, \,\perp} \, W_c)(r \, n) \,\, h_v(0) \,    \right \}   | \bar B_v \rangle \,.
\label{correlation function for Xi-T}
\end{eqnarray}
Employing the $\rm {SCET_I}$ representation (\ref{SCET representation of the transverse vector current})
of the interpolating current for the transversely polarized vector meson, it is straightforward to
identify the leading power contribution of the correlation function (\ref{correlation function for Xi-T})
\begin{eqnarray}
\widetilde{\Pi}_{\mu  \nu \rho, \perp}(p, q, \tau) &=& { n \cdot p \over 2\, \pi} \, \int d^4 x \, e^{i p \cdot x} \,
\, \int d^4 y  \,  \int d r \, e^{-i n \cdot p \, \tau \, r}  \,
\bigg  \langle 0 \bigg | {\rm T}  \bigg  \{j_{\xi \xi, \nu \rho}^{(0)}(x),  \,\, i \, {\cal L}_{\xi q_s}^{(1)}(y)\,,
\nonumber  \\
&&  \left (\bar \xi \, W_c \right)(0) \,
\gamma_5 \,\, \gamma_{\mu \perp} \,\,
(W_c^{\dagger} \,\, i  \not \! \! D_{c, \,\perp} \, W_c)(r \, n) \,\, h_v(0) \,    \bigg  \}   \bigg | \bar B_v  \bigg \rangle \,.
\label{correlation function for Xi-T at LP}
\end{eqnarray}
Performing the perturbative matching of the $\rm {SCET_I}$ matrix element onto HQET yields the
soft-collinear factorization formula
\begin{eqnarray}
\widetilde{\Pi}_{\mu \nu \rho, \perp}(p, q, \tau)  =
{\tilde{f}_B(\mu) \, m_B \over 2}\,
\sum_{m = \pm } \, \int_0^{+\infty} \, d \omega \, \tilde{J}_{\perp, m}
\left ({\mu^2 \over n \cdot p \, \omega}, {\omega \over \bar n \cdot p}, \tau \right )  \,
\phi_B^{m}(\omega, \mu) \,\, g_{\mu \rho \perp} \,\, \bar n_{\nu}.
\label{factorization formula of Pi-tilde-T-B}
\end{eqnarray}
The short-distance functions $\tilde{J}_{\perp, m}$ can be extracted from the hard-collinear contribution
to the partonic matrix element
\begin{eqnarray}
\widetilde{F}_{\mu \nu \rho, \perp}(p, q, \tau) &=& { n \cdot p \over 2\, \pi} \, \int d^4 x \, e^{i p \cdot x} \,
\, \int d^4 y  \,  \int d r \, e^{-i n \cdot p \, \tau \, r}  \,\,
\bigg \langle 0 \bigg | {\rm T}  \bigg \{ j_{\xi \xi, \nu \rho}^{(0)}(x),  \,\, i \, {\cal L}_{\xi q_s}^{(1)}(y)\,,
\nonumber \\
&&   \left (\bar \xi \, W_c \right)(0) \,
\gamma_5 \,\, \gamma_{\mu \perp} \,\,
(W_c^{\dagger} \,\, i  \not \! D_{c, \,\perp} \, W_c)(r \, n) \,\, h_v(0) \,    \bigg \}
\bigg | \bar q_s(k) \,\, h_v \bigg \rangle \,.
\label{partonic correlation function for Xi-T at LP}
\end{eqnarray}
Employing the SCET Feynman rules we can readily derive the tree-level amplitude
\begin{eqnarray}
\widetilde{F}_{\mu \nu \rho, \perp}(p, q, \tau) &=&
g_s^2 \, C_F \, \, \bar q_s(k) \, \left [ (D-4) \, \gamma_{\rho \perp} \, \gamma_{\mu \perp}
+ 2 \,\gamma_{\mu \perp} \,  \gamma_{\rho \perp}  \right ]  \,\,
\, {\not \! \bar n \over 2} \, \gamma_5 \,  h_v \,\, \bar n_{\nu} \nonumber \\
&&  \hspace{-1.5 cm} \int{d^D l \over (2 \pi)^D } \,
{ n \cdot l \, n \cdot (p-l) \, \delta(\tau - n \cdot l / n \cdot p)
\over [n \cdot l \, \bar n \cdot (l+k) + l_{\perp}^2 + i 0]
[ n \cdot (p-l) \, \bar n \cdot (p-l-k) + l_{\perp}^2  + i 0] [l^2 + i 0] } \,,
\hspace{1.0 cm}
\end{eqnarray}
which can be further computed as follows
\begin{eqnarray}
&& \widetilde{F}_{\mu \nu \rho, \perp}(p, q, \tau)  \nonumber \\
&& = \left ( - i \right )  \, {\alpha_s \, C_F \over 2 \, \pi} \, {n \cdot p \over \omega} \,
\ln (1+\eta) \, \left [ (1-\tau) \, \theta(\tau) \, \theta(1-\tau) \right ] \,\,
 \, \bar q_s(k) \,  \gamma_{\mu \perp}  \, \gamma_{\rho \perp} \, {\not \! \bar n \over 2} \, \gamma_5 \,  h_v
 \,\, \bar n_{\nu}\,.
\end{eqnarray}
Applying the perturbative matching relation for the SCET matrix element
\begin{eqnarray}
\widetilde{F}_{\mu \nu \rho, \perp}(p, q, \tau) =
(-i) \, \sum_{k=1,2,3}  \, {\cal \tilde{J}}_{\perp, \bar n}^{A, k}
\left ({\mu^2 \over n \cdot p \, \omega^{\prime}}, {\omega^{\prime} \over \bar n \cdot p}, \tau \right )  \ast
\langle  O_{\mu \rho, \perp}^{(\bar n, k)}(\omega, \omega^{\prime}) \rangle
\,\, \bar n_{\nu} \,,
\end{eqnarray}
with the light-cone HQET operators  in the momentum space defined by
\begin{eqnarray}
 O_{\mu \rho, \perp}^{(\bar n, 1)}(\omega^{\prime})  &=& {1 \over 2 \, \pi} \, \int d t \, e^{i \, t \, \omega^{\prime}} \,
\left ( \bar q_s Y_s \right )(t \, \bar n) \, \,  \left [  \, g_{\rho \mu  \perp} \,
 {\not \! \bar n \over 2}  \,\, \gamma_5 \, \right ] \,\, \left ( Y_s^{\dag} h_v \right )(0) \,, \nonumber \\
 O_{\mu \rho, \perp}^{(\bar n, 2)}(\omega^{\prime})  &=& {1 \over 2 \, \pi} \, \int d t \, e^{i \, t \, \omega^{\prime}} \,
\left ( \bar q_s Y_s \right )(t \, \bar n) \, \, \left [ \,  i \, \epsilon_{\rho \mu \perp} \,
 {\not \! \bar n \over 2}  \, \right ] \,\,  \left ( Y_s^{\dag} h_v \right )(0) \,, \nonumber \\
 O_{\mu \rho, \perp}^{(\bar n, 3)}(\omega^{\prime})  &=& {1 \over 2 \, \pi} \, \int d t \, e^{i \, t \, \omega^{\prime}} \,
\left ( \bar q_s Y_s \right )(t \, \bar n) \, \, \left [ {\not \! \bar n \over 2}  \,\,
\left ( \, { [\gamma_{\rho \perp}, \gamma_{\mu \perp}] \over 2 } \, \gamma_ 5
+ i \, \epsilon_{\rho \mu \perp} \, \right ) \, \right ] \,\, \left ( Y_s^{\dag} h_v \right )(0) \,,
\hspace{0.8 cm}
\end{eqnarray}
and implementing the infrared subtraction scheme with the evanescent operator approach
described in the previous subsections, the determined jet functions are given by
\begin{eqnarray}
\tilde{J}_{\perp, \, +} = {\cal \tilde{J}}_{\perp, \bar n}^{A, 1}
= {\alpha_s \, C_F \over 2\, \pi} \, {n \cdot p \over \omega} \,
\ln (1+\eta) \, \left [ (1-\tau) \, \theta(\tau) \, \theta(1-\tau) \right ] \,, \qquad
\tilde{J}_{\perp, \, -} =  {\cal \tilde{J}}_{\perp, n}^{A, 1}  = 0. \hspace{0.5 cm}
\end{eqnarray}

Taking advantage of the spectral representation of the factorization formula
(\ref{factorization formula of Pi-tilde-T-B}) for the vacuum-to-$B$-meson
correlation function $\widetilde{\Pi}_{\mu \nu \rho, \perp}$
\begin{eqnarray}
\widetilde{\Pi}_{\mu \nu \rho, \perp}(p, q, \tau) &=& {\alpha_s \, C_F \over 4 \, \pi} \,
\tilde{f}_B(\mu) \, m_B \, \left [ (1-\tau) \, \theta(\tau) \, \theta(1-\tau) \right ] \, \nonumber \\
&& \times \, \int_{0}^{\infty} {d \omega^{\prime} \over \omega^{\prime}  - \bar n \cdot p - i 0} \,
\left [ \int_{\omega^{\prime}}^{\infty}  d \omega \, {n \cdot p \over \omega} \, \phi_B^{+}(\omega, \mu) \right ] \,
 g_{\mu \rho \perp} \,\, \bar n_{\nu} \,,
\end{eqnarray}
with the aid of the corresponding hadronic dispersion relation
\begin{eqnarray}
\widetilde{\Pi}_{\mu \nu \rho, \perp}(p, q, \tau) &=& \bigg [- {f_{V, \perp}(\nu) \,\, m_b \over m_V^2/n \cdot p - \bar n \cdot p - i 0} \,
\left ({n \cdot p \over 2 } \right )\,\, \Xi_{\perp}(\tau, n \cdot p) \nonumber \\
&& + \int_{\omega_s}^{+\infty} \, {d \omega^{\prime} \over \omega^{\prime}  - \bar n \cdot p - i0} \,\,
\tilde{\rho}^{h}_{\perp}(\omega^{\prime}, n \cdot p, \tau) \bigg ] \, g_{\mu \rho \perp} \,\, \bar n_{\nu} \,,
\end{eqnarray}
we obtain the desired sum rules  for the  non-local  form factor $\Xi_{\perp}(\tau, n \cdot p)$ under
the parton-hadron duality approximation
\begin{eqnarray}
\Xi_{\perp}(\tau, n \cdot p) &=& - {\alpha_s \, C_F \over 2 \, \pi} \,
{U_2(\mu_{h2}, \mu) \, \, \tilde{f}_B(\mu)  \over f_{V, \perp}(\nu)} \,
{ m_B   \over  m_b} \,
\, \left [ (1-\tau) \, \theta(\tau) \, \theta(1-\tau) \right ] \,
\nonumber \\
&& \times \, \int_{0}^{\omega_s} \, d \omega^{\prime} \,
{\rm exp} \left [ - {n \cdot p \, \omega^{\prime} - m_V^2 \over n \cdot p \, \omega_M} \right ]
\int_{\omega^{\prime}}^{\infty}  \, d \omega \, {\phi_B^{+}(\omega, \mu) \over \omega} \,
+ {\cal O}(\alpha_s^2) \,.
\label{SCET sum rules for Xi-T}
\end{eqnarray}
Comparing the tree-level sum rules for the $\rm {B}$-type SCET  form factors $\Xi_{\|, \perp}(\tau, n \cdot p)$
presented in (\ref{SCET sum rules for Xi-L}) and (\ref{SCET sum rules for Xi-T}) leads to the
following relation
\begin{eqnarray}
{\Xi_{\perp}(\tau, n \cdot p) \over \Xi_{\|}(\tau, n \cdot p) }
={f_{V, \perp}(\nu) \over f_{V, \|}} \,\,
{n \cdot p \over 2 \, m_V} \,\,  + {\cal O}(\alpha_s^2) \,,
\end{eqnarray}
which is in precise agreement with the SCET factorization formulae obtained in \cite{Beneke:2005gs}.
It remains to be verified that  whether the LCSR calculations of $\Xi_{\|, \perp}(\tau, n \cdot p)$
with the $B$-meson distribution amplitudes can reproduce the already accomplished
SCET computations at ${\cal O}(\alpha_s^2)$.

\subsection{RG improvement of the hard matching coefficients}

Plugging the obtained sum rules for the ``effective" form factors (\ref{SCET sum rules for xi-L}),
(\ref{SCET sum rules for Xi-L}), (\ref{LCSR for xi-T}) and (\ref{SCET sum rules for Xi-T})
into the $\rm{SCET_I}$ factorization formulae (\ref{SCET-I factorization formulae}) gives rise to
the explicit expressions of the leading-power contributions to the seven semileptonic $B \to V$ form factors
in the heavy quark expansion, which serve as one of the major technical results of our paper.
Taking the factorization scale $\mu$ of order $\sqrt{m_b \, \Lambda}$,
the hard matching functions $C_i^{(\rm {A0})}$ and $C_i^{(\rm {B1})}$ involve the
parametrically enhanced logarithms $\ln (m_b/\Lambda)$, which need to be summed up to all orders in perturbation
theory at NLL and LL accuracy. To achieve this goal, we will apply the RG evolution equations for these short-distance
functions in the momentum space \cite{Hill:2004if,Beneke:2005gs}
\begin{eqnarray}
{d \over d \ln \mu} \, C_i^{(\rm {A0})}(n \cdot p, \mu) &=&
\left [- \Gamma_{\rm cusp}(\alpha_s) \, \ln \left ( {\mu \over n \cdot p} \right ) + \gamma(\alpha_s)\right ] \,
C_i^{(\rm {A0})}(n \cdot p, \mu) \,, \nonumber \\
{d \over d \ln \mu} \, C_i^{(\rm {B1})}(n \cdot p, \tau, \mu) &=&
- \Gamma_{\rm cusp}(\alpha_s) \, \ln \left ( {\mu \over n \cdot p} \right ) \,
C_i^{(\rm {B1})}(n \cdot p, \tau, \mu) \nonumber \\
&& + \int_0^1 d \tau^{\prime} \, \gamma_i^{(\rm {B1})}(\tau^{\prime}, \tau) \,
 C_i^{(\rm {B1})}(n \cdot p, \tau^{\prime}, \mu)\,,
\end{eqnarray}
where the anomalous dimension $\gamma(\alpha_s)$ does not depend on the Dirac structures of
the ${\rm A0}$-type SCET currents, however,  the non-local evolution kernels $\gamma_i^{(\rm {B1})}(\tau^{\prime}, \tau)$
are dependent on the spin structures of the ${\rm B1}$-type SCET currents.
The general solutions to these RG equations can be written as
\begin{eqnarray}
C_i^{(\rm {A0})}(n \cdot p, \mu) &=& U_1(n \cdot p, \mu_h, \mu) \, C_i^{(\rm {A0})}(n \cdot p, \mu_h)\,,
\label{result of U1}  \\
C_i^{(\rm {B1})}(n \cdot p, \tau, \mu) &=& {\rm Exp} \left [-S(n \cdot p, \mu_h, \mu) \right ] \,
\int_0^1 d \tau^{\prime} \, U_i^{(\rm {B1})}(\tau, \tau^{\prime}, \mu_h, \mu) \, C_i^{(\rm {B1})}(n \cdot p, \tau^{\prime}, \mu_h) \,,
\hspace{0.5 cm}
\end{eqnarray}
where the NLL  approximation of the evolution function $U_1$ and the LL expansion of the $S$ function
can be found in \cite{Beneke:2011nf} and \cite{Beneke:2005gs}, respectively.
As the tree-level expressions of the ${\rm B1}$-type hard functions
$C_i^{(\rm {B1})}(n \cdot p, \tau, \mu_h)$ displayed in the Appendix \ref{appendix: hard functions}
are independent of the $\tau$ variable, the solution to the corresponding RG equation can be further reduced as
\begin{eqnarray}
C_{i, \, \rm{LL}}^{(\rm {B1})}(n \cdot p, \tau, \mu) &=& {\rm Exp} \left [-S(n \cdot p, \mu_h, \mu) \right ] \,
U_i^{(\rm {B1})}(\tau,  \mu_h, \mu) \, C_{i, \, \rm{LO}}^{(\rm {B1})}(n \cdot p, \mu_h) \,,
\label{evolution result of the B-type hard function}
\end{eqnarray}
with the newly defined evolution function
\begin{eqnarray}
U_i^{(\rm {B1})}(\tau,  \mu_h, \mu) = \int_0^1 d \tau^{\prime} \, U_i^{(\rm {B1})}(\tau, \tau^{\prime}, \mu_h, \mu)  \,.
\end{eqnarray}
An approximate solution to $U_i^{(\rm {B1})}(\tau,  \mu_h, \mu)$ (better than 1\%) at the LL accuracy reads
\begin{eqnarray}
U_{i, \, \rm{app}}^{(\rm {B1})}(\tau,  \mu_h, \mu) =
\left ( {\alpha_s(\mu) \over \alpha_s(\mu_h) } \right )^{-\gamma_i^{(\rm B1)}(\tau)/(2 \, \beta_0)}\,,
\end{eqnarray}
with the explicit expressions of $\gamma_i^{(\rm B1)}(\tau)$ given by \cite{Beneke:2005gs}
\begin{eqnarray}
\gamma_{\|}^{(\rm B1)}(\tau) &=& -C_F + 4 \, \left (C_F - {C_A \over 2} \right ) \, {\ln \bar \tau \over \tau } \,, \nonumber \\
\gamma_{\perp}^{(\rm B1)}(\tau) &=& -C_F  \, \left [{4 \, \tau \ln \tau \over \bar \tau } + 1 \right ]
+ 4 \, \left ( C_F - {C_A \over 2} \right ) \,
\left [ { 1 + \tau \over \tau } \, \ln \bar \tau +  {\tau \, \ln  \tau \over \bar \tau }  \right ] \,.
\end{eqnarray}
The large logarithmic resummation improved SCET factorization formulae can then be deduced  by substituting
the solutions (\ref{result of U1}) and (\ref{evolution result of the B-type hard function})
into (\ref{SCET-I factorization formulae}) with
\begin{eqnarray}
i &=& \|  \hspace{0.5 cm} {\rm for} \hspace{0.3 cm}  C_{f_0}^{(\rm B1)}, \,\, C_{f_+}^{(\rm B1)}, \,\, C_{f_T}^{(\rm B1)} \,, \nonumber \\
i &=& \perp  \hspace{0.3 cm} {\rm for} \hspace{0.3 cm}  C_{V}^{(\rm B1)}, \,\, C_{T_1}^{(\rm B1)} \,.
\end{eqnarray}

\section{The higher-twist corrections to $B \to V$ form factors}
\label{section: SCET rum rules at higher twist}

We are now in a position to compute the higher-twist corrections to the semileptonic $B \to V$ form factors
from both the two-particle and three-particle $B$-meson distribution amplitudes at tree level by
employing the LCSR approach.  To this end, we will need to establish the QCD factorization formulae for the
following vacuum-to-$B$-meson correlation functions
\begin{eqnarray}
\hat{\Pi}_{\mu, \|}^{(a)}(p, q) &=& \int d^4 x \,\, e^{i p \cdot x} \,\,
\langle 0 | {\rm T} \, \left \{ j_{\|}^{V}(x), \,\,
\bar q(0) \, \Gamma_{\mu}^{(a)}  \, b(0) \right \}  |  \bar B(p+q)\rangle \,, \nonumber \\
\hat{\Pi}_{\delta \mu, \perp}^{(a)}(p, q) &=& \int d^4 x \,\, e^{i p \cdot x} \,\,
\langle 0 | {\rm T} \, \left \{ j_{\delta, \perp}^{V}(x), \,\,
\bar q(0) \, \Gamma_{\mu}^{(a)}  \, b(0) \right \}  |  \bar B(p+q)\rangle \,,
\end{eqnarray}
where the interpolating QCD currents for the longitudinally and transversely polarized
vector mesons are given by
\begin{eqnarray}
j_{\|}^{V}(x)= \bar q^{\prime}(x) \, {\not \! n \over 2} \, q(x) \,, \qquad
j_{\delta, \perp}^{V}(x) = \bar q^{\prime}(x) \, {\not \! n \over 2} \, \gamma_{\delta \perp}\, q(x) \,,
\end{eqnarray}
and the Dirac structures of the heavy-to-light transition currents under discussion are
\begin{eqnarray}
\Gamma_{\mu}^{(a)} \in \left \{ \gamma_{\mu} \,  (1-\gamma_5),  \qquad
 \, i \, \sigma_{\mu \nu} \, (1 + \gamma_5)  \, q^{\nu} \right \} \,.
\end{eqnarray}

Employing the light-cone expansion of the quark propagator in the background gluon filed
up to the gluon field strength terms without the covariant derivatives \cite{Balitsky:1987bk}
(see also \cite{Rusov:2017chr} for an update for the massive quark case)
\begin{eqnarray}
\langle 0 | {\rm T} \, \{\bar q (x), q(0) \} | 0\rangle
 \supset   i \, g_s \, \int_0^{\infty} \,\, {d^4 l \over (2 \pi)^4} \, e^{- i \, l \cdot x} \,
\int_0^1 \, d u \, \left  [ {u \, x_{\mu} \, \gamma_{\nu} \over l^2 - m^2}
 - \frac{(\not  l + m) \, \sigma_{\mu \nu}}{2 \, (l^2 - m^2)^2}  \right ]
\, G^{\mu \nu}(u \, x), \hspace{0.8 cm}
\end{eqnarray}
with $G_{\mu \nu}=G_{\mu \nu}^{a} \, T^a=[D_{\mu}, A_{\nu}]$,
and applying the general parametrization of the vacuum-to-$B$-meson matrix element of
the three-body light-ray HQET operator \cite{Braun:2017liq} (see \cite{Kawamura:2001jm} for the original
but incomplete parametrization in terms of four independent  distribution amplitudes)
\begin{eqnarray}
&& \langle 0 | \bar q_{\alpha}(z_1 \, \bar n) \, g_s \, G_{\mu \nu}(z_2 \, \bar n) \,
h_{v \, \beta}(0) | \bar B_v \rangle \nonumber \\
&& = {\tilde{f}_B(\mu) \, m_B \over 4} \,
\bigg [ (1 + \not v) \, \bigg \{ (v_{\mu} \gamma_{\nu} - v_{\nu} \gamma_{\mu})  \,
\left [\Psi_A(z_1, z_2, \mu) - \Psi_V(z_1, z_2, \mu) \right ]
- i \, \sigma_{\mu \nu} \, \Psi_V(z_1, z_2, \mu) \nonumber  \\
&& \hspace{0.4 cm}
- (\bar n_{\mu} \, v_{\nu} - \bar n_{\nu} \, v_{\mu} ) \, X_A(z_1, z_2, \mu)
+ (\bar n_{\mu} \, \gamma_{\nu} - \bar n_{\nu} \, \gamma_{\mu} ) \,
\left [ W(z_1, z_2, \mu)  + Y_A(z_1, z_2, \mu)   \right ] \nonumber \\
&& \hspace{0.4 cm} + \, i \, \epsilon_{\mu \nu \alpha \beta} \,
\bar n^{\alpha} \, v^{\beta}  \, \gamma_5 \, \tilde{X}_A(z_1, z_2, \mu)
- \, i \, \epsilon_{\mu \nu \alpha \beta} \,
\bar n^{\alpha} \, \gamma^{\beta}  \, \gamma_5 \, \tilde{Y}_A(z_1, z_2, \mu)  \nonumber \\
&&  \hspace{0.4 cm}  - \, ( \bar n_{\mu} \, v_{\nu} -  \bar n_{\nu} \, v_{\mu} ) \,
\not \bar  n \, W(z_1, z_2, \mu)
+ \, ( \bar n_{\mu} \, \gamma_{\nu} - \bar n_{\nu} \, \gamma_{\mu} ) \,
\not \bar  n \, Z(z_1, z_2, \mu)   \bigg \}  \, \gamma_5 \bigg ]_{\beta \, \alpha}  \,,
\label{def: 3-particle B-meson DAs}
\end{eqnarray}
we can readily derive the three-particle higher twist corrections
to the aforementioned correlation functions at LO in ${\cal O}(\alpha_s)$
\begin{eqnarray}
\hat{\Pi}_{\mu, \|}^{(V-A), \, \rm{3P}}(p, q) &=& - {\tilde{f}_B(\mu) \, m_B \over 2 \, n \cdot p} \,
\int_0^{\infty} \, d \omega_1 \, \int_0^{\infty} \, d \omega_2 \, \int_0^{1} \, d u  \,
{1 \over [\bar n \cdot p - \omega_1-  u \, \omega_2 + i 0]^2} \, \nonumber \\
&&   \times \, \bigg \{ \bar n_{\mu} \, \left [\rho_{\bar n, \|, \, \rm {LP}}^{(V-A), \, \rm{3P}}(u, \omega_1, \omega_2)
+ {m \over n \cdot p} \, \rho_{\bar n, \|, \, \rm {NLP}}^{(V-A), \, \rm{3P}}(u, \omega_1, \omega_2)  \right ] \nonumber \\
&& \hspace{0.8 cm} + \,  n_{\mu} \, \left [\rho_{n, \|, \, \rm {LP}}^{(V-A), \, \rm{3P}}(u, \omega_1, \omega_2)
+ {m \over n \cdot p} \, \rho_{n, \|, \, \rm {NLP}}^{(V-A), \, \rm{3P}}(u, \omega_1, \omega_2)  \right ]   \bigg \},  \nonumber  \\
\hat{\Pi}_{\delta \mu, \perp}^{(V-A), \, \rm{3P}}(p, q)  &=&   - {\tilde{f}_B(\mu) \, m_B \over 2 \, n \cdot p} \,
\int_0^{\infty} \, d \omega_1 \, \int_0^{\infty} \, d \omega_2 \, \int_0^{1} \, d u  \,
{1 \over [\bar n \cdot p - \omega_1-  u \, \omega_2 + i 0]^2}  \nonumber \\
&& \hspace{-1.8 cm}  \times \, \bigg \{ \left [g_{\delta \mu \perp} + i \, \epsilon_{\delta \mu \perp} \right ] \,
\rho_{\perp, \, \rm{LP}}^{(V-A), \, \rm{3P}}(u,\omega_1, \omega_2 )
+  {m \over n \cdot p} \,  \left [g_{\delta \mu \perp} - i \, \epsilon_{\delta \mu \perp} \right ] \,
\rho_{\perp, \, \rm{NLP}}^{(V-A), \, \rm{3P}}(u,\omega_1, \omega_2 ) \bigg \}   \,, \nonumber \\
\hat{\Pi}_{\mu, \|}^{(T + \tilde{T}), \, \rm{3P}}(p, q) &=&  {\tilde{f}_B(\mu) \, m_B \over 4 \, n \cdot p} \,
\int_0^{\infty} \, d \omega_1 \, \int_0^{\infty} \, d \omega_2 \, \int_0^{1} \, d u  \,
{1 \over [\bar n \cdot p - \omega_1-  u \, \omega_2 + i 0]^2} \, \nonumber  \\
&&  \times  \, \left (\bar n_{\mu} \, n \cdot q - n_{\mu} \, \bar n \cdot q  \right ) \,\,
\left [ \rho_{\|, \, \rm{LP}}^{(T+\tilde{T}), \, \rm{3P}}(u,\omega_1, \omega_2 )
+  {m \over n \cdot p} \,  \rho_{\|, \, \rm{NLP}}^{(T+\tilde{T}), \, \rm{3P}}(u,\omega_1, \omega_2 )  \right ] \,,  \nonumber \\
\hat{\Pi}_{\delta \mu, \perp}^{(T + \tilde{T}), \, \rm{3P}}(p, q) &=&  {\tilde{f}_B(\mu) \, m_B \over 2 \, n \cdot p} \,
\, \bar n \cdot q \,  \int_0^{\infty} \, d \omega_1 \, \int_0^{\infty} \, d \omega_2 \, \int_0^{1} \, d u  \,
{1 \over [\bar n \cdot p - \omega_1-  u \, \omega_2 + i 0]^2} \, \nonumber \\
&& \hspace{-1.8 cm} \times \, \bigg \{ \left [g_{\delta \mu \perp} + i \, \epsilon_{\delta \mu \perp} \right ] \,
\rho_{\perp, \, \rm{LP}}^{(T+\tilde{T}), \, \rm{3P}}(u,\omega_1, \omega_2 )
+  {m \, n \cdot q \over 2 \, p \cdot q} \,
\left [g_{\delta \mu \perp} - i \, \epsilon_{\delta \mu \perp} \right ] \,
\rho_{\perp, \, \rm{NLP}}^{(T+\tilde{T}), \, \rm{3P}}(u,\omega_1, \omega_2 ) \bigg \}.  \nonumber \\
\label{factorization formulae of the 3P cntribution}
\end{eqnarray}
The explicit expressions of the invariant functions entering the tree-level factorization formulae
(\ref{factorization formulae of the 3P cntribution}) can be written as follows
\begin{eqnarray}
\rho_{\bar n, \|, \, \rm {LP}}^{(V-A), \, \rm{3P}} &=&
(2\, u -1) \, (X_A - \Psi_A-2\, Y_A) + \tilde{X}_A + \Psi_V - 2 \, \tilde{Y}_A \,, \nonumber \\
\rho_{\bar n, \|, \, \rm {NLP}}^{(V-A), \, \rm{3P}}&=&
2 \,(\Psi_A - \Psi_V) + 4 \, (W+ Y_A + \tilde{Y}_A - 2\, Z) \,, \nonumber \\
\rho_{n, \|, \, \rm {LP}}^{(V-A), \, \rm{3P}}&=& 2 \, (1-u) \, (\Psi_A + \Psi_V)\,, \nonumber \\
\rho_{n, \|, \, \rm {NLP}}^{(V-A), \, \rm{3P}}&=& (\Psi_A - \Psi_V)
- (X_A + \tilde{X}_A - 2\, Y_A  - 2 \, \tilde{Y}_A )\,, \nonumber \\
\rho_{\perp, \, \rm {LP}}^{(V-A), \, \rm{3P}}&=& (2 \, u -1) \, (X_A - \Psi_A - 2 \, Y_A)
- \tilde{X}_A - \Psi_V +  2 \, \tilde{Y}_A  \,, \nonumber \\
\rho_{\perp, \, \rm {NLP}}^{(V-A), \, \rm{3P}}&=& -(\Psi_A + \Psi_V)
+ X_A - \tilde{X}_A - 2 \, (Y_A - \tilde{Y}_A)  \,, \nonumber \\
\rho_{\|, \, \rm{LP}}^{(T+\tilde{T}), \, \rm{3P}}&=&
(2\, u -1) \, (X_A + \Psi_V - 2\, Y_A) + \tilde{X}_A - \Psi_A - 2 \, \tilde{Y}_A\,, \nonumber \\
\rho_{\|, \, \rm{NLP}}^{(T+\tilde{T}), \, \rm{3P}}&=&
(\Psi_A - \Psi_V) +   X_A + \tilde{X}_A + 2 \, (Y_A + \tilde{Y}_A)
+ 4 \, (W- 2 \, Z)\,, \nonumber \\
\rho_{\perp, \, \rm{LP}}^{(T+\tilde{T}), \, \rm{3P}}&=&
(2 \, u -1) \, (X_A - \Psi_A - 2 \, Y_A)
- \tilde{X}_A - \Psi_V +  2 \, \tilde{Y}_A  \,, \nonumber \\
\rho_{\perp, \, \rm{NLP}}^{(T+\tilde{T}), \, \rm{3P}}&=&
-(\Psi_A + \Psi_V) + X_A - \tilde{X}_A - 2 \, (Y_A - \tilde{Y}_A)  \,.
\end{eqnarray}
Apparently, our results for the six invariant functions  appearing in the
factorization formulae of $\hat{\Pi}_{\mu, \|}^{(a)}$ are identical
to the corresponding coefficient functions in the QCD representations
of the vacuum-to-$B$ correlating functions defined by the pseudoscalar-meson
interpolating current and the $b \to q$ weak transition currents.
In addition, two interesting relations for the four invariant functions
entering the tree-level factorization formulae of $\hat{\Pi}_{\delta \mu, \perp}^{(a)}$
\begin{eqnarray}
\rho_{\perp, \, \rm{LP}}^{(V-A), \, \rm{3P}}=\rho_{\perp, \, \rm{LP}}^{(T+\tilde{T}), \, \rm{3P}}, \qquad
\rho_{\perp, \, \rm{NLP}}^{(V-A), \, \rm{3P}}=\rho_{\perp, \, \rm{NLP}}^{(T+\tilde{T}), \, \rm{3P}},
\end{eqnarray}
can be  established due to the equation of motion for the effective heavy quark.

We proceed to compute the higher-twist two-particle corrections to the correlation functions
$\hat{\Pi}_{\mu, \|}^{(a)}$ and $\hat{\Pi}_{\delta \mu, \perp}^{(a)}$, from
the non-vanishing partonic transverse momenta, to fulfill the non-trivial constraints due to
the classical QCD equations of motion.
Keeping the light-cone correction to the HQET matrix element of the two-body light-ray operator
up to the ${\cal O}(x^2)$ accuracy, it is straightforward to generalize the previous definition
(\ref{Definition of B-meson LCDA})  beyond the light-cone approximation
\begin{eqnarray}
&& \langle  0 | \left (\bar q_s \, Y_s \right)_{\beta} (x) \,
\left (Y_s^{\dag} \, h_v \right )_{\alpha}(0)| \bar B_v \rangle \nonumber \\
&& = - \frac{i \tilde f_B(\mu) \, m_B}{4}  \,
\int_0^{\infty} \, d \omega \, e^{- i \, \omega \, v \cdot x} \,
\bigg [  \frac{1+ \! \! \not v}{2} \, \,
\bigg \{ 2 \, \left [ \phi_{B}^{+}(\omega, \mu) + x^2 \, g_B^{+}(\omega, \mu)  \right ] \nonumber \\
&& \hspace{0.5 cm} - {\not \! x  \over v \cdot x}  \,
\left  [ \left ( \phi_{B}^{+}(\omega, \mu) - \phi_{B}^{-}(\omega, \mu)  \right )
+  x^2 \, \left ( g_{B}^{+}(\omega, \mu) - g_{B}^{-}(\omega, \mu)  \right )   \right ]  \,
\bigg \}  \, \gamma_5 \bigg ]_{\alpha \beta} \,.
\label{def: general two-particle B-meson DAs}
\end{eqnarray}
Applying the precise operator identities for the light-cone HQET operators \cite{Kawamura:2001jm}
\begin{eqnarray}
&& {\partial \over \partial x^{\mu}}  \left (\bar q_s \, Y_s \right) (x) \,
\gamma^{\mu}  \Gamma \, \left (Y_s^{\dag} \, h_v \right )(0)  \nonumber \\
&& = - i \int_0^1 \, d u \, u \, \left (\bar q_s \, Y_s \right) (x)  \,
x^{\alpha}  g_s \,\left ( Y_s^{\dag} \, G_{\alpha \mu} Y_s \right ) (u x) \,
\gamma^{\mu}  \Gamma \, \left (Y_s^{\dag} \, h_v \right )(0) \,,  \\
&& v_{\mu} \, {\partial \over \partial x_{\mu}} \, \left (\bar q_s \, Y_s \right) (x) \,
\Gamma \, \left (Y_s^{\dag} \, h_v \right )(0) \nonumber \\
&& = i \, \int_0^1 \, d u \, \bar u \, \left (\bar q_s \, Y_s \right) (x)  \,
x^{\alpha} \, g_s \,\left ( Y_s^{\dag} \, G_{\alpha \mu} Y_s \right ) (u x) \,
v^{\mu}  \, \Gamma \, \left (Y_s^{\dag} \, h_v \right )(0)  \nonumber \\
&& \hspace{0.3 cm} + \, (v \cdot \partial) \, \left (\bar q_s \, Y_s \right) (x) \,\,
 \Gamma \, \left (Y_s^{\dag} \, h_v \right )(0) \,,
\end{eqnarray}
one can express $g_B^{+}(\omega, \mu)$ and $g_B^{-}(\omega, \mu) $
in terms of the higher-twist three-particle
$B$-meson distribution amplitudes \cite{Braun:2017liq,Lu:2018cfc}
\begin{eqnarray}
-2 \, {d^2 \over d \omega^2} \, g_B^{+}(\omega, \mu) &=&
\left [ {3 \over 2} + (\omega - \bar \Lambda) \, {d \over d \omega}   \right ] \, \phi_B^{+}(\omega, \mu)
- {1 \over 2}  \, \phi_B^{-}(\omega, \mu)
+ \int_0^{\infty} \, {d \omega_2 \over \omega_2 } \, {d \over d \omega} \, \Psi_4(\omega, \omega_2, \mu) \nonumber \\
&& - \int_0^{\infty} \, {d \omega_2 \over \omega_2^2 } \, \Psi_4(\omega, \omega_2, \mu)
+ \int_0^{\omega} \, {d \omega_2 \over \omega_2^2 } \, \Psi_4(\omega-\omega_2, \omega_2, \mu) \,,
\label{the gB-plus EOM}  \\
-2 \, {d^2 \over d \omega^2} \, g_B^{-}(\omega, \mu) &=&
\left [ {3 \over 2} + (\omega - \bar \Lambda) \, {d \over d \omega}   \right ] \, \phi_B^{-}(\omega, \mu)
- {1 \over 2}  \, \phi_B^{+}(\omega, \mu)
+ \int_0^{\infty} \, {d \omega_2 \over \omega_2 } \, {d \over d \omega} \, \Psi_5(\omega, \omega_2, \mu) \nonumber \\
&& - \int_0^{\infty} \, {d \omega_2 \over \omega_2^2 } \, \Psi_5(\omega, \omega_2, \mu)
+ \int_0^{\omega} \, {d \omega_2 \over \omega_2^2 } \, \Psi_5(\omega-\omega_2, \omega_2, \mu) \,.
\label{the gB-minus EOM}
\end{eqnarray}
The resulting factorization formulae for the two-particle higher twist contributions to
$\hat{\Pi}_{\mu, \|}^{(a)}$ and $\hat{\Pi}_{\delta \mu, \perp}^{(a)}$
at tree level can be written as
\begin{eqnarray}
\hat{\Pi}_{\mu, \|}^{(V-A), \, \rm{2PHT}}(p, q) &=&
{2 \, \tilde{f}_B(\mu) \, m_B \over  n \cdot p} \, \bar n_{\mu} \,
\bigg \{ \int_0^{\infty} \, {d \omega \over (\bar n \cdot p - \omega)^2 } \,
\rho_{\|, \, 1}^{(V-A), \, \rm{2PHT}}(\omega, \mu)  \nonumber \\
&& + \, \int_0^{\infty} d \omega_1  \, \int_0^{\infty} d \omega_2  \, \int_0^{1} d u \,
{ \rho_{\|, \, 2}^{(V-A), \, \rm{2PHT}}(\omega_1, \omega_2, u, \mu)  \over (\bar n \cdot p - \omega_1 - u \, \omega_2)^2 } \,
\bigg \} \,,
\nonumber  \\
\hat{\Pi}_{\delta \mu, \perp}^{(V-A), \, \rm{2PHT}}(p, q)  &=&
{2 \, \tilde{f}_B(\mu) \, m_B \over  n \cdot p} \,
[g_{\delta \mu \perp} + i \,\epsilon_{\delta \mu \perp}  ] \,
\bigg \{ \int_0^{\infty} \, {d \omega \over (\bar n \cdot p - \omega)^2 } \,
\rho_{\perp, \, 1}^{(V-A), \, \rm{2PHT}}(\omega, \mu)  \nonumber \\
&& + \, \int_0^{\infty} d \omega_1  \, \int_0^{\infty} d \omega_2  \, \int_0^{1} d u \,
{ \rho_{\perp, \, 2}^{(V-A), \, \rm{2PHT}}(\omega_1, \omega_2, u, \mu)  \over (\bar n \cdot p - \omega_1 - u \, \omega_2)^2 } \,
\bigg \}   \,, \nonumber \\
\hat{\Pi}_{\mu, \|}^{(T+\tilde{T}), \, \rm{2PHT}}(p, q) &=&
{\tilde{f}_B(\mu) \, m_B \over  n \cdot p} \,
\left [\bar n \cdot q \, n_{\mu} - n \cdot q \, \bar n_{\mu}  \right ] \,
\bigg \{ \int_0^{\infty} \, {d \omega \over (\bar n \cdot p - \omega)^2 } \,
\rho_{\|, \, 1}^{(T+\tilde{T}), \, \rm{2PHT}}(\omega, \mu)  \nonumber \\
&& + \, \int_0^{\infty} d \omega_1  \, \int_0^{\infty} d \omega_2  \, \int_0^{1} d u \,
{ \rho_{\|, \, 2}^{(T+\tilde{T}), \, \rm{2PHT}}(\omega_1, \omega_2, u, \mu)  \over (\bar n \cdot p - \omega_1 - u \, \omega_2)^2 } \bigg \}  \,,
\nonumber  \\
\hat{\Pi}_{\delta \mu, \perp}^{(T+\tilde{T}), \, \rm{2PHT}}(p, q)  &=&
- {2 \, \tilde{f}_B(\mu) \, m_B \over  n \cdot p} \,\, \bar n \cdot q \,\,
[g_{\delta \mu \perp} + i \,\epsilon_{\delta \mu \perp}  ] \,
\bigg \{ \int_0^{\infty} \, {d \omega \over (\bar n \cdot p - \omega)^2 } \,
\rho_{\perp, \, 1}^{(T+\tilde{T}), \, \rm{2PHT}}(\omega, \mu)  \nonumber \\
&& + \, \int_0^{\infty} d \omega_1  \, \int_0^{\infty} d \omega_2  \, \int_0^{1} d u \,
{ \rho_{\perp, \, 2}^{(T+\tilde{T}), \, \rm{2PHT}}(\omega_1, \omega_2, u, \mu)  \over (\bar n \cdot p - \omega_1 - u \, \omega_2)^2 } \bigg \}   \,.
\end{eqnarray}
The explicit expressions of the newly introduced invariant functions are given by
\begin{eqnarray}
\rho_{\|, \, 1}^{(V-A), \, \rm{2PHT}} &=&
\rho_{\perp, \, 1}^{(V-A), \, \rm{2PHT}}
=\rho_{\|, \, 1}^{(T+\tilde{T}), \, \rm{2PHT}}
=\rho_{\perp, \, 1}^{(T+\tilde{T}), \, \rm{2PHT}}
= \hat{g}_B^{-}(\omega, \mu)\,,  \nonumber \\
\rho_{\|, \, 2}^{(V-A), \, \rm{2PHT}}&=&
\rho_{\perp, \, 2}^{(V-A), \, \rm{2PHT}}
=\rho_{\|, \, 2}^{(T+\tilde{T}), \, \rm{2PHT}}
=\rho_{\perp, \, 2}^{(T+\tilde{T}), \, \rm{2PHT}}
=-{1 \over 2} \, \bar u \, \Psi_5(\omega_1, \omega_2, \mu)\,,
\end{eqnarray}
with the ``genuine" two-particle twist-five two-particle  distribution amplitude
\begin{eqnarray}
\hat{g}_B^{-}(\omega, \mu) =  {1 \over 4} \, \int_{\omega}^{\infty}  \, d \rho \,
\bigg \{ (\rho - \omega) \, \left [ \phi_B^{+}(\rho) -  \phi_B^{-}(\rho) \right ]
- 2 \, (\bar \Lambda - \rho)  \, \phi_B^{-}(\rho) \bigg \}  \,.
\label{def: gBminhat}
\end{eqnarray}

Adding up the two-particle and three-particle higher twist corrections
to the vacuum-to-$B$-meson correlation functions
$\hat{\Pi}_{\mu, \|}^{(a)}$ and $\hat{\Pi}_{\delta \mu, \perp}^{(a)}$ yields
\begin{eqnarray}
\hat{\Pi}_{\mu, \|}^{(V-A), \, \rm{HT}}(p, q) &=&
- {\tilde{f}_B(\mu) \, m_B \over  2 \, n \cdot p} \,
\bigg \{ \int_0^{\infty} \, {d \omega \over (\bar n \cdot p - \omega)^2 } \,\,
\bar n_{\mu} \,\, \left [ -4 \, \rho_{\|, \, 1}^{(V-A), \rm{2PHT}}(\omega, \mu) \right ]  \nonumber \\
&&  + \, \int_0^{\infty} d \omega_1  \, \int_0^{\infty} d \omega_2  \, \int_0^{1} d u \,
{ 1 \over (\bar n \cdot p - \omega_1 - u \, \omega_2)^2 } \, \nonumber \\
&&  \hspace{0.4 cm} \bigg  [\bar n_{\mu}  \, \left ( \rho_{\bar n, \|, \, \rm{LP}}^{(V-A)}(\omega_1, \omega_2, u, \mu)
+ {m \over n \cdot p} \,\,  \rho_{\bar n, \|, \, \rm{NLP}}^{(V-A)}(\omega_1, \omega_2, u, \mu)  \right ) \nonumber \\
&& \hspace{0.6 cm} + \, n_{\mu}  \, \left ( \rho_{n, \|, \, \rm{LP}}^{(V-A)}(\omega_1, \omega_2, u, \mu)
+ {m \over n \cdot p} \,\,  \rho_{n, \|, \, \rm{NLP}}^{(V-A)}(\omega_1, \omega_2, u, \mu)  \right )   \bigg ]  \bigg \}   \,,
\nonumber \\
\hat{\Pi}_{\delta \mu, \perp}^{(V-A), \, \rm{HT}}(p, q) &=&
- {\tilde{f}_B(\mu) \, m_B \over  2 \, n \cdot p} \,
\bigg \{ (g_{\delta \mu \perp} + i \, \epsilon_{\delta \mu \perp}) \,
\int_0^{\infty} \, {d \omega \over (\bar n \cdot p - \omega)^2 } \,\,
\left [ -4 \, \rho_{\perp, \, 1}^{(V-A), \rm{2PHT}}(\omega, \mu) \right ]  \nonumber \\
&& + \, \int_0^{\infty} d \omega_1  \, \int_0^{\infty} d \omega_2  \, \int_0^{1} d u \,
{ 1 \over (\bar n \cdot p - \omega_1 - u \, \omega_2)^2 } \, \nonumber \\
&&  \hspace{0.4 cm} \bigg  [  (g_{\delta \mu \perp} + i \, \epsilon_{\delta \mu \perp}) \,
\rho_{\perp, \, \rm{LP}}^{(V-A)}(\omega_1, \omega_2, u, \mu)  \nonumber \\
&& \hspace{0.6 cm} + {m \over n \cdot p} \, (g_{\delta \mu \perp} - i \, \epsilon_{\delta \mu \perp}) \,
\rho_{\perp, \, \rm{NLP}}^{(V-A)}(\omega_1, \omega_2, u, \mu)    \bigg ]   \bigg \}   \,, \nonumber \\
\hat{\Pi}_{\mu, \|}^{(T+\tilde{T}), \, \rm{HT}}(p, q) &=&
{\tilde{f}_B(\mu) \, m_B \over  4 \, n \cdot p} \,
\left [n \cdot q \, \bar n_{\mu}   - \bar n \cdot q \, n_{\mu} \right ] \,
 \bigg \{  \int_0^{\infty} \, {d \omega \over (\bar n \cdot p - \omega)^2 } \,\,
\left [ -4 \, \rho_{\|, \, 1}^{(T+\tilde{T}), \rm{2PHT}}(\omega, \mu) \right ]  \nonumber \\
&& + \, \int_0^{\infty} d \omega_1  \, \int_0^{\infty} d \omega_2  \, \int_0^{1} d u \,
{ 1 \over (\bar n \cdot p - \omega_1 - u \, \omega_2)^2 } \, \nonumber \\
&&  \hspace{0.4 cm}  \left [  \rho_{\|, \, \rm{LP}}^{(T+\tilde{T})}(\omega_1, \omega_2, u, \mu)
+ {m \over n \cdot p} \, \rho_{\|, \, \rm{NLP}}^{(T+\tilde{T})}(\omega_1, \omega_2, u, \mu)  \right ] \bigg \}  \,, \nonumber \\
\hat{\Pi}_{\delta \mu, \perp}^{(T+\tilde{T}), \, \rm{HT}}(p, q) &=&
{\tilde{f}_B(\mu) \, m_B \over  2 \, n \cdot p} \,\,   \bar n \cdot q \,\,
\bigg \{ (g_{\delta \mu \perp} + i \, \epsilon_{\delta \mu \perp}) \,
\int_0^{\infty} \, {d \omega \over (\bar n \cdot p - \omega)^2 } \,\,
\left [ -4 \, \rho_{\perp, \, 1}^{(T+\tilde{T}), \rm{2PHT}}(\omega, \mu) \right ]  \nonumber \\
&& + \, \int_0^{\infty} d \omega_1  \, \int_0^{\infty} d \omega_2  \, \int_0^{1} d u \,
{ 1 \over (\bar n \cdot p - \omega_1 - u \, \omega_2)^2 } \, \nonumber \\
&&  \hspace{0.4 cm} \bigg  [  (g_{\delta \mu \perp} + i \, \epsilon_{\delta \mu \perp}) \,
\rho_{\perp, \, \rm{LP}}^{(T+\tilde{T})}(\omega_1, \omega_2, u, \mu)  \nonumber \\
&& \hspace{0.6 cm} + {m \,\, n \cdot q \over 2 \,\, p \cdot q} \, (g_{\delta \mu \perp} - i \, \epsilon_{\delta \mu \perp}) \,
\rho_{\perp, \, \rm{NLP}}^{(T+\tilde{T})}(\omega_1, \omega_2, u, \mu)    \bigg ]   \bigg \}    \,,
\label{LO factorization formula of the higher-twist corrections}
\end{eqnarray}
where we have introduced the following  conventions
\begin{eqnarray}
\rho_{\bar n, \|, \, \rm{LP}}^{(V-A)} &=& \Psi_5 - \tilde{\Psi}_5 \,, \qquad  \hspace{2.6 cm}
\rho_{\bar n, \|, \, \rm{NLP}}^{(V-A)} = 2 \, \Phi_6 \,, \nonumber \\
\rho_{n, \|, \, \rm{LP}}^{(V-A)} &=& 2 \, (1-u) \, \Phi_4  \,, \qquad  \hspace{2.1 cm}
\rho_{n, \|, \, \rm{NLP}}^{(V-A)} = \tilde{\Psi}_5 - \Psi_5 \,, \nonumber \\
\rho_{\perp, \, \rm{LP}}^{(V-A)} &=& \Psi_5 + \tilde{\Psi}_5  \,, \qquad \hspace{2.7 cm}
\rho_{\perp, \, \rm{NLP}}^{(V-A)}= \Psi_5 + \tilde{\Psi}_5  \,, \nonumber \\
\rho_{\|, \, \rm{LP}}^{(T+\tilde{T})} &=& 2 \, (u-1) \, \Phi_4 + \Psi_5 - \tilde{\Psi}_5 \,, \qquad  \hspace{0.1 cm}
\rho_{\|, \, \rm{NLP}}^{(T+\tilde{T})} = 2\, \Phi_6 + \Psi_5 - \tilde{\Psi}_5 \,, \nonumber \\
\rho_{\perp, \, \rm{LP}}^{(T+\tilde{T})} &=& \Psi_5 + \tilde{\Psi}_5  \,, \qquad  \hspace{2.7 cm}
\rho_{\perp, \, \rm{NLP}}^{(T+\tilde{T})} = \Psi_5 + \tilde{\Psi}_5 \,.
\end{eqnarray}
Following the standard strategy, we need to  write down the hadronic dispersion relations for
the  above-mentioned correlation functions
\begin{eqnarray}
\hat{\Pi}_{\mu, \|}^{(V-A)}(p, q) &=&
{1 \over 2} \, {f_{V, \|} \,\, m_V  \over m_V^2/n \cdot p - \bar n \cdot p - i0} \,
\left ( {n \cdot p \over 2 \, m_V} \right )^2  \,
\bigg \{ {m_B \over m_B - n \cdot p} \, n_{\mu} \, \nonumber \\
&& \hspace{0.3 cm} \left [ \left ( - {2\, m_V  \over n \cdot p} \, A_0(q^2) \right )
+ \left (  {m_B + m_V \over n \cdot p} \, A_1(q^2)
- {m_B - m_V \over m_B} \, A_2(q^2) \right )  \right ] \nonumber \\
&& - \bar n_{\mu} \,
\left [ \left (  {2\, m_V  \over n \cdot p} \, A_0(q^2) \right )
+ \left (  {m_B + m_V \over n \cdot p} \, A_1(q^2)
- {m_B - m_V \over m_B} \, A_2(q^2) \right )  \right ] \bigg \}  \nonumber \\
&& + \int d \omega^{\prime} \, {1 \over \omega^{\prime} - \bar n \cdot p - i 0} \,
\left [n_{\mu} \, \varrho_{n, \|}^{(V-A)}(\omega^{\prime}, n \cdot p)
+ \bar n_{\mu} \, \varrho_{\bar n, \|}^{(V-A)}(\omega^{\prime}, n \cdot p)  \right ],
\hspace{0.8 cm} \nonumber \\
\hat{\Pi}_{\delta \mu, \perp}^{(V-A)}(p, q) &=&
- {1 \over 2} \, {f_{V, \perp}(\nu) \,\, (n \cdot p)^2  \over m_V^2/n \cdot p - \bar n \cdot p - i0} \,  \nonumber \\
&& \bigg [ g_{\delta \mu \perp} \, \left ( {m_B + m_V \over n \cdot p} \, A_1(q^2) \right )
+  i \, \epsilon_{\delta \mu \perp} \, \left (  {m_B \over m_B +m_V} \, V(q^2) \right )  \bigg  ] \nonumber \\
&& + \int d \omega^{\prime} \, {1 \over \omega^{\prime} - \bar n \cdot p - i 0} \,
\left [ g_{\delta \mu \perp} \, \varrho_{\perp, A_1}^{(V-A)}(\omega^{\prime}, n \cdot p)
+  i \, \epsilon_{\delta \mu \perp} \, \, \varrho_{\perp, V}^{(V-A)}(\omega^{\prime}, n \cdot p)  \right ], \nonumber \\
\hat{\Pi}_{\mu, \|}^{(T+\tilde{T})}(p, q) &=&
{1 \over 2} \, {f_{V, \|} \,\, m_V  \over m_V^2/n \cdot p - \bar n \cdot p - i0} \,
\left ( {n \cdot p \over 2 \, m_V} \right )^2  \,
\left [n \cdot q \, \bar n_{\mu}   - \bar n \cdot q \, n_{\mu} \right ]  \,
\left  [ {m_B \over n \cdot p} \, T_2(q^2) - T_3(q^2) \right ]  \nonumber \\
&& +  \int d \omega^{\prime} \, {1 \over \omega^{\prime} - \bar n \cdot p - i 0} \,
\left [n \cdot q \, \bar n_{\mu}   - \bar n \cdot q \, n_{\mu} \right ]  \,
\varrho_{\|}^{(T+\tilde{T})}(\omega^{\prime}, n \cdot p)\,, \nonumber \\
\hat{\Pi}_{\delta \mu, \perp}^{(T+\tilde{T})}(p, q) &=&
{1 \over 2} \, {f_{V, \perp}(\nu) \,\, n \cdot p \, m_B \over m_V^2/n \cdot p - \bar n \cdot p - i0} \,
\left [g_{\delta \mu \perp} \,\, \left ( {m_B \over n \cdot p} \, T_2(q^2) \right )
+ i \, \epsilon_{\delta \mu \perp}  \, T_1(q^2) \right ] \nonumber \\
&& + \int d \omega^{\prime} \, {1 \over \omega^{\prime} - \bar n \cdot p - i 0} \,
\left [ g_{\delta \mu \perp} \, \varrho_{\perp, T_2}^{(T+\tilde{T})}(\omega^{\prime}, n \cdot p)
+  i \, \epsilon_{\delta \mu \perp} \, \, \varrho_{\perp, T_1}^{(T+\tilde{T})}(\omega^{\prime}, n \cdot p)  \right ].
\label{dispersion relation for the correlators: higher twist}
\end{eqnarray}
Matching the dispersion representations of the tree-level
factorization formulae (\ref{LO factorization formula of the higher-twist corrections})
with the hadronic representations of the vacuum-to-$B$-meson correlation functions
(\ref{dispersion relation for the correlators: higher twist}) and
applying the parton-hadron duality approximation leads to the desired sum rules
for the higher-twist contributions to the semileptonic $B \to V$ form factors
\begin{eqnarray}
&& - f_{V, \|} \,\, m_V \, \left ( {n \cdot p \over 2 \, m_V} \right )^2 \,\,
{\rm Exp} \left [ - {m_V^2 \over n \cdot p \, \omega_M} \right ] \,\,
\left [ {2 \, m_V \over n \cdot p} \,\, A_0^{\rm {HT}}(q^2) \right ] \nonumber \\
&& = - {\tilde{f}_B(\mu) \, m_B \over  2 \, n \cdot p} \,
\bigg \{   \int_0^{\omega_s} \, d \omega \, e^{-\omega/\omega_M} \,
\left [- 4 \,\, {d \over d  \omega} \, \rho_{\|, \, 1}^{(V-A), \rm{2PHT}}(\omega, \mu)   \right ] \nonumber \\
&& \hspace{0.5 cm} + \, \int_0^{\omega_s} \, d \omega_1 \, \int_{\omega_s-\omega_1}^{\infty} \,
{d \omega_2 \over \omega_2} \, e^{-\omega_s /\omega_M} \,
\bigg [ \left ( \rho_{\bar n, \|, \, \rm{LP}}^{(V-A)}(\omega_1, \omega_2, u, \mu)
 + {m \over n \cdot p} \,\,  \rho_{\bar n, \|, \, \rm{NLP}}^{(V-A)}(\omega_1, \omega_2, u, \mu)  \right ) \nonumber \\
&& \hspace{0.9 cm}  + \, {m_B - n \cdot p \over m_B} \,\,
\left ( \rho_{n, \|, \, \rm{LP}}^{(V-A)}(\omega_1, \omega_2, u, \mu)
+ {m \over n \cdot p} \,\,  \rho_{n, \|, \, \rm{NLP}}^{(V-A)}(\omega_1, \omega_2, u, \mu)  \right )  \bigg ]
\bigg|_{u=(\omega_s-\omega_1)/\omega_2} \nonumber \\
&& \hspace{0.5 cm} + \, \int_0^{\omega_s} \, d \omega^{\prime} \, \int_0^{\omega^{\prime}} d \omega_1 \,
\int_{\omega^{\prime}-\omega_1}^{\infty} \, { d \omega_2 \over \omega_2}  \,
{ e^{-\omega^{\prime} /\omega_M} \over \omega_M} \,\, \nonumber \\
&&  \hspace{0.5 cm}  \bigg [ \left ( \rho_{\bar n, \|, \, \rm{LP}}^{(V-A)}(\omega_1, \omega_2, u, \mu)
+ {m \over n \cdot p} \,\,  \rho_{\bar n, \|, \, \rm{NLP}}^{(V-A)}(\omega_1, \omega_2, u, \mu)  \right ) \nonumber \\
&& \hspace{0.5 cm}  + \, {m_B - n \cdot p \over m_B} \,\,
\left ( \rho_{n, \|, \, \rm{LP}}^{(V-A)}(\omega_1, \omega_2, u, \mu)
+ {m \over n \cdot p} \,\,  \rho_{n, \|, \, \rm{NLP}}^{(V-A)}(\omega_1, \omega_2, u, \mu)  \right )  \bigg ]
\bigg|_{u=(\omega^{\prime}-\omega_1)/\omega_2}  \bigg \}\,,   \\
\nonumber \\
&& - f_{V, \|} \,\, m_V \, \left ( {n \cdot p \over 2 \, m_V} \right )^2 \,\,
{\rm Exp} \left [ - {m_V^2 \over n \cdot p \, \omega_M} \right ] \,\,
\left [ {m_B + m_V \over n \cdot p} \,\, A_1^{\rm {HT}}(q^2)
-  {m_B - m_V \over m_B} \,\, A_2^{\rm {HT}}(q^2)  \right ] \nonumber \\
&& = - {\tilde{f}_B(\mu) \, m_B \over  2 \, n \cdot p} \,
\bigg \{   \int_0^{\omega_s} \, d \omega \, e^{-\omega/\omega_M} \,
\left [- 4 \,\, {d \over d  \omega} \, \rho_{\|, \, 1}^{(V-A), \rm{2PHT}}(\omega, \mu)   \right ] \nonumber \\
&& \hspace{0.5 cm} + \, \int_0^{\omega_s} \, d \omega_1 \, \int_{\omega_s-\omega_1}^{\infty} \,
{d \omega_2 \over \omega_2} \, e^{-\omega_s /\omega_M} \,
\bigg [ \left ( \rho_{\bar n, \|, \, \rm{LP}}^{(V-A)}(\omega_1, \omega_2, u, \mu)
+ {m \over n \cdot p} \,\,  \rho_{\bar n, \|, \, \rm{NLP}}^{(V-A)}(\omega_1, \omega_2, u, \mu)  \right ) \nonumber \\
&& \hspace{0.9 cm}   \textcolor{red} {-}  \, {m_B - n \cdot p \over m_B} \,\,
\left ( \rho_{n, \|, \, \rm{LP}}^{(V-A)}(\omega_1, \omega_2, u, \mu)
+ {m \over n \cdot p} \,\,  \rho_{n, \|, \, \rm{NLP}}^{(V-A)}(\omega_1, \omega_2, u, \mu)  \right )  \bigg ]
\bigg|_{u=(\omega_s-\omega_1)/\omega_2} \nonumber \\
&& \hspace{0.5 cm} + \, \int_0^{\omega_s} \, d \omega^{\prime} \, \int_0^{\omega^{\prime}} d \omega_1 \,
\int_{\omega^{\prime}-\omega_1}^{\infty} \, { d \omega_2 \over \omega_2}  \,
{ e^{-\omega^{\prime} /\omega_M} \over \omega_M} \,\, \nonumber \\
&& \hspace{0.9 cm}  \bigg [ \left ( \rho_{\bar n, \|, \, \rm{LP}}^{(V-A)}(\omega_1, \omega_2, u, \mu)
+ {m \over n \cdot p} \,\,  \rho_{\bar n, \|, \, \rm{NLP}}^{(V-A)}(\omega_1, \omega_2, u, \mu)  \right ) \nonumber \\
&& \hspace{0.9 cm} \textcolor{red} {-} \, {m_B - n \cdot p \over m_B} \,\,
\left ( \rho_{n, \|, \, \rm{LP}}^{(V-A)}(\omega_1, \omega_2, u, \mu)
+ {m \over n \cdot p} \,\,  \rho_{n, \|, \, \rm{NLP}}^{(V-A)}(\omega_1, \omega_2, u, \mu)  \right )  \bigg ]
\bigg|_{u=(\omega^{\prime}-\omega_1)/\omega_2}  \bigg \},  \hspace{1.1 cm} \\
\nonumber \\
&& - {1 \over 2} \, f_{V, \perp}(\nu) \, n \cdot p \,\,
{\rm Exp} \left [ - {m_V^2 \over n \cdot p \, \omega_M} \right ] \,\,
\left [ {m_B \over m_B + m_V } \,\, V^{\rm {HT}}(q^2)  \right ]  \nonumber \\
&& = - {\tilde{f}_B(\mu) \, m_B \over  2 \, n \cdot p} \,
\bigg \{   \int_0^{\omega_s} \, d \omega \, e^{-\omega/\omega_M} \,
\left [- 4 \,\, {d \over d  \omega} \, \rho_{\perp, \, 1}^{(V-A), \rm{2PHT}}(\omega, \mu)   \right ] \nonumber \\
&& \hspace{0.5 cm} + \, \int_0^{\omega_s} \, d \omega_1 \, \int_{\omega_s-\omega_1}^{\infty} \,
{d \omega_2 \over \omega_2} \, e^{-\omega_s /\omega_M} \, \nonumber \\
&& \hspace{0.9 cm} \left ( \rho_{\perp, \, \rm{LP}}^{(V-A)}(\omega_1, \omega_2, u, \mu)
- {m \over n \cdot p} \,\,  \rho_{\perp, \, \rm{NLP}}^{(V-A)}(\omega_1, \omega_2, u, \mu)  \right )
\bigg|_{u=(\omega_s-\omega_1) / \omega_2} \nonumber \\
&& \hspace{0.5 cm}  + \, \int_0^{\omega_s} \, d \omega^{\prime} \, \int_0^{\omega^{\prime}} d \omega_1 \,
\int_{\omega^{\prime}-\omega_1}^{\infty} \, { d \omega_2 \over \omega_2}  \,
{ e^{-\omega^{\prime} /\omega_M} \over \omega_M} \,\,  \nonumber \\
&& \hspace{0.9 cm} \left ( \rho_{\perp, \, \rm{LP}}^{(V-A)}(\omega_1, \omega_2, u, \mu)
- {m \over n \cdot p} \,\,  \rho_{\perp, \, \rm{NLP}}^{(V-A)}(\omega_1, \omega_2, u, \mu)  \right )
\bigg|_{u=(\omega^{\prime}-\omega_1) / \omega_2}  \,\, \bigg \},  \\
\nonumber \\
&& - {1 \over 2} \, f_{V, \perp}(\nu) \, n \cdot p \,\,
{\rm Exp} \left [ - {m_V^2 \over n \cdot p \, \omega_M} \right ] \,\,
\left [ {m_B + m_V \over n \cdot p} \,\, A_1^{\rm {HT}}(q^2)  \right ]  \nonumber \\
&& = - {\tilde{f}_B(\mu) \, m_B \over  2 \, n \cdot p} \,
\bigg \{   \int_0^{\omega_s} \, d \omega \, e^{-\omega/\omega_M} \,
\left [- 4 \,\, {d \over d  \omega} \, \rho_{\perp, \, 1}^{(V-A), \rm{2PHT}}(\omega, \mu)   \right ] \nonumber \\
&& \hspace{0.5 cm} + \, \int_0^{\omega_s} \, d \omega_1 \, \int_{\omega_s-\omega_1}^{\infty} \,
{d \omega_2 \over \omega_2} \, e^{-\omega_s /\omega_M} \, \nonumber \\
&& \hspace{0.9 cm} \left ( \rho_{\perp, \, \rm{LP}}^{(V-A)}(\omega_1, \omega_2, u, \mu)
\textcolor{red} {+} {m \over n \cdot p} \,\,  \rho_{\perp, \, \rm{NLP}}^{(V-A)}(\omega_1, \omega_2, u, \mu)  \right )
\bigg|_{u=(\omega_s-\omega_1) / \omega_2} \nonumber \\
&& \hspace{0.5 cm}  + \, \int_0^{\omega_s} \, d \omega^{\prime} \, \int_0^{\omega^{\prime}} d \omega_1 \,
\int_{\omega^{\prime}-\omega_1}^{\infty} \, { d \omega_2 \over \omega_2}  \,
{ e^{-\omega^{\prime} /\omega_M} \over \omega_M} \,\,  \nonumber \\
&& \hspace{0.9 cm} \left ( \rho_{\perp, \, \rm{LP}}^{(V-A)}(\omega_1, \omega_2, u, \mu)
\textcolor{red} {+} {m \over n \cdot p} \,\,  \rho_{\perp, \, \rm{NLP}}^{(V-A)}(\omega_1, \omega_2, u, \mu)  \right )
\bigg|_{u=(\omega^{\prime}-\omega_1) / \omega_2}  \,\, \bigg \},  \\
\nonumber \\
&& {1 \over 2} \, f_{V, \|} \, m_V \, \left ({ n \cdot p \over 2 \, m_V} \right )^2 \,\,
{\rm Exp} \left [ - {m_V^2 \over n \cdot p \, \omega_M} \right ] \,\,
\left [ {m_B \over n \cdot p} \,\, T_2^{\rm {HT}}(q^2)
- T_3^{\rm {HT}}(q^2)   \right ]  \nonumber \\
&& = {\tilde{f}_B(\mu) \, m_B \over  4 \, n \cdot p} \,
\bigg \{   \int_0^{\omega_s} \, d \omega \, e^{-\omega/\omega_M} \,
\left [- 4 \,\, {d \over d  \omega} \, \rho_{\|, \, 1}^{(T+\tilde{T}), \rm{2PHT}}(\omega, \mu)   \right ] \nonumber \\
&& \hspace{0.5 cm} + \, \int_0^{\omega_s} \, d \omega_1 \, \int_{\omega_s-\omega_1}^{\infty} \,
{d \omega_2 \over \omega_2} \, e^{-\omega_s /\omega_M} \, \nonumber \\
&& \hspace{0.9 cm} \left ( \rho_{\|, \, \rm{LP}}^{(T+\tilde{T})}(\omega_1, \omega_2, u, \mu)
+ {m \over n \cdot p} \,\,  \rho_{\|, \, \rm{NLP}}^{(T+\tilde{T})}(\omega_1, \omega_2, u, \mu)  \right )
\bigg|_{u=(\omega_s-\omega_1) / \omega_2} \nonumber \\
&& \hspace{0.5 cm}  + \, \int_0^{\omega_s} \, d \omega^{\prime} \, \int_0^{\omega^{\prime}} d \omega_1 \,
\int_{\omega^{\prime}-\omega_1}^{\infty} \, { d \omega_2 \over \omega_2}  \,
{ e^{-\omega^{\prime} /\omega_M} \over \omega_M} \,\,  \nonumber \\
&& \hspace{0.9 cm} \left ( \rho_{\|, \, \rm{LP}}^{(T+\tilde{T})}(\omega_1, \omega_2, u, \mu)
+ {m \over n \cdot p} \,\,  \rho_{\|, \, \rm{NLP}}^{(T+\tilde{T})}(\omega_1, \omega_2, u, \mu)  \right )
\bigg|_{u=(\omega^{\prime}-\omega_1) / \omega_2}  \,\,  \bigg \} \,, \\
\nonumber \\
&& {1 \over 2} \, f_{V, \perp}(\nu) \, n \cdot p  \,\,
{\rm Exp} \left [ - {m_V^2 \over n \cdot p \, \omega_M} \right ] \,\,
T_1^{\rm {HT}}(q^2) \nonumber \\
&& = {\tilde{f}_B(\mu) \, m_B \over  2 \, n \cdot p} \,
\bigg \{   \int_0^{\omega_s} \, d \omega \, e^{-\omega/\omega_M} \,
\left [- 4 \,\, {d \over d  \omega} \, \rho_{\perp, \, 1}^{(T+\tilde{T}), \rm{2PHT}}(\omega, \mu)   \right ] \nonumber \\
&& \hspace{0.5 cm} + \, \int_0^{\omega_s} \, d \omega_1 \, \int_{\omega_s-\omega_1}^{\infty} \,
{d \omega_2 \over \omega_2} \, e^{-\omega_s /\omega_M} \, \nonumber \\
&& \hspace{0.9 cm} \left ( \rho_{\perp, \, \rm{LP}}^{(T+\tilde{T})}(\omega_1, \omega_2, u, \mu)
- {m \,\, n \cdot q \over n \cdot p \, \bar n \cdot q } \,\,
\rho_{\perp, \, \rm{NLP}}^{(T+\tilde{T})}(\omega_1, \omega_2, u, \mu)  \right )
\bigg|_{u=(\omega_s-\omega_1) / \omega_2} \nonumber \\
&& \hspace{0.5 cm}  + \, \int_0^{\omega_s} \, d \omega^{\prime} \, \int_0^{\omega^{\prime}} d \omega_1 \,
\int_{\omega^{\prime}-\omega_1}^{\infty} \, { d \omega_2 \over \omega_2}  \,
{ e^{-\omega^{\prime} /\omega_M} \over \omega_M} \,\,  \nonumber \\
&& \hspace{0.9 cm} \left ( \rho_{\perp, \, \rm{LP}}^{(T+\tilde{T})}(\omega_1, \omega_2, u, \mu)
- {m \,\, n \cdot q \over n \cdot p \, \bar n \cdot q } \,\,
\rho_{\perp, \, \rm{NLP}}^{(T+\tilde{T})}(\omega_1, \omega_2, u, \mu)  \right )
\bigg|_{u=(\omega^{\prime}-\omega_1) / \omega_2}  \,\,  \bigg \} \,, \\
\nonumber \\
&& {1 \over 2} \, f_{V, \perp}(\nu) \, n \cdot p  \,\,
{\rm Exp} \left [ - {m_V^2 \over n \cdot p \, \omega_M} \right ] \,\,
\left [  {m_B \over n \cdot p} \,\, T_2^{\rm {HT}}(q^2) \right ]  \nonumber \\
&& = {\tilde{f}_B(\mu) \, m_B \over  2 \, n \cdot p} \,
\bigg \{   \int_0^{\omega_s} \, d \omega \, e^{-\omega/\omega_M} \,
\left [- 4 \,\, {d \over d  \omega} \, \rho_{\perp, \, 1}^{(T+\tilde{T}), \rm{2PHT}}(\omega, \mu)   \right ] \nonumber \\
&& \hspace{0.5 cm} + \, \int_0^{\omega_s} \, d \omega_1 \, \int_{\omega_s-\omega_1}^{\infty} \,
{d \omega_2 \over \omega_2} \, e^{-\omega_s /\omega_M} \, \nonumber \\
&& \hspace{0.9 cm} \left ( \rho_{\perp, \, \rm{LP}}^{(T+\tilde{T})}(\omega_1, \omega_2, u, \mu)
\textcolor{red} {+} {m \,\, n \cdot q \over n \cdot p \, \bar n \cdot q } \,\,
\rho_{\perp, \, \rm{NLP}}^{(T+\tilde{T})}(\omega_1, \omega_2, u, \mu)  \right )
\bigg|_{u=(\omega_s-\omega_1) / \omega_2} \nonumber \\
&& \hspace{0.5 cm}  + \, \int_0^{\omega_s} \, d \omega^{\prime} \, \int_0^{\omega^{\prime}} d \omega_1 \,
\int_{\omega^{\prime}-\omega_1}^{\infty} \, { d \omega_2 \over \omega_2}  \,
{ e^{-\omega^{\prime} /\omega_M} \over \omega_M} \,\,  \nonumber \\
&& \hspace{0.9 cm} \left ( \rho_{\perp, \, \rm{LP}}^{(T+\tilde{T})}(\omega_1, \omega_2, u, \mu)
\textcolor{red} {+} {m \,\, n \cdot q \over n \cdot p \, \bar n \cdot q } \,\,
\rho_{\perp, \, \rm{NLP}}^{(T+\tilde{T})}(\omega_1, \omega_2, u, \mu)  \right )
\bigg|_{u=(\omega^{\prime}-\omega_1) / \omega_2}  \,\,  \bigg \} \,.
\end{eqnarray}

Several comments on the subleading power contributions from the higher-twist $B$-meson distribution amplitudes
are in order.

\begin{itemize}

\item{The two-particle higher-twist corrections
preserve the large recoil symmetry relations for the soft contributions
to the semileptonic $B \to V$ form factors.
In addition, the twist-four $B$-meson distribution amplitude $g_B^{+}(\omega, \mu)$
will not appear in the tree-level sum rules due to the fact that
$\hat{\Pi}_{\mu, \|}^{(a)}$ and $\hat{\Pi}_{\delta \mu, \perp}^{(a)}$
are defined with the leading-power interpolating currents
for the longitudinally and transversely polarized vector mesons. }

\item{The three-particle  higher-twist $B$-meson distribution amplitudes can generate
the large-recoil symmetry breaking effects for the soft form factors already at tree level.
In particular, the two form-factor relations  presented in (\ref{exact form factor relations at LP}),
which are valid up to all orders in ${\cal O}(\alpha_s)$ at leading power in $\Lambda/m_b$,
will be violated by the subleading power corrections due to the light-quark mass contributions.  }

\end{itemize}

\section{Numerical analysis}
\label{section: numerical analysis}

The major objective of this section is the numerical  exploration of the resummation improved
LCSR for the semileptonic  $B \to V$ form factors including the subleading power corrections
from the higher-twist $B$-meson distribution amplitudes up to the twist-six accuracy.
Applying the $z$-series parametrization, we will further extrapolate the obtained LCSR predictions for
these  QCD  form factors at large hadronic recoil to the whole kinematical region.
Phenomenological applications of our results to the semileptonic $B \to (\rho,  \omega) \, \ell \nu_{\ell}$  decays
and the rare exclusive $B \to K^{\ast} \, \nu_{\ell} \, \bar \nu_{\ell}$ decays will be also discussed
with an emphasis on the determination of the CKM matrix element $|V_{ub}|$,
the normalized differential branching fractions,
and the $q^2$-binned $K^{\ast}$ longitudinal polarization fractions.

\subsection{Theory inputs}

The fundamental ingredients entering the derived sum rules for  $B \to V$ form factors include
the two-particle and three-particle $B$-meson distribution amplitudes up to the twist-six accuracy,
the decay constants of the $B$-meson and the light vector mesons as well as the intrinsic sum rule parameters.
We will employ two phenomenological models for the involved $B$-meson distribution amplitudes consistent
with the classical QCD equations of motion as constructed in \cite{Braun:2017liq,Lu:2018cfc},
whose explicit expressions will be collected in Appendix \ref{appendix: B-meson DAs} for completeness.
Two independent HQET parameters, $\lambda_B(\mu)$ and $R(\mu)=\lambda_E^2(\mu)/\lambda_H^2(\mu)$,
are introduced to parameterize the shapes of these non-perturbative distribution amplitudes
(see \cite{Beneke:2018wjp} for more discussions on the alternative parametrizations of the twist-two
LCDA).  Applying the Lange-Neubert evolution equation for $\phi_B^{+}(\omega, \mu)$ \cite{Lange:2003pk},
the RG evolution  of the inverse moment $\lambda_B(\mu)$ at the one-loop accuracy can be written as
\cite{Beneke:2011nf,Bell:2013tfa}
\begin{eqnarray}
{\lambda_B (\mu_0) \over \lambda_B (\mu)} =
1 + {\alpha_s(\mu_0) \, C_F \over 4 \, \pi} \, \ln {\mu \over \mu_0} \,
\left [ 2- 2 \,  \ln {\mu \over \mu_0} - 4 \, \sigma_B^{(1)}(\mu_0)  \right ]
+ {\cal O}(\alpha_s^2)\,,
\end{eqnarray}
with the inverse-logarithmic moment $\sigma_B^{(1)}$ given by
\begin{eqnarray}
\sigma_B^{(1)} (\mu)= \lambda_B (\mu) \, \int_0^{\infty} \, {d \omega \over \omega} \,
\ln \left ( {\mu \over \omega} \right )    \,\,  \phi_B^{+}(\omega, \mu)  \,.
\end{eqnarray}
The NLO determination of $\sigma_B^{(1)}(\mu_0)=1.4 \pm 0.4$ from the method of QCD sum rules
\cite{Braun:2003wx} will be taken in the subsequent numerical analysis.
The one-loop evolution equations for $\lambda_E^2(\mu)$ and $\lambda_H^2(\mu)$
defined by the  matrix elements of the dimension-five  HQET operators \cite{Grozin:1996hk,Nishikawa:2011qk}
\begin{eqnarray}
{d \over d \ln \mu} \, \left(
                         \begin{array}{c}
                           \lambda_E^2(\mu) \\
                           \lambda_H^2(\mu) \\
                         \end{array}
                       \right)
+ {\alpha_s(\mu) \over 4 \, \pi}  \, \gamma_{EH}  \, \left(
                         \begin{array}{c}
                           \lambda_E^2(\mu) \\
                           \lambda_H^2(\mu) \\
                         \end{array}
                       \right) = 0
\,,
\label{RGE of lambdaE and lamdbaH}
\end{eqnarray}
where the anomalous dimension matrix $\gamma_{EH}$ reads
\begin{eqnarray}
\gamma_{EH} =  \left(
                 \begin{array}{cc}
                   {8 \over 3} \, C_F + {3 \over 2}  \, N_c \,\,  &  \,\,  {4 \over 3} \, C_F - {3 \over 2}  \, N_c \\
                   {4 \over 3} \, C_F - {3 \over 2}  \, N_c \,\,  & \,\,  {8 \over 3} \, C_F + {5 \over 2}  \, N_c \\
                 \end{array}
               \right)
\,.
\end{eqnarray}
Diagonalizing this renormalization mixing matrix, one can readily obtain the solution
to the RG equation (\ref{RGE of lambdaE and lamdbaH})  in the LL approximation \cite{Grozin:1996hk}
\begin{eqnarray}
\left(
                         \begin{array}{c}
                           \lambda_E^2(\mu) \\
                           \lambda_H^2(\mu) \\
                         \end{array}
                       \right)
= \hat{V} \,  \left [  \left ( {\alpha_s(\mu) \over \alpha_s(\mu_0)} \right )^{\gamma_i^{(0)}/(2 \, \beta_0)} \right ]_{\rm diag}
  \,  \hat{V}^{-1} \,\,\,
\left(
                         \begin{array}{c}
                           \lambda_E^2(\mu_0) \\
                           \lambda_H^2(\mu_0) \\
                         \end{array}
                       \right)\,,
\label{solution to RGE of lambda-E and lambda-H}
\end{eqnarray}
where $\hat{V}$ the matrix that diagonalize  $\gamma_{EH}$, so that
\begin{eqnarray}
\hat{V}^{-1} \, \gamma_{EH} \, \hat{V} = [\gamma_i^{(0)}]_{\rm diag}\,,
\end{eqnarray}
with the eigenvalues of the one-loop anomalous dimension matrix
\begin{eqnarray}
\gamma_{\pm}^{(0)}= \left ( {8 \over 3} \, C_F + 2 \, N_c  \right ) \pm
{1 \over 6} \, \sqrt{64 \, C_F^2 - 144 \, N_C \, C_F + 90 \, N_C^2}
= {1 \over 9} \, \left ( 86  \pm \sqrt{{1565 \over 2}} \right )\,.
\end{eqnarray}
It is evident that the RG evolution of the ratio $R(\mu)=\lambda_E^2(\mu)/\lambda_H^2(\mu)$
at LL accuracy can be readily deduced from (\ref{solution to RGE of lambda-E and lambda-H}).
We further employ the QCD sum rule estimate for $R(\mu_0)=0.5 \pm 0.1$
at the  reference scale $\mu_0=1 \, {\rm GeV}$ by combining the results from \cite{Grozin:1996pq} at
the LO approximation  and from \cite{Nishikawa:2011qk} including the higher-order perturbative and non-perturbative corrections.

Following the standard strategy \cite{Beneke:2011nf}, the HQET $B$-meson decay constant $\tilde{f}_B(\mu)$ will be expressed
in terms of the QCD decay constant $f_B$ by virtue of  the matching relation (\ref{HQET relation for fB}).
The Lattice QCD determination  $f_B=(192.0 \pm 4.3) \, {\rm MeV}$ with $N_f=2+1$
from the Flavour Lattice Averaging Group (FLAG) \cite{Aoki:2016frl}
will be adopted in the following.
The longitudinal  decay constants of the light vector mesons can be extracted from the leptonic
decays $V^{0} \to e^{+} \, e^{-}$ and from the tau lepton decays $\tau^{+} \to V^{+} \, \nu_{\tau}$.
Including the flavour mixing of $\rho^{0}-\omega-\phi$ due to the QCD and QED interactions
gives rise to \cite{Straub:2015ica}
\begin{eqnarray}
f_{\rho, \, \|}=(213 \pm 5) \, {\rm MeV}, & \qquad  &
f_{\rho, \, \perp}(1 \, {\rm GeV})=(160 \pm 7) \, \rm{MeV}\,,  \nonumber \\
f_{\omega, \, \|}=(197 \pm 8) \, {\rm MeV}, & \qquad  &
f_{\omega, \, \perp}(1 \, {\rm GeV})=(148 \pm 13) \, \rm{MeV}\,,  \nonumber  \\
f_{K^{\ast}, \, \|}=(204 \pm 7) \, {\rm MeV}, & \qquad  &
f_{K^{\ast}, \, \perp}(1 \, {\rm GeV})=(159 \pm 6) \, \rm{MeV}\,,
\end{eqnarray}
where the renormalization-scale dependent transverse decay constants of the vector mesons at
$\mu_0=1 \, {\rm GeV}$ are also displayed by making use of the ratios
$f_{V, \, \perp}(2 \, {\rm GeV})/f_{V, \, \|}$ computed from the Lattice QCD simulation
with $2+1$ flavours of domain wall quarks and the Iwasaki gauge action \cite{Allton:2008pn}.
The RG evolution of $f_{V, \, \perp}(\nu)$ at the NLL  accuracy can be determined
by solving the equation (\ref{RGE for fVT}) straightforwardly.

We proceed to discuss the determinations of the Borel masses and the threshold parameters
for the light vector-meson channels entering both the leading-power and the subleading-power LCSR of
the semileptonic $B \to V$ form factors.
The interval of the Borel mass for the $\rho$-meson channel $M^2_{\rho}= (1.5 \pm 0.5) \, {\rm GeV^2}$
extracted from the two-point QCD sum rules \cite{Ball:1998sk} will be employed in the numerical calculations.
Taking into account the SU(3) symmetry breaking effects for the improved LCSR of $B \to V$ form factors,
we will employ the relations proposed in \cite{Ball:1998sk}  for the determinations of the Borel masses
for the $\omega$ and $K^{\ast}$  channels
\begin{eqnarray}
M^2_{\omega} - M^2_{\rho} = m^2_{\omega} - m^2_{\rho} \,, \qquad
M^2_{K^{\ast}} - M^2_{\rho} = m^2_{K^{\ast}} - m^2_{\rho} \,.
\label{approximate relations for Borel masses}
\end{eqnarray}
The continuum threshold for the longitudinally polarized $\rho$-meson channel
$s_{0, \, \rho}^{\|}= (1.5 \pm 0.1) \, {\rm GeV^2}$  \cite{Ball:1998sk,Khodjamirian:2006st}
is determined by the requirement that the QCD sum rule
prediction of the vector decay constant $f_{\rho, \|}$ at ${\cal O}(\alpha_s)$ can reproduce the
corresponding experimentally measured value.
By contrast, a lower value of the threshold parameter for the transversely polarized $\rho$-meson channel
$s_{0, \, \rho}^{\perp}= (1.2 \pm 0.1) \, {\rm GeV^2}$ will be adopted to incorporate the contaminating contributions
of the $b_1(1235)$ channel to the QCD sum rules for the transverse decay constant $f_{\rho, \perp}$ effectively
(see \cite{Ball:1996tb} for more discussions).
The continuum threshold parameters for the $\omega$ and $K^{\ast}$  channels will be fixed by applying the
approximate relations in analogy to  (\ref{approximate relations for Borel masses}).

The bottom-quark mass in the ${\rm \overline{MS}}$ scheme
$\overline{m_b} (\overline{m_b})=4.193^{+0.022}_{-0.035} \,\, {\rm GeV}$
determined from  non-relativistic QCD sum rules at  next-to-next-to-next-to-leading order (NNNLO)
\cite{Beneke:2014pta} (see also \cite{Dehnadi:2015fra} for alternative determinations with relativistic
QCD sum rules and \cite{Mateu:2017hlz} for the NNNLO determination from the  bottomonium spectrum)
will be employed in the numerical analysis.
We further employ the intervals for the light quark masses in the ${\rm \overline{MS}}$ scheme
at a renormalization scale of $2 \, {\rm GeV}$ from \cite{Tanabashi:2018oca}
\begin{eqnarray}
m_u(2 \, {\rm GeV}) &=& (2.15 \pm 0.15) \, {\rm MeV}  \,, \qquad
m_d(2 \, {\rm GeV}) = (4.70 \pm 0.20) \, {\rm MeV}  \,, \nonumber \\
m_s(2 \, {\rm GeV}) &=& (93.8 \pm 1.5 \pm 1.9) \, {\rm MeV}  \,.
\end{eqnarray}

Following the discussions displayed in \cite{Lu:2018cfc},
the factorization scale $\mu$ entering the obtained LCSR for $B \to V$ form factors will be
varied in the interval  $1 \, {\rm GeV} \leq \mu \leq 2 \, {\rm GeV}$ around the default
value $\mu=1.5 \, {\rm GeV}$. The renormalization scale for the QCD tensor current will be taken
as $\nu=m_b$ varying in the range $[m_b/2, 2 \, m_b]$.
In addition,  the initial scales for the RG evolutions of the hard matching coefficients
$C_i^{(\rm A0)}(n \cdot p, \mu)$ and $C_i^{(\rm B1)}(n \cdot p, \tau, \mu)$,
the HQET decay constant $\tilde{f}_B(\mu)$ and the transverse decay constant $f_{V, \perp}(\nu)$
will be chosen as $\mu_{h1}=\mu_{h2}=\nu_h \in [m_b/2, 2 \, m_b]$ around $m_b$.

\subsection{Theory predictions for $B \to V$ form factors}

We are now in a position to investigate the numerical impacts of the perturbative QCD
corrections and the higher twist contributions to the semileptonic $B \to V$
form factors applying the SCET based formulation of the LCSR approach.
To this end, we first need to determine the inverse moment $\lambda_B(\mu_0)$ of the leading-twist $B$-meson
distribution amplitude, which  serves as the principle theory input for the precision description
for  exclusive $B$-meson decay amplitudes in QCD generally.
The non-perturbative calculations of  $\lambda_B(\mu_0)$ from the method of HQET sum rules \cite{Braun:2003wx}
and  the complementary indirect extractions from measuring the integrated  branching fractions of
the radiative leptonic $B$-meson decays \cite{Beneke:2011nf,Wang:2016qii,Wang:2018wfj,Beneke:2018wjp}
provided us meaningful  but still loose constraints of this key quantity at present.
Due to the limited knowledge of $\lambda_B(\mu_0)$, we prefer to perform an independent determination
by matching our prediction for the vector form factor $V_{B \to \rho}(q^2)$ at the maximal hadronic recoil
with the corresponding result from the improved NLO LCSR with the $\rho$-meson distribution amplitudes \cite{Straub:2015ica}.
Proceeding with this matching procedure immediately gives rise to the following constraints
\begin{eqnarray}
\lambda_B(\mu_0) = \left\{
\begin{array}{l}
343^{+22}_{-20} \,\, {\rm MeV}  \,, \qquad  \hspace{1.5 cm}
(\rm Exponential \,\, Model) \vspace{1.0 cm} \\
370^{+24}_{-22} \,\, {\rm MeV}  \,,
 \qquad  \hspace{1.5 cm}
(\rm Local \,\, Duality \,\, Model)
\end{array}
 \hspace{0.5 cm} \right.
\,
\label{results of lambdaB}
\end{eqnarray}
which can be further traded into the intervals of the HQET parameter $\bar \Lambda$
\begin{eqnarray}
\bar \Lambda = \left\{
\begin{array}{l}
515^{+33}_{-30} \,\, {\rm MeV}  \,, \qquad  \hspace{1.5 cm}
(\rm Exponential \,\, Model) \vspace{1.0 cm} \\
463^{+30}_{-28} \,\, {\rm MeV}  \,.
 \qquad  \hspace{1.5 cm}
(\rm Local \,\, Duality \,\, Model)
\end{array}
 \hspace{0.5 cm} \right.
\,
\label{results of Lambda}
\end{eqnarray}
It is interesting to notice that the extracted values of $\lambda_B(\mu_0)$ are in nice agreement with the
previous determination by matching the distinct LCSR for the vector $B \to \pi$ form factor $f_{B \to \pi}^{+}(q^2)$ with the analogous
prescription \cite{Wang:2015vgv} and are also consistent with the implications of experimental data
for the two-body charmless hadronic $B$-meson decays from the QCD factorization approach \cite{Beneke:2003zv}.
In the following we will take the exponential model of the $B$-meson distribution amplitudes as our default choice
to explore the phenomenological aspects of the newly derived SCET sum rules,
and the systematic uncertainty due to the model dependence of the $B$-meson LCDA will be taken into account
in the final theory predictions for the semileptonic $B \to V$ form factors.

\begin{figure}
\begin{center}
\includegraphics[width=0.55 \columnwidth]{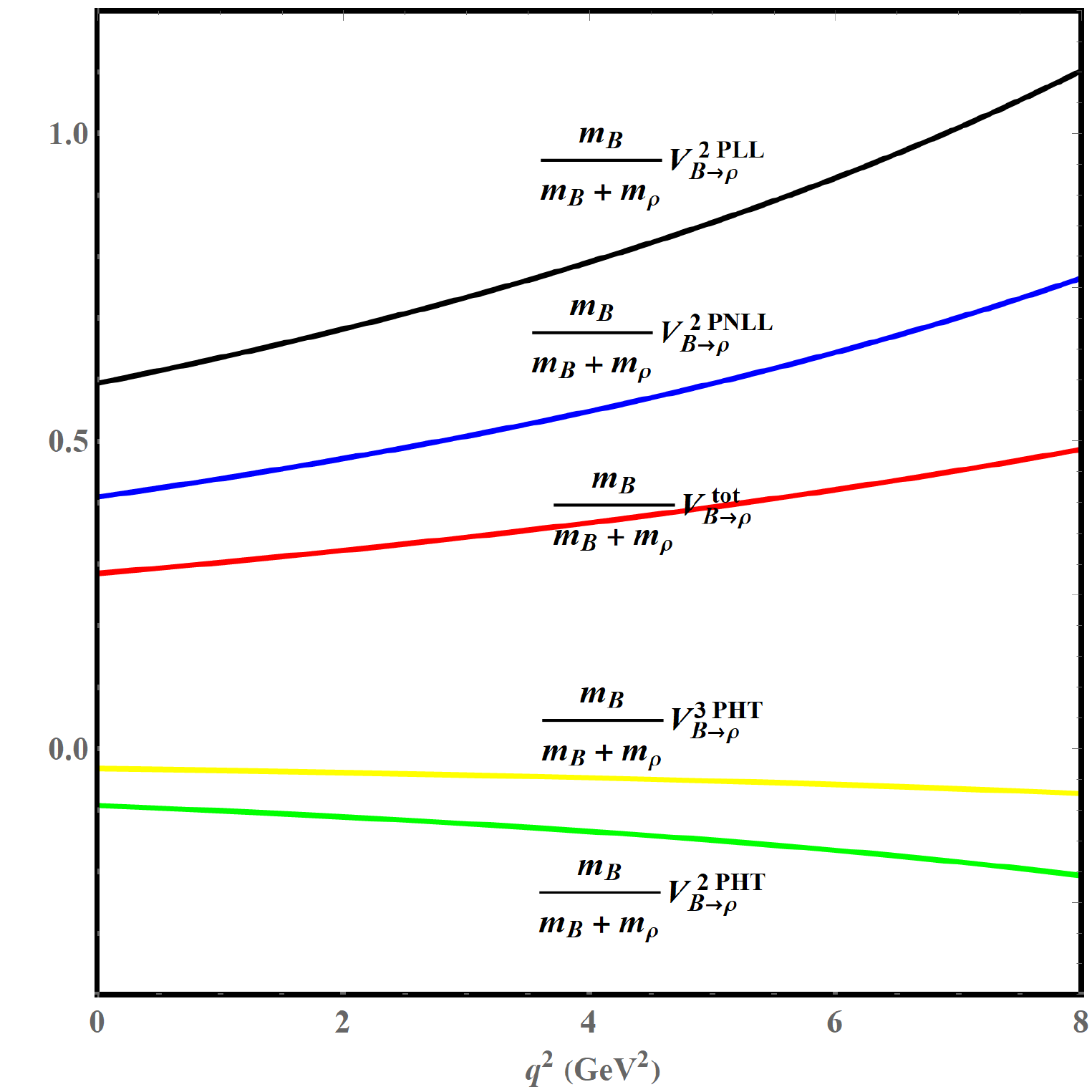}
\includegraphics[width=0.55 \columnwidth]{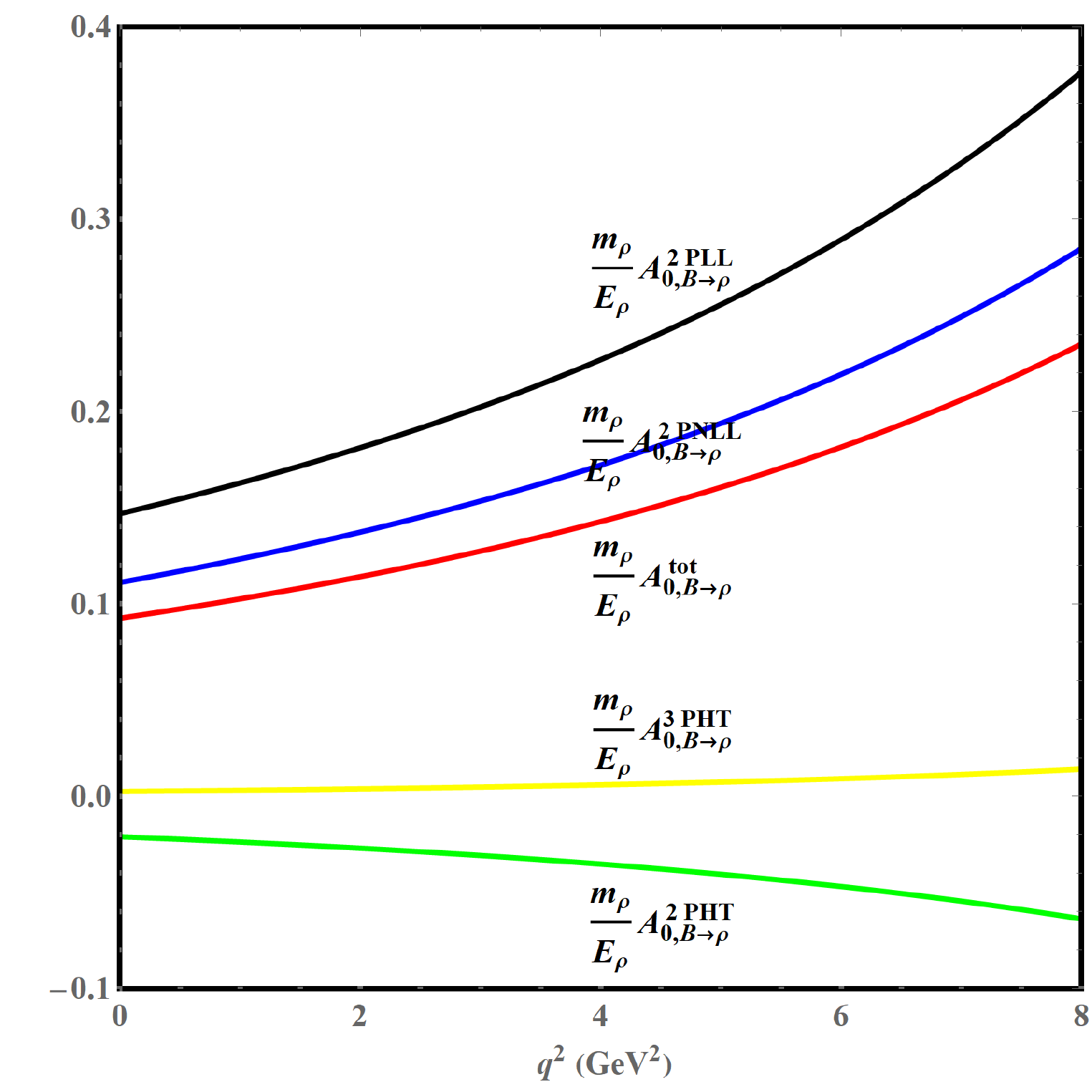}
\vspace*{0.1cm}
\caption{Breakdown of  the various terms contributing to the  two $B \to \rho$ form factors $V_{B \to \rho}(q^2)$
and $A_{0, \, B \to \rho}(q^2)$ from the SCET sum rules  with the exponential model
of the $B$-meson distribution amplitudes at $q^2 \leq 8 \, {\rm GeV^2}$.
The individual contributions correspond to the two-particle  leading-twist effects  at LL
(black curves) and at NLL (blue curves), the two-particle higher-twist corrections (green curves)
and the three-particle higher-twist effects (yellow curves).}
\label{fig: Breakdown of B to rho form factors}
\end{center}
\end{figure}

To develop a transparent understanding of the higher-order perturbative corrections and
the higher-twist contributions from the two-particle and three-particle
$B$-meson distribution amplitudes computed in this work, we display in
figure \ref{fig: Breakdown of B to rho form factors} the
numerical effects of distinct pieces contributing to the final sum rules for the two
$B \to \rho$ form factors $V_{B \to \rho}(q^2)$ and $A_{0, \, B \to \rho}(q^2)$
at large hadronic recoil.
It is evident that the NLL QCD radiative corrections to the leading-twist contributions can
give rise to approximately $(25 -30) \, \%$ reduction of the corresponding
resummation improved tree-level predictions.
In particular,  the two-particle twist-five contributions to both the two $B \to \rho$
form factors at LO in QCD generate sizeable corrections, numerically $(20 -30) \, \%$,
to the leading-power predictions at NLL, in analogy to the earlier observation
for $B \to \pi, K$ form factors \cite{Lu:2018cfc} (see also \cite{Gubernari:2018wyi}
for  independent calculations of the higher-twist effects up to the twist-four $B$-meson
distribution amplitudes). By contrast, the genuine three-particle higher-twist corrections
yield approximately ${\cal O} (10 \, \%)$ and ${\cal O} (2\, \%)$ enhancement of
the leading-twist calculations for the transverse and longitudinal $B \to \rho$ form factors, respectively.
We have verified that the preceding observed patterns for the higher-order and higher-twist corrections
are also satisfied for the SCET sum rules predictions of the remaining
$B \to V$ (with $V=\rho, \omega, K^{\ast}$) form factors.

\begin{figure}
\begin{center}
\includegraphics[width=0.35 \columnwidth]{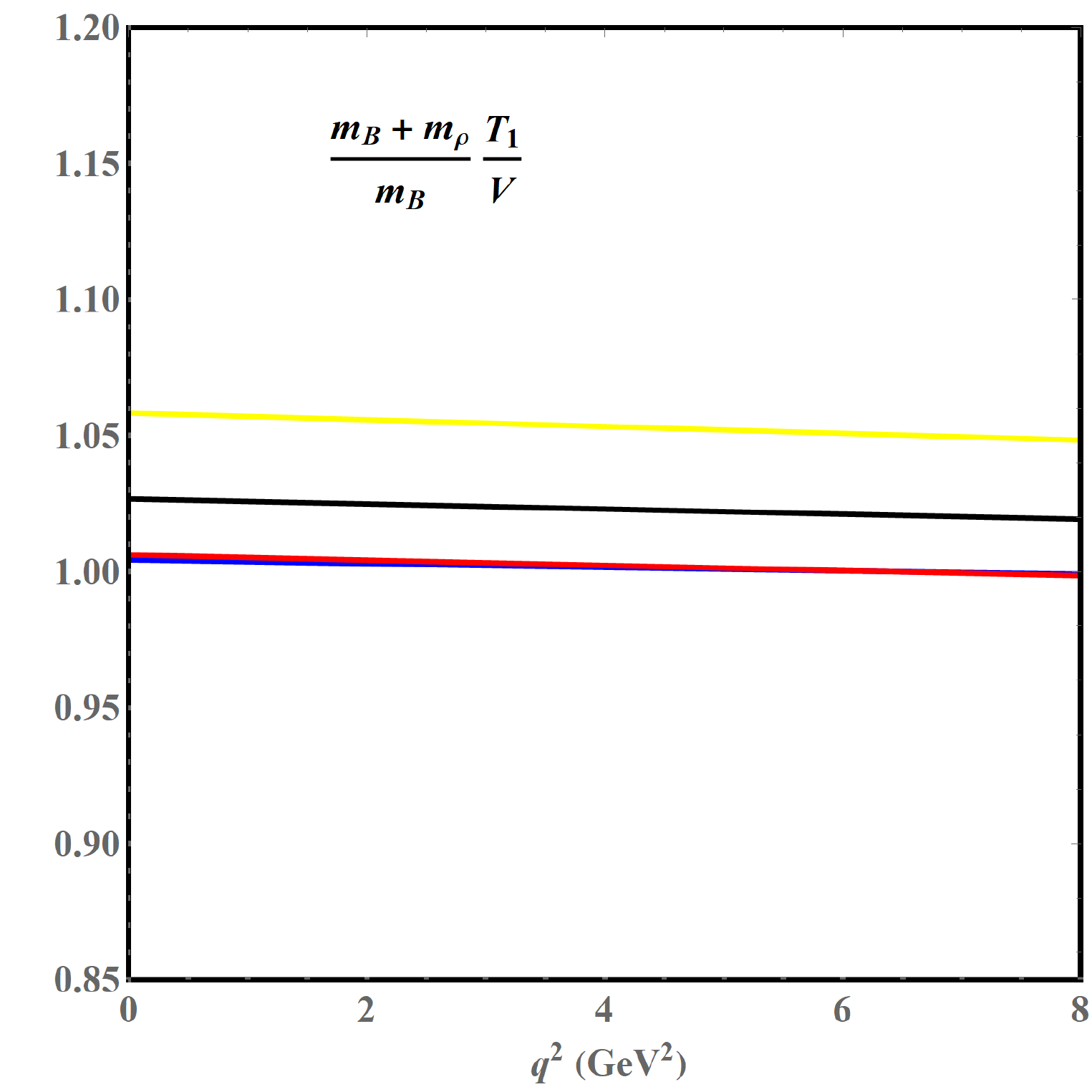}
\hspace{0.2 cm}
\includegraphics[width=0.35 \columnwidth]{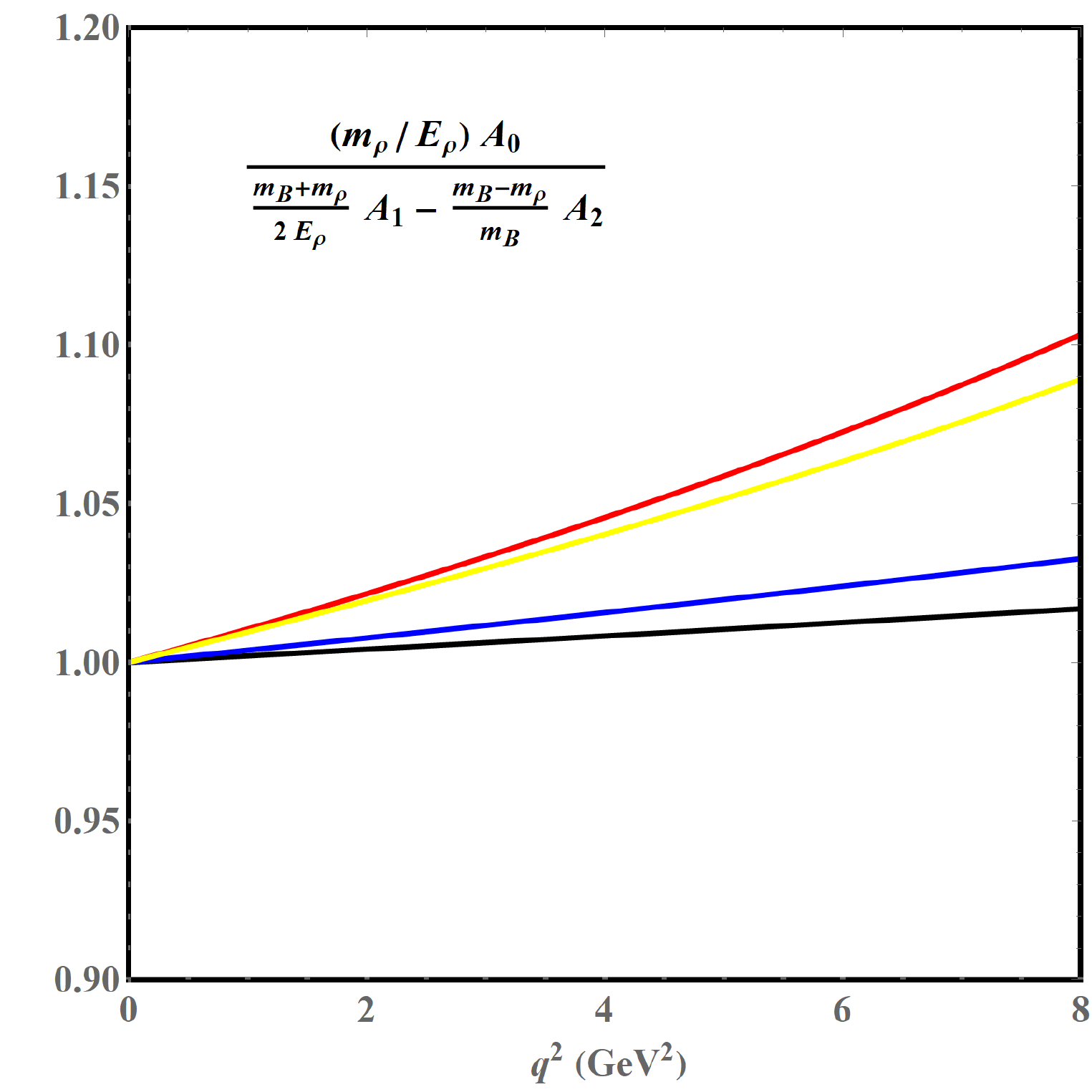}
\\
\includegraphics[width=0.35 \columnwidth]{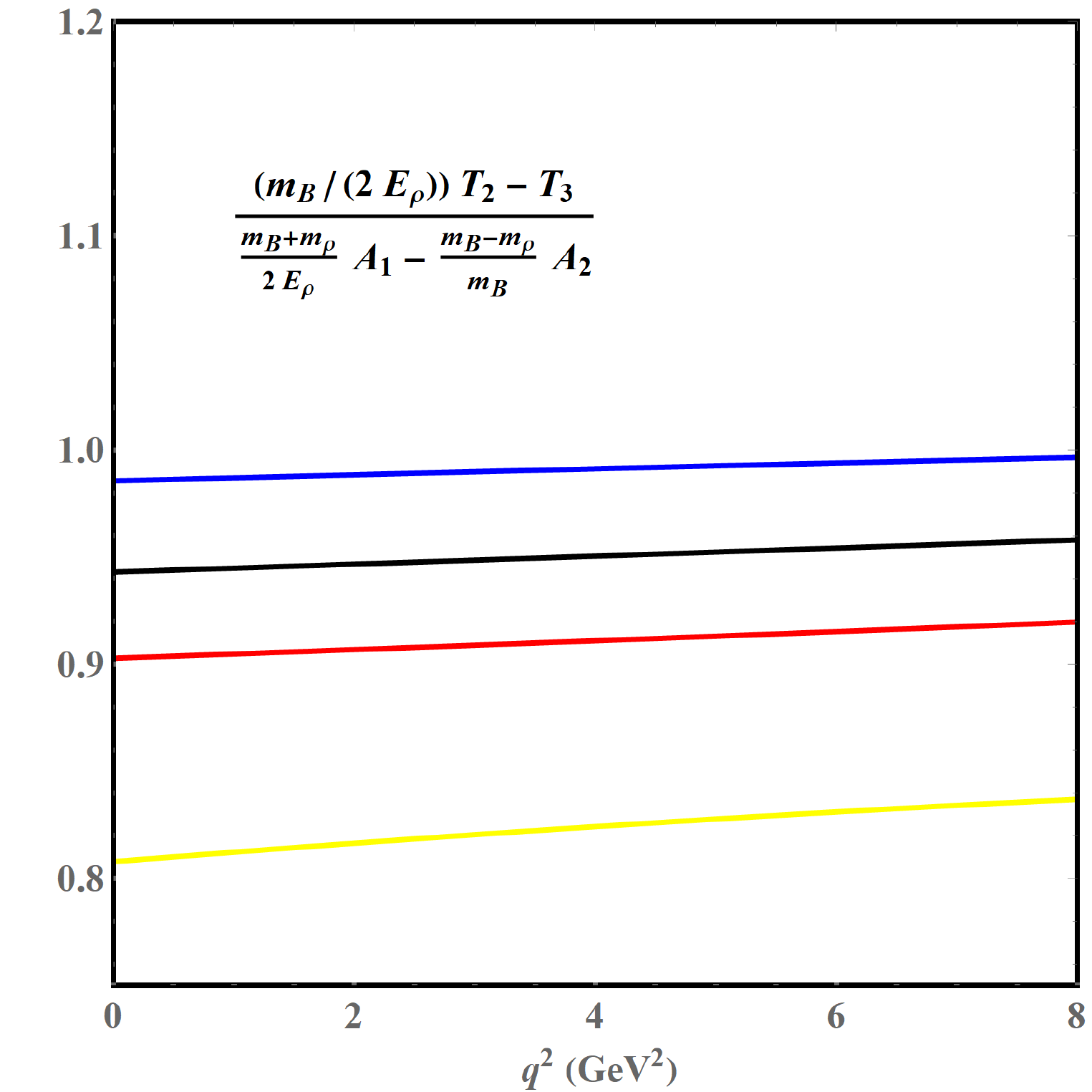}
\hspace{0.2 cm}
\includegraphics[width=0.35 \columnwidth]{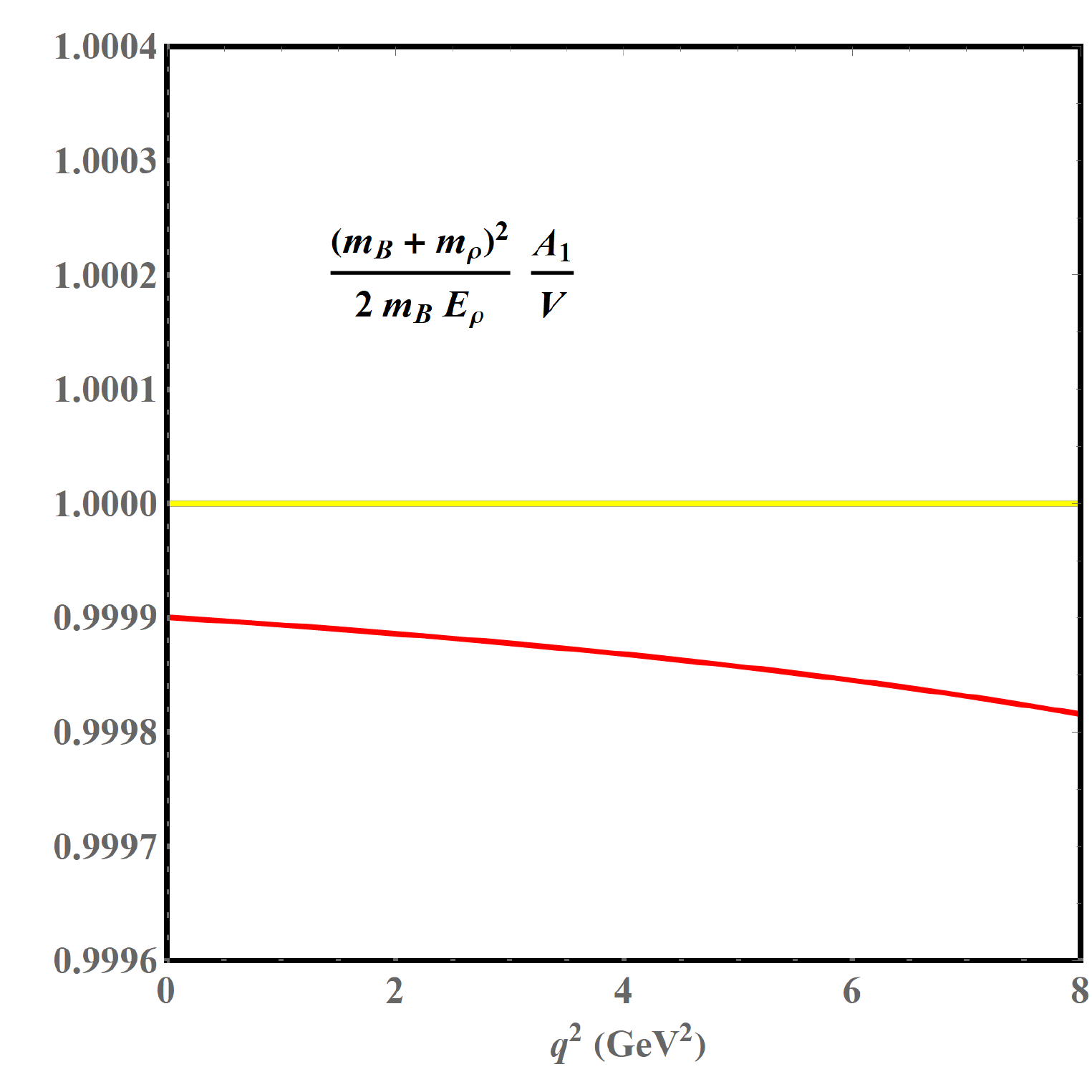}
\\
\includegraphics[width=0.35 \columnwidth]{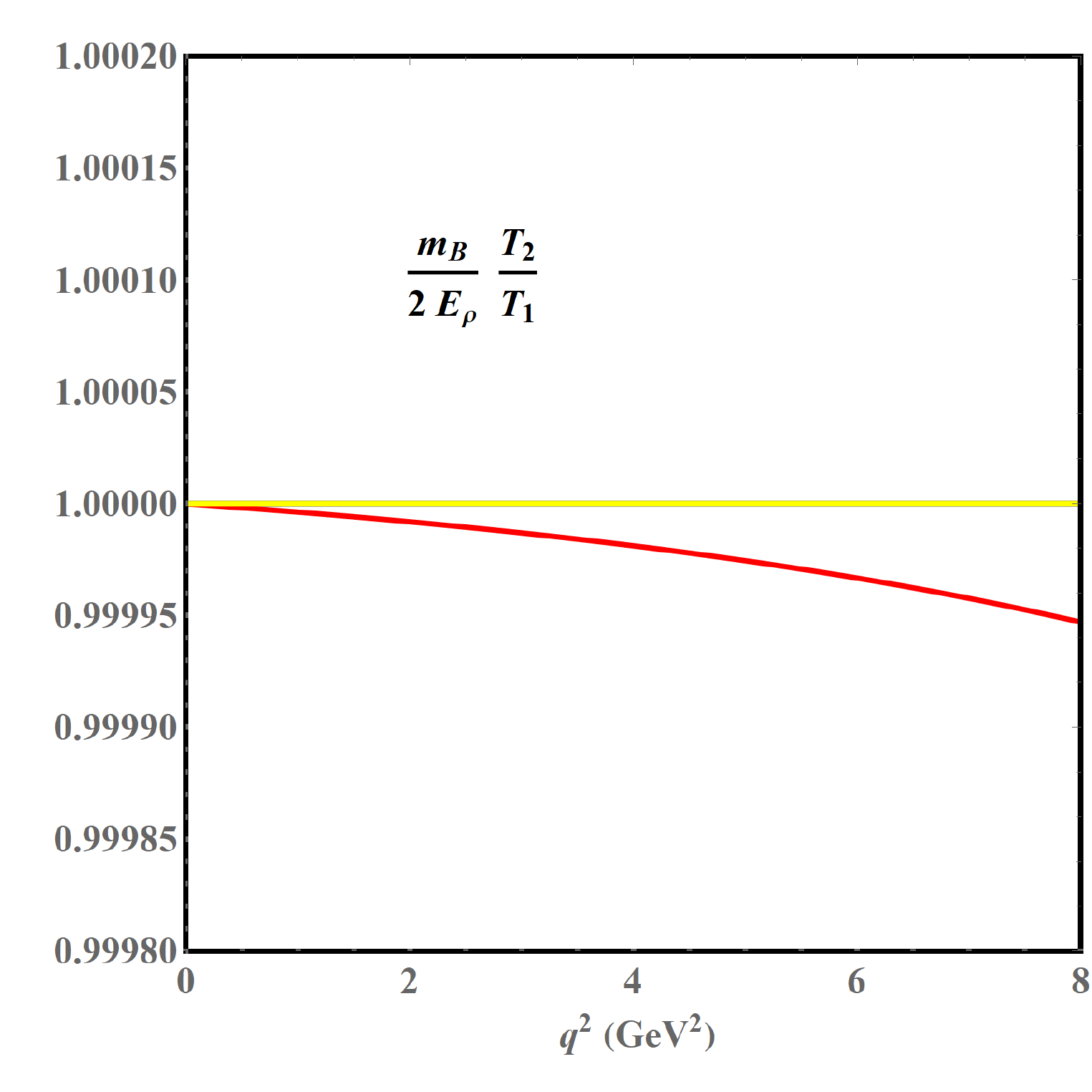}
\vspace*{0.1cm}
\caption{Theory predictions for the $B \to V$ form-factor ratios from the improved SCET
sum rules derived in this work and from the QCD factorization approach
with the so-called  physical form factor scheme \cite{Beneke:2000wa,Beneke:2005gs}.
Black curves: the leading-power contributions at the LL accuracy;
blue curves: the leading-power contributions at the NLL accuracy;
red curves: full results including both the leading-power effects at NLL and
the subleading-twist corrections  at LO up to the twist-six accuracy. The yellow curves are obtained
from the leading-power computations with the QCD factorization approach at NLO.}
\label{fig: comparision of the LCSR and QCDF results}
\end{center}
\end{figure}

Now we proceed to investigate an interesting issue of  QCD computations for the heavy-to-light $B$-meson
decay form factors at large recoil from the SCET factorization approach and the LCSR method with the light-meson
distribution amplitudes. Generally theory predictions for the form-factor ratios from these two different approaches
are in reasonable agreement with each other, however, the obtained results for the following $B \to V$ form-factor ratios
\begin{eqnarray}
\mathcal{R}_1={m_B + m_{V} \over m_B} \, {T_{1} \over V}, \qquad
\mathcal{R}_2={m_B/(2 \, E) \, T_{2} - T_{3} \over
(m_B + m_{V})/(2 \, E)\, A_{1}
-  (m_B - m_{V})/m_B\, A_{2}}\,,
\end{eqnarray}
differ in  both the magnitude and sign of the large-recoil symmetry breaking effects
\cite{Beneke:2000wa,Beneke:2005gs,Bell:2010mg}.
It is our purpose to address whether such discrepancies are due to the yet higher-order
corrections in both $\alpha_s$ and $\Lambda/m_b$ or due to the systematic uncertainties
of the method of QCD sum rules.
To achieve this goal, we display our predictions for the form-factor ratios from the improved
SCET sum rules with the $B$-meson distribution amplitudes
in figure \ref{fig: comparision of the LCSR and QCDF results}, including the corresponding NLL results
from the QCD factorization approach for a comparison.
It can be observed that our predictions for all the $B \to \rho$ form-factor ratios,
particularly the sign of the symmetry breaking,  are in agreement
with the SCET results displayed in figure 6 of \cite{Beneke:2005gs}.
We are therefore led to conclude that the above-mentioned discrepancies between
the two different QCD calculations mainly arise from the parton-hadron duality approximation for
the $B$-meson channel implemented in the construction of the sum rules
for the heavy-to-light form factors with the vector-meson distribution amplitudes.
This can be also understood from the fact that the traditional LCSR for the semileptonic
$B \to V$  form factors in the heavy-quark limit will introduce new non-perturbative  quantities,
for instance $\phi_{\perp}^{\prime}(1)$ and $\Phi_{\|}^{\prime}(1)$ \cite{Ball:1997rj},
which cannot be constructed from a finite number of Gegenbauer moments of the corresponding
vector-meson distribution amplitudes and whose field-theoretical definitions
are absent in SCET \cite{Beneke:2007zz}.
On the contrary, the new LCSR for  the ${\rm A0}$- and ${\rm B1}$- type SCET form factors
with the $B$-meson distribution amplitudes involve the two quantities
$\phi_B^{-}(0, \mu)$ and $\lambda_B^{-1}(\mu)$, which are identical in the Wandzura-Wilczek approximation
(namely, neglecting the effect of the three-particle $B$-meson LCDA $\Psi_A-\Psi_V$) \cite{Wandzura:1977qf}
and are well defined parameters in the SCET framework.

In addition, we notice that the magnitudes of the large-recoil symmetry violations  predicted from
the SCET sum rules with $B$-meson distribution amplitudes are generally smaller than those predicted by
the QCD factorization approach (see also \cite{Beneke:2005gs} for a similar observation).
To identify the underlying mechanism responsible for such discrepancy, we  write down explicitly the
separate terms  generating the symmetry correction to  $\mathcal{R}_1$
\begin{eqnarray}
\mathcal{R}_{1, \, {\rm LCSR}} &=& 1 + {[(-0.179)-(-0.154)]\times 0.560 \over 0.285} \bigg |_{C_{i}^{(\rm A0)}}
+ {[(-1)+0.203]\times (-0.0192) \over 0.285} \bigg |_{C_{i}^{(\rm B1)}}  \nonumber  \\
&& +  {[(-0.0922) -(-0.0922) ] \over 0.285 } \bigg|_{\rm 2PHT}
+  { [(-0.03220)- (-0.03219)] \over 0.285 } \bigg|_{\rm3PHT} \nonumber \\
&=&  1 + (-0.049)\big |_{C_{i}^{(\rm A0)}} + (+0.054) \big |_{C_{i}^{(\rm B1)}}
+ (-3.5 \times 10^{-5}) \big|_{\rm3PHT}   \,, \nonumber \\
\mathcal{R}_{1, \, {\rm QCDF}} &=& 1 + (-0.023)\big |_{C_{i}^{(\rm A0)}}
+ (+0.086) \, [1 + {\cal O}(\alpha_s)] \big |_{C_{i}^{(\rm B1)}}  \,,
\end{eqnarray}
where the expression of $\mathcal{R}_{1, \, {\rm QCDF}}$ is borrowed
directly from (124) of \cite{Beneke:2005gs} by dropping out the NLO correction
to the hard-spectator scattering contribution.
It is then evident that the negligible symmetry breaking effect from the
$B$-meson LCSR calculation is due to the strong cancellation between
the ${\rm A0}$- and ${\rm B1}$-type SCET matrix elements weighted by the corresponding hard matching
coefficients.
By contrast,  the symmetry violation predicted in QCD factorization is numerically dominated by
the hard-spectator scattering and will be further enhanced by the higher-order perturbative
correction and the RG resummation effect as indicated in \cite{Beneke:2005gs}.
More specifically, the strong cancellation mechanics from the $B$-meson LCSR computation
can be attributed to the following reasoning.

\begin{itemize}

\item{The LCSR prediction of the non-local SCET form factor $\Xi_{\perp}(\tau, n \cdot p)$
is approximately $40 \%$ smaller than the QCD factorization result at tree level.
This observation implies that approximating the collinear dynamics of the energetic vector meson
by the static QCD decay constants is numerically insufficient for the theory description
of the semileptonic $B \to V$ form factors beyond the heavy quark limit.}

\item{At the one-loop accuracy, the NLO corrections to the hard functions of the ${\rm A0}$-type
SCET operators must be multiplied with  the LO sum rule result of the ${\rm A0}$-type
form factor $\xi_{\perp}(n \cdot p)$ instead of the physics QCD form factor
\begin{eqnarray}
\xi_{\perp}^{\rm FF}(n \cdot p) \equiv {m_B \over m_B+m_V} \, V(n \cdot p)\,,
\end{eqnarray}
which has to be taken as an hadronic input in QCD factorization.
The SCET sum rule prediction of $\xi_{\perp}(m_B)$
with $B$-meson distribution amplitudes at tree level is approximately twice of
the complete  result for $\xi_{\perp}^{\rm FF}(m_B)$ including both the NLO QCD correction
and the higher-twist correction. As a consequence, our prediction for the symmetry correction due to
$C_{i, \, \rm{NLL}}^{(\rm A0)} \,\, \xi_{\perp}$
is enhanced by a factor of two when compared with the  corresponding QCD factorization computation.}

\end{itemize}

\begin{figure}
\begin{center}
\includegraphics[width=0.45 \columnwidth]{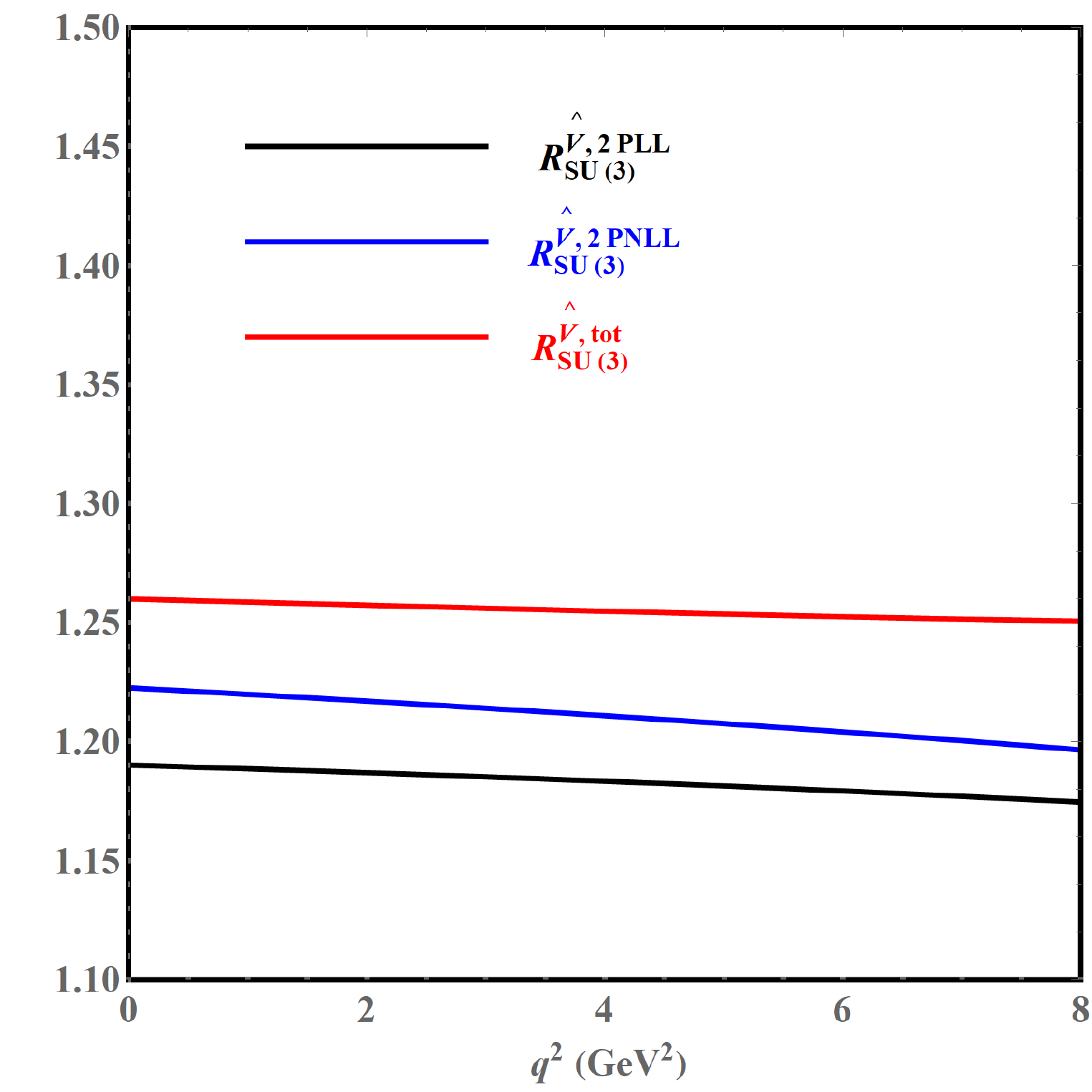}
\hspace{0.2 cm}
\includegraphics[width=0.45 \columnwidth]{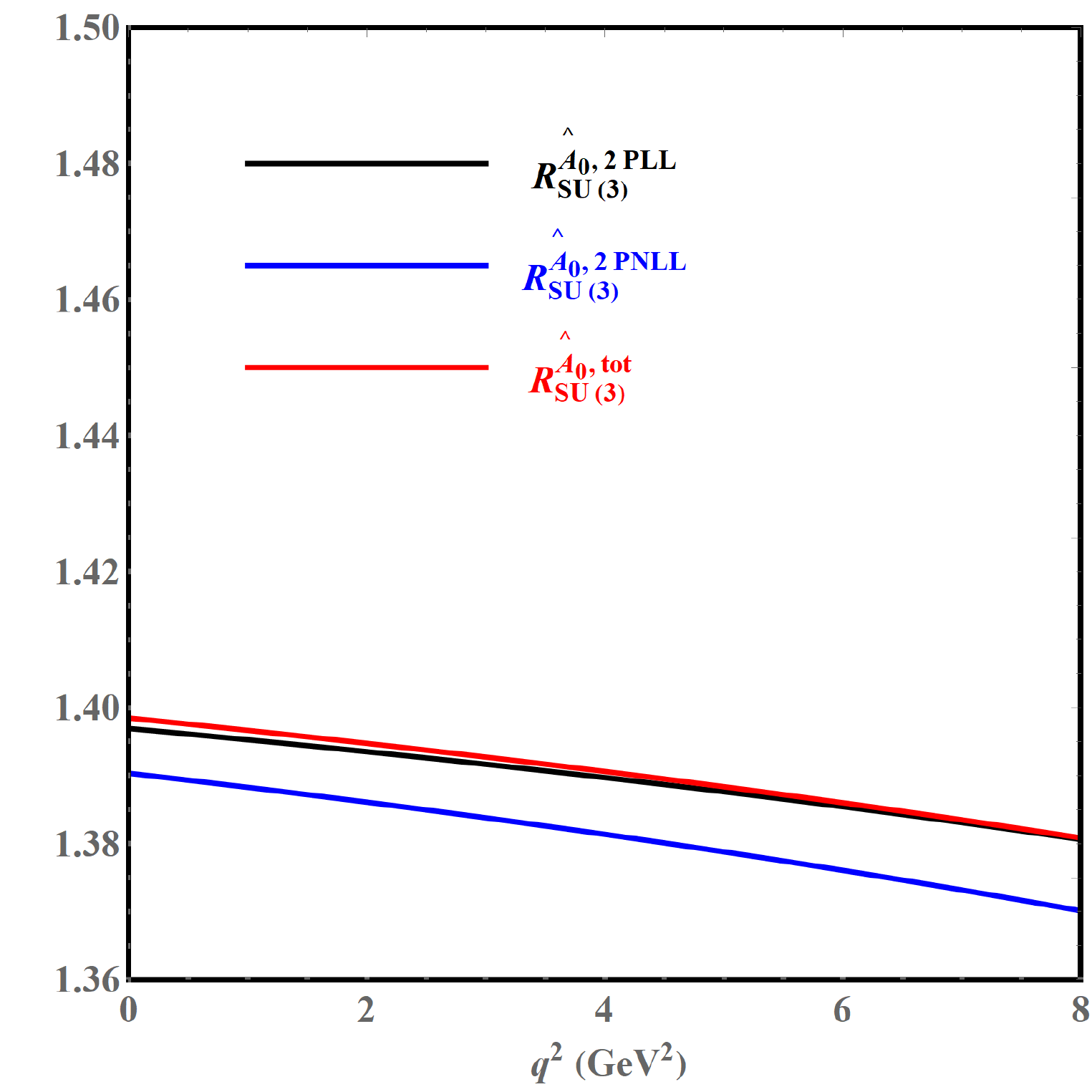}
\vspace*{0.1cm}
\caption{The SU(3) symmetry breaking effects for the transverse form factor
${m_B \over m_B+m_V} \, V(q^2)$  and for the longitudinal form factor ${m_V \over E_V} \, A_0(q^2)$
at large hadronic recoil, computed from the SCET sum rules with $B$-meson distribution amplitudes.}
\label{fig: SU(3) symmtey breaking}
\end{center}
\end{figure}

We now explore the SU(3)  flavour  symmetry breaking effects between $B \to \rho$ and
$B \to K^{\ast}$ form factors from the $B$-meson LCSR numerically.
In our theoretical framework they originate from the explicit corrections proportional to the
light-quark masses and the light vector meson masses,
the differences in the values of the threshold parameters and Boreal masses,
and the discrepancies  in the longitudinal and transverse decay constants for $\rho$ and $K^{\ast}$.
Additional sources of the  SU(3) flavour  symmetry  violations due to the electromagnetic corrections
and the process-dependent systematic uncertainties (e.g., the parton-hadron duality ansatz) are not
taken into account.
For the phenomenological convenience we introduce the  following quantity to character the SU(3) symmetry corrections
\begin{eqnarray}
R_{\rm SU(3)}^{i}(q^2)= {F_{B \to K^{\ast}}^{i}(q^2) \over F_{B \to \rho}^{i}(q^2)} \,,
\end{eqnarray}
where $F_{B \to V}^{i}$ represent the seven QCD form-factor combinations
appearing in (\ref{SCET-I factorization formulae}) generally.
It is evident from figure \ref{fig: SU(3) symmtey breaking} the SU(3) flavour symmetry breaking effects
for the transverse and longitudinal $B \to V$ form factors are approximately $25 \%$ and $40 \%$
in the large recoil region $0 \, {\rm GeV^2} \leq q^2 \leq 8 \, {\rm GeV^2}$, respectively.
This pattern can be understood from the fact that  the leading-power light-quark mass effect
does not contribute to the  SCET form factor $\xi_{\perp}(n \cdot p)$ at the one-loop approximation
as  demonstrated in (\ref{result of F-T-C}).
Our predictions for the SU(3) flavour symmetry corrections are in excellent agreement with the
previous computations based upon the  LCSR method with the vector-meson distribution amplitudes \cite{Ball:2004rg},
but are significantly larger than the updated results presented in  \cite{Straub:2015ica},
which predicted remarkably small flavour symmetry violations  (approximately $2 \, \%$ and $15 \, \%$
for the transverse and longitudinal $B \to V$ form factors at the maximal hadronic recoil).
We further notice that the subleading power higher-twist corrections are of minor numerical importance for generating
the SU(3) flavour symmetry breaking effects.

\begin{figure}
\begin{center}
\includegraphics[width=0.45 \columnwidth]{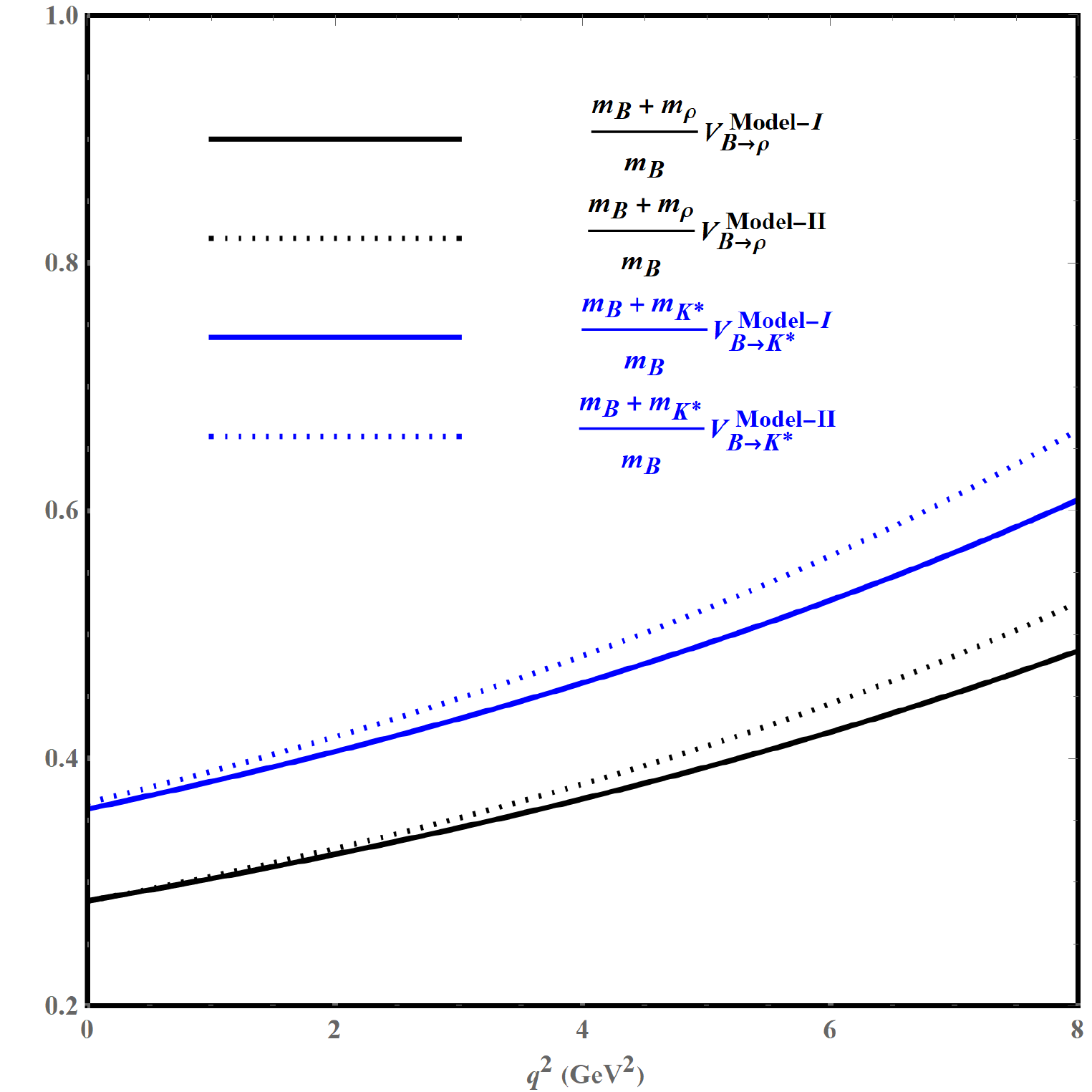}
\hspace{0.2 cm}
\includegraphics[width=0.45 \columnwidth]{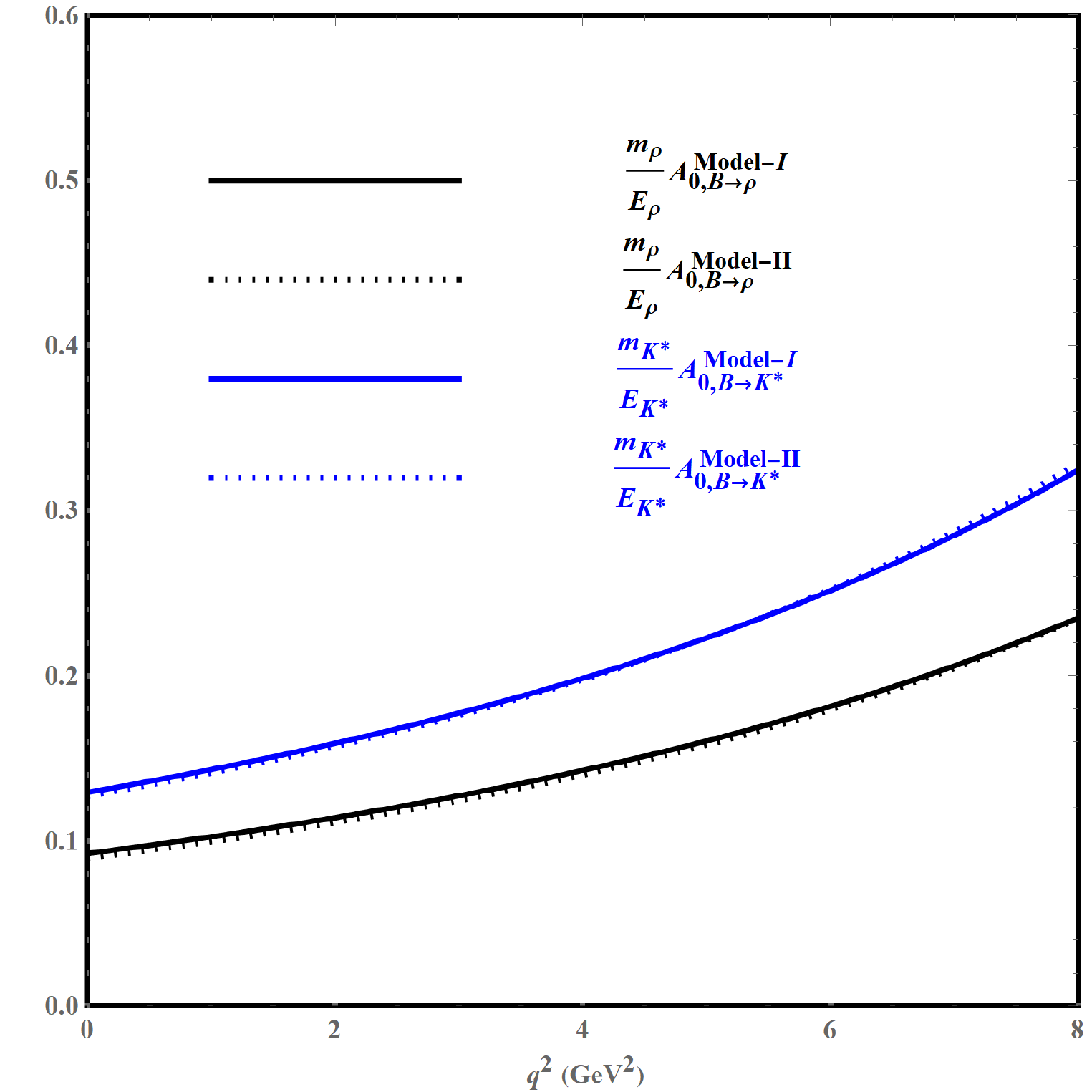}
\vspace*{0.1cm}
\caption{Model dependence of the transverse and longitudinal $B \to \rho, \, K^{\ast}$
form factors on the shapes of $B$-meson distribution amplitudes at
$0  \leq q^2 \leq 8 \, {\rm GeV^2}$. The superscripts ``Model-I" and ``Model-II" refer to
the exponential model and the local duality model of the two-particle and three-particle $B$-meson LCDA
displayed in Appendix \ref{appendix: B-meson DAs}. }
\label{fig: model dependence of B to V form factors}
\end{center}
\end{figure}

As already demonstrated in \cite{Wang:2015vgv}, the knowledge of the complete functional forms
of the $B$-meson distribution amplitudes is in demand for the evaluation for the heavy-to-light
$B$-meson decay form factors. To reduce the model dependence of our predictions,
the inverse moment $\lambda_B(\mu_0)$ for a given model of the $B$-meson LCDA has been determined
by reproducing the alternative LCSR prediction for $V_{B \to \rho}(q^2=0)$  with the vector-meson LCDA
as described in the previous  paragraphs.
In other words, we  aim at predicting the momentum-transfer dependence of the transverse $B \to \rho$
form factor $V_{B \to \rho}(q^2)$ merely. However, both the normalization factors at the maximal recoil
and the $q^2$-shapes for all the remaining form factors will be obtained from the derived SCET sum rules
subsequently. It can be observed from figure \ref{fig: model dependence of B to V form factors} that
the model dependence of our predictions on the precise $\omega$-behaviours of the $B$-meson distribution amplitudes
is drastically reduced by implementing the above-mentioned prescription, in analogy to the earlier
observation for the semileptonic $B \to \pi, K$  form factors computed in the same framework \cite{Wang:2015vgv,Lu:2018cfc}.

Evidently, the obtained LCSR for the SCET  form factors $\xi_{i}(n \cdot p)$ and $\Xi_{i}(\tau, n \cdot p)$
(with $i=\|, \, \perp$) cannot be constructed without demonstrating the soft-collinear factorization for the various
vacuum-to-$B$-meson correlation functions under discussion in the first place,
which can be validated with the light-cone operator product expansion (OPE) technique only at the large hadronic recoil.
To extrapolate the SCET sum rule predictions for the $B \to V$ form factors toward the high $q^2$ region,
we will employ the model-independent $z$-series parametrizations \cite{Bourrely:1980gp} motivated by
the analytical properties and the asymptotic behaviours of the heavy-to-light form factors.
The complex cut $q^2$-plane will then be mapped onto the unit disc $z(q^2, t_0)\leq 1$ under the conformal transformation
\begin{eqnarray}
z(q^2, t_0) = \frac{\sqrt{t_{+}-q^2}-\sqrt{t_{+}-t_0}}{\sqrt{t_{+}-q^2}+\sqrt{t_{+}-t_0}}  \,,
\end{eqnarray}
where two parameters $t_{+}$ and $t_0$ are given by \cite{Lu:2018cfc} (see also \cite{Khodjamirian:2017fxg})
\begin{eqnarray}
t_{+}=(m_B +m_V)^2\,, \qquad  t_0=(m_B - m_V)^2 \,.
\end{eqnarray}

\begin{figure}
\begin{center}
\includegraphics[width=0.28 \columnwidth]{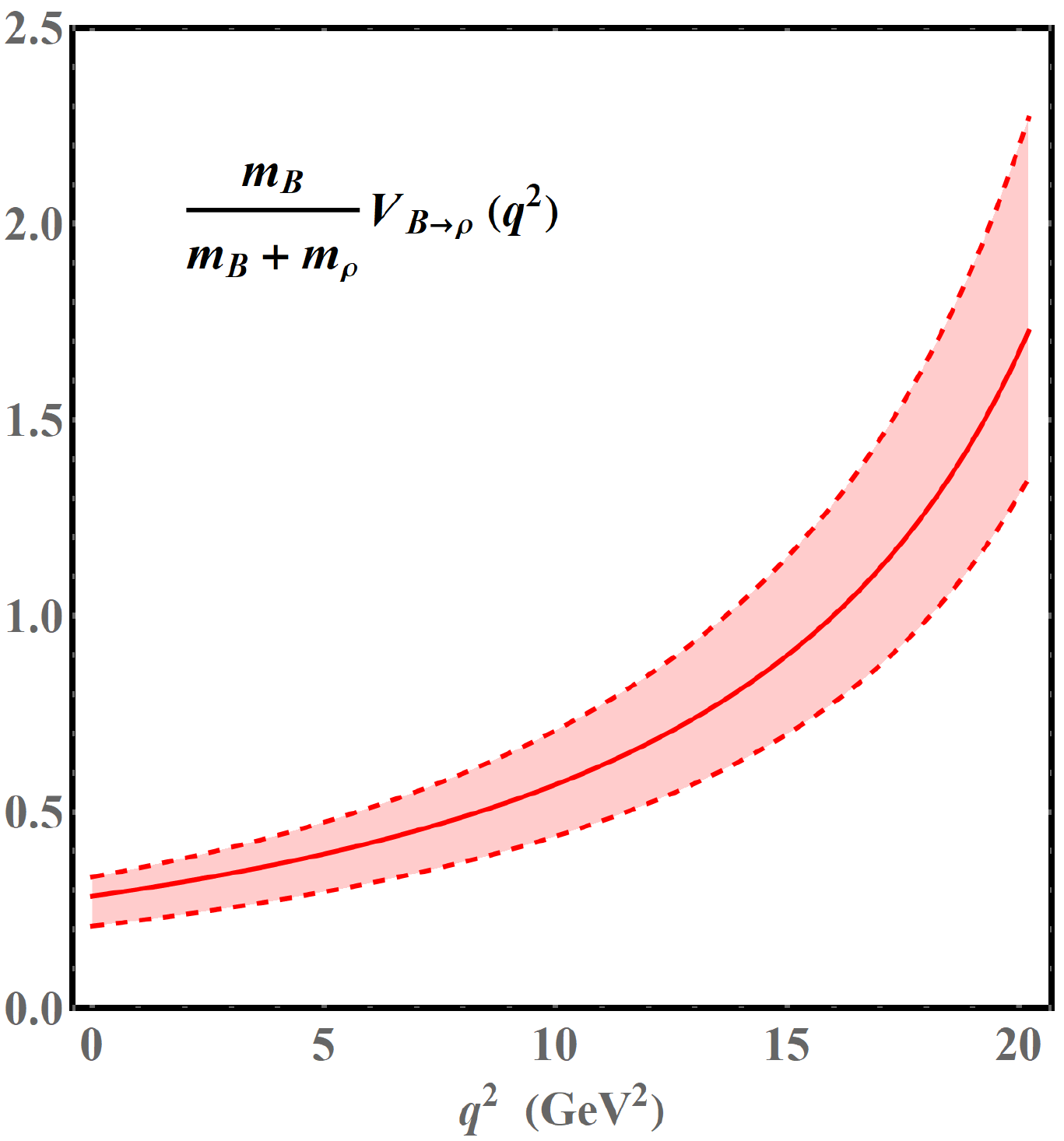}
\hspace{0.8 cm}
\includegraphics[width=0.28 \columnwidth]{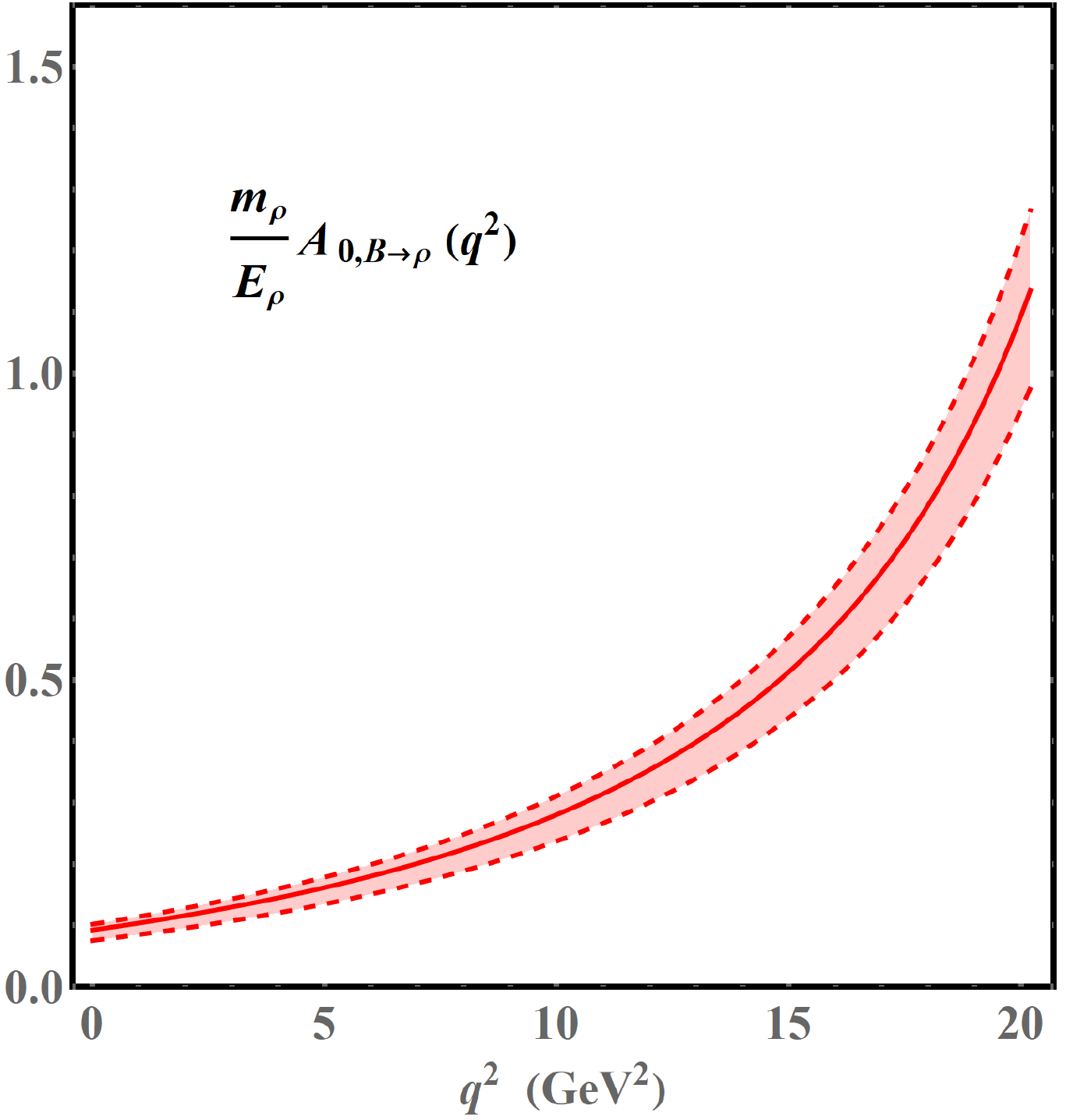}
\\
\includegraphics[width=0.28 \columnwidth]{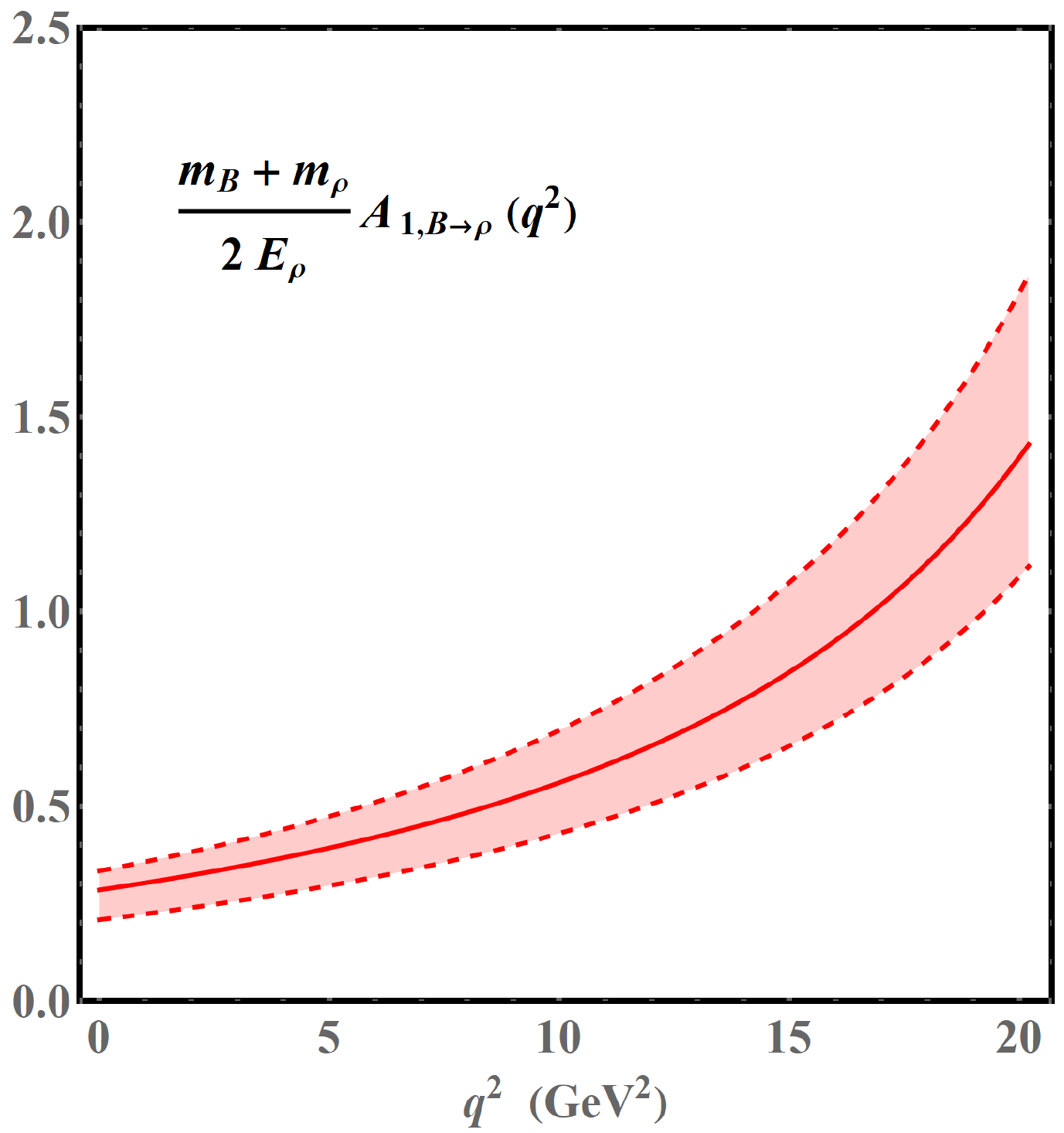}
\hspace{0.8 cm}
\includegraphics[width=0.28 \columnwidth]{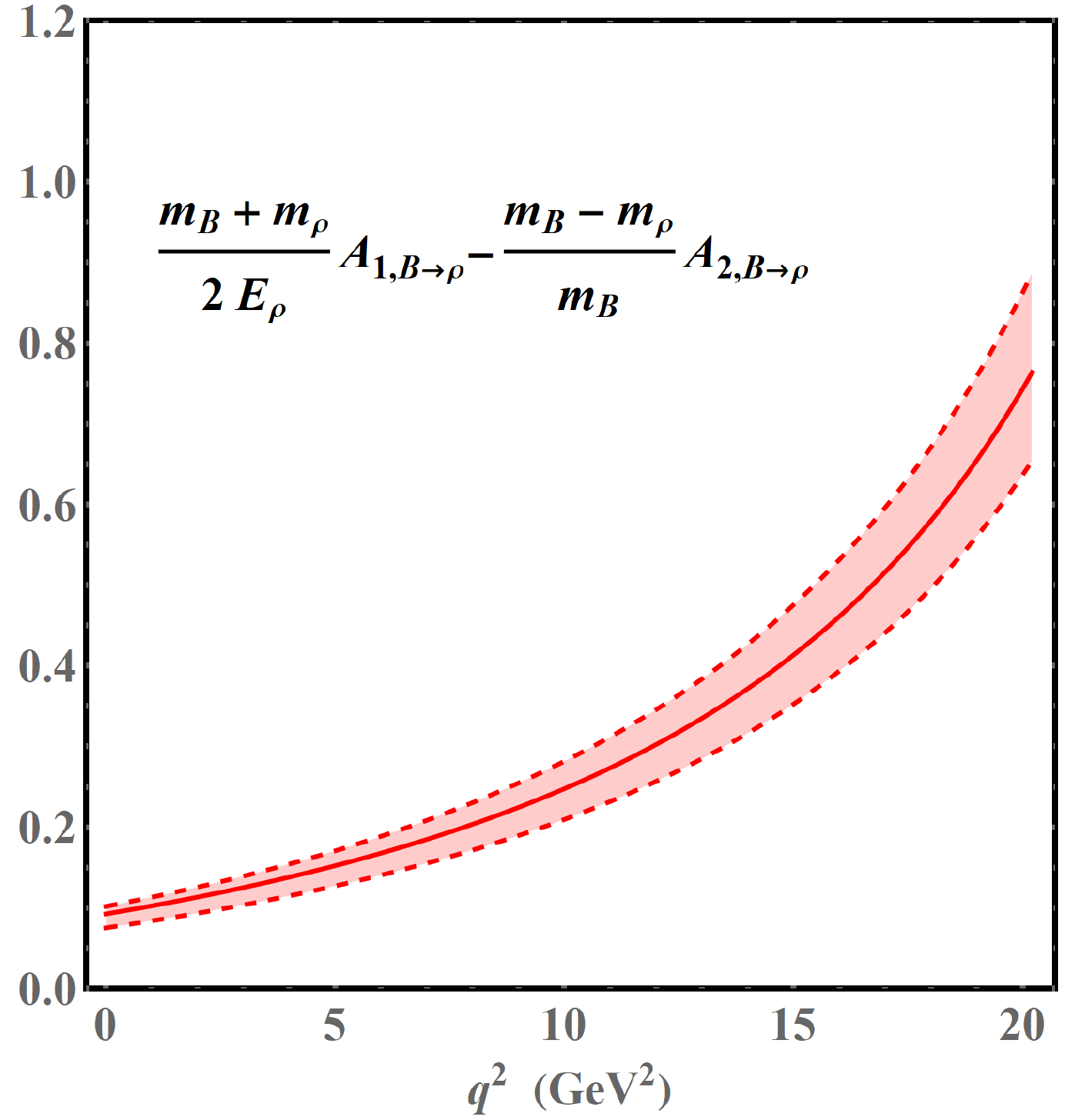}
\\
\includegraphics[width=0.28 \columnwidth]{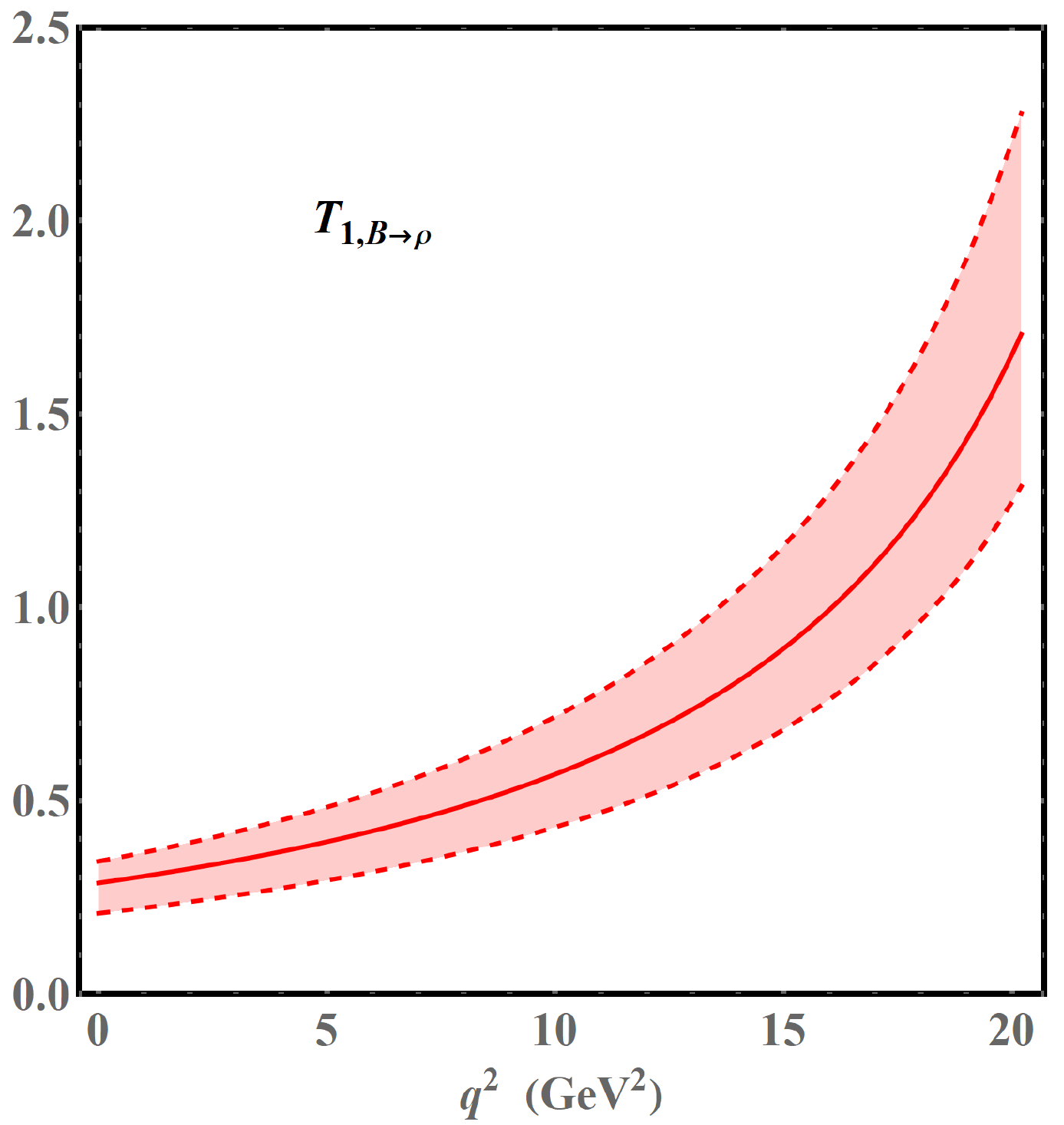}
\hspace{0.8 cm}
\includegraphics[width=0.28 \columnwidth]{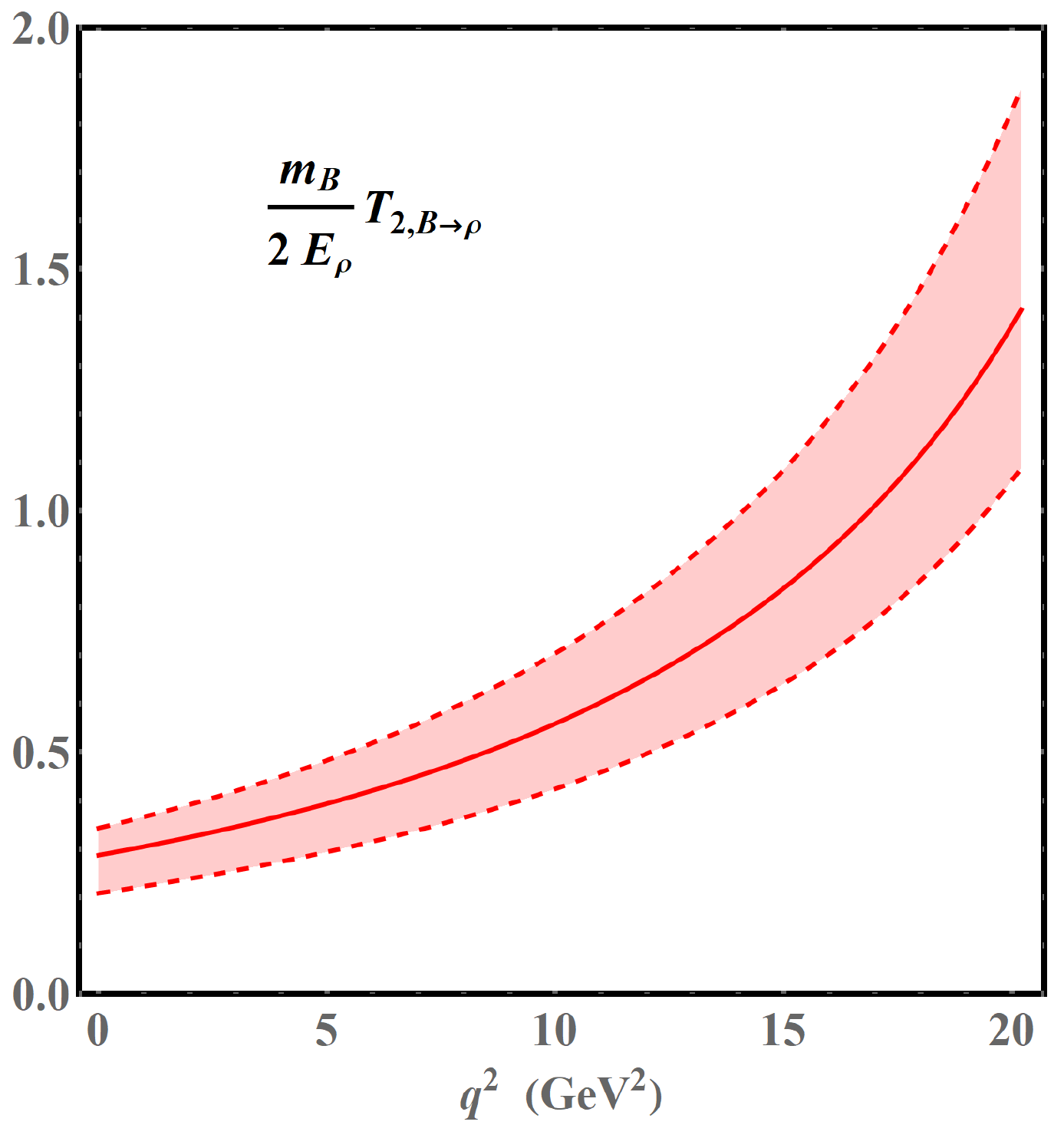}
\\
\includegraphics[width=0.28 \columnwidth]{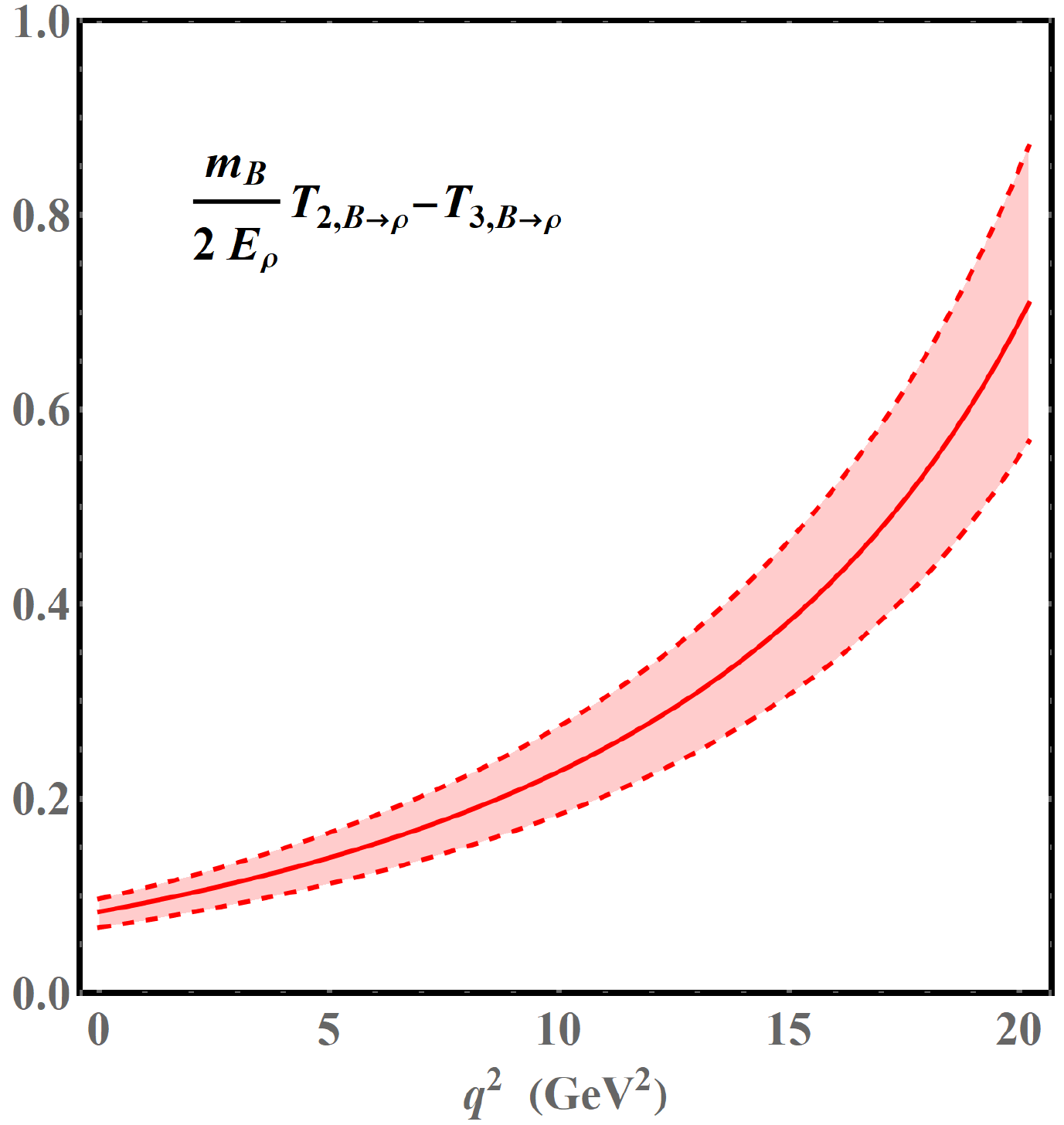}
\vspace*{0.1cm}
\caption{Theory predictions of the semileptonic $B \to \rho$ form factors obtained from the
SCET sum rules with the $B$-meson distribution amplitudes with an extrapolation to the entire
kinematical region by applying the $z$-series parametrizations (\ref{z-series expansion}).}
\label{fig: B to rho form factors}
\end{center}
\end{figure}

\begin{figure}
\begin{center}
\includegraphics[width=0.28 \columnwidth]{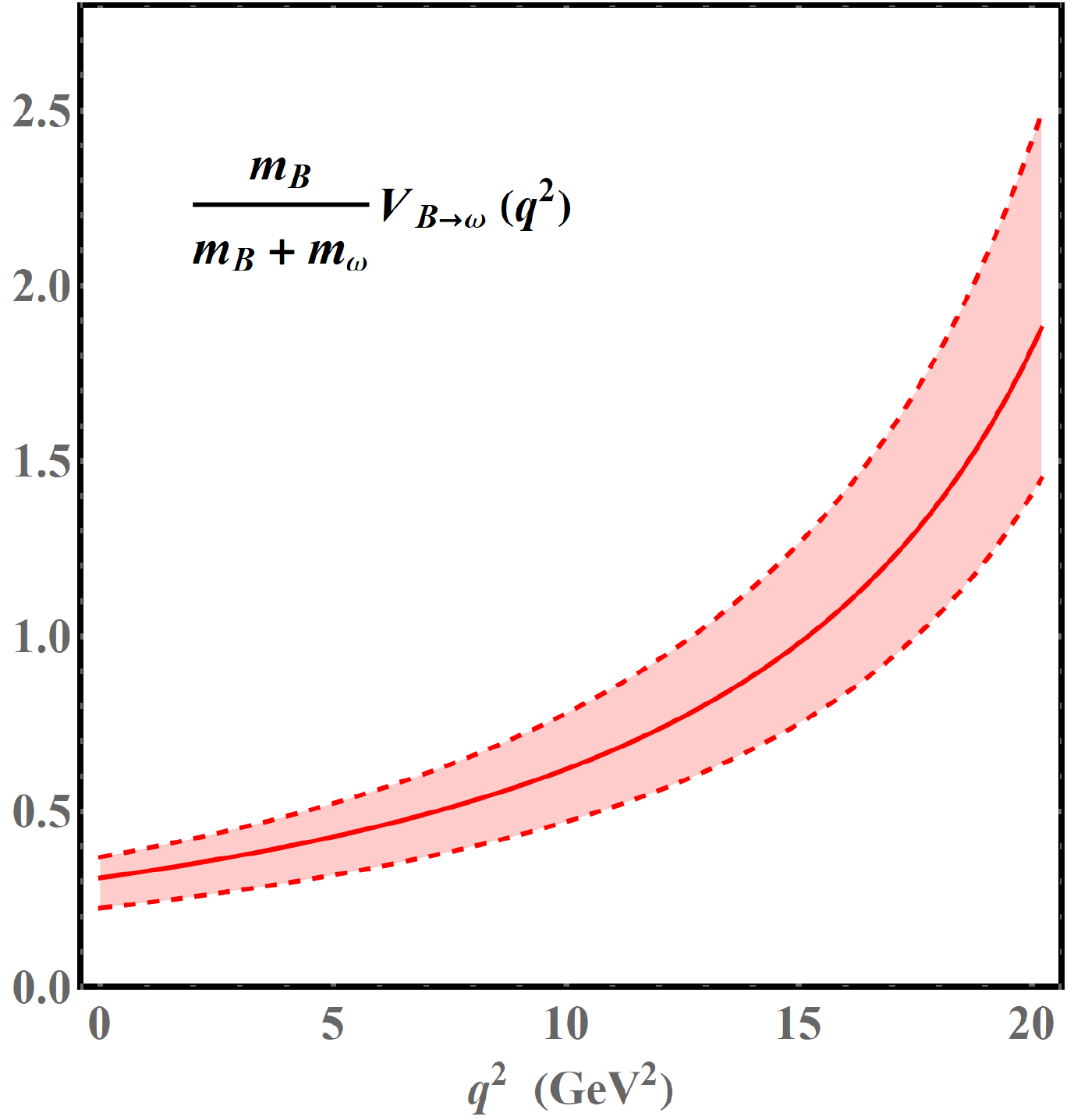}
\hspace{0.8 cm}
\includegraphics[width=0.28 \columnwidth]{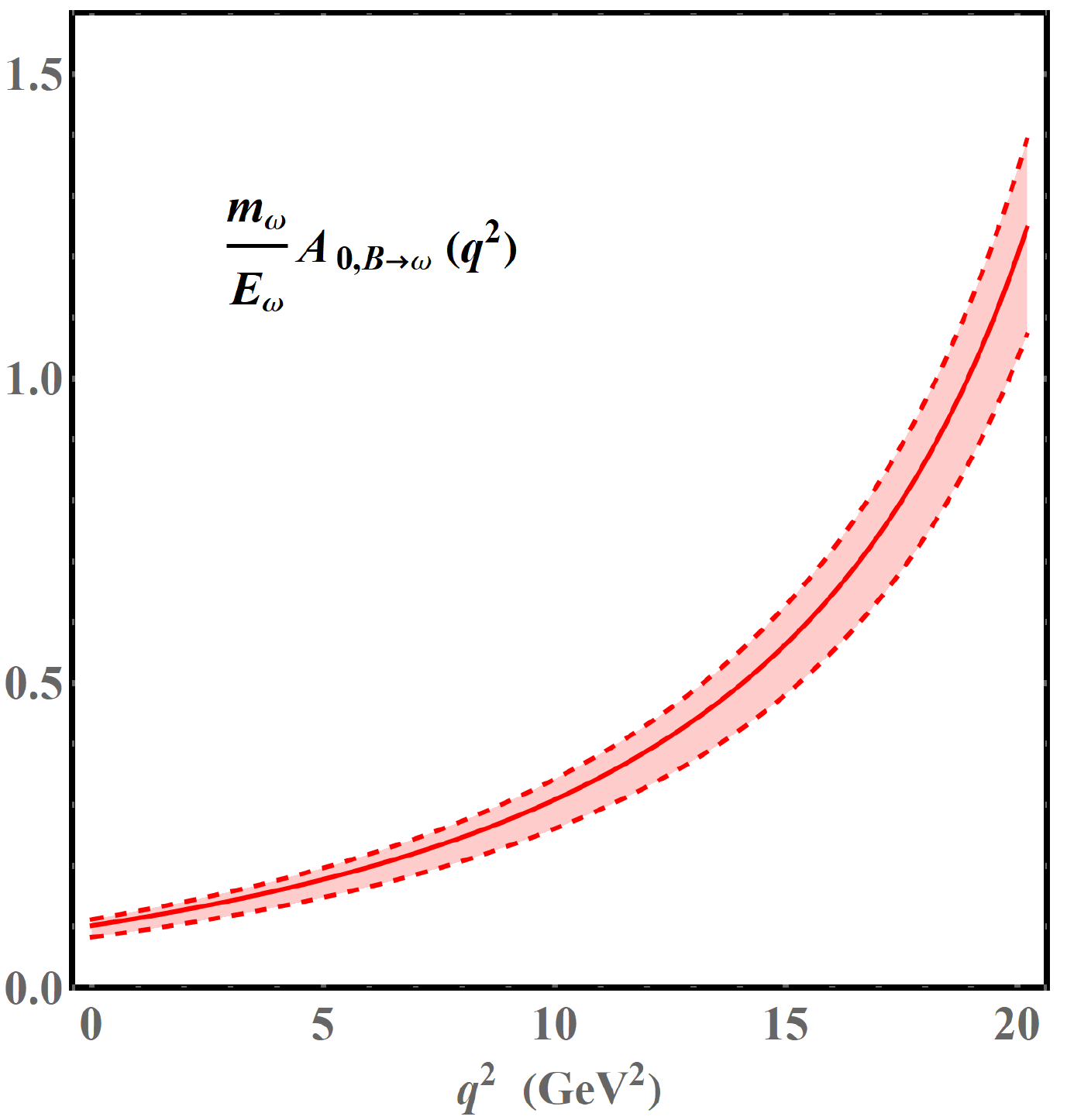}
\\
\includegraphics[width=0.28 \columnwidth]{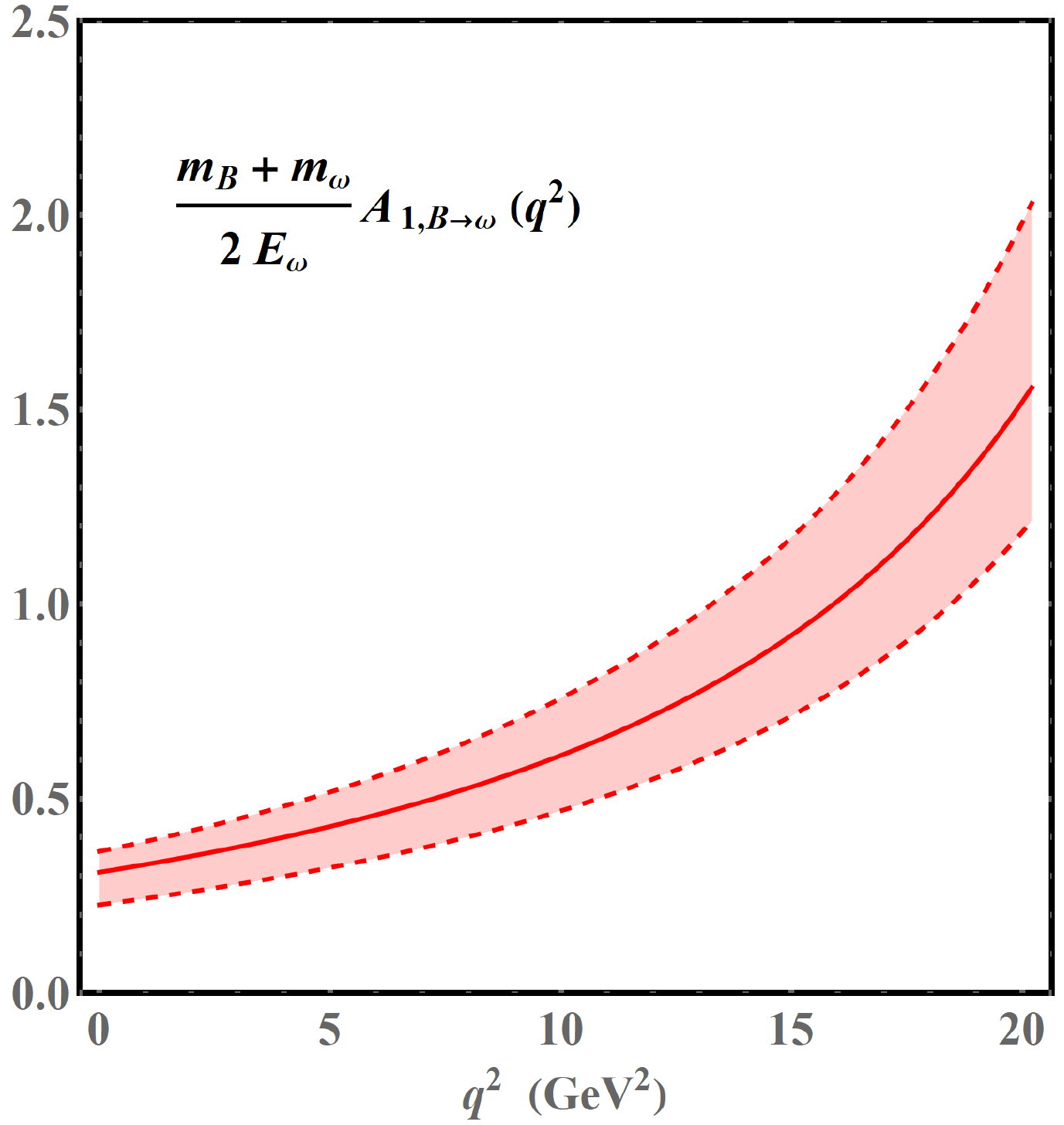}
\hspace{0.8 cm}
\includegraphics[width=0.28 \columnwidth]{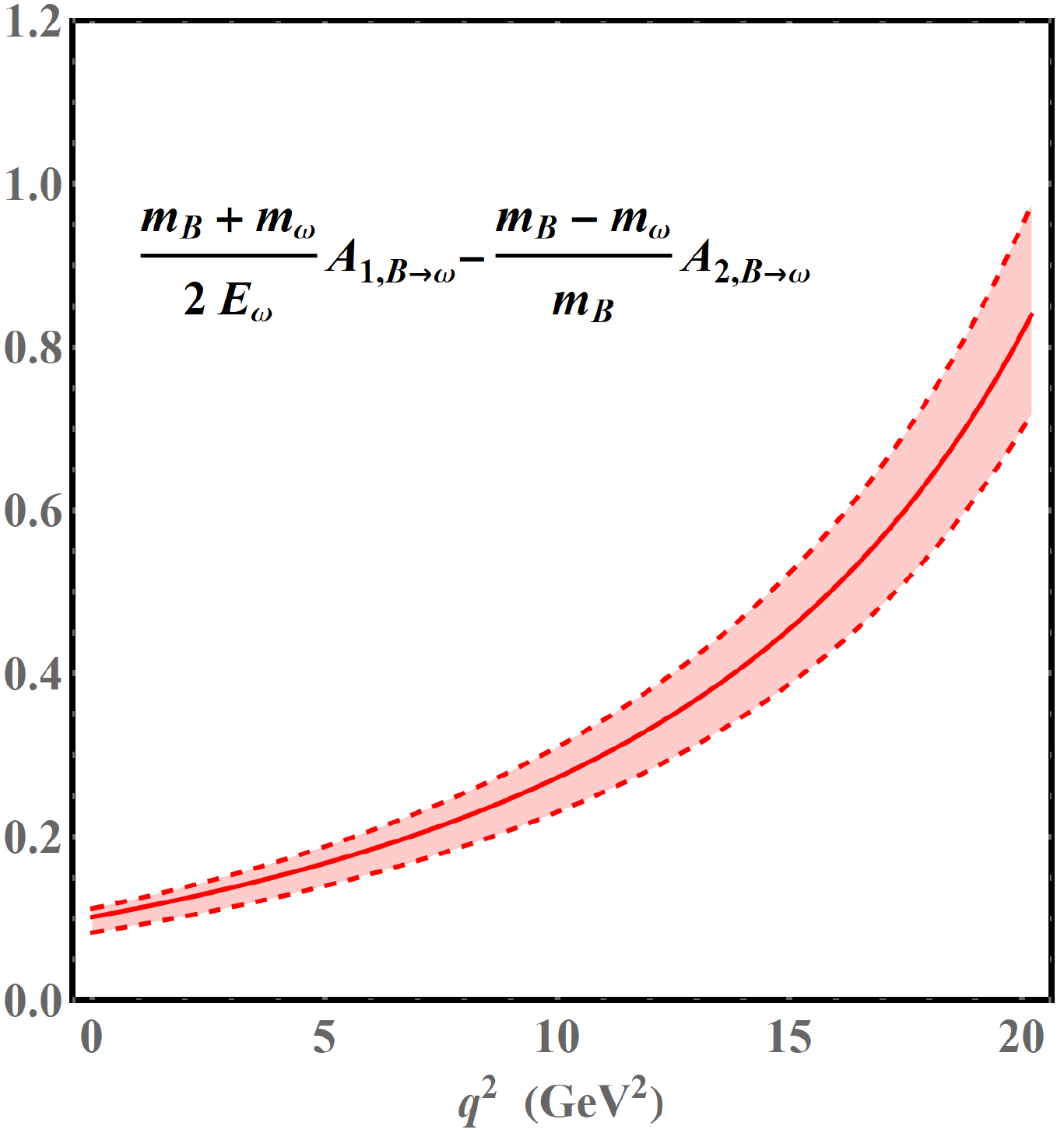}
\\
\includegraphics[width=0.28 \columnwidth]{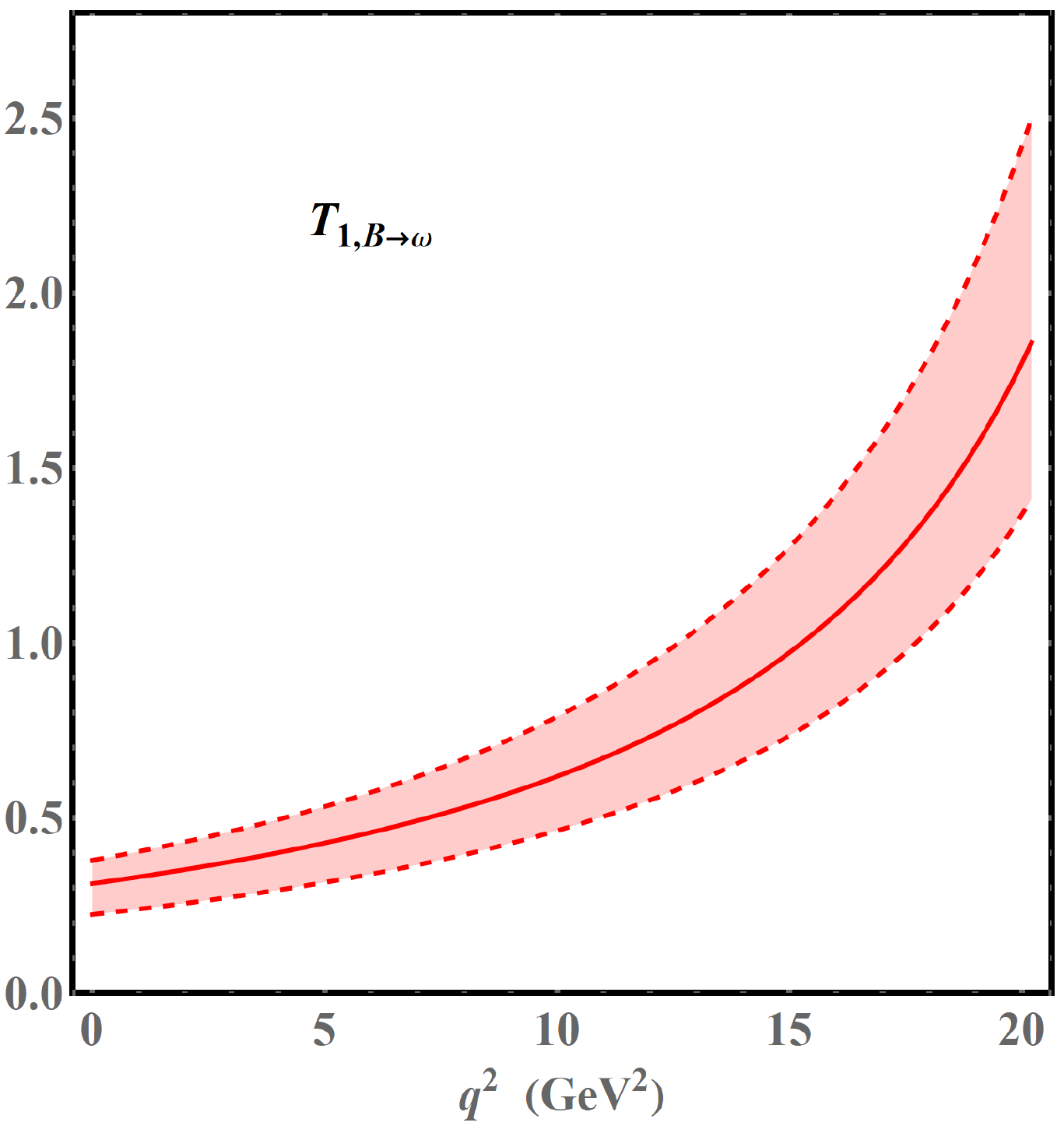}
\hspace{0.8 cm}
\includegraphics[width=0.28 \columnwidth]{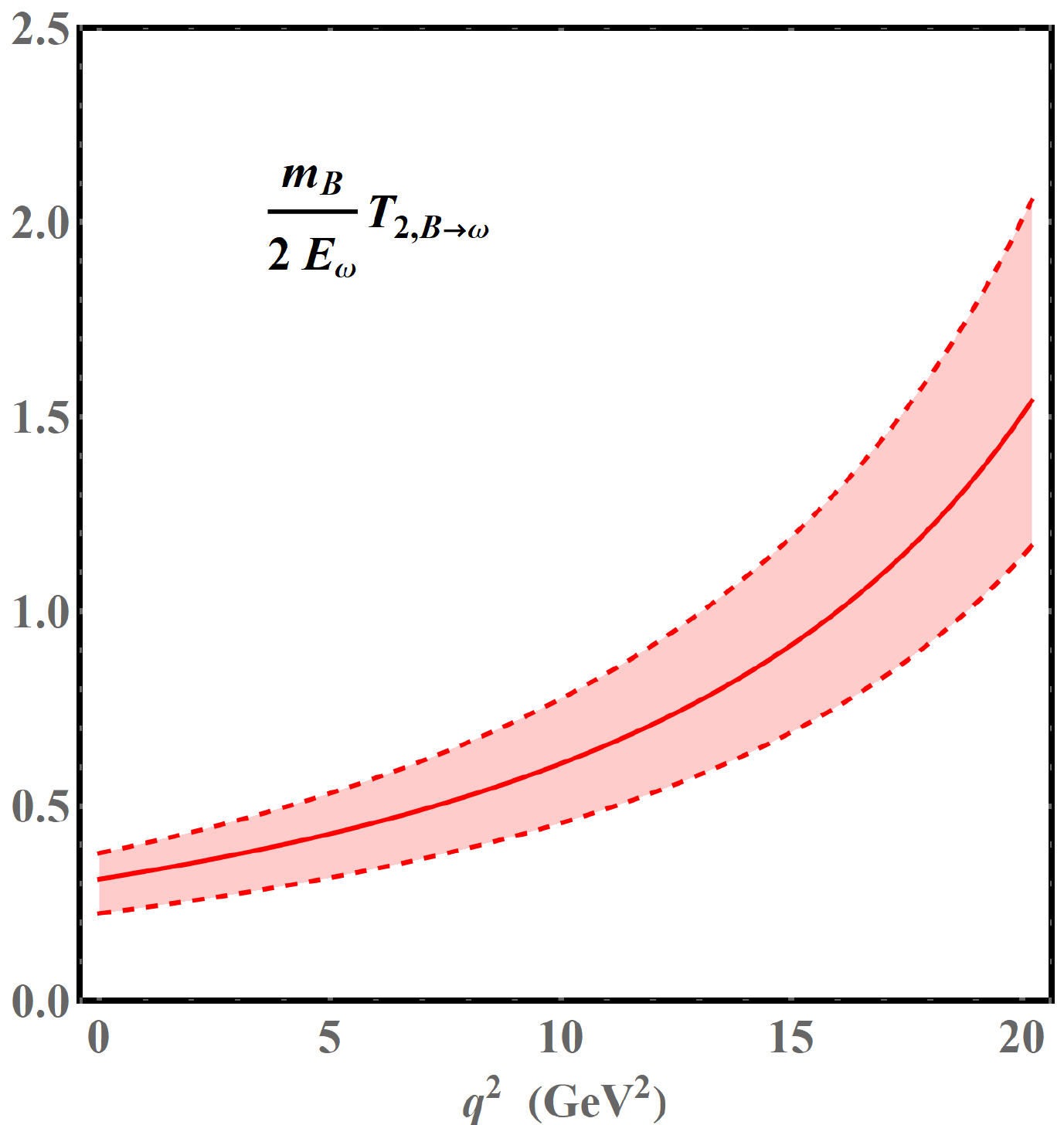}
\\
\includegraphics[width=0.28 \columnwidth]{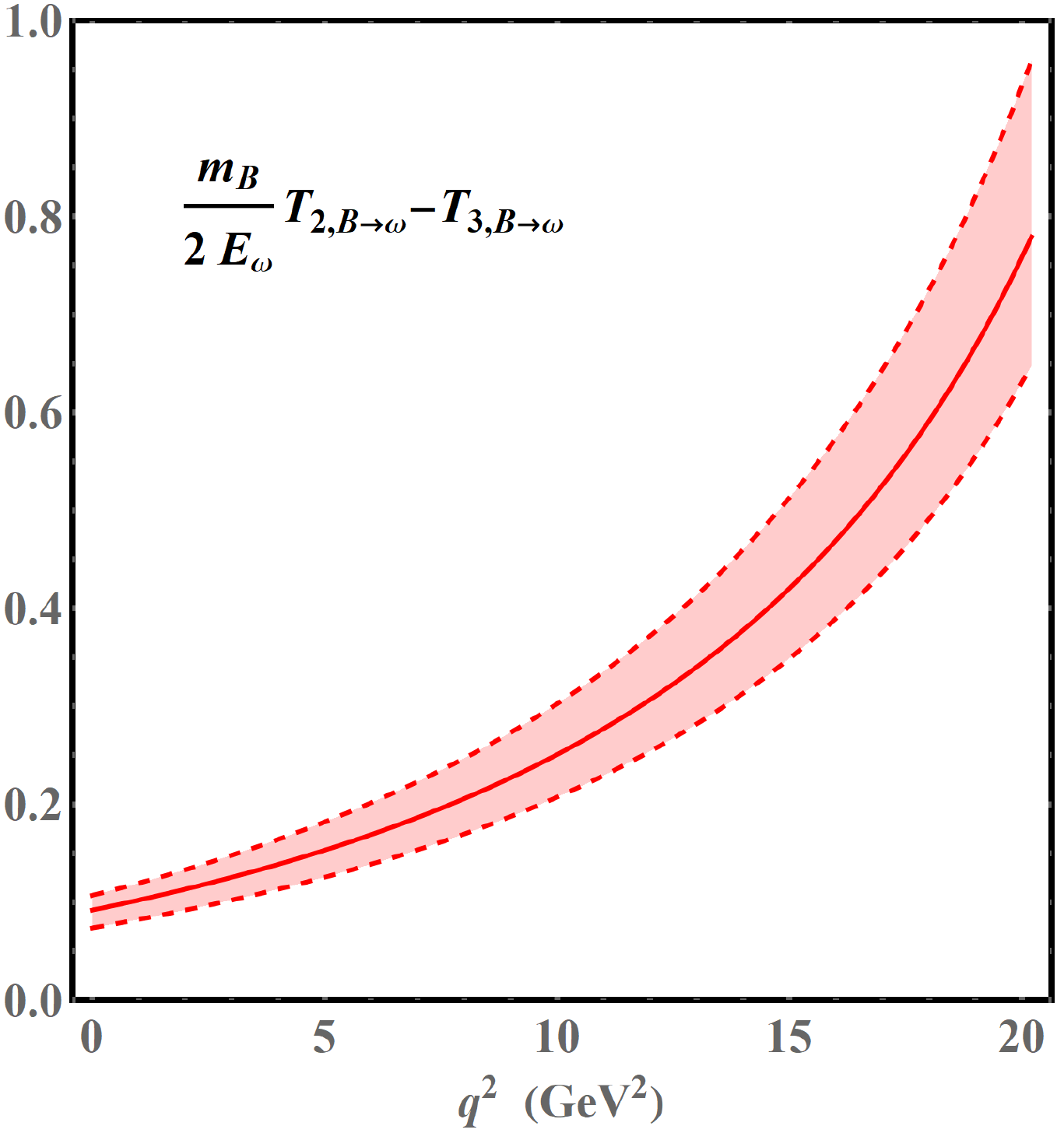}
\vspace*{0.1cm}
\caption{Theory predictions of the semileptonic $B \to \omega$ form factors obtained from the
SCET sum rules with the $B$-meson distribution amplitudes with an extrapolation to the entire
kinematical region by applying the $z$-series parametrizations (\ref{z-series expansion}).}
\label{fig: B to omega form factors}
\end{center}
\end{figure}

\begin{figure}
\begin{center}
\includegraphics[width=0.27 \columnwidth]{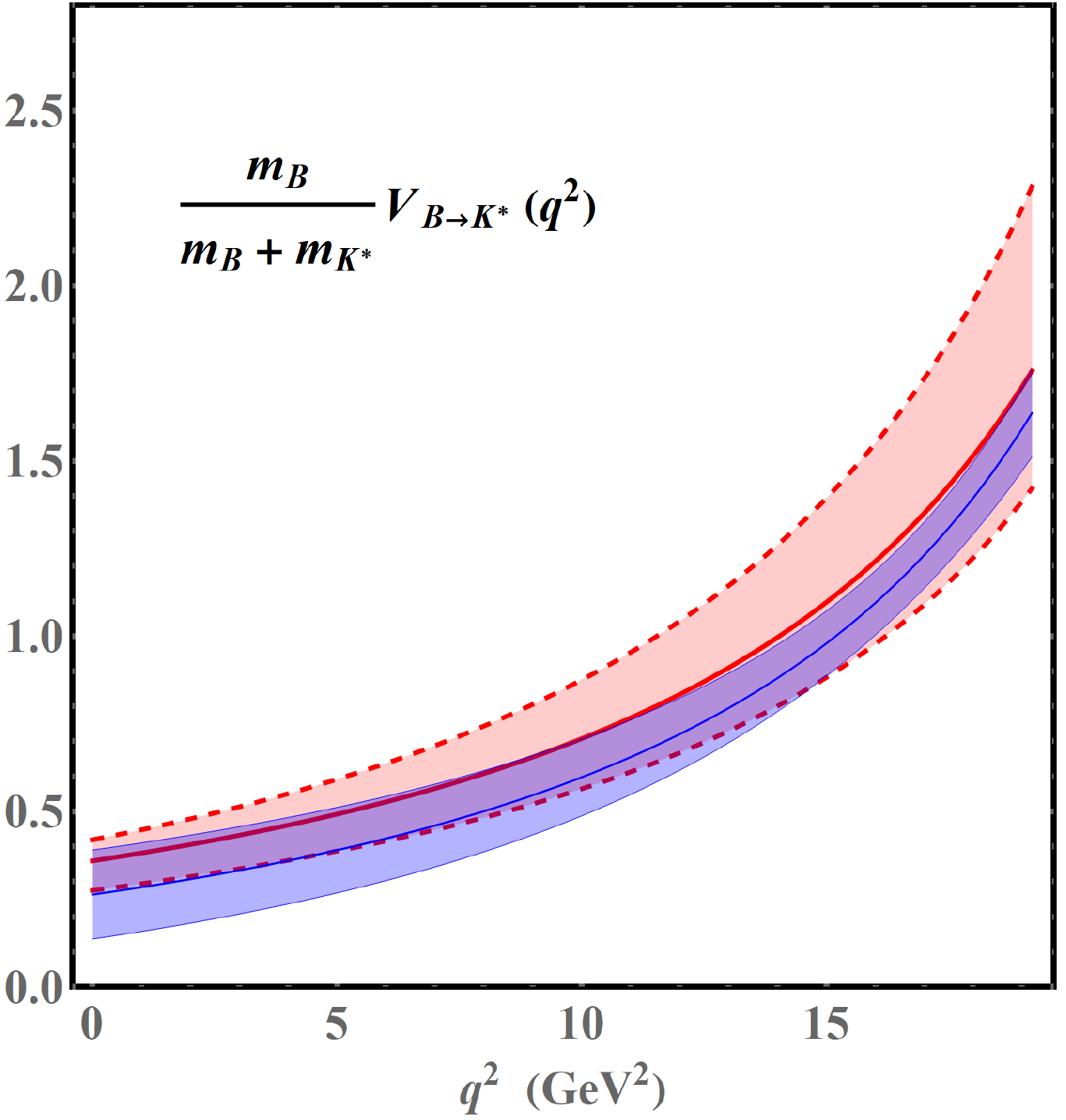}
\hspace{0.8 cm}
\includegraphics[width=0.27 \columnwidth]{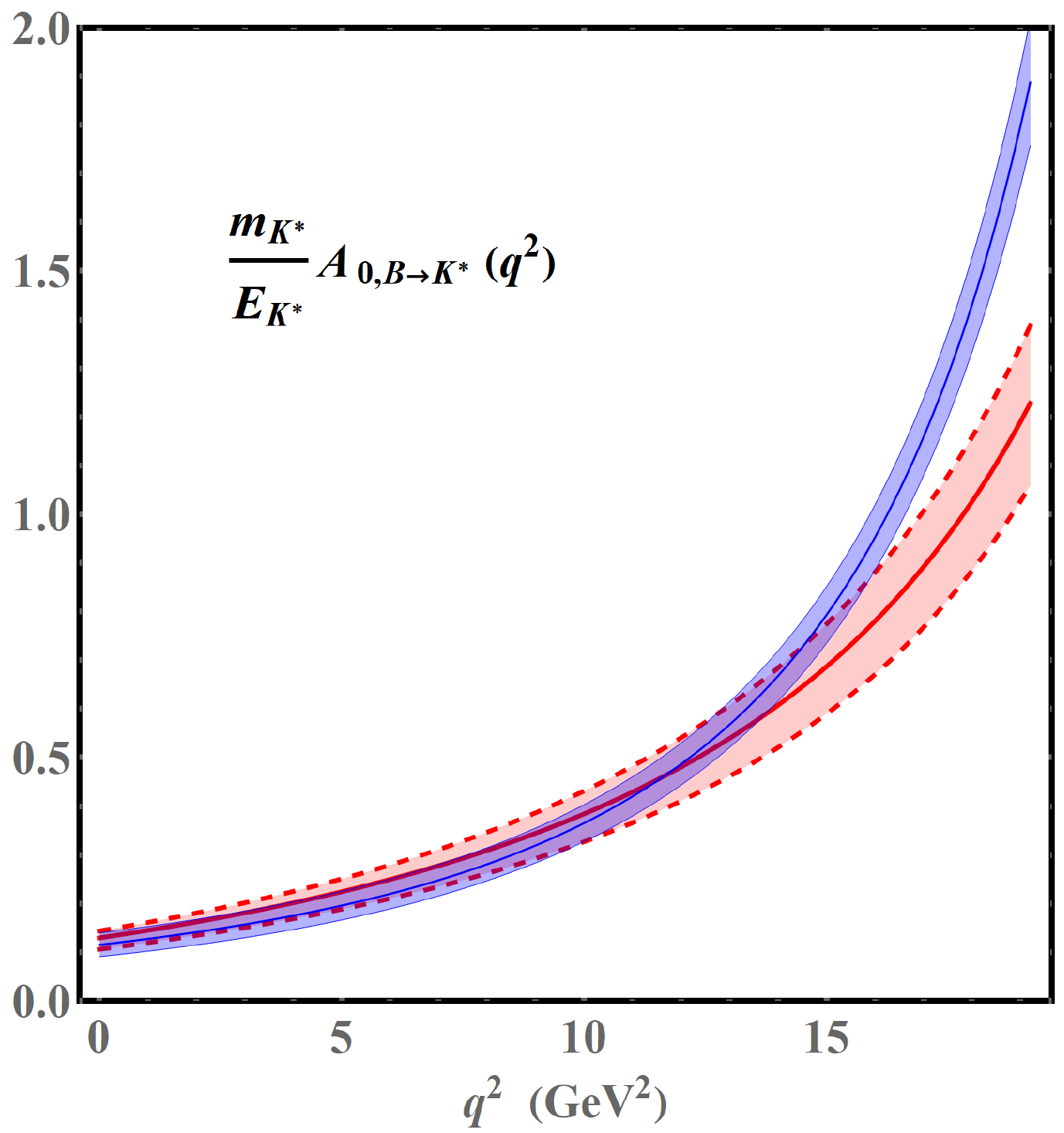}
\\
\includegraphics[width=0.27 \columnwidth]{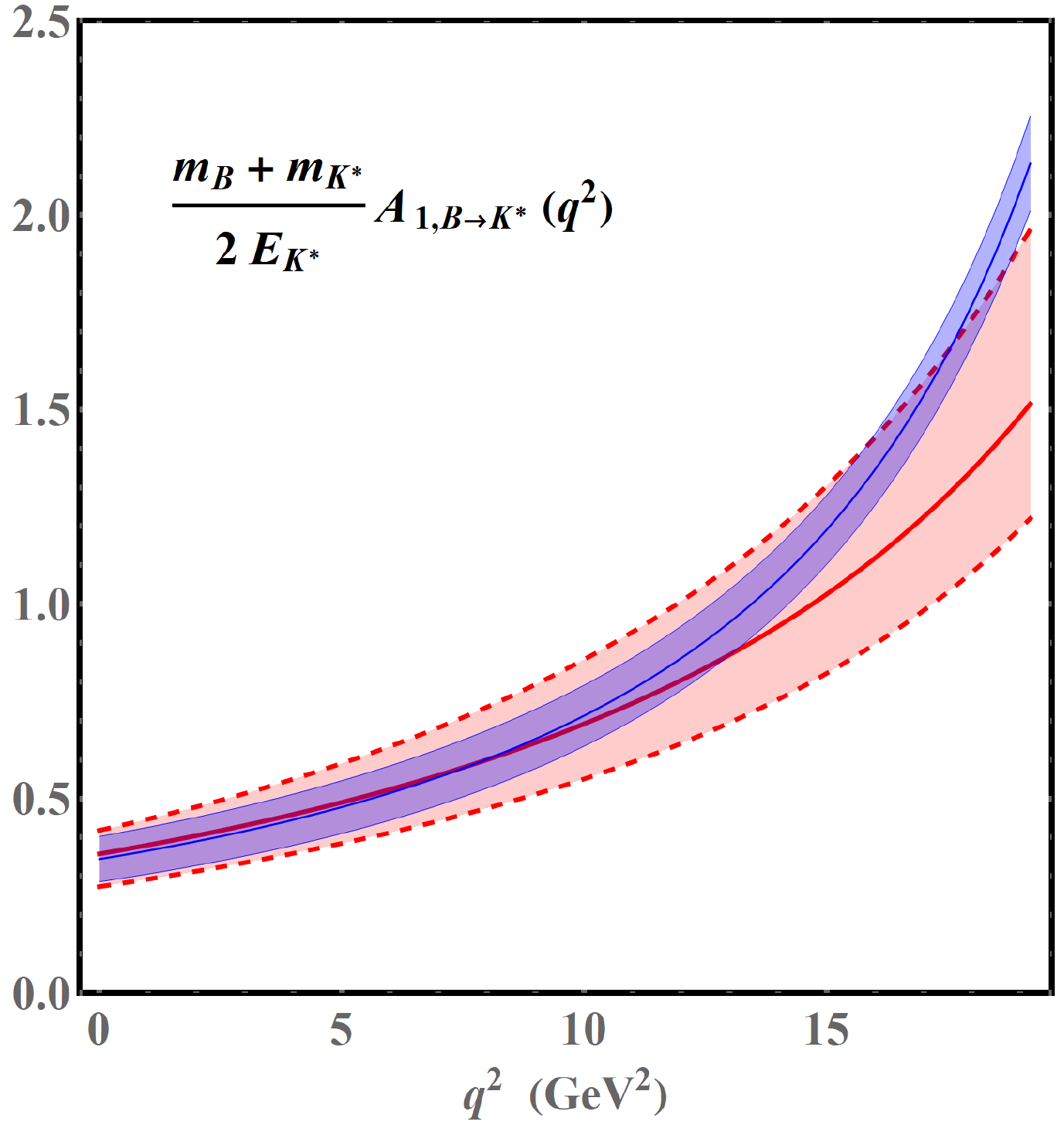}
\hspace{0.8 cm}
\includegraphics[width=0.27 \columnwidth]{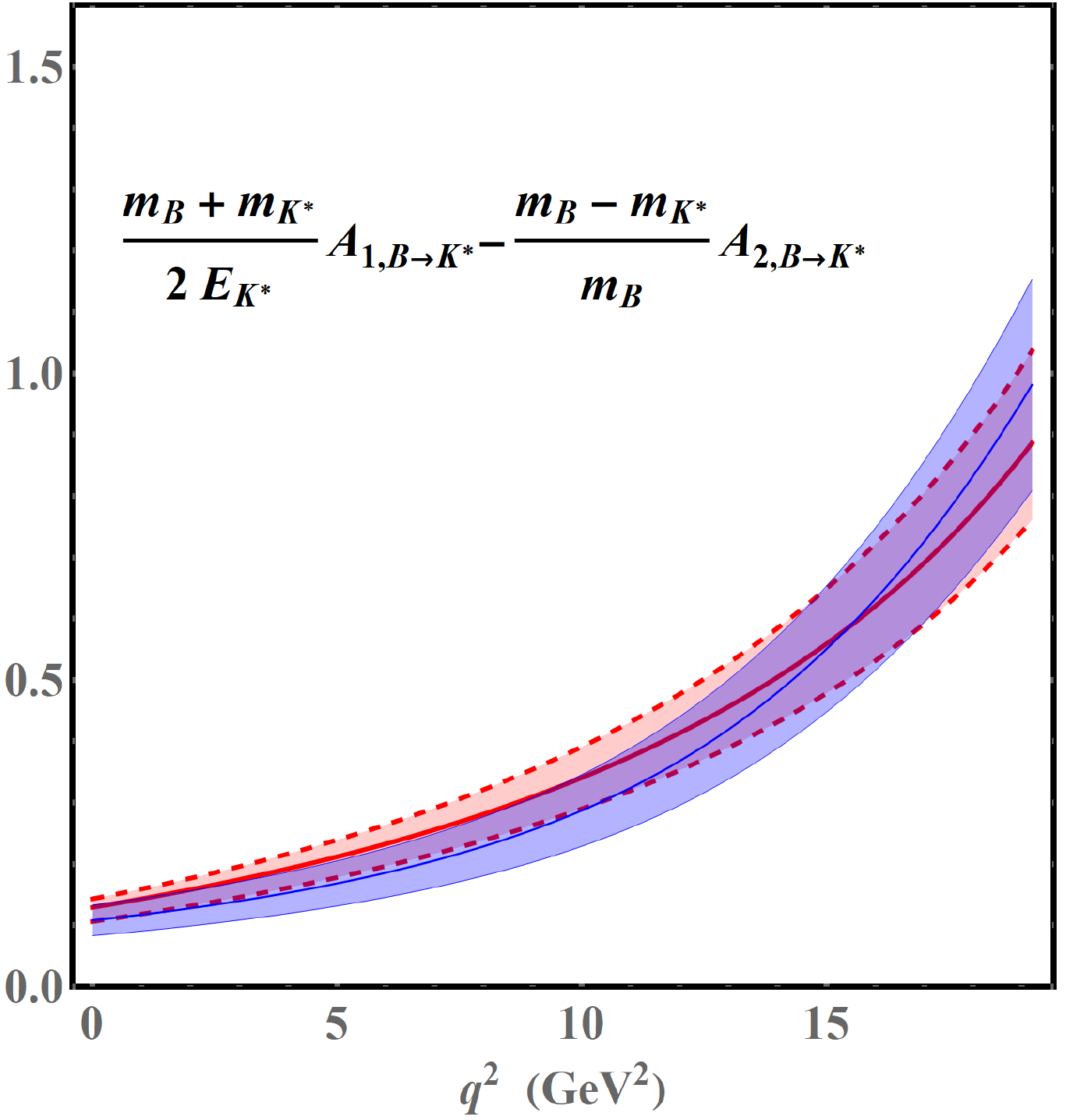}
\\
\includegraphics[width=0.27 \columnwidth]{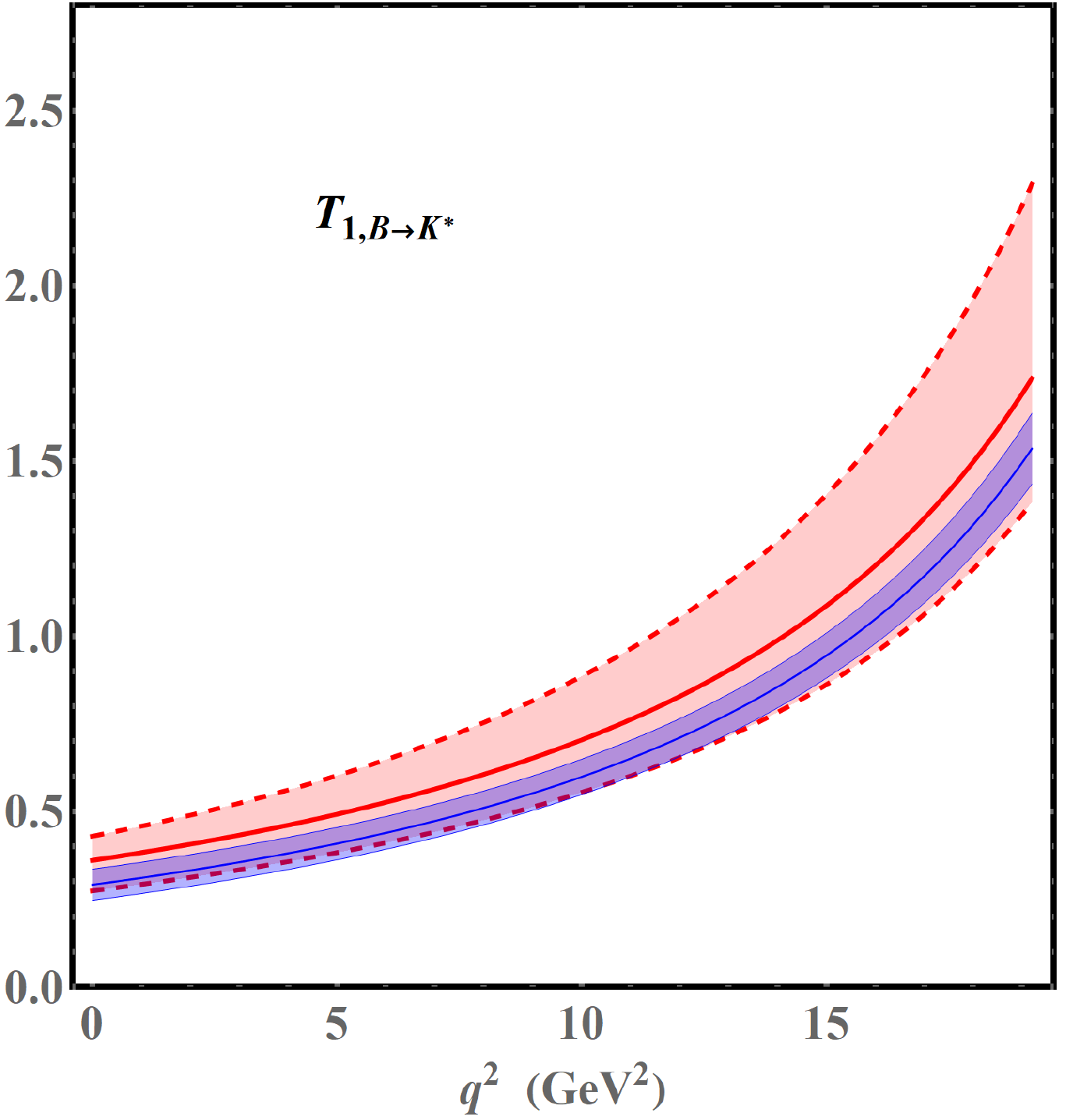}
\hspace{0.8 cm}
\includegraphics[width=0.27 \columnwidth]{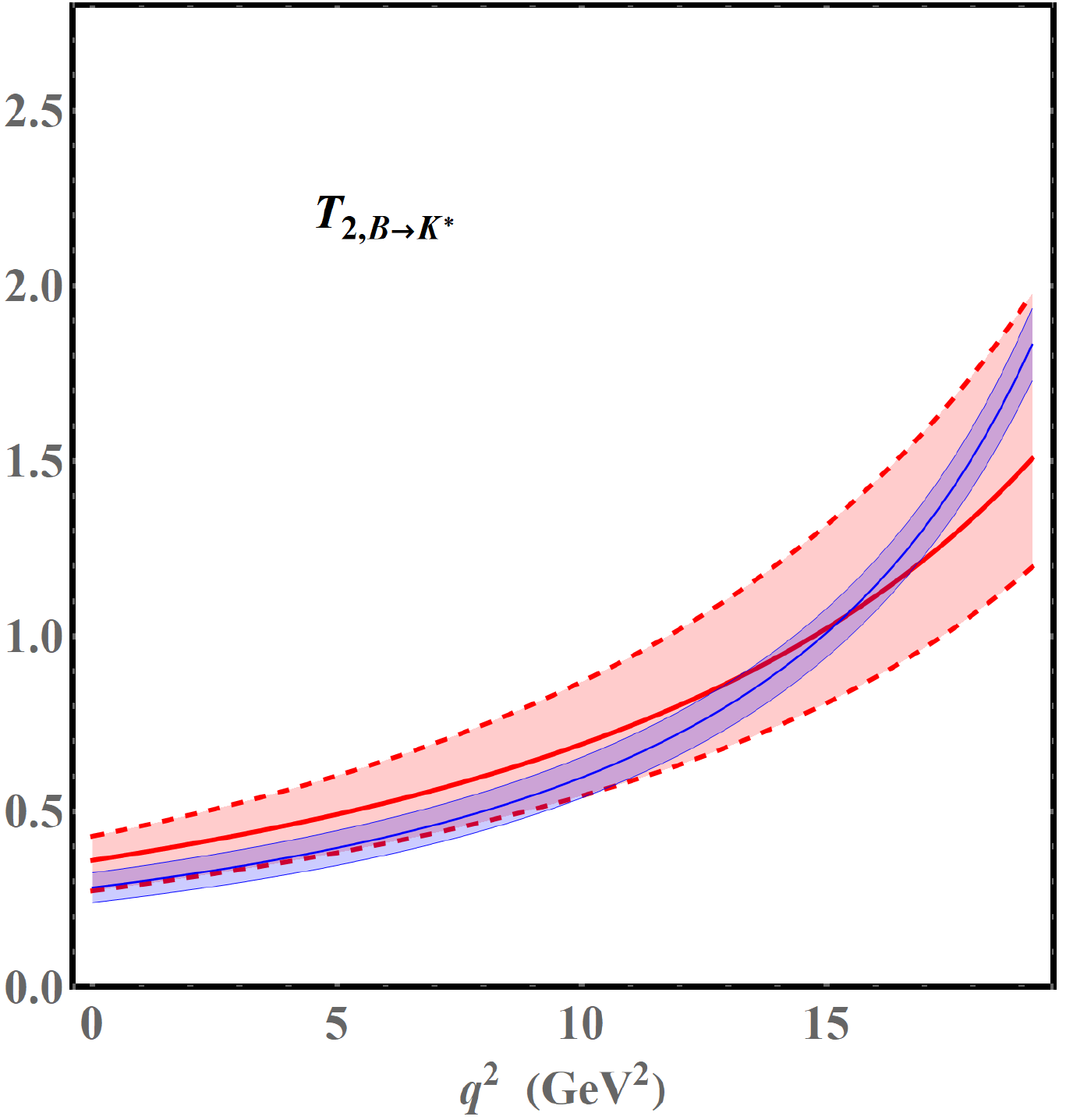}
\\
\includegraphics[width=0.27 \columnwidth]{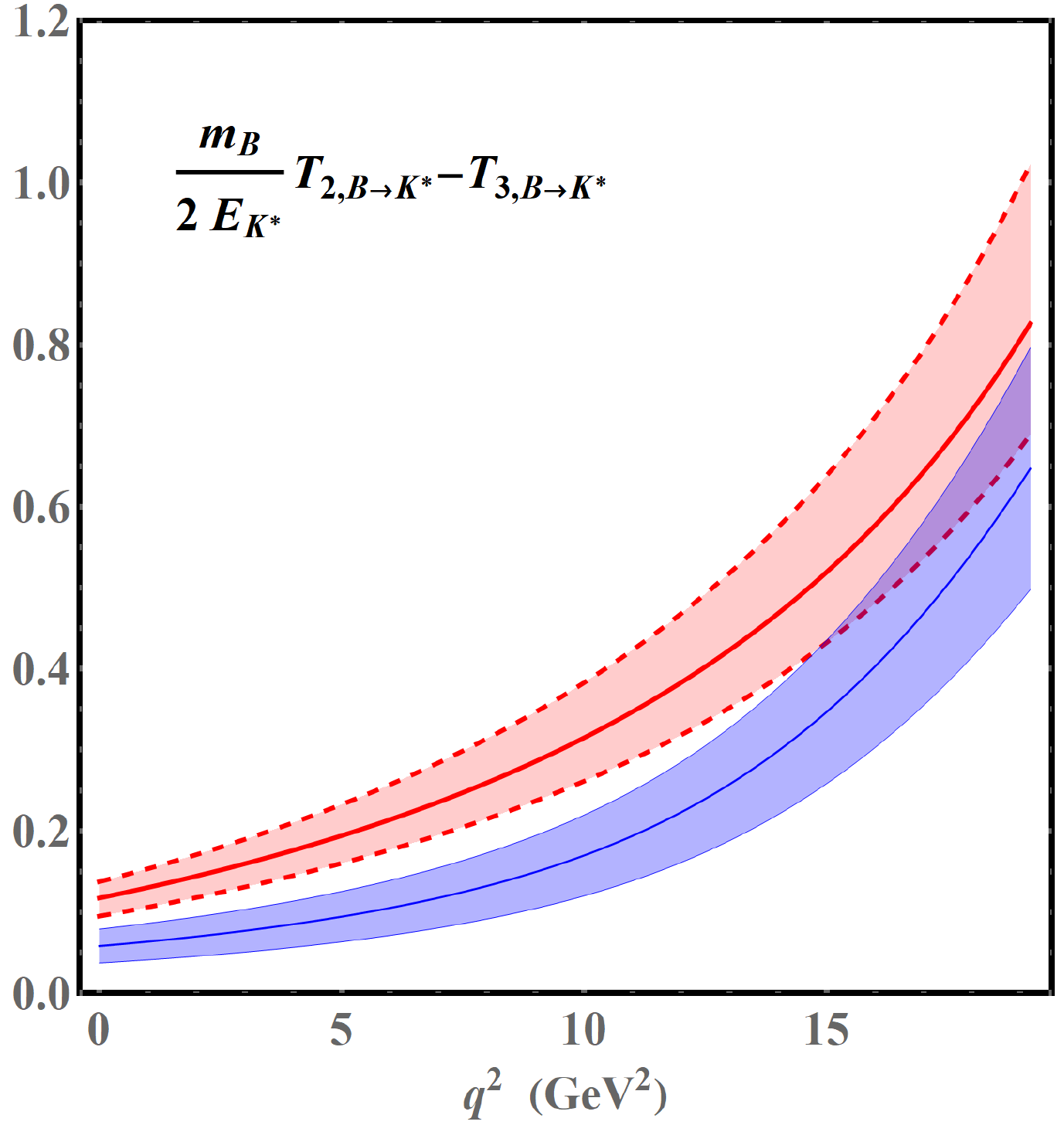}
\vspace*{0.1cm}
\caption{Theory predictions of the electro-weak penguin $B \to K^{\ast}$ decay form factors obtained from the
SCET sum rules with the $B$-meson distribution amplitudes with an extrapolation to the entire
kinematical region by applying the $z$-series parametrizations (\ref{z-series expansion}).
We also display the Lattice QCD predictions for these form factors with $2+1$ flavours of sea quarks
\cite{Horgan:2013hoa,Horgan:2015vla} as indicated by the blue bands for a comparison.}
\label{fig: B to kstar form factors}
\end{center}
\end{figure}

For the phenomenological applications we will adopt the Bourrely-Caprini-Lellouch (BCL)
version of the $z$-series expansion \cite{Bourrely:2008za}
(see \cite{Boyd:1995cf} for an alternative version and \cite{Khodjamirian:2011ub} for more discussions
in the context of the semileptonic $B \to \pi$ form factors)
\begin{eqnarray}
F_{B \to V}^{i}(q^2) = {F_{B \to V}^{i}(0) \over 1-q^2/m_{i, \, \rm pole}^2} \,
\left \{1 + \sum_{k=1}^N \, b_k^i \, \left [z(q^2, t_0)^k - z(0, t_0)^k \right ]  \right \}  \,.
\label{z-series expansion}
\end{eqnarray}
The adopted values of the various resonance masses from the Particle Data Group (PDG) \cite{Tanabashi:2018oca}
and from the heavy-hadron chiral perturbation theory \cite{Bardeen:2003kt} are summarized
in Table \ref{table: inputs of resonance masses}.
\begin{table}[t!bph]
\begin{center}
\begin{tabular}{|c|c|c|c|}
  \hline
  \hline
   &  &   & \\
  $F_{B \to V}^{i}(q^2)$ & \,\,\, $J^P$ \,\,\, & $b \to d$ \, (in  ${\rm GeV}$) & $b \to s$ \, (in ${\rm GeV}$) \\
   &  &   & \\
  \hline
 &  &   & \\
 $\mathcal{V}(q^2)$, \, $\mathcal{T}_1 (q^2)$  & $1^{-}$ & 5.325 & 5.415 \\
 &  &   & \\
 $\mathcal{A}_0 (q^2)$ & $0^{-}$ & 5.279 & 5.366 \\
  &  &   & \\
 $\mathcal{A}_1 (q^2)$,  \, $\mathcal{A}_{12} (q^2)$,  \, $\mathcal{T}_2 (q^2)$ \, $\mathcal{T}_{23}(q^2)$
 & $1^{+}$ & 5.724 & 5.829 \\
  &  &   & \\
  \hline
  \hline
\end{tabular}
\end{center}
\caption{Theory summary of the resonance masses with different quantum numbers entering the $z$-series
expansions of the QCD $B \to V$ form factors (\ref{z-series expansion}).
The calligraphic form factors represent the linear combinations of the conventionally defined form factors,
appearing in the ${\rm SCET_I}$ factorization formulae (\ref{SCET-I factorization formulae}) directly. }
\label{table: inputs of resonance masses}
\end{table}
For the practical purpose we will truncate the $z$-series expansion (\ref{z-series expansion})
at $N=1$ for the sake of fitting the coefficients $b_k^i$, keeping in mind that
$|z(q^2, t_0)|^2 \leq 0.04$ in the whole kinematic region
(see \cite{Bharucha:2010im} for further  discussions on the systematic uncertainties due to the
truncation-scheme dependence and \cite{Bigi:2016mdz} on the implementation of the strong unitary constraints).

It is straightforward to implement  the matching procedure for the semileptonic $B \to V$
form factors by employing the improved LCSR calculations at $-6 \,{\rm GeV}^2 \leq q^2 \leq 8 \, {\rm GeV}^2 $
and the $z$-series parametrizations (\ref{z-series expansion}).
Our predictions for the twenty-one form factors responsible for
the exclusive $B \to \rho, \, \omega, \, K^{\ast}$ transitions in
the entire kinematic region are displayed in figures
\ref{fig: B to rho form factors}, \ref{fig: B to omega form factors} and \ref{fig: B to kstar form factors},
where the theory uncertainties are obtained by adding all the separate uncertainties in quadrature
and the updated Lattice QCD results of $B \to \, K^{\ast}$ form factors with physical-mass bottom quarks and
$2+1$ flavours of sea quarks \cite{Horgan:2015vla} are also shown for a comparison.
Generally these two different QCD techniques lead to consistent  form-factor predictions  at  large 
hadronic recoil, with the exception of the longitudinal form factor
$m_B / (2 \, E_{K^{\ast}}) \, T_{2, \, B \to K^{\ast}}(q^2)-T_{3, \, B \to K^{\ast}}(q^2)$.
Such discrepancy  may be attributed to the fact that the form factor $T_3(q^2)$ cannot be isolated directly from
the helicity form factor $T_{23}(q^2)$  at large $q^2$ in the Lattice QCD simulations  \cite{Horgan:2013hoa},
due to the phase-space suppression.
We further collect the fitted results for the shape parameters $b_k^i$ and
the normalization constants $F_{B \to V}^{i}(0)$ entering the $z$-expansion (\ref{z-series expansion})
with numerically important uncertainties in Tables
\ref{table: results of  the shape parameters and normalizations of the vector B to rho FF},
\ref{table: results of  the shape parameters and normalizations of the tensor B to rho FF},
\ref{table: results of  the shape parameters and normalizations of the vector B to omega FF},
\ref{table: results of  the shape parameters and normalizations of the tensor B to omega FF},
\ref{table: results of  the shape parameters and normalizations of the vector B to kstar FF},
\ref{table: results of  the shape parameters and normalizations of the tensor B to kstar FF}.
Several remarks on the obtained numerical results are in order.

\begin{itemize}

\item{It is evident that the dominant theory uncertainties of the resulting predictions for
$F_{B \to V}^{i}(0)$ and $b_k^i$ originate from the model dependence of the $B$-meson distribution amplitudes
at a reference scale (including the (logarithmic)-inverse moments $\lambda_B$ and $\sigma_1$),
the factorization scale $\mu$ and the QCD renormalization scale $\nu$ for the tensor weak transition currents.
Consequently, it is of interest to perform the non-perturbative determination of  the momentum-dependence of
the leading-twist $B$-meson LCDA with the Lattice QCD technique and to compute the yet higher-order perturbative
QCD corrections to the ${\rm A0}$- and ${\rm B}$-type  SCET form factors with the method of sum rules. }

\item{Our theory predictions for the QCD $B \to \rho, \, \, K^{\ast}$ form factors in the whole kinematic region
are in reasonable agreement with the previous calculations applying the same framework \cite{Gubernari:2018wyi}.
However, the leading-twist contributions to  $B \to M$ form factors were only computed at LO
in the strong coupling $\alpha_s$ \cite{Gubernari:2018wyi} without implementing
the  summation of enhanced logarithms of $m_b/\Lambda$.
In addition,  the higher-twist corrections from the three-particle
$B$-meson distribution amplitudes were also estimated at the twist-four accuracy \cite{Gubernari:2018wyi},
implying the  violation of the QCD equation of motion (\ref{the gB-minus EOM}) already at the classical level.
It needs further to be pointed out that a comprehensive study of the higher-twist $B$-meson LCDA
up to the twist-six accuracy, including both the off light-cone corrections
and the four-body light-ray HQET operator effects, still remains as an interesting problem for the future improvement. }

\end{itemize}

\begin{table}[t!bph]
\begin{center}
\begin{tabular}{|c|c|c|c|c|c|c|c|c|}
  \hline
  \hline
  &  &  &  &  &  &  &  &  \\
  Parameters & Central \,\, value & $\lambda_B$ & $\sigma_1$ & $\mu$ & $\nu$ & $M^2$ & $s_0$ & $\phi_B^{\pm}(\omega)$ \\
  &  &  &  &  &  &  &  &  \\
  \hline
 \text{} & \text{} & \text{} & \text{} & \text{} & \text{} & \text{} & \text{} & \text{} \\
 $ \mathcal{V}_{B\rightarrow \rho }\text{(0)}$ & $ 0.285$ & $_{-0.027}^{+0.027}$ &
   $_{-0.016}^{+0.016}$ & $_{-0.055}^{+0.002}$ & - & $_{-0.000}^{+0.001}$ &
   $_{-0.018}^{+0.016}$ & $_{-0.027}^{+0.028}$ \\
 \text{} & \text{} & \text{} & \text{} & \text{} & \text{} & \text{} & \text{} & \text{} \\
 $ b_{1,\rho }^{\mathcal{V}}$ & $ -3.72$ & $_{-0.09}^{+0.14}$ & $_{-0.06}^{+0.08}$ &
   $_{-0.76}^{+0.45}$ & - & $_{-0.07}^{+0.03}$ & $_{-0.09}^{+0.08}$ & $_{-1.43}^{+0.00}$
   \\
 \text{} & \text{} & \text{} & \text{} & \text{} & \text{} & \text{} & \text{} & \text{} \\
 \hline
 \text{} & \text{} & \text{} & \text{} & \text{} & \text{} & \text{} & \text{} & \text{} \\
 $ \mathcal{A}_{0,B\rightarrow \rho }\text{(0)}$ & $ 0.093$ & $_{-0.006}^{+0.006}$ &
   $_{-0.004}^{+0.004}$ & $_{-0.011}^{+0.000}$ & - & $_{-0.000}^{+0.000}$ &
   $_{-0.003}^{+0.002}$ & $_{-0.010}^{+0.004}$ \\
 \text{} & \text{} & \text{} & \text{} & \text{} & \text{} & \text{} & \text{} & \text{} \\
 $ b_{1,\rho }^{\mathcal{A}_0}$ & $ -11.9$ & $_{-0.1}^{+0.1}$ & $_{-0.1}^{+0.1}$ &
   $_{-0.6}^{+0.2}$ & - & $_{-0.4}^{+0.2}$ & $_{-0.1}^{+0.1}$ & $_{-0.9}^{+0.0}$ \\
 \text{} & \text{} & \text{} & \text{} & \text{} & \text{} & \text{} & \text{} & \text{} \\
 \hline
 \text{} & \text{} & \text{} & \text{} & \text{} & \text{} & \text{} & \text{} & \text{} \\
 $ \mathcal{A}_{1,B\rightarrow \rho }\text{(0)}$ & $ 0.285$ & $_{-0.027}^{+0.027}$ &
   $_{-0.016}^{+0.016}$ & $_{-0.055}^{+0.002}$ & - & $_{-0.001}^{+0.001}$ &
   $_{-0.018}^{+0.016}$ & $_{-0.027}^{+0.028}$ \\
 \text{} & \text{} & \text{} & \text{} & \text{} & \text{} & \text{} & \text{} & \text{} \\
 $ b_{1,\rho }^{\mathcal{A}_1}$ & $ -4.63$ & $_{-0.10}^{+0.15}$ & $_{-0.06}^{+0.08}$ &
   $_{-0.78}^{+0.46}$ & - & $_{-0.07}^{+0.03}$ & $_{-0.09}^{+0.08}$ & $_{-1.48}^{+0.00}$
   \\
 \text{} & \text{} & \text{} & \text{} & \text{} & \text{} & \text{} & \text{} & \text{} \\
 \hline
 \text{} & \text{} & \text{} & \text{} & \text{} & \text{} & \text{} & \text{} & \text{} \\
 $ \mathcal{A}_{12,B\rightarrow \rho }\text{(0)}$ & $ 0.093$ & $_{-0.006}^{+0.006}$ &
   $_{-0.004}^{+0.004}$ & $_{-0.011}^{+0.000}$ & - & $_{-0.000}^{+0.000}$ &
   $_{-0.003}^{+0.002}$ & $_{-0.010}^{+0.004}$ \\
 \text{} & \text{} & \text{} & \text{} & \text{} & \text{} & \text{} & \text{} & \text{} \\
 $ b_{1,\rho }^{\mathcal{A}_{12}}$ & $ -10.8$ & $_{-0.0}^{+0.1}$ & $_{-0.0}^{+0.0}$ &
   $_{-0.5}^{+0.3}$ & - & $_{-0.2}^{+0.1}$ & $_{-0.1}^{+0.1}$ & $_{-1.4}^{+0.0}$ \\
 \text{} & \text{} & \text{} & \text{} & \text{} & \text{} & \text{} & \text{} & \text{} \\
  \hline
  \hline
\end{tabular}
\end{center}
\caption{Theory summary of the fitted results for the shape parameters and normalizations
of the (axial)-vector $B \to \rho$ form factors with the numerically sizeable uncertainties
by varying  different input parameters.}
\label{table: results of  the shape parameters and normalizations of the vector B to rho FF}
\end{table}

\begin{table}[t!bph]
\begin{center}
\begin{tabular}{|c|c|c|c|c|c|c|c|c|}
  \hline
  \hline
  &  &  &  &  &  &  &  &  \\
  Parameters & Central \,\, value & $\lambda_B$ & $\sigma_1$ & $\mu$ & $\nu$ & $M^2$ & $s_0$ & $\phi_B^{\pm}(\omega)$ \\
  &  &  &  &  &  &  &  &  \\
  \hline
 \text{} & \text{} & \text{} & \text{} & \text{} & \text{} & \text{} & \text{} & \text{} \\
 $ \mathcal{T}_{1,B\rightarrow \rho }\text{(0)}$ & $ 0.287$ & $_{-0.027}^{+0.027}$ &
   $_{-0.016}^{+0.016}$ & $_{-0.055}^{+0.002}$ & $_{-0.016}^{+0.026}$ &
   $_{-0.000}^{+0.001}$ & $_{-0.018}^{+0.016}$ & $_{-0.027}^{+0.028}$ \\
 \text{} & \text{} & \text{} & \text{} & \text{} & \text{} & \text{} & \text{} & \text{} \\
 $ b_{1,\rho }^{\mathcal{T}_1}$ & $ -3.57$ & $_{-0.10}^{+0.15}$ & $_{-0.06}^{+0.08}$ &
   $_{-0.73}^{+0.44}$ & $_{-0.07}^{+0.04}$ & $_{-0.06}^{+0.03}$ & $_{-0.08}^{+0.08}$ &
   $_{-1.43}^{+0.00}$ \\
 \text{} & \text{} & \text{} & \text{} & \text{} & \text{} & \text{} & \text{} & \text{} \\
 \hline
 \text{} & \text{} & \text{} & \text{} & \text{} & \text{} & \text{} & \text{} & \text{} \\
 $ \mathcal{T}_{2,B\rightarrow \rho }\text{(0)}$ & $ 0.287$ & $_{-0.027}^{+0.027}$ &
   $_{-0.016}^{+0.016}$ & $_{-0.055}^{+0.002}$ & $_{-0.016}^{+0.026}$ &
   $_{-0.000}^{+0.001}$ & $_{-0.018}^{+0.016}$ & $_{-0.027}^{+0.028}$ \\
 \text{} & \text{} & \text{} & \text{} & \text{} & \text{} & \text{} & \text{} & \text{} \\
 $ b_{1,\rho }^{\mathcal{T}_2}$ & $ -4.48$ & $_{-0.10}^{+0.16}$ & $_{-0.07}^{+0.08}$ &
   $_{-0.75}^{+0.46}$ & $_{-0.07}^{+0.04}$ & $_{-0.06}^{+0.03}$ & $_{-0.09}^{+0.08}$ &
   $_{-1.47}^{+0.00}$ \\
 \text{} & \text{} & \text{} & \text{} & \text{} & \text{} & \text{} & \text{} & \text{} \\
 \hline
 \text{} & \text{} & \text{} & \text{} & \text{} & \text{} & \text{} & \text{} & \text{} \\
 $ \mathcal{T}_{23,B\rightarrow \rho }\text{(0)}$ & $ 0.084$ & $_{-0.006}^{+0.006}$ &
   $_{-0.004}^{+0.004}$ & $_{-0.011}^{+0.000}$ & $_{-0.004}^{+0.007}$ &
   $_{-0.000}^{+0.001}$ & $_{-0.003}^{+0.002}$ & $_{-0.006}^{+0.007}$ \\
 \text{} & \text{} & \text{} & \text{} & \text{} & \text{} & \text{} & \text{} & \text{} \\
 $ b_{1,\rho }^{\mathcal{T}_{23}}$ & $ -11.3$ & $_{-0.0}^{+0.1}$ & $_{-0.0}^{+0.0}$ &
   $_{-0.6}^{+0.4}$ & $_{-0.1}^{+0.0}$ & $_{-0.2}^{+0.1}$ & $_{-0.1}^{+0.1}$ &
   $_{-1.5}^{+0.0}$ \\
 \text{} & \text{} & \text{} & \text{} & \text{} & \text{} & \text{} & \text{} & \text{} \\
  \hline
  \hline
\end{tabular}
\end{center}
\caption{Theory summary of the fitted results for the shape parameters and normalizations
of the tensor $B \to \rho$ form factors with the numerically sizeable uncertainties
by varying  different input parameters.}
\label{table: results of  the shape parameters and normalizations of the tensor B to rho FF}
\end{table}

\begin{table}[t!bph]
\begin{center}
\begin{tabular}{|c|c|c|c|c|c|c|c|c|}
  \hline
  \hline
  &  &  &  &  &  &  &  &  \\
  Parameters & Central \,\, value & $\lambda_B$ & $\sigma_1$ & $\mu$ & $\nu$ & $M^2$ & $s_0$ & $\phi_B^{\pm}(\omega)$ \\
  &  &  &  &  &  &  &  &  \\
  \hline
 \text{} & \text{} & \text{} & \text{} & \text{} & \text{} & \text{} & \text{} & \text{} \\
 $ \mathcal{V}_{B\rightarrow \omega }\text{(0)}$ & $ 0.311$ & $_{-0.030}^{+0.030}$ &
   $_{-0.018}^{+0.017}$ & $_{-0.060}^{+0.002}$ & - & $_{-0.000}^{+0.001}$ &
   $_{-0.019}^{+0.018}$ & $_{-0.030}^{+0.030}$ \\
 \text{} & \text{} & \text{} & \text{} & \text{} & \text{} & \text{} & \text{} & \text{} \\
 $ b_{1,\omega }^{\mathcal{V}}$ & $ -3.73$ & $_{-0.09}^{+0.14}$ & $_{-0.06}^{+0.08}$ &
   $_{-0.76}^{+0.45}$ & - & $_{-0.07}^{+0.04}$ & $_{-0.09}^{+0.08}$ & $_{-1.44}^{+0.00}$
   \\
 \text{} & \text{} & \text{} & \text{} & \text{} & \text{} & \text{} & \text{} & \text{} \\
 \hline
 \text{} & \text{} & \text{} & \text{} & \text{} & \text{} & \text{} & \text{} & \text{} \\
 $ \mathcal{A}_{0,B\rightarrow \omega }\text{(0)}$ & $ 0.102$ & $_{-0.007}^{+0.007}$ &
   $_{-0.004}^{+0.004}$ & $_{-0.012}^{+0.000}$ & - & $_{-0.000}^{+0.001}$ &
   $_{-0.003}^{+0.003}$ & $_{-0.011}^{+0.004}$ \\
 \text{} & \text{} & \text{} & \text{} & \text{} & \text{} & \text{} & \text{} & \text{} \\
 $ b_{1,\omega }^{\mathcal{A}_0}$ & $ -11.9$ & $_{-0.1}^{+0.1}$ & $_{-0.1}^{+0.1}$ &
   $_{-0.6}^{+0.2}$ & - & $_{-0.4}^{+0.2}$ & $_{-0.1}^{+0.1}$ & $_{-0.9}^{+0.0}$ \\
 \text{} & \text{} & \text{} & \text{} & \text{} & \text{} & \text{} & \text{} & \text{} \\
 \hline
 \text{} & \text{} & \text{} & \text{} & \text{} & \text{} & \text{} & \text{} & \text{} \\
 $ \mathcal{A}_{1,B\rightarrow \omega }\text{(0)}$ & $ 0.310$ & $_{-0.030}^{+0.030}$ &
   $_{-0.018}^{+0.017}$ & $_{-0.060}^{+0.002}$ & - & $_{-0.000}^{+0.001}$ &
   $_{-0.019}^{+0.018}$ & $_{-0.030}^{+0.030}$ \\
 \text{} & \text{} & \text{} & \text{} & \text{} & \text{} & \text{} & \text{} & \text{} \\
 $ b_{1,\omega }^{\mathcal{A}_1}$ & $ -4.64$ & $_{-0.10}^{+0.15}$ & $_{-0.06}^{+0.08}$ &
   $_{-0.78}^{+0.46}$ & - & $_{-0.07}^{+0.04}$ & $_{-0.09}^{+0.08}$ & $_{-1.48}^{+0.00}$
   \\
 \text{} & \text{} & \text{} & \text{} & \text{} & \text{} & \text{} & \text{} & \text{} \\
 \hline
 \text{} & \text{} & \text{} & \text{} & \text{} & \text{} & \text{} & \text{} & \text{} \\
 $ \mathcal{A}_{12,B\rightarrow \omega }\text{(0)}$ & $ 0.102$ & $_{-0.007}^{+0.007}$ &
   $_{-0.004}^{+0.004}$ & $_{-0.012}^{+0.000}$ & - & $_{-0.000}^{+0.001}$ &
   $_{-0.003}^{+0.003}$ & $_{-0.011}^{+0.004}$ \\
 \text{} & \text{} & \text{} & \text{} & \text{} & \text{} & \text{} & \text{} & \text{} \\
 $ b_{1,\omega }^{\mathcal{A}_{12}}$ & $ -10.9$ & $_{-0.0}^{+0.1}$ & $_{-0.0}^{+0.0}$ &
   $_{-0.5}^{+0.3}$ & - & $_{-0.2}^{+0.1}$ & $_{-0.1}^{+0.1}$ & $_{-1.4}^{+0.0}$ \\
 \text{} & \text{} & \text{} & \text{} & \text{} & \text{} & \text{} & \text{} & \text{} \\
  \hline
  \hline
\end{tabular}
\end{center}
\caption{Theory summary of the fitted results for the shape parameters and normalizations
of the (axial)-vector $B \to \omega$ form factors with the numerically sizeable uncertainties
by varying  different input parameters.}
\label{table: results of  the shape parameters and normalizations of the vector B to omega FF}
\end{table}

\begin{table}[t!bph]
\begin{center}
\begin{tabular}{|c|c|c|c|c|c|c|c|c|}
  \hline
  \hline
  &  &  &  &  &  &  &  &  \\
  Parameters & Central \,\, value & $\lambda_B$ & $\sigma_1$ & $\mu$ & $\nu$ & $M^2$ & $s_0$ & $\phi_B^{\pm}(\omega)$ \\
  &  &  &  &  &  &  &  &  \\
  \hline
 \text{} & \text{} & \text{} & \text{} & \text{} & \text{} & \text{} & \text{} & \text{} \\
 $ \mathcal{T}_{1,B\rightarrow \omega }\text{(0)}$ & $ 0.312$ & $_{-0.030}^{+0.030}$ &
   $_{-0.018}^{+0.018}$ & $_{-0.061}^{+0.002}$ & $_{-0.017}^{+0.028}$ &
   $_{-0.000}^{+0.001}$ & $_{-0.020}^{+0.018}$ & $_{-0.029}^{+0.031}$ \\
 \text{} & \text{} & \text{} & \text{} & \text{} & \text{} & \text{} & \text{} & \text{} \\
 $ b_{1,\omega }^{\mathcal{T}_1}$ & $ -3.58$ & $_{-0.10}^{+0.15}$ & $_{-0.06}^{+0.08}$ &
   $_{-0.73}^{+0.44}$ & $_{-0.07}^{+0.04}$ & $_{-0.06}^{+0.03}$ & $_{-0.08}^{+0.08}$ &
   $_{-1.43}^{+0.00}$ \\
 \text{} & \text{} & \text{} & \text{} & \text{} & \text{} & \text{} & \text{} & \text{} \\
 \hline
 \text{} & \text{} & \text{} & \text{} & \text{} & \text{} & \text{} & \text{} & \text{} \\
 $ \mathcal{T}_{2,B\rightarrow \omega }\text{(0)}$ & $ 0.312$ & $_{-0.030}^{+0.030}$ &
   $_{-0.018}^{+0.018}$ & $_{-0.061}^{+0.002}$ & $_{-0.017}^{+0.028}$ &
   $_{-0.000}^{+0.001}$ & $_{-0.020}^{+0.018}$ & $_{-0.029}^{+0.031}$ \\
 \text{} & \text{} & \text{} & \text{} & \text{} & \text{} & \text{} & \text{} & \text{} \\
 $ b_{1,\omega }^{\mathcal{T}_2}$ & $ -4.49$ & $_{-0.10}^{+0.16}$ & $_{-0.06}^{+0.08}$ &
   $_{-0.75}^{+0.46}$ & $_{-0.07}^{+0.04}$ & $_{-0.06}^{+0.03}$ & $_{-0.09}^{+0.08}$ &
   $_{-1.47}^{+0.00}$ \\
 \text{} & \text{} & \text{} & \text{} & \text{} & \text{} & \text{} & \text{} & \text{} \\
 \hline
 \text{} & \text{} & \text{} & \text{} & \text{} & \text{} & \text{} & \text{} & \text{} \\
 $ \mathcal{T}_{23,B\rightarrow \omega }\text{(0)}$ & $ 0.092$ & $_{-0.007}^{+0.007}$ &
   $_{-0.004}^{+0.004}$ & $_{-0.012}^{+0.000}$ & $_{-0.005}^{+0.008}$ &
   $_{-0.000}^{+0.000}$ & $_{-0.003}^{+0.003}$ & $_{-0.007}^{+0.008}$ \\
 \text{} & \text{} & \text{} & \text{} & \text{} & \text{} & \text{} & \text{} & \text{} \\
 $ b_{1,\omega }^{\mathcal{T}_{23}}$ & $ -11.3$ & $_{-0.0}^{+0.1}$ & $_{-0.0}^{+0.0}$ &
   $_{-0.7}^{+0.4}$ & $_{-0.1}^{+0.0}$ & $_{-0.2}^{+0.1}$ & $_{-0.1}^{+0.1}$ &
   $_{-1.5}^{+0.0}$ \\
 \text{} & \text{} & \text{} & \text{} & \text{} & \text{} & \text{} & \text{} & \text{} \\
  \hline
  \hline
\end{tabular}
\end{center}
\caption{Theory summary of the fitted results for the shape parameters and normalizations
of the tensor $B \to \omega$ form factors with the numerically sizeable uncertainties
by varying  different input parameters.}
\label{table: results of  the shape parameters and normalizations of the tensor B to omega FF}
\end{table}

\begin{table}[t!bph]
\begin{center}
\begin{tabular}{|c|c|c|c|c|c|c|c|c|}
  \hline
  \hline
  &  &  &  &  &  &  &  &  \\
  Parameters & Central \,\, value & $\lambda_B$ & $\sigma_1$ & $\mu$ & $\nu$ & $M^2$ & $s_0$ & $\phi_B^{\pm}(\omega)$ \\
  &  &  &  &  &  &  &  &  \\
  \hline
 \text{} & \text{} & \text{} & \text{} & \text{} & \text{} & \text{} & \text{} & \text{} \\
 $ \mathcal{V}_{B\rightarrow K^*}\text{(0)}$ & $ 0.359$ & $_{-0.032}^{+0.032}$ &
   $_{-0.019}^{+0.019}$ & $_{-0.062}^{+0.001}$ & - & $_{-0.004}^{+0.010}$ &
   $_{-0.017}^{+0.016}$ & $_{-0.027}^{+0.038}$ \\
 \text{} & \text{} & \text{} & \text{} & \text{} & \text{} & \text{} & \text{} & \text{} \\
 $ b_{1,K^*}^{\mathcal{V}}$ & $ -3.94$ & $_{-0.06}^{+0.11}$ & $_{-0.04}^{+0.06}$ &
   $_{-0.73}^{+0.42}$ & - & $_{-0.07}^{+0.04}$ & $_{-0.08}^{+0.08}$ & $_{-1.41}^{+0.00}$
   \\
 \text{} & \text{} & \text{} & \text{} & \text{} & \text{} & \text{} & \text{} & \text{} \\
 \hline
 \text{} & \text{} & \text{} & \text{} & \text{} & \text{} & \text{} & \text{} & \text{} \\
 $ \mathcal{A}_{0,B\rightarrow K^*}\text{(0)}$ & $ 0.129$ & $_{-0.008}^{+0.008}$ &
   $_{-0.005}^{+0.005}$ & $_{-0.016}^{+0.001}$ & - & $_{-0.002}^{+0.004}$ &
   $_{-0.003}^{+0.003}$ & $_{-0.011}^{+0.006}$ \\
 \text{} & \text{} & \text{} & \text{} & \text{} & \text{} & \text{} & \text{} & \text{} \\
 $ b_{1,K^*}^{\mathcal{A}_0}$ & $ -12.4$ & $_{-0.1}^{+0.1}$ & $_{-0.1}^{+0.1}$ &
   $_{-0.6}^{+0.2}$ & - & $_{-0.3}^{+0.2}$ & $_{-0.1}^{+0.1}$ & $_{-0.9}^{+0.0}$ \\
 \text{} & \text{} & \text{} & \text{} & \text{} & \text{} & \text{} & \text{} & \text{} \\
 \hline
 \text{} & \text{} & \text{} & \text{} & \text{} & \text{} & \text{} & \text{} & \text{} \\
 $ \mathcal{A}_{1,B\rightarrow K^*}\text{(0)}$ & $ 0.358$ & $_{-0.032}^{+0.031}$ &
   $_{-0.019}^{+0.018}$ & $_{-0.062}^{+0.001}$ & - & $_{-0.005}^{+0.010}$ &
   $_{-0.017}^{+0.016}$ & $_{-0.026}^{+0.039}$ \\
 \text{} & \text{} & \text{} & \text{} & \text{} & \text{} & \text{} & \text{} & \text{} \\
 $ b_{1,K^*}^{\mathcal{A}_1}$ & $ -4.81$ & $_{-0.07}^{+0.12}$ & $_{-0.05}^{+0.06}$ &
   $_{-0.75}^{+0.43}$ & - & $_{-0.07}^{+0.04}$ & $_{-0.08}^{+0.08}$ & $_{-1.48}^{+0.00}$
   \\
 \text{} & \text{} & \text{} & \text{} & \text{} & \text{} & \text{} & \text{} & \text{} \\
 \hline
 \text{} & \text{} & \text{} & \text{} & \text{} & \text{} & \text{} & \text{} & \text{} \\
 $ \mathcal{A}_{12,B\rightarrow K^*}\text{(0)}$ & $ 0.129$ & $_{-0.008}^{+0.008}$ &
   $_{-0.005}^{+0.005}$ & $_{-0.016}^{+0.001}$ & - & $_{-0.002}^{+0.004}$ &
   $_{-0.003}^{+0.003}$ & $_{-0.011}^{+0.006}$ \\
 \text{} & \text{} & \text{} & \text{} & \text{} & \text{} & \text{} & \text{} & \text{} \\
 $ b_{1,K^*}^{\mathcal{A}_{12}}$ & $ -11.3$ & $_{-0.0}^{+0.1}$ & $_{-0.0}^{+0.0}$ &
   $_{-0.5}^{+0.3}$ & - & $_{-0.2}^{+0.1}$ & $_{-0.1}^{+0.1}$ & $_{-1.5}^{+0.0}$ \\
 \text{} & \text{} & \text{} & \text{} & \text{} & \text{} & \text{} & \text{} & \text{} \\
  \hline
  \hline
\end{tabular}
\end{center}
\caption{Theory summary of the fitted results for the shape parameters and normalizations
of the (axial)-vector $B \to K^{\ast}$ form factors with the numerically sizeable uncertainties
by varying  different input parameters.}
\label{table: results of  the shape parameters and normalizations of the vector B to kstar FF}
\end{table}

\begin{table}[t!bph]
\begin{center}
\begin{tabular}{|c|c|c|c|c|c|c|c|c|}
  \hline
  \hline
  &  &  &  &  &  &  &  &  \\
  Parameters & Central \,\, value & $\lambda_B$ & $\sigma_1$ & $\mu$ & $\nu$ & $M^2$ & $s_0$ & $\phi_B^{\pm}(\omega)$ \\
  &  &  &  &  &  &  &  &  \\
  \hline
 \text{} & \text{} & \text{} & \text{} & \text{} & \text{} & \text{} & \text{} & \text{} \\
 $ \mathcal{T}_{1,B\rightarrow K^*}\text{(0)}$ & $ 0.361$ & $_{-0.032}^{+0.032}$ &
   $_{-0.019}^{+0.019}$ & $_{-0.062}^{+0.001}$ & $_{-0.019}^{+0.032}$ &
   $_{-0.004}^{+0.011}$ & $_{-0.017}^{+0.016}$ & $_{-0.026}^{+0.039}$ \\
 \text{} & \text{} & \text{} & \text{} & \text{} & \text{} & \text{} & \text{} & \text{} \\
 $ b_{1,K^*}^{\mathcal{T}_1}$ & $ -3.78$ & $_{-0.07}^{+0.11}$ & $_{-0.05}^{+0.06}$ &
   $_{-0.71}^{+0.41}$ & $_{-0.06}^{+0.04}$ & $_{-0.07}^{+0.04}$ & $_{-0.08}^{+0.07}$ &
   $_{-1.41}^{+0.00}$ \\
 \text{} & \text{} & \text{} & \text{} & \text{} & \text{} & \text{} & \text{} & \text{} \\
 \hline
 \text{} & \text{} & \text{} & \text{} & \text{} & \text{} & \text{} & \text{} & \text{} \\
 $ \mathcal{T}_{2,B\rightarrow K^*}\text{(0)}$ & $ 0.361$ & $_{-0.032}^{+0.032}$ &
   $_{-0.019}^{+0.019}$ & $_{-0.062}^{+0.001}$ & $_{-0.019}^{+0.032}$ &
   $_{-0.004}^{+0.011}$ & $_{-0.017}^{+0.016}$ & $_{-0.026}^{+0.039}$ \\
 \text{} & \text{} & \text{} & \text{} & \text{} & \text{} & \text{} & \text{} & \text{} \\
 $ b_{1,K^*}^{\mathcal{T}_2}$ & $ -4.67$ & $_{-0.07}^{+0.12}$ & $_{-0.05}^{+0.07}$ &
   $_{-0.72}^{+0.42}$ & $_{-0.07}^{+0.04}$ & $_{-0.07}^{+0.04}$ & $_{-0.08}^{+0.08}$ &
   $_{-1.46}^{+0.00}$ \\
 \text{} & \text{} & \text{} & \text{} & \text{} & \text{} & \text{} & \text{} & \text{} \\
 \hline
 \text{} & \text{} & \text{} & \text{} & \text{} & \text{} & \text{} & \text{} & \text{} \\
 $ \mathcal{T}_{23,B\rightarrow K^*}\text{(0)}$ & $ 0.117$ & $_{-0.008}^{+0.008}$ &
   $_{-0.005}^{+0.005}$ & $_{-0.016}^{+0.000}$ & $_{-0.006}^{+0.010}$ &
   $_{-0.001}^{+0.003}$ & $_{-0.003}^{+0.003}$ & $_{-0.006}^{+0.011}$ \\
 \text{} & \text{} & \text{} & \text{} & \text{} & \text{} & \text{} & \text{} & \text{} \\
 $ b_{1,K^*}^{\mathcal{T}_{23}}$ & $ -11.8$ & $_{-0.0}^{+0.1}$ & $_{-0.0}^{+0.0}$ &
   $_{-0.7}^{+0.4}$ & $_{-0.1}^{+0.0}$ & $_{-0.2}^{+0.1}$ & $_{-0.1}^{+0.1}$ &
   $_{-1.5}^{+0.0}$ \\
 \text{} & \text{} & \text{} & \text{} & \text{} & \text{} & \text{} & \text{} & \text{} \\
  \hline
  \hline
\end{tabular}
\end{center}
\caption{Theory summary of the fitted results for the shape parameters and normalizations
of the tensor $B \to K^{\ast}$ form factors with the numerically sizeable uncertainties
by varying  different input parameters.}
\label{table: results of  the shape parameters and normalizations of the tensor B to kstar FF}
\end{table}

\subsection{Semileptonic $B \to (\rho,  \omega) \, \ell \bar \nu_{\ell}$ decays}

Having at our disposal the theory predictions for all the  $B \to \rho,  \, \omega$ form factors
in QCD, we proceed to  investigate the phenomenological aspects of the semileptonic
$B \to (\rho,  \omega) \, \ell \,  \bar \nu_{\ell}$ decays, which provide  a complementary way for
the exclusive determination of the CKM  matrix element $|V_{ub}|$.
However, we will not explore the further applications of the calculated form factors to
the more challenging  radiative and electroweak penguin $B$-meson decays, which are crucial to
the precision extraction of the CKM  matrix element $|V_{td}|$ \cite{Bosch:2004nd}
and to the intensive hunting of new physics beyond the Standard Model (SM)
\cite{Beneke:2004dp,Khodjamirian:2010vf,Khodjamirian:2012rm},
due to the appearance of the complex non-local hadronic matrix elements
even at leading power in the heavy quark expansion.
The differential decay rate of $B \to V \, \ell \, \bar \nu_{\ell}$ can be readily written as
\begin{eqnarray}
 \frac{d^2 \Gamma(B \to V \, \ell \, \bar \nu_{\ell})}{d q^2 \, d \cos \theta} &=&
{G_F^2 \,|V_{ub}|^2 \over 256 \, \pi^3 \, m_B^3} \, {q^2 \over c_V^2} \, \lambda^{1/2}(m_B^2, m_V^2, q^2) \,
\bigg \{ \sin^2 \theta \,\, |H_0(q^2)|^2
+ (1 - \cos \theta)^2 \, {|H_{+}(q^2)|^2  \over 2}  \nonumber \\
&&  + \, (1 + \cos \theta)^2 \, {|H_{-}(q^2)|^2  \over 2} \bigg \}  \,,
\end{eqnarray}
where the three  helicity amplitudes $H_{i}(q^2)$ ($i=\pm, \, 0$) can be expressed in terms of the semileptonic
$B \to V$ form factors
\begin{eqnarray}
H_{\pm}(q^2) &=& (m_B +m_V) \, \left [ A_1(q^2) \mp { 2\, m_B \, |\vec{p}_{V}| \over (m_B+m_V)^2} \, V(q^2)\right ] \,,
\nonumber  \\
H_{0}(q^2) &=&  {m_B +m_V \over 2 \, m_V \, \sqrt{q^2}} \,
\bigg [ (m_B^2 - m_V^2 -q^2) \, A_1(q^2)
- {4 \, m_B^2 \, |\vec{p}_{V}|^2 \over (m_B +m_V)^2} \, A_2(q^2) \bigg ]   \,,
\end{eqnarray}
with the momentum $|\vec{p}_{V}|$ of the light-vector meson in the $B$-meson rest frame given by
\begin{eqnarray}
|\vec{p}_{V}| =  {1 \over 2 \, m_B} \, \lambda^{1/2}(m_B^2, m_V^2, q^2) \,, \qquad
\lambda(a,b,c)=a^2+b^2+c^2-2 ab - 2 ac -2 bc \,.
\end{eqnarray}

\begin{figure}
\begin{center}
\includegraphics[width=0.55 \columnwidth]{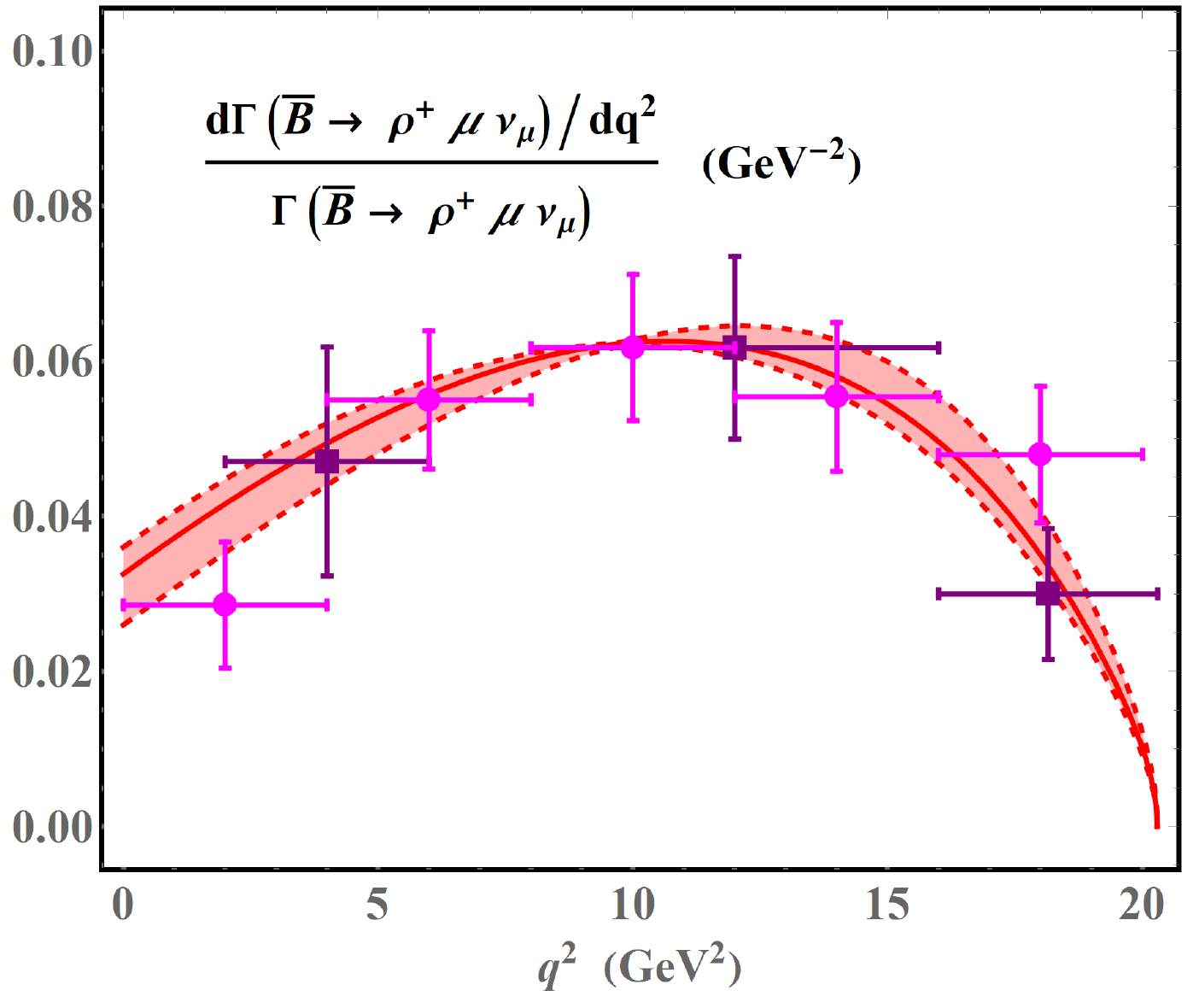}
\vspace{0.2 cm}
\includegraphics[width=0.55 \columnwidth]{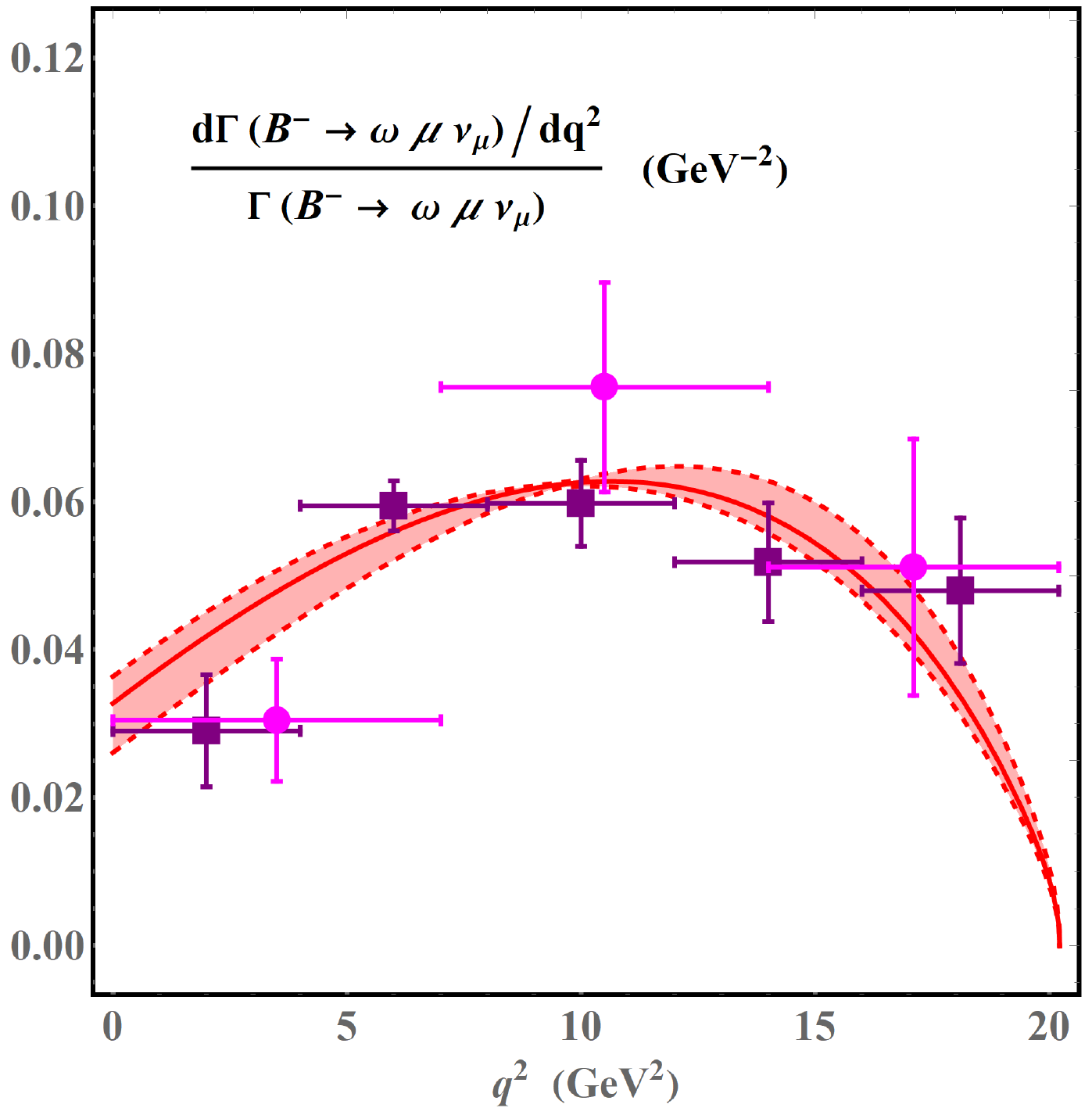}
\vspace*{0.1cm}
\caption{The normalized differential $q^2$ distributions of the semileptonic $B \to V \, \ell \, \bar \nu_{\ell}$
($V=\rho, \, \omega$) decays with the heavy-to-light form factors computed in this work (red band).
For a comparison, we also display the experimental measurements for the binned distributions
from the BaBar \cite{delAmoSanchez:2010af,Lees:2012vv} and Belle \cite{Sibidanov:2013rkk} Collaborations.}
\label{fig: differential B to V l nu distribution}
\end{center}
\end{figure}

For the determination of the CKM matrix element $|V_{ub}|$ we introduce the  standard  quantity
\begin{eqnarray}
\Delta \zeta_{V}(q_1^2, q_2^2)= {1 \over |V_{ub}|^2} \,
\int_{q_1^2}^{q_2^2} \, d q^2  \, {d \Gamma(B \to V \, \ell \, \bar \nu_{\ell}) \over dq^2 }\,,
\end{eqnarray}
which can be computed by performing the phase-space integration over the obtained hadronic $B \to V$ form factors.
The resulting predictions for $\Delta \zeta_{V}$ with the  theoretical uncertainties from varying the
input parameters are given by
\begin{eqnarray}
\Delta \zeta_{\rho}(0, 16 \, {\rm GeV}^2) &=&
\left ( 14.35 \,{}^{+2.59}_{-2.41} \, \big|_{\lambda_B}\,{}^{+1.49}_{-1.46} \, \big|_{\sigma_B^{(1)}}
\,{}^{+0.00}_{-3.63} \, \big|_{\mu}
\,{}^{+0.29}_{-1.02} \, \big|_{\mu_{h1}}
\,{}^{+0.71}_{-1.60} \, \big|_{\mu_{h2}}
\,{}^{+1.14}_{-1.24} \, \big|_{s_0}
\,{}^{+4.43}_{-1.16} \, \big|_{\phi_B^{\pm}} \right ) \,\,\, {\rm ps}^{-1} \nonumber   \\
&=& 14.35^{+5.71}_{-5.41} \,\,\, {\rm ps}^{-1} \,,  \nonumber \\
\Delta \zeta_{\omega}(0, 12 \, {\rm GeV}^2) &=&
\left ( 6.25 \,{}^{+1.08}_{-1.01} \, \big|_{\lambda_B}\,{}^{+0.62}_{-0.61} \, \big|_{\sigma_B^{(1)}}
\,{}^{+0.00}_{-1.59} \, \big|_{\mu}
\,{}^{+0.11}_{-0.41} \, \big|_{\mu_{h1}}
\,{}^{+0.28}_{-0.65} \, \big|_{\mu_{h2}}
\,{}^{+0.48}_{-0.52} \, \big|_{s_0}
\,{}^{+1.63}_{-0.66} \, \big|_{\phi_B^{\pm}} \right ) \,\,\, {\rm ps}^{-1} \nonumber   \\
&=& 6.25^{+2.26}_{-2.37} \,\,\, {\rm ps}^{-1} \,,
\end{eqnarray}
where the subdominant uncertainties from  variations of the remaining parameters have been taken
into account in the final combined uncertainties.
Employing the experimental measurements of the partial branching fractions for $B \to \rho \, \ell \, \bar \nu_{\ell}$
\cite{delAmoSanchez:2010af,Sibidanov:2013rkk} and $B \to \omega \, \ell \, \bar \nu_{\ell}$ \cite{Lees:2012vv,Sibidanov:2013rkk}
we can readily obtain the following intervals of $|V_{ub}|$
\begin{eqnarray}
|V_{ub}| = \bigg ( 3.05\,{}^{+0.67}_{-0.52} \big |_{\rm th.}\,{}^{+0.19}_{-0.20} \big |_{\rm exp.} \bigg )
\times 10^{-3} \,,  \qquad  [{\rm from} \,\, B \to \rho \ell \nu_{\ell} ] \nonumber \\
|V_{ub}| = \bigg ( 2.54 \,{}^{+0.56}_{-0.40} \big |_{\rm th.}\,{}^{+0.18}_{-0.19} \big |_{\rm exp.} \bigg )
\times 10^{-3} \,.  \qquad  [{\rm from} \,\, B \to \omega \ell \nu_{\ell} ]
\end{eqnarray}
Apparently,  the extracted values of $|V_{ub}|$ from the semileptonic $B \to \omega \, \ell \, \bar \nu_{\ell}$ decay
are significantly lower than from the exclusive channel  $B \to \rho \, \ell \, \bar \nu_{\ell}$ as already observed in
\cite{Sibidanov:2013rkk}.
Furthermore, the central values of both determinations of $|V_{ub}|$ from $B \to V \, \ell \, \bar \nu_{\ell}$
are somewhat smaller than the corresponding result derived from the ``golden" channel
$B \to \pi \, \ell \, \bar \nu_{\ell}$ \cite{Tanabashi:2018oca}
\begin{eqnarray}
|V_{ub}|_{\rm PDG} = \bigg ( 3.70 \pm 0.12 \big |_{\rm th.}\, \pm 0.10 \big |_{\rm exp.} \bigg )
\times 10^{-3} \,.
\end{eqnarray}
Exploring the underlying mechanisms responsible for such discrepancy from both the theoretical and experimental
aspects will be certainly in demand for resolving the $|V_{ub}|$ puzzle.

We further display in figure \ref{fig: differential B to V l nu distribution}
the normalized differential $q^2$ distribution of $B \to V \, \ell \, \bar \nu_{\ell}$
in the entire kinematic region by applying the computed form factors from the LCSR technique with the
aid of the $z$-series expansion. Due to the strong cancellation of the theory uncertainties for the
normalized physical quantities the resulting $q^2$ shapes of
\begin{eqnarray}
{\cal N}(B \to V \, \ell \, \bar \nu_{\ell}) \equiv {1 \over \Gamma(B \to V \, \ell \, \bar \nu_{\ell})} \,
{d \Gamma(B \to V \, \ell \, \bar \nu_{\ell} )  \over dq^2 } \,,  \qquad (V=\rho, \, \omega)
\end{eqnarray}
are more accurate than the predicted form factors shown in figures
\ref{fig: B to rho form factors}, \ref{fig: B to omega form factors} and \ref{fig: B to kstar form factors}.
Our predictions for ${\cal N}(B \to V \, \ell \, \bar \nu_{\ell})$ are also in nice agreement with the experimental measurements
from the BaBar \cite{delAmoSanchez:2010af,Lees:2012vv} and Belle \cite{Sibidanov:2013rkk} Collaborations.

\subsection{Rare exclusive $B \to K^{\ast} \, \nu_{\ell} \, \bar \nu_{\ell}$ decays}

\begin{figure}
\begin{center}
\includegraphics[width=0.55 \columnwidth]{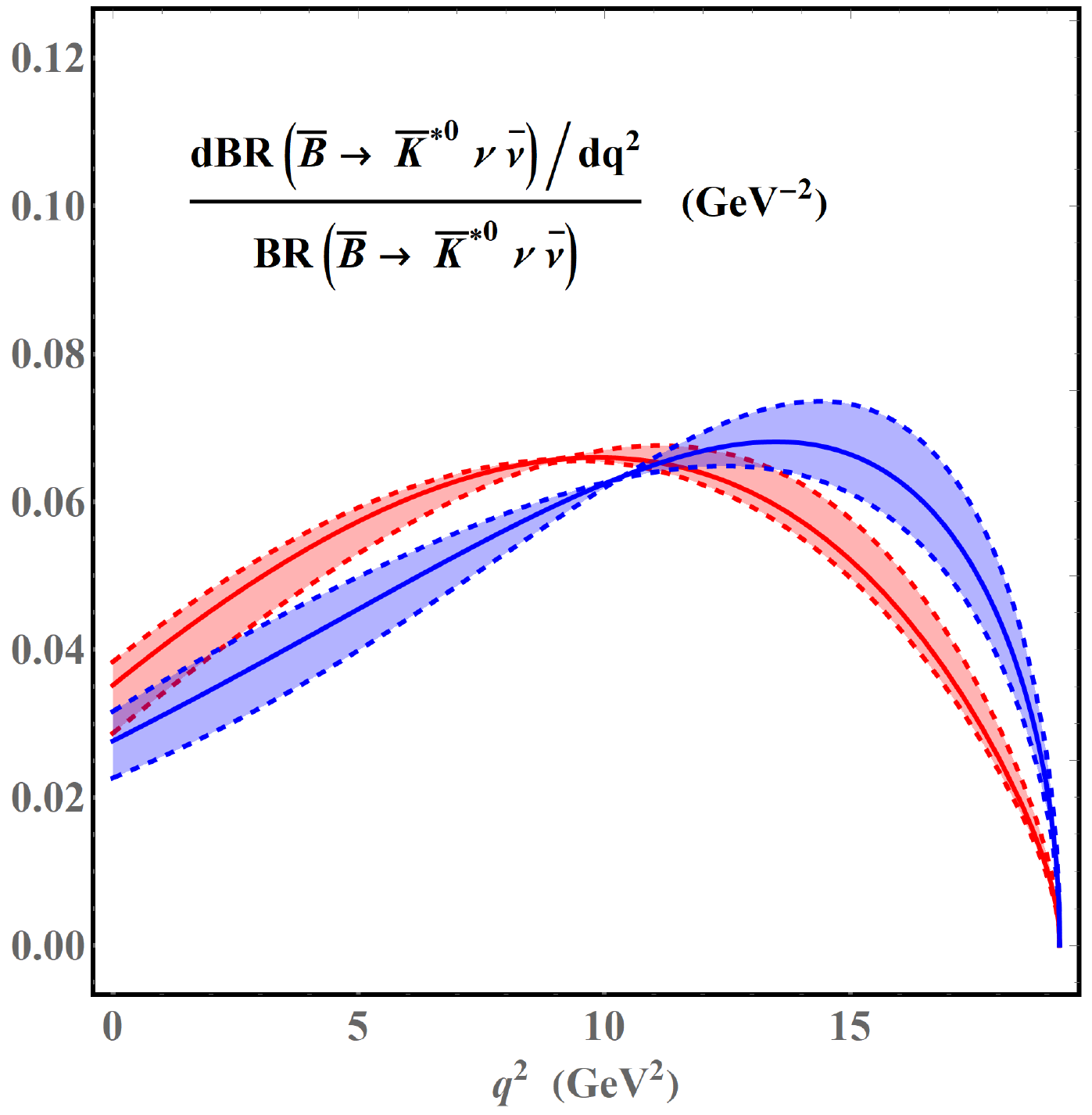}
\vspace*{0.1cm}
\caption{Theory predictions for the normalized differential $q^2$ distribution of
$B \to K^{\ast} \, \nu_{\ell} \, \bar \nu_{\ell}$ with the semileptonic $B \to K^{\ast}$
form factors computed from the improved LCSR approach (pink band) and from the
Lattice QCD simulation \cite{Horgan:2015vla} (blue band).}
\label{fig: differential B to kstar nu nu distribution}
\end{center}
\end{figure}

Thanks to the high-luminosity Belle II experiment, the exclusive rare
$B \to K^{\ast} \, \nu_{\ell} \, \bar \nu_{\ell}$ decays are expected to be observed with
$10 \, {\rm ab}^{-1}$ of data and the corresponding branching fraction will be further determined
at  ${\cal O}(10) \%$ accuracy with $50 \, {\rm ab}^{-1}$ of data \cite{Kou:2018nap}.
We are therefore well motivated to explore the phenomenological aspects of $B \to K^{\ast} \, \nu_{\ell} \, \bar \nu_{\ell}$
for understanding the strong interaction dynamics of $B \to K^{\ast}$ form factors and
for searching the exotic particle $X$ in the dark matter context.
It is straightforward to derive the differential decay width formula for the exclusive process
$B \to K^{\ast} \, \nu_{\ell} \, \bar \nu_{\ell}$ \cite{Buras:2014fpa}
\begin{eqnarray}
{d \Gamma(B \to K^{\ast} \, \nu_{\ell} \, \bar \nu_{\ell})  \over d q^2}
&=&  {G_F^2 \, \alpha_{\rm em}^2  \over 256 \, \pi^5}  \,
\frac{\lambda^{3/2}(m_B^2, m_{K^{\ast}}^2, q^2)}{m_B^3 \, \sin^4 \, \theta_W} \, |V_{tb} \, V_{ts}^{\ast}|^2  \,
\left [ X_t \left ({m_t^2 \over m_W^2}, \, {m_H^2 \over m_t^2}, \, \sin \, \theta_W, \, \mu \right ) \right ]^2  \,
\nonumber \\
&& \times \, \left [  H_V(q^2) +  H_{A_1}(q^2) +  H_{A_{12}}(q^2)  \right ] \,,
\end{eqnarray}
where the three invariant functions $H_{i}(q^2)$ can be further expressed by the
semileptonic  $B \to K^{\ast}$ form factors
\begin{eqnarray}
H_V(q^2) &=&  {2 \, q^2  \over (m_B + m_{K^{\ast}})^2 } \,
\left [ V(q^2) \right ]^2   \,,  \qquad
H_{A_1}(q^2) =  {2 \, q^2 \, (m_B+m_{K^{\ast}})^2 \over \lambda(m_B^2, m_{K^{\ast}}^2, q^2) } \,
\left [ A_1(q^2) \right ]^2   \,, \nonumber \\
H_{A_{12}}(q^2) &=&  {64 \,m_{K^{\ast}}^2 \, m_B^2  \over \lambda(m_B^2, m_{K^{\ast}}^2, q^2) } \,
\left [ A_{12}(q^2) \right ]^2   \,,
\end{eqnarray}
with the helicity form factor $A_{12}$ introduced in \cite{Horgan:2013hoa}
\begin{eqnarray}
A_{12}(q^2)= \frac{(m_B+m_{K^{\ast}})^2 \, (m_B^2-m_{K^{\ast}}^2-q^2) \, A_1(q^2)
- \lambda(m_B^2, m_{K^{\ast}}^2, q^2) \, A_2(q^2)}
{16 \, m_B \, m_{K^{\ast}}^2 \, (m_B+ m_{K^{\ast}})}  \,.
\end{eqnarray}
The short-distance Wilson coefficient $X_t$ can be expanded perturbatively in terms of the SM
coupling constants
\begin{eqnarray}
X_t = X_t^{(0)} + {\alpha_s \over 4\, \pi} \, X_t^{(1)} + {\alpha_{\rm em} \over 4\, \pi} \,X_t^{\rm EW \, (1)}  + ...  \,,
\end{eqnarray}
where the LO contribution $X_t^{(0)}$ \cite{Inami:1980fz}, the NLO QCD correction $X_t^{(1)}$
\cite{Buchalla:1998ba,Buchalla:1992zm,Misiak:1999yg} and the two-loop electroweak correction
$X_t^{\rm EW \, (1)}$ \cite{Brod:2010hi} are already available analytically.
We display the theory prediction for the normalized differential branching fraction of
$B \to K^{\ast} \, \nu_{\ell} \, \bar \nu_{\ell}$ in figure \ref{fig: differential B to kstar nu nu distribution},
including the results obtained from the Lattice QCD calculations of $B \to K^{\ast}$ form factors \cite{Horgan:2015vla}
for a comparison.
In general we find a fair agreement of the two different calculations in the physical  $q^2$ range
of $B \to K^{\ast} \, \nu_{\ell} \, \bar \nu_{\ell}$ , albeit with the weak mismatch of the peak regions.

\begin{table}[t!bph]
\begin{center}
\begin{tabular}{|c|c|c|}
  \hline
  \hline
  &   &    \\
  $[q_1^2, \, q_2^2]$  \,\, (${\rm in \,\, GeV}^2$) & $10^6 \times \Delta {\cal BR}(q_1^2, q_2^2)$
  & $\,\,\,  \Delta \, F_L  (q_1^2, q_2^2) \,\,\, $  \\
  &   &    \\
  \hline
  &   &    \\
  $[0.0, 1.0]$ & $0.41^{+0.10}_{-0.14}$  &  $0.95^{+0.01}_{-0.01}$  \\
  &   &    \\
  $[1.0, 2.5]$ & $0.72^{+0.19}_{-0.25}$  &  $0.84^{+0.02}_{-0.02}$  \\
  &   &    \\
  $[2.5, 4.0]$ & $0.83^{+0.25}_{-0.29}$  &  $0.74^{+0.03}_{-0.02}$  \\
  &   &    \\
  $[4.0, 6.0]$ & $1.25^{+0.42}_{-0.43}$  &  $0.65^{+0.03}_{-0.03}$  \\
  &   &    \\
  $[6.0, 8.0]$ & $1.36^{+0.52}_{-0.48}$  &  $0.56^{+0.03}_{-0.03}$  \\
  &   &    \\
  $[8.0, 12.0]$ & $2.84^{+1.26}_{-1.00}$  &  $0.46^{+0.02}_{-0.03}$  \\
  &   &    \\
  $[12.0, 16.0]$ & $2.46^{+1.30}_{-0.88}$  &  $0.36^{+0.02}_{-0.02}$  \\
  &   &    \\
  $\left [16.0, \,\, (m_B-m_{K^{\ast}})^2 \right ]$ & $1.01^{+1.91}_{-1.25}$  &  $0.31^{+0.01}_{-0.01}$  \\
  &   &    \\
  $\left [0, \,\, (m_B-m_{K^{\ast}})^2 \right ]$ & $10.88^{+4.62}_{-3.77}$  &  $0.52^{+0.03}_{-0.04}$  \\
  &   &    \\
  \hline
  \hline
\end{tabular}
\end{center}
\caption{Theory predictions for the $q^2$-binned observables $\Delta {\cal BR}$
and $\Delta \, F_L$ of the electroweak penguin
$B \to K^{\ast} \, \nu_{\ell} \, \bar \nu_{\ell}$  decays with the heavy-to-light form
factors computed from the $B$-meson LCSR approach and the $z$-series expansion.}
\label{table: binned distributions of B to Kstar nu nu}
\end{table}

We can proceed to  define the differential longitudinal $K^{\ast}$ polarization fraction  $F_L$ of
the electroweak penguin $B \to K^{\ast} \, \nu_{\ell} \, \bar \nu_{\ell}$  decays
\begin{eqnarray}
F_L(q^2) = \frac{H_{A_{12}}(q^2)}{H_V(q^2)+H_{A_1}(q^2)+H_{A_{12}}(q^2)} \,.
\end{eqnarray}
In addition, we define the following two $q^2$-binned observables for the comparison with the future Belle II data
\begin{eqnarray}
\Delta \, {\cal BR} (q_1^2, q_2^2) &=& \tau_{B_0} \,
\int_{q_1^2}^{q_2^2} \, d q^2 \,  {d \Gamma(B \to K^{\ast} \, \nu_{\ell} \, \bar \nu_{\ell})  \over d q^2} \,,
\nonumber \\
\Delta \, F_L  (q_1^2, q_2^2) &=&  \frac{\int_{q_1^2}^{q_2^2} \, d q^2 \, \lambda^{3/2}(m_B^2, m_{K^{\ast}}^2, q^2) \, H_{A_{12}}(q^2)}
{\int_{q_1^2}^{q_2^2} \, d q^2 \,\lambda^{3/2}(m_B^2, m_{K^{\ast}}^2, q^2) \,
\left [ H_V(q^2) +  H_{A_1}(q^2) +  H_{A_{12}}(q^2) \right ]} \,.
\end{eqnarray}
Our predictions for these two quantities with the choices of the $q^2$-intervals following \cite{Kou:2018nap}
are presented in Table \ref{table: binned distributions of B to Kstar nu nu}.
Apparently,  the theory uncertainties of the binned longitudinal $K^{\ast}$ polarization fractions
are much reduced compared with the resulting predictions of  $\Delta {\cal BR}$,
due to the less sensitivity of the form-factor ratios to the precise shapes of the
two-particle and three-particle $B$-meson distribution amplitudes.

\section{Conclusion}
\label{section: summary}

In this paper we have presented improved QCD calculations of the twenty-one $B \to V$
($V=\rho\,, \omega \,,  K^{\ast}$) form factors by first implementing the hard-collinear factorization
for the weak transition currents \cite{Hill:2004if,Becher:2004kk,Beneke:2005gs} and
then computing the resulting $\rm {SCET_{I}}$ matrix elements from the LCSR technique
with the HQET $B$-meson distribution amplitudes.
The hard-collinear functions entering the factorization formulae for the $\rm {SCET_{I}}$  vacuum-to-$B$-meson
correlation functions under discussion were determined at NLO in QCD for the leading-twist
two-particle contributions, by employing the evanescent-operator approach with the dimensional regularization scheme.
In particular, we demonstrated explicitly that the light-quark-mass terms appearing in the sum rules for the
${\rm A0}$-type SCET form factors $\xi_{\|, \, \perp}(n \cdot p)$,
with the generic power counting scheme $m \sim \Lambda$,
are not suppressed by any powers of $\Lambda/m_b$
in the heavy quark expansion as expected in  \cite{Leibovich:2003jd}.
We further computed the  higher-twist corrections to the semileptonic
$B \to V$ form factors from the two-particle and three-particle $B$-meson distribution amplitudes,
up to the twist-six accuracy, with the same LCSR method at tree level in QCD.
It turned out that the twist-five two-particle corrections to both the longitudinal and transverse
$B \to V$ form factors were numerically dominant, in analogy to the previous observation for the  semileptonic  $B \to P$
form factors \cite{Lu:2018cfc,Gubernari:2018wyi}.
However, we also noticed that the genuine three-particle higher-twist corrections were non-negligible for the transverse
$B \to V$ form factors, in contrast to the pattern appeared in the longitudinal vector form factors.
We proceeded to investigate the long-standing puzzles of the large-recoil symmetry breaking effects for the
$B \to V$ form-factor ratios predicted by the QCD factorization approach and by the sum rule technique \cite{Beneke:2000wa}.
Following the standard strategy \cite{Khodjamirian:2011ub,Wang:2015vgv},
we extrapolated the LCSR calculations of the heavy-to-light form factors,
with the two distinct models for the $B$-meson distribution amplitudes, toward the large $q^2$ region
by virtue of the well-motivated $z$-series expansion.

We explored the phenomenological applications of the obtained predictions for $B$-meson decay form factors
to the semileponic $B \to V \ell \nu_{\ell}$ decays as well as the electroweak penguin
$B \to K^{\ast} \nu_{\ell} \bar \nu_{\ell}$ decays.
The newly extracted interval of the CKM matrix element
$|V_{ub}|=\left ( 3.05\,{}^{+0.67}_{-0.52} \big |_{\rm th.}\,{}^{+0.19}_{-0.20} \big |_{\rm exp.}  \right )
\times 10^{-3}$
from the exclusive process $B \to \rho \ell \nu_{\ell}$  is in  agreement with the previous determination
from $B \to \pi  \ell \nu_{\ell}$ applying the same computational framework,
but the analogous determination  from $B \to \omega \ell \nu_{\ell}$ yields somewhat smaller values
of $|V_{ub}|$, albeit with the sizeable theory uncertainties mainly due to the  limited knowledge of the
small $\omega$-behaviours  of the two-particle $B$-meson distribution amplitude $\phi_B^{\pm}(\omega, \mu)$.
The two $q^2$-binned observables $\Delta \, {\cal BR} $ and  $\Delta \, F_L $
for the exclusive rare $B \to K^{\ast} \, \nu_{\ell} \, \bar \nu_{\ell}$ decays were further predicted
for the sake of hunting for new physics beyond the SM at the Belle II experiment \cite{Kou:2018nap}.

Future developments of QCD calculations of $B \to V$ form factors at large hadronic recoil can be pushed forward
both conceptually and technically.
First,  it will be of wide interest to carry out  the two-particle and three-particle higher-twist contributions
to the semileptonic $B \to V$ form factors, up to the twist-six accuracy,
at NLO in QCD in order to verify explicitly that the higher-Fock state contributions of both the $B$-meson
and the energetic vector meson generate the leading-power effects in the heavy quark expansion as demonstrated
in \cite{Beneke:2003pa}.
The computational challenges of determining the hard-collinear functions entering the $\rm {SCET_{I}}$
factorization formulae for the vacuum-to-$B$-meson correlation functions originate from the non-trivial
mixing of the different light-ray HQET operators under renormalization  beyond the twist-four accuracy \cite{Braun:2017liq},
making the infrared subtraction of the perturbative matching procedure  tediously in dimensional regularization.
In addition, constructing the meaningful  constraints for the higher-twist $B$-meson distribution amplitudes from the QCD
equations of motion beyond the LO in QCD are complicated by the appearance of the  light-cone divergences.
Second, improving theory calculations for the higher-dimensional local HQET matrix elements $\lambda_E^2(\mu)$
and $\lambda_H^2(\mu)$ will be of value for reducing the parametric uncertainties on the phenomenological aspect,
in view of the significant discrepancies of the  two independent QCD sum rule calculations presented in
\cite{Grozin:1996pq} and \cite{Nishikawa:2011qk} \footnote{Here, the NLO QCD correction to the dimension-five
quark-gluon mixing condensate and the LO contribution to the dimension-six four-quark condensate are  included,
in addition to the perturbative and non-perturbative contributions already  estimated in \cite{Grozin:1996pq}.}.
However, evaluating the NLO QCD correction to the leading-power contribution in the framework of
QCD sum rules will  also necessitate challenging computations of the two-point HQET diagrams at three loops.
Third, implementing a complete NLL QCD resummation for the enhanced logarithms due to the RG evolutions of both the short-distance
Wilson coefficients and the leading-twist $B$-meson distribution amplitude \cite{Braun:2019wyx} is
essential to the precision calculations of the semileptonic $B \to V$ form factors, but
it is also a technically demanding task to achieve this goal in a full analytical form.
It will be probably more promising to derive the NLL resummation improved SCET factorization formula
for the radiative leptonic $B$-meson decays in this respect.
To summarize, we expect interesting extensions of our work for deepening our understanding of the factorization
and resummation properties of heavy-to-light $B$-meson decay form factors in QCD.

\subsection*{Acknowledgements}

C.D.L is supported in part by the National Natural Science Foundation
of China (NSFC) with Grant No. 11521505 and 11621131001.
The work of Y.L.S is supported by the Natural Science Foundation of Shandong Province,
China under Grant No. ZR2015AQ006.
Y.M.W acknowledges support from the National Youth Thousand Talents Program,
the Youth Hundred Academic Leaders Program of Nankai University, and the NSFC with
Grant No. 11675082 and 11735010.
The work of Y.B.W is supported in part by the NSFC with Grant No. 11847238.
This work was also supported in part by the Mainz Institute for Theoretical Physics (MITP)
of the Johannes Gutenberg University.


\appendix

\section{Hard functions for the SCET currents at ${\cal O}(\alpha_s)$}
\label{appendix: hard functions}

Here we collect the hard matching coefficient functions for the A0-type
and B-type ${\rm SCET_{I}}$ currents entering the factorization formulae
(\ref{SCET-I factorization formulae}) for the semileptonic $B \to V$ form factors.
\begin{eqnarray}
C_{f_+}^{\rm (A0)} &=& 1 + {\alpha_s \, C_F \over 4 \, \pi}  \,
\bigg \{ -2 \, \ln^2 \left ({r \over \hat \mu} \right )
+ 5 \, \ln \left ({r \over \hat \mu} \right ) - 2 \, {\rm Li}_2 (1-r)
- 3 \, \ln r -{\pi^2 \over 12} - 6 \bigg \},
\\
C_{f_0}^{\rm (A0)} &=& 1 + {\alpha_s \, C_F \over 4 \, \pi}  \,
\bigg  \{ -2 \, \ln^2 \left ({r \over \hat \mu} \right )
+ 5 \, \ln \left ({r \over \hat \mu} \right ) - 2 \, {\rm Li}_2 (1-r)
- {3 - 5 \, r \over 1 - r} \, \ln r \nonumber \\
&&  - {\pi^2 \over 12}  - 4 \bigg \} \,,
\\
C_{f_T}^{\rm (A0)} &=&   1 + {\alpha_s \, C_F \over 4 \, \pi}  \,
\bigg \{ -2 \, \ln \hat{\nu}  - 2 \, \ln^2 \left ({r \over \hat \mu} \right )
+ 5 \, \ln \left ({r \over \hat \mu} \right ) - 2 \, {\rm Li}_2 (1-r)
- {3 -  r \over 1 - r} \, \ln r  \nonumber \\
&& - {\pi^2 \over 12} - 6 \bigg \} \,,
\\
C_{V}^{\rm (A0)} &=& 1 + {\alpha_s \, C_F \over 4 \, \pi}  \,
\bigg \{ -2 \, \ln^2 \left ({r \over \hat \mu} \right )
+ 5 \, \ln \left ({r \over \hat \mu} \right ) - 2 \, {\rm Li}_2 (1-r)
- {3 - 2 \, r \over 1 - r} \, \ln r \nonumber \\
&&  - {\pi^2 \over 12} - 6 \bigg \} \,,
\\
C_{T_1}^{\rm (A0)} &=& 1 + {\alpha_s \, C_F \over 4 \, \pi}  \,
\bigg \{ -2 \, \ln \hat{\nu} -2 \, \ln^2 \left ({r \over \hat \mu} \right )
+ 5 \, \ln \left ({r \over \hat \mu} \right ) - 2 \, {\rm Li}_2 (1-r) - 3 \, \ln r \nonumber \\
&&   -{\pi^2 \over 12} - 6 \bigg \} \,,
\\
C_{f_+}^{\rm (B1)} &=&  \left (-2 +{1 \over r} \right ) + {\cal O}(\alpha_s) \,,
\qquad
C_{f_0}^{\rm (B1)} =  \left (-{1 \over r} \right ) + {\cal O}(\alpha_s) \,,
\\
C_{f_T}^{\rm (B1)} &=&  \left ({1 \over r} \right ) + {\cal O}(\alpha_s) \,,
\qquad
C_{V}^{\rm (B1)} =  0 + {\cal O}(\alpha_s) \,,
\qquad
C_{T_1}^{\rm (B1)} =  -1  + {\cal O}(\alpha_s) \,,
\end{eqnarray}
where we have introduced the variables
\begin{eqnarray}
r = {n \cdot p \over m_b}\,, \qquad
\hat \mu ={\mu \over m_b}\,, \qquad
\hat \nu ={\nu \over m_b}\,.
\end{eqnarray}

\section{$B$-meson distribution amplitudes}
\label{appendix: B-meson DAs}

In this appendix we  collect the explicit expressions for two different models of the two-particle
and three-particle $B$-meson distribution amplitudes employed in our numerical study of the obtained sum rules for $B \to V$
form factors.

\begin{itemize}

\item{Exponential model:
\begin{eqnarray}
\phi_B^{+, \, \rm exp}(\omega, \mu) &=& {\omega \over \omega_0^2} \,  e^{-\omega/\omega_0} \,, \nonumber \\
\phi_B^{-, \, \rm exp}(\omega, \mu) &=& {1 \over \omega_0} \,  e^{-\omega/\omega_0}
- {\lambda_E^2 - \lambda_H^2 \over 9 \, \omega_0^3}  \,
\left [ 1 - 2 \, \left ( {\omega \over \omega_0} \right )
+ {1 \over 2} \, \left ( {\omega \over \omega_0} \right )^2 \right ]
\,  e^{-\omega/\omega_0} \,, \nonumber \\
\hat{g}_B^{-,  \, \rm exp}(\omega, \mu) &=& \omega \,
\left \{ {3 \over 4}  - {\lambda_E^2 - \lambda_H^2 \over 12 \, \omega_0^2} \,
\left [ 1 - \left ( {\omega \over \omega_0} \right )
+ {1 \over 3} \, \left ( {\omega \over \omega_0} \right )^2 \right ] \right \}
\,  e^{-\omega/\omega_0}  \,, \nonumber \\
\Phi_3^{\rm exp}(\omega_1, \omega_2, \mu) &=&
{\lambda_E^2 - \lambda_H^2 \over 6 \, \omega_0^5} \, \omega_1 \, \omega_2^2 \,
e^{-(\omega_1 + \omega_2)/\omega_0} \,, \nonumber \\
\Phi_4^{\rm exp}(\omega_1, \omega_2, \mu) &=&
{\lambda_E^2 + \lambda_H^2 \over 6 \, \omega_0^4} \, \omega_2^2 \,
e^{-(\omega_1 + \omega_2)/\omega_0} \,, \nonumber \\
\Psi_4^{\rm exp}(\omega_1, \omega_2, \mu) &=&
{\lambda_E^2 \over 3 \, \omega_0^4} \, \omega_1 \, \omega_2 \,
e^{-(\omega_1 + \omega_2)/\omega_0} \,, \nonumber \\
\tilde{\Psi}_4^{\rm exp}(\omega_1, \omega_2, \mu) &=&
{\lambda_H^2 \over 3 \, \omega_0^4} \, \omega_1 \, \omega_2 \,
e^{-(\omega_1 + \omega_2)/\omega_0} \,, \nonumber \\
\Phi_5^{\rm exp}(\omega_1, \omega_2, \mu)
&=& {\lambda_E^2 + \lambda_H^2 \over 3 \, \omega_0^3} \, \omega_1 \,
e^{-(\omega_1 + \omega_2)/\omega_0} \,, \nonumber \\
\Psi_5^{\rm exp}(\omega_1, \omega_2, \mu)
&=& - {\lambda_E^2 \over 3 \, \omega_0^3} \, \omega_2 \,
e^{-(\omega_1 + \omega_2)/\omega_0} \,, \nonumber \\
\tilde{\Psi}_5^{\rm exp}(\omega_1, \omega_2, \mu)
&=& - {\lambda_H^2 \over 3 \, \omega_0^3} \, \omega_2 \,
e^{-(\omega_1 + \omega_2)/\omega_0} \,, \nonumber \\
\Phi_6^{\rm exp}(\omega_1, \omega_2, \mu)
&=& {\lambda_E^2 - \lambda_H^2 \over 3 \, \omega_0^2} \,
e^{-(\omega_1 + \omega_2)/\omega_0} \,.
\end{eqnarray}
The classical QCD equations of motion further imply the following relations of the HQET parameters
appearing in this specifical model \cite{Braun:2017liq}
\begin{eqnarray}
\omega_0 =  \lambda_B = {2 \over 3} \, \bar \Lambda\,, \qquad
2 \, \bar \Lambda^2 = 2 \, \lambda_E^2 + \lambda_H^2 \,,
\end{eqnarray}
so that only two of the three non-perturbative parameters $\lambda_B$,
$\lambda_E^2$ and $\lambda_H^2$ are independent of each other at tree level.
}

\item{Local-duality model:
\begin{eqnarray}
\phi_B^{+, \rm LD}(\omega, \mu) &=& {5 \over 8 \, \omega_0^5}  \, \omega(2 \, \omega_0 - \omega)^3 \,
\theta(2 \, \omega_0 - \omega)\,,  \nonumber  \\
\phi_B^{-, \rm LD}(\omega, \mu) &=& {5 (2 \, \omega_0 - \omega)^2 \over 192\, \omega_0^5}  \,
\bigg \{6 \, (2 \, \omega_0 - \omega)^2 - {7 \, (\lambda_E^2 - \lambda_H^2) \over \omega_0^2} \,
(15 \, \omega^2 - 20 \, \omega \, \omega_0 + 4 \, \omega_0^2) \bigg \}   \, \nonumber \\
&& \times \, \theta(2 \, \omega_0 - \omega)\,,  \nonumber  \\
\hat{g}_B^{-, \rm LD}(\omega, \mu) &=& {\omega \, (2 \, \omega_0 - \omega)^3 \over \omega_0^5} \,
\bigg  \{ - {35 \, (\lambda_E^2 - \lambda_H^2) \over 1536} \,
\left [ 4 - 12 \, \left ( {\omega \over \omega_0} \right )
+ 11 \, \left ( {\omega \over \omega_0} \right )^2  \right ] \nonumber \\
&&  + {5 \over 256} \, (2 \, \omega_0 - \omega)^2   \bigg \} \,\,
\theta(2 \, \omega_0 - \omega) \,, \nonumber \\
\Phi_3^{\rm LD}(\omega_1, \omega_2, \mu) &=&  {105 \, (\lambda_E^2 - \lambda_H^2) \over 8 \, \omega_0^7} \,
\omega_1 \, \omega_2^2 \, \left (\omega_0 - {\omega_1 + \omega_2 \over 2} \right )^2  \,
\theta(2 \, \omega_0 - \omega_1 - \omega_2)  \,,  \nonumber  \\
\Phi_4^{\rm LD}(\omega_1, \omega_2, \mu) &=&  {35 \, (\lambda_E^2 + \lambda_H^2) \over 4 \, \omega_0^7} \,
\omega_2^2 \, \left (\omega_0 - {\omega_1 + \omega_2 \over 2} \right )^3  \,
\theta(2 \, \omega_0 - \omega_1 - \omega_2)  \,,  \nonumber  \\
\Psi_4^{\rm LD}(\omega_1, \omega_2, \mu) &=&  {35 \, \lambda_E^2 \over 2 \, \omega_0^7} \,
\omega_1 \, \omega_2 \, \left (\omega_0 - {\omega_1 + \omega_2 \over 2} \right )^3  \,
\theta(2 \, \omega_0 - \omega_1 - \omega_2)  \,,  \nonumber  \\
\tilde{\Psi}_4^{\rm LD}(\omega_1, \omega_2, \mu) &=&  {35 \, \lambda_H^2 \over 2 \, \omega_0^7} \,
\omega_1 \, \omega_2 \, \left (\omega_0 - {\omega_1 + \omega_2 \over 2} \right )^3  \,
\theta(2 \, \omega_0 - \omega_1 - \omega_2)  \,, \nonumber \\
\Phi_5^{\rm LD}(\omega_1, \omega_2, \mu)  &=& {35 \over 64} \, (\lambda_E^2 + \lambda_H^2) \,
{\omega_1 \over \omega_0^7} \, (2 \, \omega_0 - \omega_1-\omega_2)^4 \,
\theta(2 \, \omega_0 - \omega_1 - \omega_2)  \,, \nonumber \\
\Psi_5^{\rm LD}(\omega_1, \omega_2, \mu)  &=& - {35 \over 64} \, \lambda_E^2 \,
{\omega_2 \over \omega_0^7} \, (2 \, \omega_0 - \omega_1-\omega_2)^4 \,
\theta(2 \, \omega_0 - \omega_1 - \omega_2)  \,, \nonumber \\
\tilde{\Psi}_5^{\rm LD}(\omega_1, \omega_2, \mu)  &=& - {35 \over 64} \, \lambda_H^2 \,
{\omega_2 \over \omega_0^7} \, (2 \, \omega_0 - \omega_1-\omega_2)^4 \,
\theta(2 \, \omega_0 - \omega_1 - \omega_2)  \,, \nonumber \\
\Phi_6^{\rm LD}(\omega_1, \omega_2, \mu)  &=& {7 \over 64} \, (\lambda_E^2 - \lambda_H^2) \,
{1 \over \omega_0^7} \, (2 \, \omega_0 - \omega_1-\omega_2)^5 \,
\theta(2 \, \omega_0 - \omega_1 - \omega_2)  \,.
\end{eqnarray}
Analogously, the HQET parameters for the local-duality model also satisfy nontrivial constraints
due to QCD equations of motion \cite{Braun:2017liq}
\begin{eqnarray}
\omega_0 = {5 \over 2} \, \lambda_B = 2 \, \bar \Lambda\,, \qquad
6 \, {\bar \Lambda}^2 = 7 \, (2 \, \lambda_E^2 + \lambda_H^2)\,.
\end{eqnarray}
}

\end{itemize}



\begin{thebibliography}{99}




\bibitem{Horgan:2013hoa}
  R.~R.~Horgan, Z.~Liu, S.~Meinel and M.~Wingate,
  Phys.\ Rev.\ D {\bf 89} (2014) no.9,  094501
  [arXiv:1310.3722 [hep-lat]].





\bibitem{Horgan:2015vla}
  R.~R.~Horgan, Z.~Liu, S.~Meinel and M.~Wingate,
  PoS LATTICE {\bf 2014} (2015) 372
  [arXiv:1501.00367 [hep-lat]].



\bibitem{Bazavov:2009bb}
  A.~Bazavov {\it et al.} [MILC Collaboration],
  Rev.\ Mod.\ Phys.\  {\bf 82} (2010) 1349
  [arXiv:0903.3598 [hep-lat]].





\bibitem{Beneke:2000wa}
  M.~Beneke and T.~Feldmann,
  Nucl.\ Phys.\ B {\bf 592} (2001) 3
  [hep-ph/0008255].






\bibitem{Bauer:2002aj}
  C.~W.~Bauer, D.~Pirjol and I.~W.~Stewart,
  Phys.\ Rev.\ D {\bf 67} (2003) 071502
  [hep-ph/0211069].




\bibitem{Beneke:2002ph}
  M.~Beneke, A.~P.~Chapovsky, M.~Diehl and T.~Feldmann,
  Nucl.\ Phys.\ B {\bf 643} (2002) 431
  [hep-ph/0206152].





\bibitem{Beneke:2003pa}
  M.~Beneke and T.~Feldmann,
  Nucl.\ Phys.\ B {\bf 685} (2004) 249
  [hep-ph/0311335].






\bibitem{Ball:1997rj}
  P.~Ball and V.~M.~Braun,
  Phys.\ Rev.\ D {\bf 55} (1997) 5561
  [hep-ph/9701238].





\bibitem{Ball:1998kk}
  P.~Ball and V.~M.~Braun,
  Phys.\ Rev.\ D {\bf 58} (1998) 094016
  [hep-ph/9805422].




\bibitem{Ball:2004rg}
  P.~Ball and R.~Zwicky,
  Phys.\ Rev.\ D {\bf 71} (2005) 014029
  [hep-ph/0412079].





\bibitem{Straub:2015ica}
  A.~Bharucha, D.~M.~Straub and R.~Zwicky,
  JHEP {\bf 1608} (2016) 098
  [arXiv:1503.05534 [hep-ph]].



\bibitem{Meissner:2013hya}
  U.~G.~Mei{\ss}ner and W.~Wang,
  Phys.\ Lett.\ B {\bf 730} (2014) 336
  [arXiv:1312.3087 [hep-ph]].






\bibitem{Hambrock:2015aor}
  C.~Hambrock and A.~Khodjamirian,
  Nucl.\ Phys.\ B {\bf 905} (2016) 373
  [arXiv:1511.02509 [hep-ph]].




\bibitem{Cheng:2017sfk}
  S.~Cheng, A.~Khodjamirian and J.~Virto,
  Phys.\ Rev.\ D {\bf 96} (2017)   051901
  [arXiv:1709.00173 [hep-ph]].




\bibitem{Cheng:2018ouz}
  W.~Cheng, X.~G.~Wu, R.~Y.~Zhou and H.~B.~Fu,
  Phys.\ Rev.\ D {\bf 98} (2018)   096013
  [arXiv:1808.10775 [hep-ph]].





\bibitem{Khodjamirian:2005ea}
  A.~Khodjamirian, T.~Mannel and N.~Offen,
  Phys.\ Lett.\ B {\bf 620} (2005) 52
  [hep-ph/0504091].




\bibitem{Khodjamirian:2006st}
  A.~Khodjamirian, T.~Mannel and N.~Offen,
  Phys.\ Rev.\ D {\bf 75} (2007) 054013
  [hep-ph/0611193].



\bibitem{Khodjamirian:2010vf}
  A.~Khodjamirian, T.~Mannel, A.~A.~Pivovarov and Y.-M.~Wang,
  JHEP {\bf 1009} (2010) 089
  [arXiv:1006.4945 [hep-ph]].




\bibitem{Khodjamirian:2012rm}
  A.~Khodjamirian, T.~Mannel and Y.~M.~Wang,
  JHEP {\bf 1302} (2013) 010
  [arXiv:1211.0234 [hep-ph]].





\bibitem{Wang:2015vgv}
  Y.~M.~Wang and Y.~L.~Shen,
  Nucl.\ Phys.\ B {\bf 898} (2015) 563
  [arXiv:1506.00667 [hep-ph]].




\bibitem{Wang:2017jow}
  Y.~M.~Wang, Y.~B.~Wei, Y.~L.~Shen and C.~D.~L\"{u},
  JHEP {\bf 1706} (2017) 062
  [arXiv:1701.06810 [hep-ph]].




\bibitem{Lu:2018cfc}
  C.~D.~L\"{u}, Y.~L.~Shen, Y.~M.~Wang and Y.~B.~Wei,
  JHEP {\bf 1901} (2019) 024
  [arXiv:1810.00819 [hep-ph]].




\bibitem{Beneke:1997zp}
  M.~Beneke and V.~A.~Smirnov,
  Nucl.\ Phys.\ B {\bf 522} (1998) 321
  [hep-ph/9711391].





\bibitem{Smirnov:2002pj}
  V.~A.~Smirnov,
  {\it Applied asymptotic expansions in momenta and masses,}
  Springer Tracts Mod.\ Phys.\  {\bf 177} (2002) 1.




\bibitem{DeFazio:2005dx}
  F.~De Fazio, T.~Feldmann and T.~Hurth,
  Nucl.\ Phys.\ B {\bf 733} (2006) 1
   Erratum: [Nucl.\ Phys.\ B {\bf 800} (2008) 405]
  [hep-ph/0504088].





\bibitem{DeFazio:2007hw}
  F.~De Fazio, T.~Feldmann and T.~Hurth,
  JHEP {\bf 0802} (2008) 031
  [arXiv:0711.3999 [hep-ph]].



\bibitem{Keum:2000wi}
  Y.~Y.~Keum, H.~N.~Li and A.~I.~Sanda,
  Phys.\ Rev.\ D {\bf 63} (2001) 054008
  [hep-ph/0004173].




\bibitem{Lu:2000em}
  C.~D.~L\"{u}, K.~Ukai and M.~Z.~Yang,
  Phys.\ Rev.\ D {\bf 63} (2001) 074009
  [hep-ph/0004213].




\bibitem{Nandi:2007qx}
  S.~Nandi and H.~n.~Li,
  Phys.\ Rev.\ D {\bf 76} (2007) 034008
  [arXiv:0704.3790 [hep-ph]].






\bibitem{Li:2010nn}
  H.~n.~Li, Y.~L.~Shen, Y.~M.~Wang and H.~Zou,
  Phys.\ Rev.\ D {\bf 83} (2011) 054029
  [arXiv:1012.4098 [hep-ph]].




\bibitem{Li:2012nk}
  H.~n.~Li, Y.~L.~Shen and Y.~M.~Wang,
  Phys.\ Rev.\ D {\bf 85} (2012) 074004
  [arXiv:1201.5066 [hep-ph]].




\bibitem{Li:2012md}
  H.~N.~Li, Y.~L.~Shen and Y.~M.~Wang,
  JHEP {\bf 1302} (2013) 008
  [arXiv:1210.2978 [hep-ph]].






\bibitem{Li:2013xna}
  H.~N.~Li, Y.~L.~Shen and Y.~M.~Wang,
  JHEP {\bf 1401} (2014) 004
  [arXiv:1310.3672 [hep-ph]].





\bibitem{Collins:1981uk}
  J.~C.~Collins and D.~E.~Soper,
  Nucl.\ Phys.\ B {\bf 193} (1981) 381
   Erratum: [Nucl.\ Phys.\ B {\bf 213} (1983) 545].





\bibitem{Collins:1984kg}
  J.~C.~Collins, D.~E.~Soper and G.~F.~Sterman,
  Nucl.\ Phys.\ B {\bf 250} (1985) 199.



\bibitem{Collins:2011zzd}
  J.~Collins,
  { \it Foundations of Perturbative QCD},
  Camb.\ Monogr.\ Part.\ Phys.\ Nucl.\ Phys.\ Cosmol.\  {\bf 32} (2011) 1.




\bibitem{Li:2014xda}
  H.~n.~Li and Y.~M.~Wang,
  JHEP {\bf 1506} (2015) 013
  [arXiv:1410.7274 [hep-ph]].




\bibitem{Wang:2015qqr}
  Y.~M.~Wang,
  EPJ Web Conf.\  {\bf 112} (2016) 01021
  [arXiv:1512.08374 [hep-ph]].




\bibitem{Dugan:1990df}
  M.~J.~Dugan and B.~Grinstein,
  Phys.\ Lett.\ B {\bf 256} (1991) 239.





\bibitem{Herrlich:1994kh}
  S.~Herrlich and U.~Nierste,
  Nucl.\ Phys.\ B {\bf 455} (1995) 39
  [hep-ph/9412375].




\bibitem{Hill:2004if}
  R.~J.~Hill, T.~Becher, S.~J.~Lee and M.~Neubert,
  JHEP {\bf 0407} (2004) 081
  [hep-ph/0404217].






\bibitem{Beneke:2005gs}
  M.~Beneke and D.~S.~Yang,
  Nucl.\ Phys.\ B {\bf 736} (2006) 34
  [hep-ph/0508250].



\bibitem{Braun:2017liq}
  V.~M.~Braun, Y.~Ji and A.~N.~Manashov,
  JHEP {\bf 1705} (2017) 022
  [arXiv:1703.02446 [hep-ph]].




\bibitem{Kawamura:2001jm}
  H.~Kawamura, J.~Kodaira, C.~F.~Qiao and K.~Tanaka,
  Phys.\ Lett.\ B {\bf 523} (2001) 111;
   Erratum: [Phys.\ Lett.\ B {\bf 536} (2002) 344]
  [hep-ph/0109181].




\bibitem{Beneke:2008ei}
  M.~Beneke, T.~Huber and X.-Q.~Li,
  Nucl.\ Phys.\ B {\bf 811} (2009) 77
  [arXiv:0810.1230 [hep-ph]].




\bibitem{Becher:2004kk}
  T.~Becher and R.~J.~Hill,
  JHEP {\bf 0410} (2004) 055
  [hep-ph/0408344].






\bibitem{Lange:2003pk}
  B.~O.~Lange and M.~Neubert,
  Nucl.\ Phys.\ B {\bf 690} (2004) 249
   Erratum: [Nucl.\ Phys.\ B {\bf 723} (2005) 201]
  [hep-ph/0311345].








\bibitem{Bauer:2000yr}
  C.~W.~Bauer, S.~Fleming, D.~Pirjol and I.~W.~Stewart,
  Phys.\ Rev.\ D {\bf 63} (2001) 114020
  [hep-ph/0011336].





\bibitem{Beneke:2004rc}
  M.~Beneke, Y.~Kiyo and D.~S.~Yang,
  Nucl.\ Phys.\ B {\bf 692} (2004) 232
  [hep-ph/0402241].











\bibitem{Beneke:2002ni}
  M.~Beneke and T.~Feldmann,
  Phys.\ Lett.\ B {\bf 553} (2003) 267
  [hep-ph/0211358].






\bibitem{Burdman:2000ku}
  G.~Burdman and G.~Hiller,
  Phys.\ Rev.\ D {\bf 63} (2001) 113008
  [hep-ph/0011266].



\bibitem{Grozin:1996pq}
  A.~G.~Grozin and M.~Neubert,
  Phys.\ Rev.\ D {\bf 55} (1997) 272
  [hep-ph/9607366].




\bibitem{Beneke:2018rbh}
  M.~Beneke, M.~Garny, R.~Szafron and J.~Wang,
  JHEP {\bf 1811} (2018) 112
  [arXiv:1808.04742 [hep-ph]].



\bibitem{Wang:2015ndk}
  Y.~M.~Wang and Y.~L.~Shen,
  JHEP {\bf 1602} (2016) 179
  [arXiv:1511.09036 [hep-ph]].



\bibitem{Beneke:2011nf}
  M.~Beneke and J.~Rohrwild,
  Eur.\ Phys.\ J.\ C {\bf 71} (2011) 1818
  [arXiv:1110.3228 [hep-ph]].




\bibitem{Ji:1991pr}
  X.~D.~Ji and M.~J.~Musolf,
  Phys.\ Lett.\ B {\bf 257} (1991) 409.





\bibitem{Broadhurst:1991fz}
  D.~J.~Broadhurst and A.~G.~Grozin,
  Phys.\ Lett.\ B {\bf 267} (1991) 105
  [hep-ph/9908362].


\bibitem{Wang:2016qii}
  Y.~M.~Wang,
  JHEP {\bf 1609} (2016) 159
  [arXiv:1606.03080 [hep-ph]].



\bibitem{Braun:2019wyx}
  V.~M.~Braun, Y.~Ji and A.~N.~Manashov,
  arXiv:1905.04498 [hep-ph].




\bibitem{Reinders:1984sr}
  L.~J.~Reinders, H.~Rubinstein and S.~Yazaki,
  Phys.\ Rept.\  {\bf 127} (1985) 1.




\bibitem{Wang:2017ijn}
  Y.~M.~Wang and Y.~L.~Shen,
  JHEP {\bf 1712} (2017) 037
  [arXiv:1706.05680 [hep-ph]].



\bibitem{Bell:2010mg}
  G.~Bell, M.~Beneke, T.~Huber and X.~Q.~Li,
  Nucl.\ Phys.\ B {\bf 843} (2011) 143
  [arXiv:1007.3758 [hep-ph]].



\bibitem{Balitsky:1987bk}
  I.~I.~Balitsky and V.~M.~Braun,
  Nucl.\ Phys.\ B {\bf 311} (1989) 541.




\bibitem{Rusov:2017chr}
  A.~V.~Rusov,
  Eur.\ Phys.\ J.\ C {\bf 77}  442 (2017)
  [arXiv:1705.01929 [hep-ph]].




\bibitem{Beneke:2018wjp}
  M.~Beneke, V.~M.~Braun, Y.~Ji and Y.~B.~Wei,
  JHEP {\bf 1807} (2018) 154
  [arXiv:1804.04962 [hep-ph]].



\bibitem{Bell:2013tfa}
  G.~Bell, T.~Feldmann, Y.~M.~Wang and M.~W.~Y.~Yip,
  JHEP {\bf 1311} (2013) 191
  [arXiv:1308.6114 [hep-ph]].


\bibitem{Braun:2003wx}
  V.~M.~Braun, D.~Y.~Ivanov and G.~P.~Korchemsky,
  Phys.\ Rev.\ D {\bf 69} (2004) 034014
  [hep-ph/0309330].



\bibitem{Grozin:1996hk}
  A.~G.~Grozin and M.~Neubert,
  Nucl.\ Phys.\ B {\bf 495} (1997) 81
  [hep-ph/9701262].



\bibitem{Nishikawa:2011qk}
  T.~Nishikawa and K.~Tanaka,
  Nucl.\ Phys.\ B {\bf 879} (2014) 110
  [arXiv:1109.6786 [hep-ph]].



\bibitem{Aoki:2016frl}
  S.~Aoki {\it et al.},
  Eur.\ Phys.\ J.\ C {\bf 77} (2017)  112
  [arXiv:1607.00299 [hep-lat]].



\bibitem{Allton:2008pn}
  C.~Allton {\it et al.} [RBC-UKQCD Collaboration],
  Phys.\ Rev.\ D {\bf 78} (2008) 114509
  [arXiv:0804.0473 [hep-lat]].



\bibitem{Ball:1998sk}
  P.~Ball, V.~M.~Braun, Y.~Koike and K.~Tanaka,
  Nucl.\ Phys.\ B {\bf 529} (1998) 323
  [hep-ph/9802299].




\bibitem{Ball:1996tb}
  P.~Ball and V.~M.~Braun,
  Phys.\ Rev.\ D {\bf 54} (1996) 2182
  [hep-ph/9602323].



\bibitem{Beneke:2014pta}
  M.~Beneke, A.~Maier, J.~Piclum and T.~Rauh,
  Nucl.\ Phys.\ B {\bf 891} (2015) 42
  [arXiv:1411.3132 [hep-ph]].



\bibitem{Dehnadi:2015fra}
  B.~Dehnadi, A.~H.~Hoang and V.~Mateu,
  JHEP {\bf 1508} (2015) 155
  [arXiv:1504.07638 [hep-ph]].




\bibitem{Mateu:2017hlz}
  V.~Mateu and P.~G.~Ortega,
  JHEP {\bf 1801} (2018) 122
  [arXiv:1711.05755 [hep-ph]].



\bibitem{Tanabashi:2018oca}
  M.~Tanabashi {\it et al.} [Particle Data Group],
  Phys.\ Rev.\ D {\bf 98} (2018)   030001.




\bibitem{Wang:2018wfj}
  Y.~M.~Wang and Y.~L.~Shen,
  JHEP {\bf 1805} (2018) 184
  [arXiv:1803.06667 [hep-ph]].



\bibitem{Beneke:2003zv}
  M.~Beneke and M.~Neubert,
  Nucl.\ Phys.\ B {\bf 675} (2003) 333
  [hep-ph/0308039].



\bibitem{Gubernari:2018wyi}
  N.~Gubernari, A.~Kokulu and D.~van Dyk,
  JHEP {\bf 1901} (2019) 150
  [arXiv:1811.00983 [hep-ph]].






\bibitem{Beneke:2007zz}
  M.~Beneke,
  eConf C {\bf 0610161} (2006) 030
   [Nucl.\ Phys.\ Proc.\ Suppl.\  {\bf 170} (2007) 57]
  [hep-ph/0612353].



\bibitem{Wandzura:1977qf}
  S.~Wandzura and F.~Wilczek,
  Phys.\ Lett.\  {\bf 72B} (1977) 195.



\bibitem{Bourrely:1980gp}
  C.~Bourrely, B.~Machet and E.~de Rafael,
  Nucl.\ Phys.\ B {\bf 189} (1981) 157.




\bibitem{Khodjamirian:2017fxg}
  A.~Khodjamirian and A.~V.~Rusov,
  JHEP {\bf 1708} (2017) 112
  [arXiv:1703.04765 [hep-ph]].




\bibitem{Bourrely:2008za}
  C.~Bourrely, I.~Caprini and L.~Lellouch,
  Phys.\ Rev.\ D {\bf 79} (2009) 013008;
   Erratum: [Phys.\ Rev.\ D {\bf 82} (2010) 099902]
  [arXiv:0807.2722 [hep-ph]].



\bibitem{Boyd:1995cf}
  C.~G.~Boyd, B.~Grinstein and R.~F.~Lebed,
  Phys.\ Lett.\ B {\bf 353} (1995) 306
  [hep-ph/9504235].



\bibitem{Khodjamirian:2011ub}
  A.~Khodjamirian, T.~Mannel, N.~Offen and Y.-M.~Wang,
  Phys.\ Rev.\ D {\bf 83} (2011) 094031
  [arXiv:1103.2655 [hep-ph]].







\bibitem{Bardeen:2003kt}
  W.~A.~Bardeen, E.~J.~Eichten and C.~T.~Hill,
  Phys.\ Rev.\ D {\bf 68} (2003) 054024
  [hep-ph/0305049].


\bibitem{Bharucha:2010im}
  A.~Bharucha, T.~Feldmann and M.~Wick,
  JHEP {\bf 1009} (2010) 090
  [arXiv:1004.3249 [hep-ph]].



\bibitem{Bigi:2016mdz}
  D.~Bigi and P.~Gambino,
  Phys.\ Rev.\ D {\bf 94} (2016)   094008
  [arXiv:1606.08030 [hep-ph]].



\bibitem{Bosch:2004nd}
  S.~W.~Bosch and G.~Buchalla,
  JHEP {\bf 0501} (2005) 035
  [hep-ph/0408231].



\bibitem{Beneke:2004dp}
  M.~Beneke, T.~Feldmann and D.~Seidel,
  Eur.\ Phys.\ J.\ C {\bf 41} (2005) 173
  [hep-ph/0412400].



\bibitem{delAmoSanchez:2010af}
  P.~del Amo Sanchez {\it et al.} [BaBar Collaboration],
  Phys.\ Rev.\ D {\bf 83} (2011) 032007
  [arXiv:1005.3288 [hep-ex]].




\bibitem{Sibidanov:2013rkk}
  A.~Sibidanov {\it et al.} [Belle Collaboration],
  Phys.\ Rev.\ D {\bf 88} (2013)   032005
  [arXiv:1306.2781 [hep-ex]].



\bibitem{Lees:2012vv}
  J.~P.~Lees {\it et al.} [BaBar Collaboration],
  Phys.\ Rev.\ D {\bf 86} (2012) 092004
  [arXiv:1208.1253 [hep-ex]].




\bibitem{Kou:2018nap}
  E.~Kou {\it et al.} [Belle-II Collaboration],
  ``{\it The Belle II Physics Book},''
  arXiv:1808.10567 [hep-ex].



\bibitem{Buras:2014fpa}
  A.~J.~Buras, J.~Girrbach-Noe, C.~Niehoff and D.~M.~Straub,
  JHEP {\bf 1502} (2015) 184
  [arXiv:1409.4557 [hep-ph]].




\bibitem{Inami:1980fz}
  T.~Inami and C.~S.~Lim,
  Prog.\ Theor.\ Phys.\  {\bf 65} (1981) 297;
   Erratum: [Prog.\ Theor.\ Phys.\  {\bf 65} (1981) 1772].





\bibitem{Buchalla:1998ba}
  G.~Buchalla and A.~J.~Buras,
  Nucl.\ Phys.\ B {\bf 548} (1999) 309
  [hep-ph/9901288].



\bibitem{Buchalla:1992zm}
  G.~Buchalla and A.~J.~Buras,
  Nucl.\ Phys.\ B {\bf 398} (1993) 285.




\bibitem{Misiak:1999yg}
  M.~Misiak and J.~Urban,
  Phys.\ Lett.\ B {\bf 451} (1999) 161
  [hep-ph/9901278].



\bibitem{Brod:2010hi}
  J.~Brod, M.~Gorbahn and E.~Stamou,
  Phys.\ Rev.\ D {\bf 83} (2011) 034030
  [arXiv:1009.0947 [hep-ph]].




\bibitem{Leibovich:2003jd}
  A.~K.~Leibovich, Z.~Ligeti and M.~B.~Wise,
  Phys.\ Lett.\ B {\bf 564} (2003) 231
  [hep-ph/0303099].







\end{thebibliography}
\end{document}